\documentclass[final,5p,times,twocolumn,authoryear]{elsarticle}

\journal{Journal of High Energy Astrophysics}

\PassOptionsToPackage{cmyk}{xcolor}
\usepackage[tbtags,fleqn]{amsmath}
\usepackage{amssymb,bm,dcolumn,booktabs,threeparttable,multirow,fleqn}
\usepackage[dvipsnames]{xcolor}
\definecolor{LinkBlue}{RGB}{6,69,173}
\definecolor{DarkBlue}{RGB}{11,0,128}
\usepackage[colorlinks=true,linkcolor=LinkBlue,urlcolor=LinkBlue,
	citecolor=LinkBlue,citecolor=LinkBlue,hyperfootnotes=true]{hyperref}
\usepackage{graphicx,paralist}
\usepackage{aas_macros}
\setdefaultitem{}{\textbullet}{$\star$}{}

\journal{The Journal of High Energy Astrophysics}
\usepackage{bm}
\usepackage{fleqn}
\usepackage{graphicx,fleqn}
\usepackage{booktabs}
\usepackage{setspace}
\usepackage{cellspace}
\usepackage{latexsym,color,multirow,array}
\usepackage{amssymb}
\usepackage{times}
\usepackage[T1]{fontenc}
\usepackage{natbib}
\usepackage{amsmath,amssymb,amsthm,mathrsfs,amsfonts,dsfont}
\usepackage{color}

\begin{document}

\begin{frontmatter}

\title{Tori sequences  as remnants of  multiple accreting  periods of   Kerr SMBHs}

\author{D. Pugliese$^*$}
\cortext[dip]{Corresponding author}
\ead{d.pugliese.physics@gmail.com}
\author{Z. Stuchl\'{\i}k}
\ead{zdenek.stuchlik@physics.cz}
\address{Institute of Physics, and Research Centre of Theoretical Physics and Astrophysics, \\
Faculty of Philosophy \& Science,
  Silesian University in Opava,\\
 Bezru\v{c}ovo n\'{a}m\v{e}st\'{i} 13, CZ-74601 Opava, Czech Republic
}

\begin{abstract}
Super-massive  black holes ({SMBHs})  hosted in  active galactic nuclei ({AGNs})   can be  characterized  by multi-accreting periods   as the  attractors  interact with the  environment during their life-time.
These multi-accretion   episodes should leave  traces in the  matter orbiting the attractor. Counterrotating and even misaligned structures orbiting  around the SMBHs would be  consequences of these episodes.
 Our task in this work is to consider  situations where such accretions occur and to trace  their remnants    represented by  several toroidal  accreting fluids, corotating  or counterrotating  relative to the   central Kerr attractor, and  created  in various regimes during  the evolution of matter configurations  around SMBHs. We  focus   particularly on the emergence of matter instabilities,  i.e.,  tori collisions, accretion  onto the central Kerr black hole, or  creation of jet-like structures (proto-jets).
 Each orbiting configuration is governed by the   general relativistic hydrodynamic Boyer  condition  of equilibrium configurations  of rotating perfect fluid.
We prove that sequences  of configurations and hot points, where an instability occurs, characterize the Kerr SMBHs, depending mainly on   their spin-mass ratios.
The occurrence of  tori accretion or  collision are strongly constrained {{{by the fluid  rotation with respect to the central black hole and the relative rotation with respect to each other}}}. Our investigation provides characteristic of  attractors  where  traces of multi-accreting episodes can be found and observed.
\end{abstract}

\begin{keyword}
Accretion disks, accretion, jets, black hole physics, hydrodynamics\MSC[2010] 00-01\sep  99-00
\end{keyword}

\end{frontmatter}


\section{Introduction}
\newcommand{\ti}[1]{\mbox{\tiny{#1}}}
\newcommand{\im}{\mathop{\mathrm{Im}}}
\def\be{\begin{equation}}
\def\ee{\end{equation}}
\def\bea{\begin{eqnarray}}

\newcommand{\mso}{mso}
\newcommand{\mbo}{mbo}
\def\eea{\end{eqnarray}}
\newcommand{\tb}[1]{\textbf{\texttt{#1}}}
\newcommand{\ttb}[1]{\textbf{#1}}
\newcommand{\rtb}[1]{\textcolor[rgb]{1.00,0.00,0.00}{\tb{#1}}}
\newcommand{\btb}[1]{\textcolor[rgb]{0.00,0.00,1.00}{\tb{#1}}}
\newcommand{\pp}{\textbf{()}}
\newcommand{\non}[1]{\not{#1}}
\newcommand*{\doi}[1]{\href{http://dx.doi.org/#1}{doi: #1}}
\newcommand{\il}{~}
\newcommand{\rc}{\rho_{\ti{C}}}
\newcommand{\dd}{\mathcal{D}}
\newcommand{\lie}{\mathcal{L}}
\newcommand{\Mie}{\mathcal{M}}
  \newcommand{\downmapsto}{\rotatebox[origin=c]{-90}{$\small\mathbf{\mapsto}$}\mkern2mu}
  \newcommand{\upmapsto}{\rotatebox[origin=c]{90}{$\small\mathbf{\mapsto}$}\mkern2mu}
  \newcommand{\cc}{\mathrm{C}}
    \newcommand{\oo}{\mathrm{O}}
\newcommand{\Tem}{T^{\rm{em}}}
\newcommand{\subsubsubsection}[1]{\paragraph{#1}\mbox{}\\}
\setcounter{secnumdepth}{4}
\setcounter{tocdepth}{4}
Galactic cores and  active galactic nuclei (\textbf{AGNs}) provide a rich scenario  to observe  super-massive  black holes (\textbf{SMBHs})    interacting  with their galactic environments.
There are several observational evidences supporting the existence
of such objects in AGNs.
To cite two of most recent studies  on  \textbf{SMBHs} in  their host  galaxies, we point out the    analysis in  \cite{Tadhunter:2017qji} exploring the link between galaxy collisions and super-massive black hole feeding, while in  \cite{Regan:2017vre}
  the  link between  galaxy collapse and rapid \textbf{SMBHs} formation is faced.
Both these studies  show the existence  of an  intense and strong   relation between the galaxy dynamics and its super-massive guest,  especially in the accretion processes characterizing the strong attractors.
 It can be   expected that, during their  life-time,  \textbf{SMBHs}   would be  influenced by the  galaxy dynamics due  to a series of multi-accreting  episodes  as a consequence of  interaction with  the galactic environment made up by  stars  and   dusts.
  These activities may leave traces  in the form of matter  remnants    orbiting the  central attractor.

Chaotical, discontinuous   accretion episodes  can  produce sequences of orbiting toroidal structures  with strongly   different features as,  for example, different rotation orientations with respect to the central Kerr \textbf{BH} where corotating and counterrotating accretion stages can be mixed \citep{Dyda:2014pia,Aligetal(2013),Carmona-Loaiza:2015fqa,Lovelace:1996kx,Gafton:2015jja},  or    disks  strongly misaligned with respect to the central \textbf{SMBH} spin  may appear \citep{Nixon:2013qfa,Dogan:2015ida,Bonnerot:2015ara,Aly:2015vqa}. Eventually, the scenario envisaged by these studies raises a series of issues and  indications  about the different stages of the  attractor accretion  periods   binding it on its intrinsic rotation.
 Motivated by these facts,  in this work  we  investigate structured toroidal  disks, so called ringed accretion disks (\textbf{RAD}),    which may be formed   during several accretion regimes occurred in the lifetime of non-isolated  Kerr \textbf{BHs}.  These configurations were first introduced in \cite{pugtot} and then  detailed in  \cite{ringed,open}.  They feature a system made up by several axis-symmetrical matter configurations orbiting in the equatorial plane of a single central Kerr \textbf{BH}.
Evidences of these special configurations are expected to be found in  the associated  X-ray spectra emission in \textbf{AGNs}.

The phenomenology associated with these toroidal complex structures may be indeed  very wide. This new  complex  scenario enables to re-interpret the phenomena analyzed so far in the single-torus framework.
Observational evidence is expected  by the spectral features of \textbf{AGNs} X-ray emission shape, due to X-ray obscuration and absorption by one of the tori, providing  therefore  a fingerprint of the tori as a radially stratified emission profile  \citep{KS10,S11etal,Schee:2008fc}. Relatively indistinct excesses of the relativistically broadened emission-line components were predicted in different  works, arising in a well-confined radial distance in the accretion structure originating by a series of episodic accretion events. Furthermore, the
radially oscillating tori can   be related to the high-frequency quasi periodic oscillations  (QPOs) observed in non-thermal
X-ray emission from compact objects.
More generally, instabilities of such configurations, we expect, may reveal crucial significance
for the high energy astrophysics related especially to accretion onto  \textbf{BH}, and the extremely energetic
phenomena occurring in quasars and \textbf{AGNs} that could be observable by the planned X-ray observatory ATHENA\footnote{http://the-athena-x-ray-observatory.eu/}.

The investigation of these \textbf{RAD} configurations, however, is influences   by significant methodological issues.
The question of how to treat  this scenario, and  how to model the dynamics of toroidal sequences,  is undoubtedly   challenging.
A major  methodological  challenge comes from the need to study different evolutive periods of the  \textbf{BH} in its environment.
Conveniently,  we may  consider  following three periods of ringed accretion disk life: $(i)$  formation of  tori, $(ii)$  the accretion periods   onto the central Kerr  attractor  and  $(iii)$ the tori interaction (emergence of tori collisions).
In the current analysis of dynamical one-torus system of   both general relativistic hydrodynamic (GR-HD)  and general relativistic magnetohydrodynamic  (GR-MHD) set-ups, the geometrically  thick disks considered in this work  are  often adopted as initial configurations for the analysis-- \citep{Igumenshchev,Fragile:2007dk,DeVilliers,Porth:2016rfi}.
We can  therefore  adopt an  analogue    approach  for the investigation of  the case of a central  Kerr \textbf{BH} and several  tori orbiting in its equatorial plane.
However, in a dynamical process the timing problem of how to depict the  different periods   is definitely  challenging, and requires  a certain number of assumptions on the history of the  \textbf{BH} in interaction  with the environment.  Therefore, fixing  a minimum model set-up  inevitably will focus the analysis  on a single,  very special situation. It would be necessary to fix:   $1.$ the attractor through its dimensionless spin, $2.$  the accretion era we are  willing to describe, and  eventually, $3.$  number of tori,  $4.$  fluid  rotation law,  $5.$ relative tori location,  $6.$ location of  the inner torus with respect to the attractor.
The immediate approach  would  be to let the chance in the choice of a specific scenario which even  in a  rich variant of the model will provide necessarily only a partially focused description of one hypothetical model.   The problem remains of how to fix, in such objectively complex scenario, the initial conditions  of multi-tori orbiting  a spinning \textbf{BH}.

In fact, recent results presented by \cite{dsystem,Letter} show
that it is not even immediate to choice the spin class for the
central Kerr \textbf{BH}: the dimensionless spin of the Kerr black hole
strongly constrains the possible couple of orbiting tori in number
of orbiting tori, in location and relative range of variation for
fluid specific  angular momentum.
In other words, to fix the initial data for a  restricted dynamical scenario, one needs to have in advance the  answer to the very question one firstly  wants to address within the simulation.

In this respect,  our  analysis stands  also as a guideline to this  choice providing a  detailed  answer to these questions.
Results found here will  be the guide for the set-up of any more complex  dynamical system.
Moreover, we are able to trace some evolutive
lines for an initial configuration of a system composed by
an attractor and general $n$ fluid configurations, corotating or counterrotating relative to the attractor, at any time of the system evolution, from the formation to the occurrence of accretion.
Fixing an initial set up,  and distinguishing  a very restricted number of classes of configurations and attractors, we discuss  the final state of a dynamical evolution from an initial configuration. There are  few evolutive lines with different final states; the occurrence of these  paths  however will be finally  established by a dynamical  analysis.
The set-up for the ringed accretion disk model was drew in \cite{ringed}, while first proposal of these configurations was in \cite{pugtot}. In \cite{ringed},  constraints and discussion on perturbations were provided.
Then in \cite{open} sequences of unstable configurations were discussed, the investigation was focused on the unstable phases of multi accreting toroidal structures.
The paper \cite{dsystem} focused on the case of  two tori as ``seed'' for larger configurations, and paper \cite{Letter} explicitly addressed collisions and energy release in colliding tori.
In this  article, we discuss the situation where several equilibrium and accreting or proto-jet  (open critical) configurations  are formed around a Kerr-\textbf{SMBH} in \textbf{AGN} environments.
Here we take full advantage of the symmetry of the Kerr geometry, considering  a stationary and axisymmetric, full general relativity (GR) model for a single  thick accretion disk with a toroidal shape. Each torus is featured as an
opaque (large optical depth) and super-Eddington (high  matter accretion rates) disk
model a radiation pressure supported accretion disk cooled by advection with low viscosity--
 \citep{Pac-Wii,Igumenshchev,Fragile:2007dk,DeVilliers,Porth:2016rfi}.
More precisely, the individual toroidal  (thick disk) configurations are  barotropic models 
where the effects of strong gravitational fields are dominant with respect to the  dissipative ones and predominant to determine  the unstable phases of the systems \citep{F-D-02,Igumenshchev,AbraFra,pugtot,Pac-Wii,Kovar:2016kqh},
As a consequence of this, during the evolution of dynamical processes, the functional form of the angular
momentum and entropy distribution depends on the initial conditions of the system and on
the details of the dissipative processes.
The tori are governed by ``Boyer's condition'' of the analytic theory of equilibrium configurations of rotating perfect fluids  \citep{Boy:1965:PCPS:}. The toroidal structures of orbiting barotropic perfect fluid are determined by an \emph{effective potential} reflecting the spacetime geometry and the centrifugal force through the  distribution of the specific angular momentum $\ell(r)$ of the orbiting fluid
   \citep{Abra83,AbraFra,2010A&A...521A..15A,AJS78,Stuchlik:2004wk,Abramowicz:1996ap,Lei:2008ui}.
The equipressure surfaces, $K=$constant,   could be closed, determining equilibrium configurations, or open (related to  proto-jets configurations  \citep{open}). The special case of cusped or critical  equipotential surfaces allows for the accretion onto the central black hole  \citep{Pac-Wii}. The outflow of matter through the cusp occurs due to an instability in the balance of the gravitational and inertial forces and the pressure gradients in the fluid, i.e., by the so called Paczynski mechanism of violation of mechanical equilibrium of the tori  \citep{Pac-Wii}.

The plan of this article is as follows:
Sec.\il(\ref{Sec:Kerr-2-Disk}) introduces the ringed accretion disk model: we  discuss the main features of geometrically thick accretion disk orbiting   a central Kerr \textbf{BH},  and we then proceed to consider the case  of  several   tori orbiting in the equatorial plane of the  central attractor. Concepts and notation used throughout this works are also introduced. The introduction of new model also requires the use of an extended notation; for easy of reference we have summarized  main notation in  Sec.\il(\ref{Sec:Kerr-2-Disk}) and we will make reference also to Table\il(\ref{Table:nature-Att}) listing main Kerr \textbf{BHs} spins used in this work.
 In Sec.\il(\ref{Sec:criticalII}) we consider the case when all the configurations around the attractor rotate with  the same orientation, i.e.,  all  are  corotating or counterrotating with respect to the central  Kerr \textbf{BH}  (\emph{$\ell$corotating}). In fact, many of the results and constraints on orbiting tori depend mainly on the fluids  relative rotation as  well as on   each torus rotation with respect to the central Kerr attractor.
The limiting  case of Schwarzschild  \textbf{BH}  is also considered. In static spacetimes,  all tori may be considered  as corotating, regardless of the  fluids relative rotation.
In general, any  configuration may be  in one of three possible states: non-accreting or equilibrium  $\textbf{(C)}$, accretion $\textbf{(A)}$  or proto-jet  $\textbf{(J)}$.
Accordingly, we developed our analysis as follows:
in Sec.\il(\ref{Sec:J-Jc}) we  consider the  proto-jets sequences $\mathbf{(J-J)}$.
Sequences  $\mathbf{(A-J)}$, formed  by at last a  configuration in accretion and a proto-jet,
 are studied in Sec.\il(\ref{Sec:A-J}), for the case where the  open     topology
is the outer one (farthest from the attractor).
In Sec.\il(\ref{Sec:innerJ-A}),
the case in which the inner  configuration of the couple has   proto-jet topology is  investigated.
 Section\il(\ref{Sec:C-J-clna-v}) describes sequences formed by a non-accreting torus   and a proto-jet.
In Section\il(\ref{Sec:J-Csmooth}), we focus on the case where the proto-jet is the closest to the attractor.
In Sec.\il(\ref{Sec:A-C-Asmooth}),
some remarks on the sequences with  $\mathbf{(C-A)}$ configurations  are addressed.
 Analogously, Section\il(\ref{Sec:ell-cont-double}) deals with the $\ell$counterrotating couples (tori having  different relative orientation of rotation). This case turns far more articulated than the $\ell$corotating one, and we address the analysis by considering the
  proto-jet-proto-jet $\mathbf{(J-J)}$ systems in Sec.\il(\ref{Sec:J-J-cont}),
 in Sec.\il(\ref{Sec:L-A-system})   proto-jet-accretion $\mathbf{(J-A)}$ systems are discussed, and in
 Sec.\il(\ref{Sec:A-A}) we address the  accretion-accretion $\mathbf{(A-A)}$ systems, with a special case where two $\ell$counterrotating tori  are accreting onto the central \textbf{BH}.
 The case where there is  an equilibrium disk and an accreting torus $\mathbf{(C-A)}$  is investigated in Sec.\il(\ref{Sec:C-A-SYStEMS}). This section ends with the study of the
 equilibrium disk-proto-jet  $\mathbf{(C-J)}$ systems in Sec.\il(\ref{Sec:C-J}).
{Section\il(\ref{Sec:P-O-E-RADs}) provides indications on possible    observational evidences  for the \textbf{R}inged \textbf{A}ccretion \textbf{D}isks (\textbf{RADs}), discussing  the phenomenology expected to be associated with these macrostructures.}

 Our analysis required a certain number of sideline results, fixing  the location  of the accretion disk edges  in the spacetime regions confined by marginally bounded, marginally stable and marginally circular (photon) orbits.
 It is clear that  the problem to assess  the  location of the inner edge of a single torus is in fact  a very relevant issue of the accretion disk theory--see
    \citep{Krolik:2002ae,BMP98,2010A&A...521A..15A,Agol:1999dn,Paczynski:2000tz,Sla-Stu:2005:CLAQG:}.
 Acknowledging the  importance of this issue,  we report in the \ref{App:location} the direct procedure provided for this part of our analysis,   with comments on the   results.
 We note that it was necessary to consider both the triplets of radii (marginal  orbits) for both corotating  and counterrotating matter, consequently, we  separated  our discussion  in two parts in  \ref{Sec:possi-conc} and  \ref{Sec:coorr}. In the investigation of  Sections (\ref{Sec:criticalII}) and (\ref{Sec:ell-cont-double}), we make direct reference to the results and quantities of \ref{App:location}. Finally, some of these results have also been used  in \cite{open};  here we propose proof of those results.
 This article  closes in Sec.\il(\ref{Sec:Conclusions})  where comments  and future perspectives are presented.
\section{Orbiting Axi-symmetric tori in a Kerr spacetime}\label{Sec:Kerr-2-Disk}
We consider  axially symmetric  configurations  orbiting in  the equatorial plane of a central Kerr   \textbf{BH} with mass parameter $M$  and dimensionless spin  $a/M\in[0,1]$
The Kerr  metric tensor can be
written as
\begin{eqnarray}\nonumber
&& ds^2=-dt^2+\frac{\rho^2}{\Delta}dr^2+\rho^2
d\theta^2+(r^2+a^2)\sin^2\theta
d\phi^2+
\\\label{alai}&&\frac{2M}{\rho^2}r(dt-a\sin^2\theta d\phi)^2\ ,\\
&&\nonumber
 \rho^2\equiv r^2+a^2\cos\theta^2, \quad \Delta\equiv r^2-2 M r+a^2,
\end{eqnarray}
where
\( \{t,r,\theta ,\phi \}\) are  the Boyer-Lindquist (BL)  coordinates;
the horizons $r_-<r_+$,  and the outer static limit $r_{\epsilon}^+$ are respectively given by:
\bea
r_{\pm}\equiv M\pm\sqrt{M^2-a^2};\quad r_{\epsilon}^{+}\equiv M+\sqrt{M^2- a^2 \cos\theta^2}.
\eea
The extreme Kerr black hole  has spin-mass ratio $a/M=1$, while  the non-rotating  limiting case $a=0$ is the   Schwarzschild metric.
In general there is $r_+<r_{\epsilon}^+$ on   $\theta\neq0$  and  $r_{\epsilon}^+=2M$  in the equatorial plane ($\theta=\pi/2$).
Metric tensor  (\ref{alai}) is independent of $\phi$ and $t$, as consequence of this  the covariant
components $p_{\phi}$ and $p_{t}$ of a particle four--momentum are
conserved along the   geodesics\footnote{We adopt the
geometrical  units $c=1=G$ and  the $(-,+,+,+)$ signature, Greek indices run in $\{0,1,2,3\}$.  The   four-velocity  satisfy $u^a u_a=-1$. The radius $r$ has unit of
mass $[M]$, and the angular momentum  units of $[M]^2$, the velocities  $[u^t]=[u^r]=1$
and $[u^{\varphi}]=[u^{\theta}]=[M]^{-1}$ with $[u^{\varphi}/u^{t}]=[M]^{-1}$ and
$[u_{\varphi}/u_{t}]=[M]$. For the seek of convenience, we always consider the
dimensionless  energy and effective potential $[V_{eff}]=1$ and an angular momentum per
unit of mass $[L]/[M]=[M]$.}
and we can introduce the constants of motion
\be\label{Eq:after}
{E} \equiv -g_{\alpha \beta}\xi_{t}^{\alpha} p^{\beta},\quad L \equiv
g_{\alpha \beta}\xi_{\phi}^{\alpha}p^{\beta}\ ,
\ee
 where  $\xi_{t}=\partial_{t} $  is
the Killing field representing the stationarity of the Kerr geometry and  $\xi_{\phi}=\partial_{\phi} $
is the
rotational Killing field.
Thus $E$, is interpreted  as the total energy
of timelike test particle
 coming from radial infinity, as measured  by  a static observer at infinity, while  $L$  is  the axial component of the angular momentum  of the particle.
 Line element  (\ref{alai}) is  also invariant under the application of any two different transformations: $x^\alpha\rightarrow-x^\alpha$
  for one of the coordinates $(t,\phi)$, or the metric parameter $a$, and  therefore  the    test particle dynamics is invariant under the mutual transformation of the parameters
$(a,L)\rightarrow(-a,-L)$. This makes possible to   limit the  analysis of the test particle circular motion to the case of  positive values of $a$
for corotating  $(L>0)$ and counterrotating   $(L<0)$ orbits with respect to the black hole.

To start of exploration of the  accretion sequences we consider
 a one-species particle perfect  fluid (simple fluid),  described by  the  energy momentum tensor
\be\label{E:Tm}
T_{\alpha \beta}=(\varrho +p) u_{\alpha} u_{\beta}+\  p g_{\alpha \beta},
\ee
where $\varrho$ and $p$ are  the total energy density and
pressure, respectively, as measured by an observer moving with the fluid whose four-velocity $u^{\alpha}$  is
a timelike flow vector field.  Then set up the
 problem symmetries, assuming to be   $\partial_t \mathbf{Q}=0$ and
$\partial_{\varphi} \mathbf{Q}=0$, for   a generic  tensor $\mathbf{Q}$.
Consequently
 the  fluid dynamics  is described by the \emph{continuity  equation} and the \emph{Euler equation} as follows
\bea\label{E:1a0}
\begin{aligned}
u^\alpha\nabla_\alpha\varrho+(p+\varrho)\nabla^\alpha u_\alpha=0,\,
\\
(p+\varrho)u^\alpha \nabla_\alpha u^\gamma+ \ h^{\beta\gamma}\nabla_\beta p=0,
\end{aligned}
\eea
where  $h_{\alpha \beta}=g_{\alpha \beta}+ u_\alpha u_\beta$ and $\nabla_\alpha g_{\beta\gamma}=0$ \citep{pugtot,Pugliese:2012ub}, moreover we assume
fluid toroidal  configurations centered on  the  plane $\theta=\pi/2$, and  defined by the constraint
$u^r=0$, with a   barotropic equation of state $p=p(\varrho)$. No
motion is assumed also in the $\theta$ angular direction ($u^{\theta}=0$).
The  continuity equation
is  identically satisfied as consequence of these conditions  and
 the Euler  equation  in (\ref{E:1a0})   reads
\bea\label{Eq:scond-d}
&&
\frac{\partial_{\mu}p}{\varrho+p}=-{\partial_{\mu }W}+\frac{\Omega \partial_{\mu}\ell}{1-\Omega \ell}
,\quad W\equiv\ln V_{eff}(\ell)
\\\nonumber
&&\mbox{where}\quad V_{eff}(\ell)=u_t= \pm\sqrt{\frac{g_{\phi t}^2-g_{tt} g_{\phi \phi}}{g_{\phi \phi}+2 \ell g_{\phi t} +\ell^2g_{tt}}}\\
&&\mbox{and }\quad \ell=\frac{L}{E}.
\eea
%
The function $W$ in Eq.\il(\ref{Eq:scond-d}) is Paczynski-Wiita  (P-W) potential,  $\Omega$ is the relativistic angular frequency of the fluid relative to the distant observer, and  $V_{eff}(\ell)$ provides an \emph{effective potential} for the fluid, assumed here  to be  characterized by a  conserved and constant specific angular momentum $\ell$  (see also \cite{Lei:2008ui,Abramowicz:2008bk}).

\begin{figure}
\centering
\includegraphics[width=1\columnwidth]{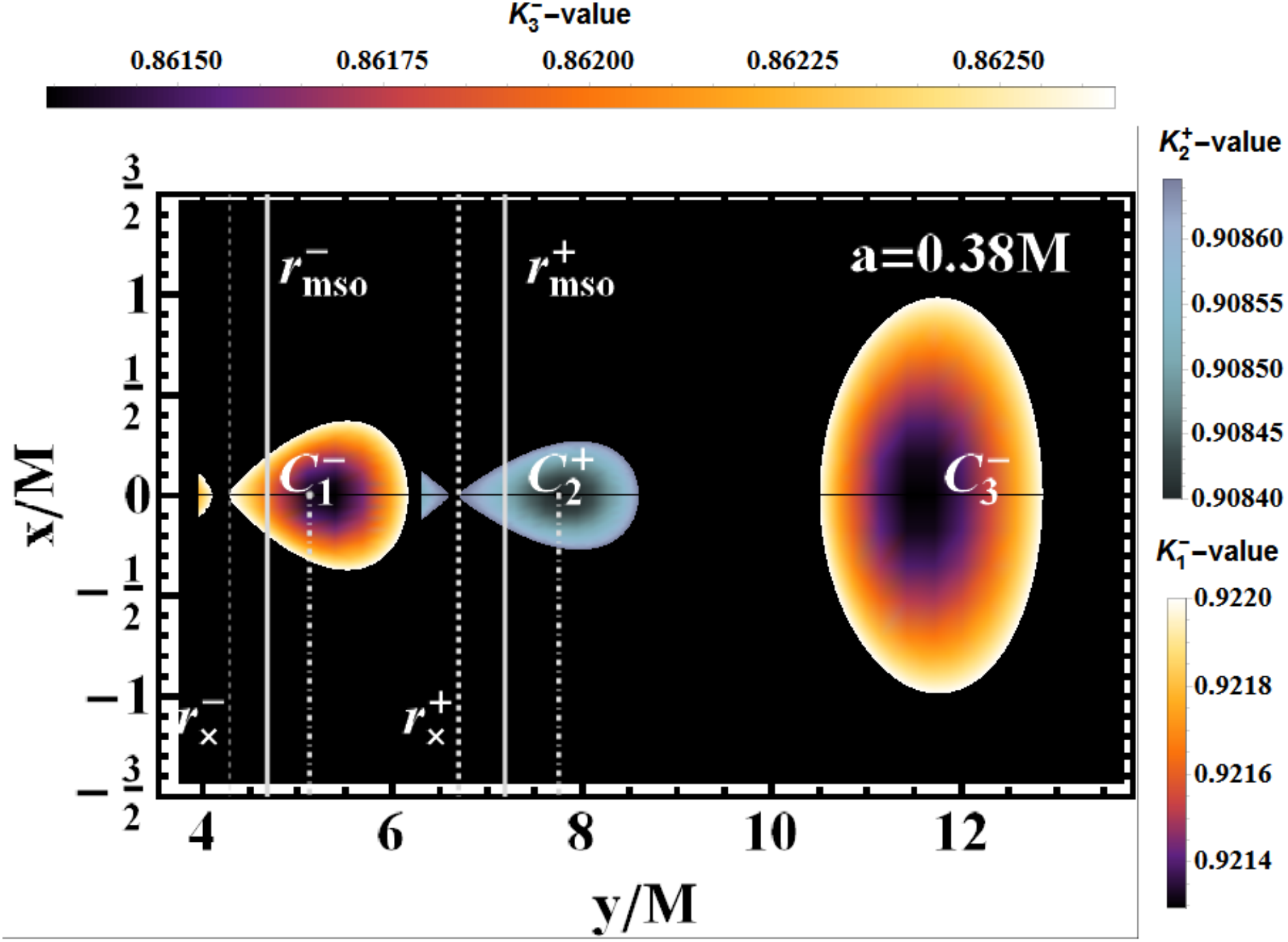}\\
\includegraphics[width=1\columnwidth]{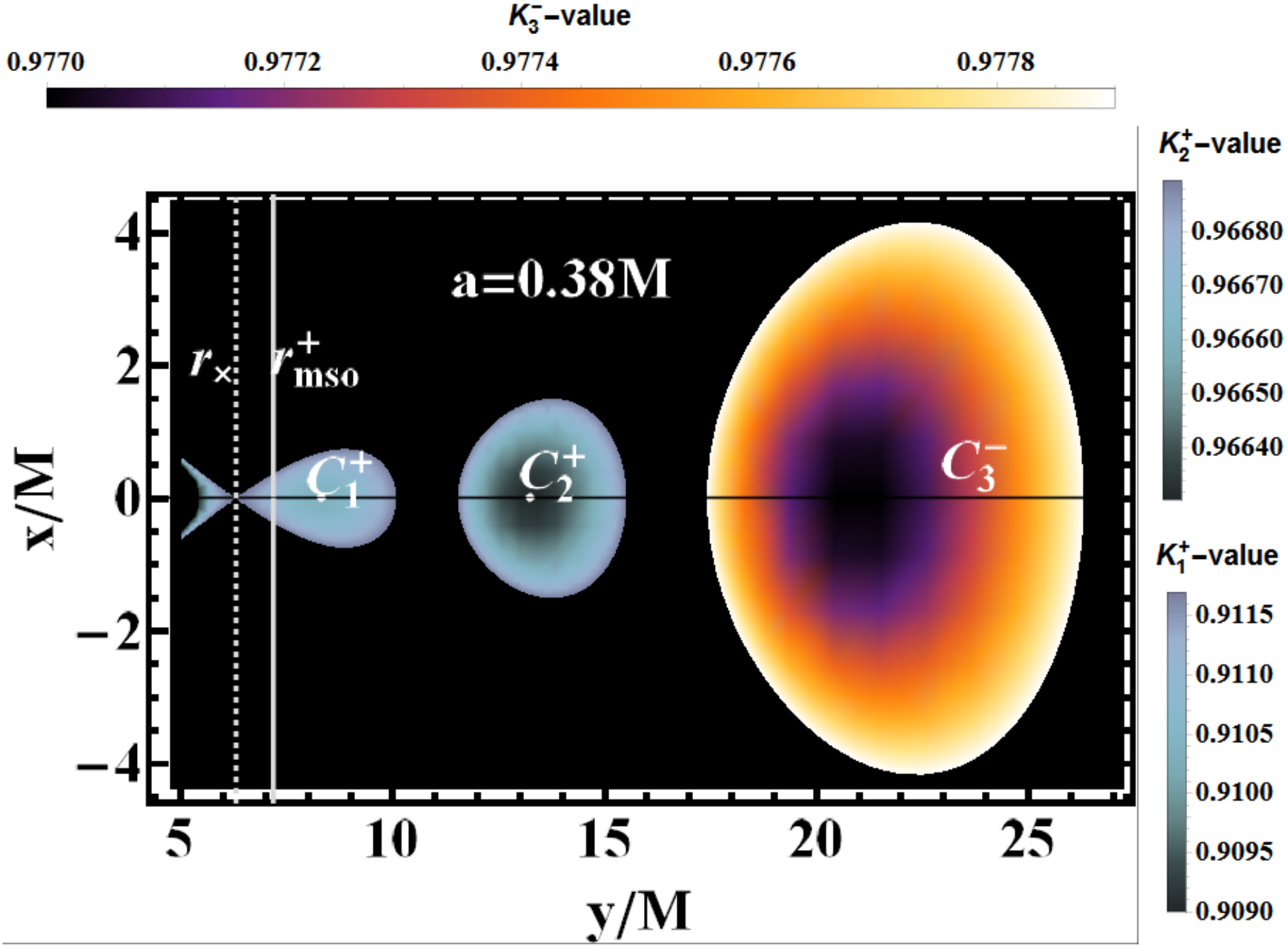}%
\\
\includegraphics[width=1\columnwidth]{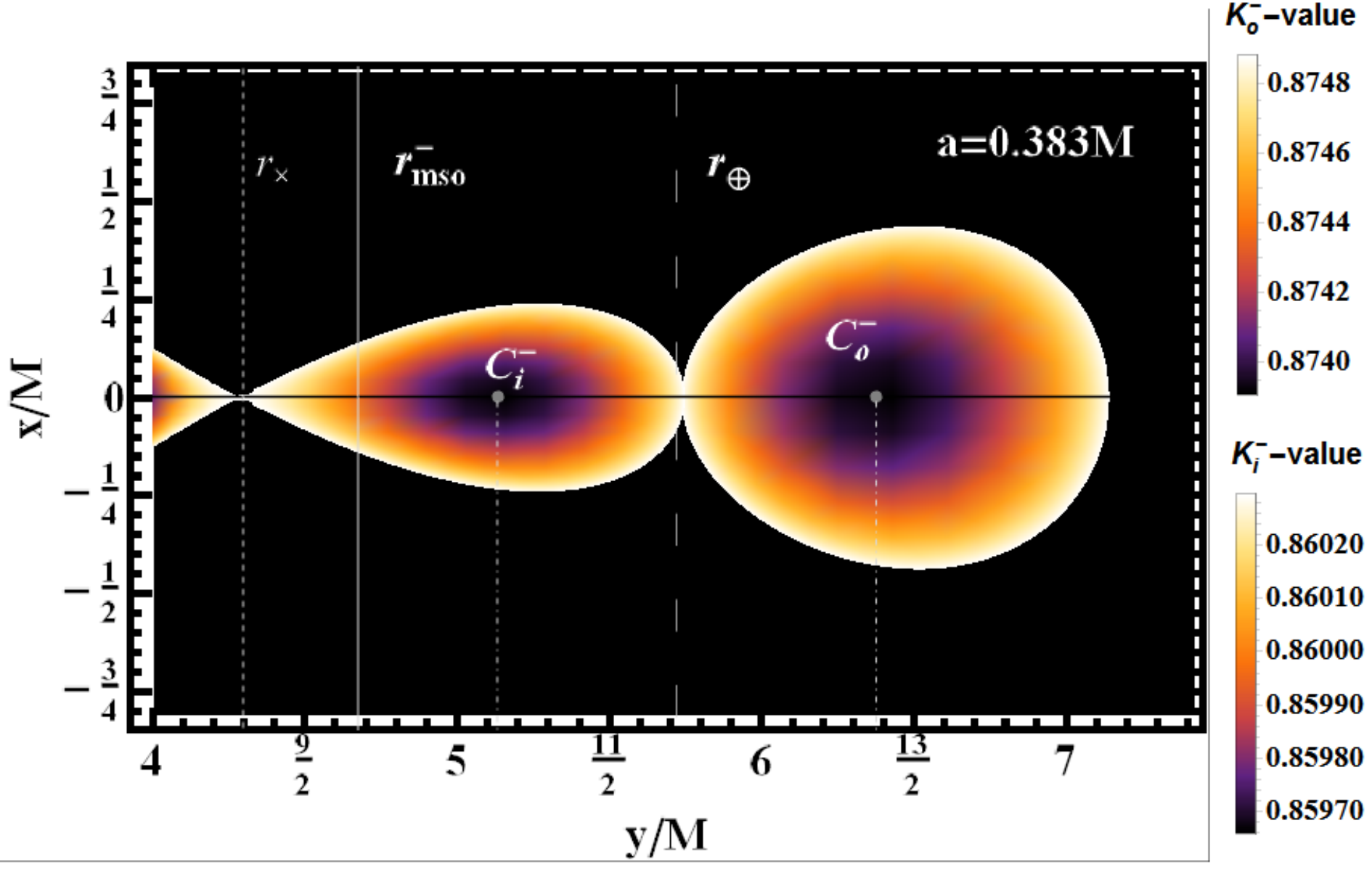}
\\
\includegraphics[width=1\columnwidth]{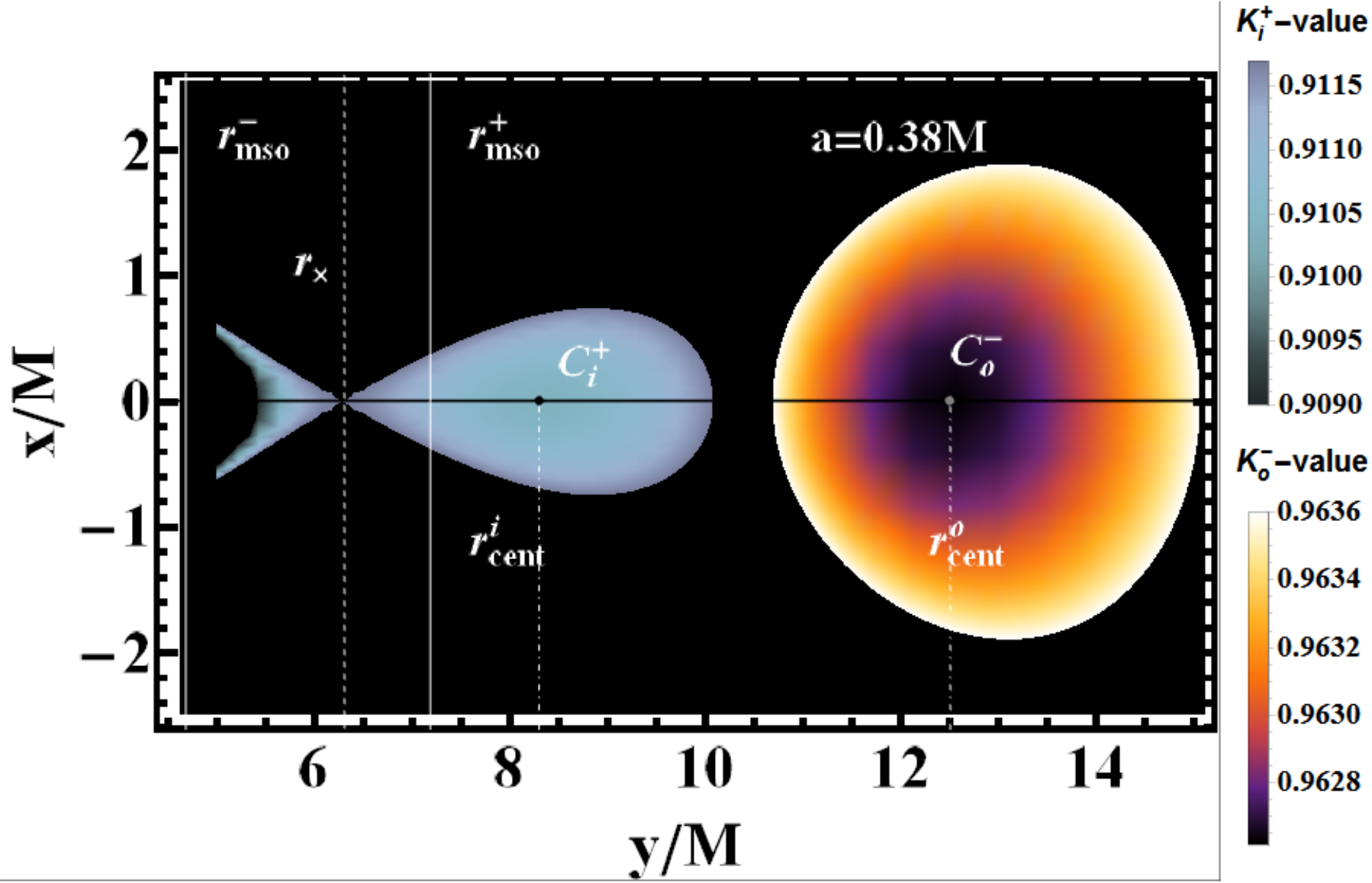}
\caption{Density plots. \emph{Upper-first} panel: $\cc^-_{x}<\cc_{x}^+
 <\cc^-$ configurations; \emph{Upper-second} panel:
$ \cc^+_{x}<\cc^+<\cc^-$ configurations.
\emph{Third} panel: colliding corotating tori
 $ \cc_{x}^-<\cc^-$, \emph{Bottom} panel: couple
$ \cc_{x}^+<\cc^-$.  $(x, y)$ are Cartesian coordinates.
\label{Fig:rssi-na} }
\end{figure}

Similarly to the case of the test particle dynamics,
the  function  $V_{eff}(\ell)$  in Eq.\il(\ref{Eq:scond-d})  is invariant under the mutual transformation of  the parameters
$(a,\ell)\rightarrow(-a,-\ell)$, therefore we can limit the analysis to  positive values of $a>0$,
for \emph{corotating}  $(\ell>0)$ and \emph{counterrotating}   $(\ell<0)$ fluids and    we adopt the notation $(\pm)$  for  counterrotating or corotating matter  respectively.
Therefore,
the accretion tori  corotate $(-)$ or counterrotate $(+)$ with respect to the  Kerr  \textbf{BH}, for $\ell_{\mp} a\gtrless0$ respectively.
As a consequence of this,
considering the case of two orbiting tori, $(i)$ and $(o)$  respectively, we need to introduce   the concept  of
 \emph{$\ell$corotating} tori,  $\ell_{i}\ell_{o}>0$ (es: Figs \il\ref{Fig:rssi-na})-\emph{Third panel}, and \emph{$\ell$counterrotating} tori,   $\ell_{i}\ell_{o}<0$--see Figs\il\ref{Fig:rssi-na}-\emph{Bottom-panel}.  The   tori can be both corotating, $\ell a>0$, or counterrotating,  $\ell a<0$, with respect to the central Kerr attractor  f\cite{ringed}.
 The configurations  are regulated by  the balance of the    hydrostatic  and   centrifugal  factors due to the fluid  rotation and by the curvature  effects  of the  Kerr background, encoded in the effective potential function $V_{eff}$. The set of these configurations (macro-configurations) is studied as ringed accretion disks in \cite{ringed,open}. Examples of integrations for these   configurations are shown in  Figs\il\ref{Fig:rssi-na} (see also Fig.\il\ref{Fig:CblaPlot}).

\begin{figure}[h!]
\begin{center}
\begin{tabular}{cc}
\includegraphics[width=1\columnwidth]{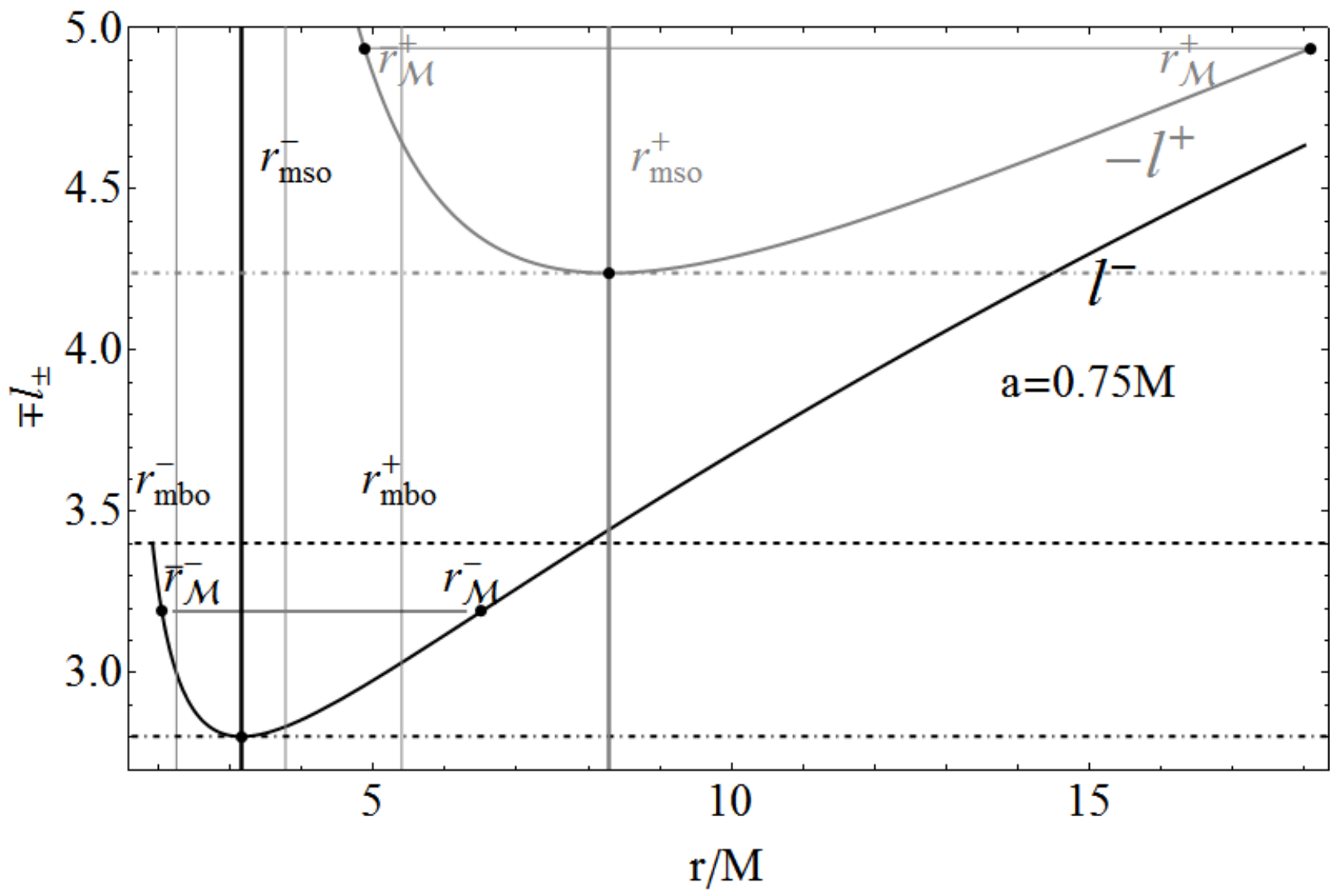}\\
\includegraphics[width=1\columnwidth]{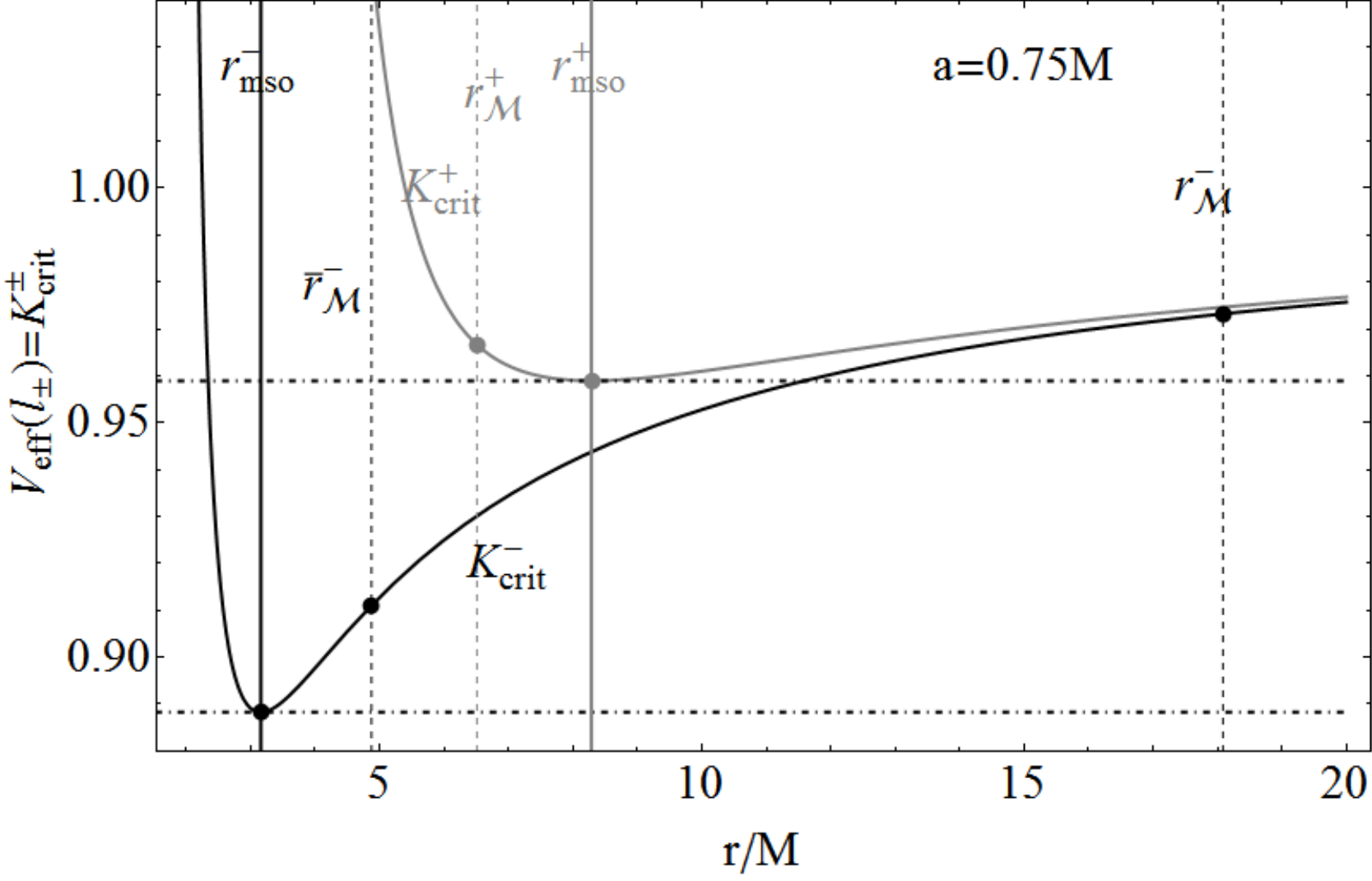}
\end{tabular}
\caption{{Spacetime spin $a=0.75M$.  The outer horizon is  at $r_+=1.66144M$.  Curves of the  fluid specific angular momentum $\pm\ell^{\mp}$ (upper panel) and $K_{crit}^{\pm}= V_{eff}(\ell_{\pm})$ of critical points of the effective potentials at different angular momenta (bottom panel) as functions of $r/M$. $r^{\pm}_{\mathcal{M}}$ the  maximum points    of
derivative $\partial_r(\mp \ell^{\pm})$ respectively. Corotating case $(-)$, black curves, and counterrotating case $(+)$, gray curves, are shown.  Minimum of the curves, signed by points, set the vales of  the functions $(\ell(r), K_{crit}(r))$  valuated at marginally stable orbits $r_{mso}^{\pm}$: toroidal configurations associated with critical points  of the pressure are possible only for $\mp \ell_{\pm}\geq\mp\ell_{mso}^{\pm}$ and $K^{\pm}\geq K_{mso}^{\pm}$ respectively.}}\label{Figs:IPCPl75}
\end{center}
\end{figure}

\begin{figure}[h!]
\begin{center}
\begin{tabular}{cc}
\includegraphics[width=.71\columnwidth]{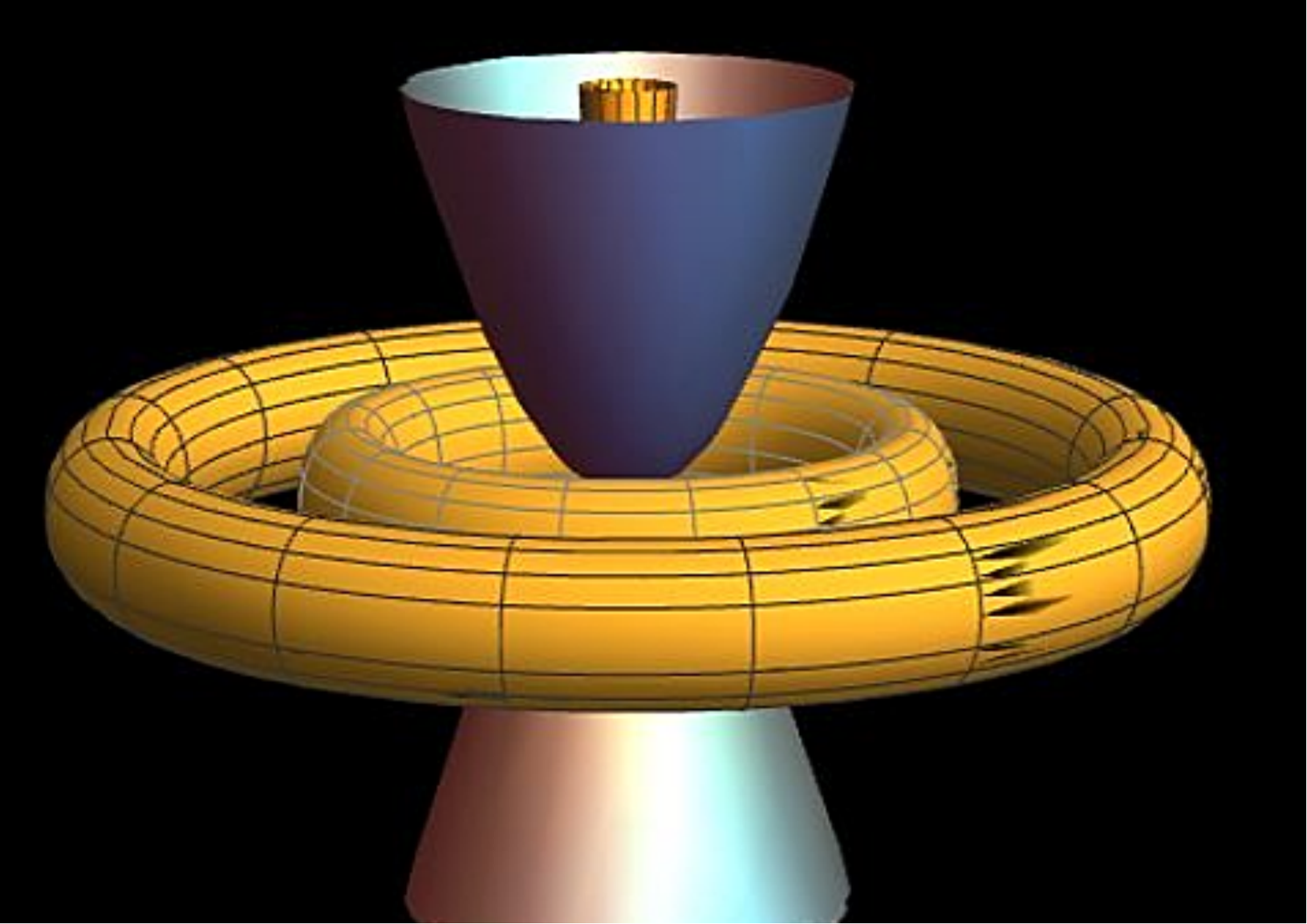}
\\
\includegraphics[width=.731\columnwidth]{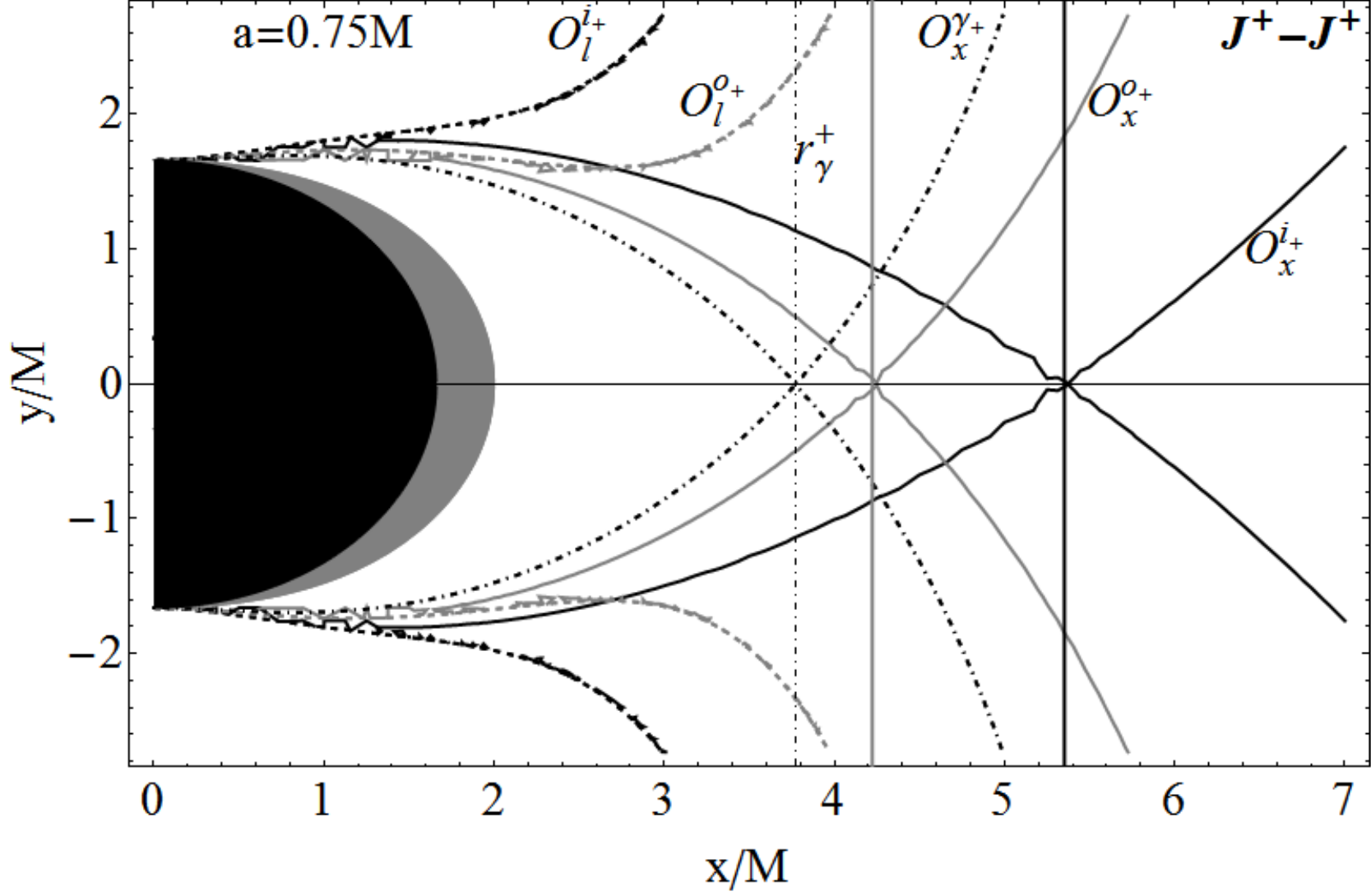}
\end{tabular}
\caption{Upper panel: Pictorial representation of a ringed accretion disk with  open $\oo_{x}$ surfaces.  Bottom panel: Spacetime spin $a=0.75M$, $\ell$corotating sequences, $\ell_i\ell_j>0$, of counterrotating open configurations $\mathbf{J^+-J^+}$  $\ell_i a<0$ $\forall i j$. Decomposition including open-crossed sub-configurations ($\gamma$-surface) $\oo^{\gamma}_x$, open cusped with angular momentum $\ell_{\gamma}$.  The outer horizon is  at $r_+=1.66144M$, black region is $r<r_+$, gray region is $r<r_{\epsilon}^+$, where $r_{\epsilon}^+$ is the static limit, and $r_{\gamma}^+$ is the photon orbit on $\Sigma_{\pi/2}$. For $O^{\oo_{+}}_x$ there is  $\ell_o=-5.62551$ and $ K_o=1.28775$, for
$O^{i_+}_x$  there is $\ell_i=-4.66487$ and $ K_i=1.00272$. The 
 open surfaces $\oo_{l}^{i_+}$ and $\oo_l^{\oo^+}$ are limiting solutions without critical points.}\label{Fig:CblaPlot}
\end{center}
\end{figure}

The procedure adopted  in the present article
borrows from the  Boyer theory on the equipressure surfaces applied to a   torus  \citep{Boy:1965:PCPS:},
where   the Boyer surfaces are given by the surfaces of constant pressure  or\footnote{{More generally $\Sigma_{\mathbf{Q}}$ is the  surface $\mathbf{Q}=$constant for any quantity or set of quantities $\mathbf{Q}$.}}  $\Sigma_{i}=$constant for \(i\in(p,\varrho, \ell, \Omega) \) \citep{Boy:1965:PCPS:,Raine}, where the angular frequency  is indeed $\Omega=\Omega(\ell)$ and $\Sigma_i=\Sigma_{j}$ for \({i, j}\in(p,\varrho, \ell, \Omega) \).  Many features of the tori dynamics and morphology like their thickness, their stretching in the equatorial plane, and the location of the tori are predominantly determined by the geometric properties of spacetime via the effective potential $V_{eff}$.  The boundary of any stationary, barotropic, perfect fluid body is determined by an equipotential surface,  i.e., the surface of constant pressure that is orthogonal to the gradient of the effective potential.
 The toroidal surfaces  are the equipotential surfaces of the effective potential  $V_{eff}(\ell) $, considered  as function of $r$ and $\theta$,  solutions $ \ln(V_{eff})=\rm{c}=\rm{constant}$ or $V_{eff}=K=$constant.
The  couple of parameters $(\ell,K)$ uniquely identifies each Boyer surface.
According to  Eq.\il(\ref{Eq:scond-d}), the maximum  of the hydrostatic pressure corresponds to the minimum $r_{min}$ of the  effective potential $V_{eff}$, and it is the torus center $r_{cent}$.   The fluids instability points  are located at the minima of the pressure and therefore maxima $r_{Max}$ of $V_{eff}$.  Equation  $\partial_r V_{eff}=0$  can be solved for the specific angular momentum of the fluid  $\ell(r)$--Fig.\il\ref{Figs:IPCPl75}-\emph{bottom}. This curve provides information about the center of the torus  $r_{cent}$ and possible critical points  $r_{Max}$, while we can calculate the values of the curve  $K$ as  $K_{crit}=V_{eff}(\ell(r))$ specifying the solution topology. These solutions, if they exist, represent non equilibrium configurations which may be closed $\cc_{x}$, for an accreting torus with accretion point $r_{x}$, or open $\oo_{x}$ which are associated to some ``proto-jet'' matter configurations\footnote{{ The role of ``proto-jet'' configurations, which in fact correspond  to limiting topologies for the  closed or closed cusped solutions associated with equilibrium or accretion,  is still  under investigation. More generally, in this model  the open surfaces   have been    always associated with the jet emission along the attractor symmetry axis--see for a general discussion \citep{KJA78,AJS78,Sadowski:2015jaa,Lasota:2015bii,Lyutikov(2009),Madau(1988),Sikora(1981),Stuchlik-AA-2000}.
  }},
 with critical point $r_{J}$ (see Fig.\il\ref{Fig:CblaPlot} and also \cite{open}). In general we use the notation $\pp$ and $\pp_{x}$ to indicate any equilibrium or critical configuration  without any further specification of its topology. Finally $\cc$ stands  for a closed equilibrium configuration whose  (stress) inner and outer edges are $r_{in}$ and $r_{out}$ respectively.
Then there is  $r_{in}\in\Delta r_{crit}$, where $\Delta r_{crit}\equiv[r_{Max}, r_{min}]$  and
$r_{out}>r_{min}$. The range $\lambda^i=r^i_{out}-r^i_{in}$ is the elongation on the equatorial plane of a  $\cc$ disk and,  in a tori couple,   $\bar{\lambda}=r^o_{in}-r^i_{out}$ is the spacing between the outer $\cc_o$ and inner $\cc_i$  torus of the couple. For colliding  tori there is $\bar{\lambda}=0$.

\begin{figure}[h!]
\begin{center}
\begin{tabular}{cc}
\includegraphics[width=1\columnwidth]{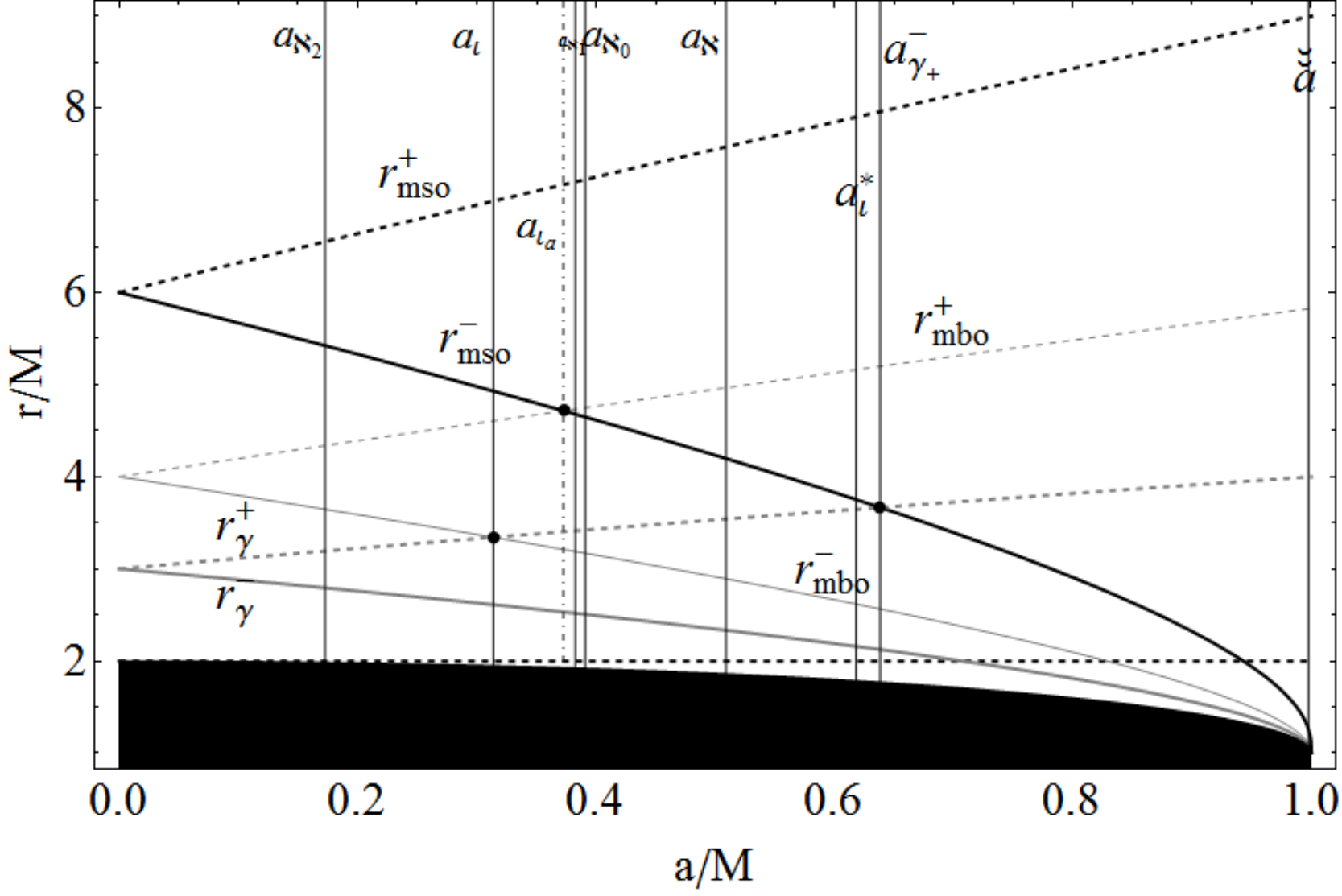}\\
 \includegraphics[width=1\columnwidth]{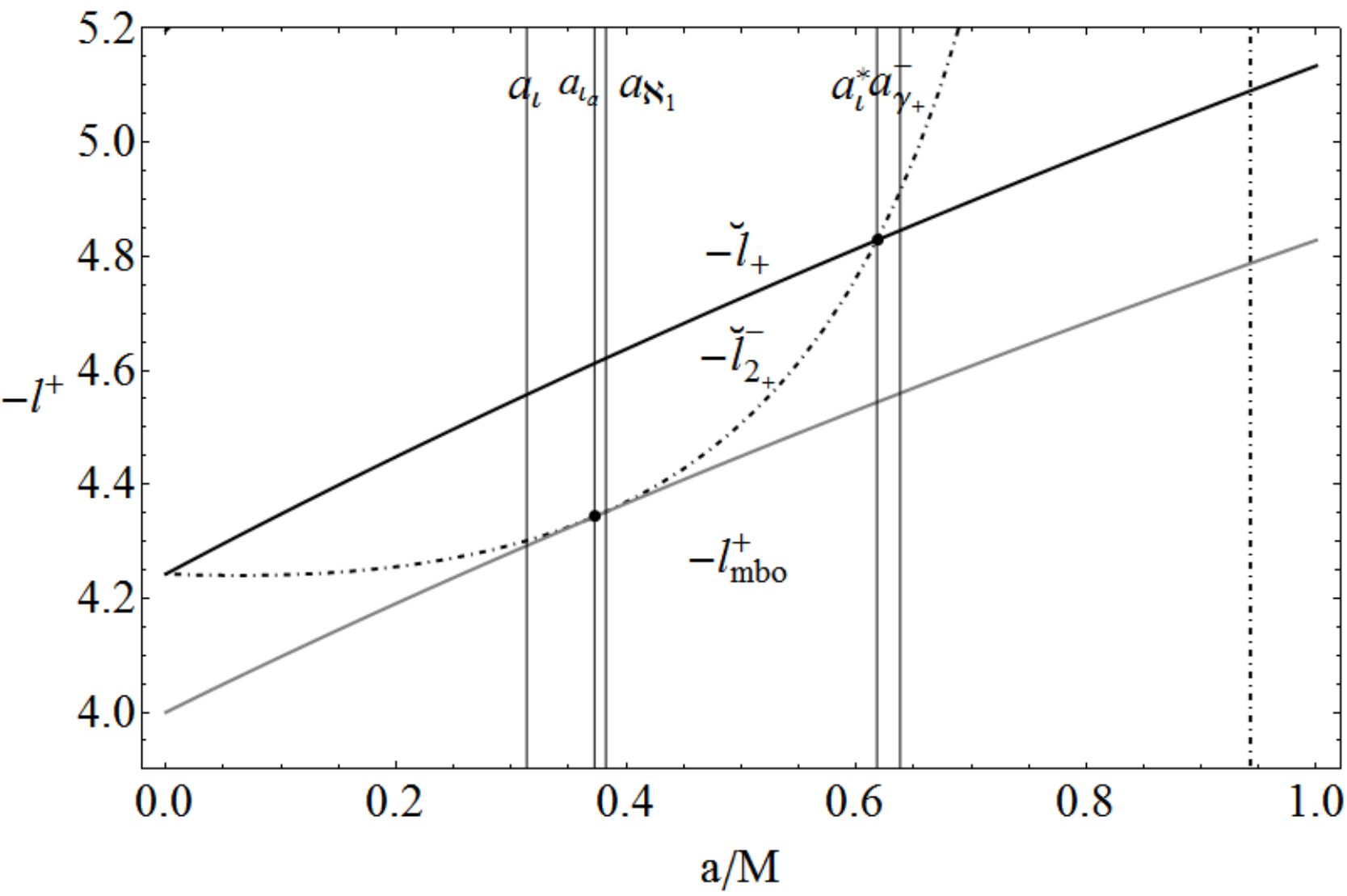}
\end{tabular}
\caption{
Upper panel: the notable radii
$r_{\mathcal{N}}^{\pm}=\{r_{\gamma}^{\pm},r_{mbo}^{\pm},r_{mso}^{\pm}\}$, for the corotating $(-)$ and
counterrotating $(+)$ orbits. Radii $r_{\gamma}^{\pm}$ are the photon orbits, $r_{mbo}^{\pm}$ the marginally bounded orbits
and $r_{mso}^{\pm}$  the marginally stable orbit. Black region is
$r<r_+$, $r_+$ being the  outer horizon.
Bottom panel:
specific angular  momentum  for counterrotating $\ell^+<0$  orbits on  the marginally bounded orbit $-\ell_{mbo}^+$
and  the curves $-\breve{\ell}_{2_+}^-$, as evaluated in $r_{mso}^-$. There is
$\breve{\ell}^-_{2_+}:\;V_{eff}(\ell_2^+, r_{mso}^-)=1$ where   $-\ell_2^{+}\in]-\ell_{mbo}^+,-\breve{\ell}_2^+[$.
The specific angular  momentum $\breve{\ell}_+:\; V_{eff}(\breve{\ell}_+,r_{mso}^+)=1$.}\label{Fig:Plotaaleph1IIa}
\end{center}
\end{figure}

Tori  are  strongly   constrained   by  the Kerr geometry \emph{geodesic structure}\footnote{It is worth specifying that this strong dependence of the model on the geometric properties  of spacetime induced by the central attractor enables us to  apply to a certain extent the   results found here to  different models of accretion disks \citep{AbraFra}.}: this comprises the  \emph{notable radii}
$r_{\mathcal{N}}^{\pm}=\{r_{\gamma}^{\pm},r_{mbo}^{\pm},r_{mso}^{\pm}\}$ made by   the \emph{{marginally stable circular orbit}}, $r_{\mso}^{\pm}$, the \emph{{marginally bounded circular orbit}}, $r_{\mbo}^{\pm}$ and  the \emph{{marginal circular orbit}} (photon orbit) $r_{\gamma}^{\pm}$-Fig.\il\ref{Fig:Plotaaleph1IIa}-\emph{Upper}-- \citep{Pugliese:2011xn,Pugliese:2013zma}.

It is simple to see that,
consistently   with  most of the axi-symmetric  accretion tori  models,
the \emph{(stress) inner edge} $r_{in}$ of the  accreting  torus is at  $r_{x}\in]r_{\mbo},r_{\mso}]$, while the torus  outer Roche lobe   is centered at   $r_{cent}>r_{\mso}$
 \citep{Krolik:2002ae,BMP98,2010A&A...521A..15A,Agol:1999dn}.
From now on given $r_{\bullet}$,
  we adopt  the  notation for any function $\mathbf{Q}(r):\;\mathbf{Q}_{\bullet}\equiv\mathbf{Q}(r_{\bullet})$, thus for example $\ell_{\mso}^+\equiv \ell^+(r_{\mso}^+)$.

\begin{figure}[h!]
\begin{center}
\begin{tabular}{c}
\includegraphics[width=1\columnwidth]{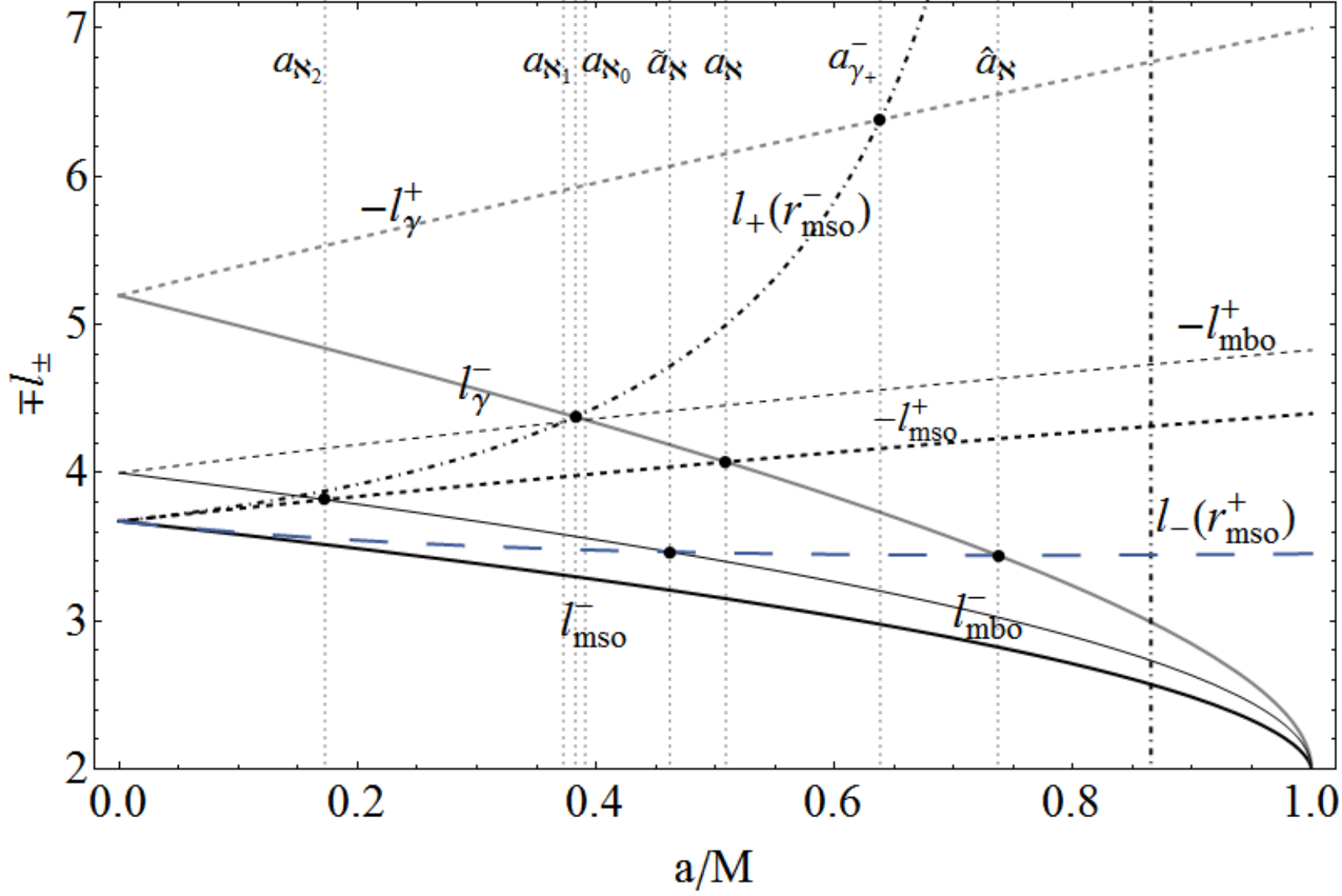}
\end{tabular}
\caption{Specific angular  momentum  for corotating $\ell^->0$ and counterrotating $-\ell^+>0$ orbits on the notable radii
$r_{\mathcal{N}}^{\pm}=\{r_{\gamma}^{\pm},r_{mbo}^{\pm},r_{mso}^{\pm}\}$, for the corotating $(-)$ and
counterrotating $(+)$ orbits.
Where $r_{\gamma}^{\pm}$ are the photon orbits, $r_{mbo}^{\pm}$ the marginally bounded orbits
and $r_{mso}^{\pm}$  the marginally stable orbit. Spins are shown where  a cross  occurs.}\label{Fig:Plotaaleph1II}
\end{center}
\end{figure}

 These radii stand as one of the main effects of the  presence of  strong curvature  of the background geometry.
 In fact, let
indexes $i\in\{1,2,3\}$ refer to the following ranges of angular momentum $\ell\in \mathbf{Li}$.
We find that
1. for fluid specific angular momentum $\ell$ in
$
\mp \mathbf{L1}^{\pm}\equiv[\mp \ell_{\mso}^{\pm},\mp\ell_{\mbo}^{\pm}[$,     topologies $(\cc_1, \cc_{x})$ are possible,  with accretion point in  $r^{\pm}_{x}\in]r^{\pm}_{\mbo},r^{\pm}_{\mso}]$.  
2. For
$\mp \mathbf{L2}^{\pm}\equiv[\mp \ell_{\mbo}^{\pm},\mp\ell_{\gamma}^{\pm}[ $,    topologies    $(\cc_2, \oo_{x})$ are possible,  with unstable point  $r^{\pm}_{J}\in]r^{\pm}_{\gamma},r^{\pm}_{\mbo}]$.   
3. For
$\mp \mathbf{L3}^{\pm}\equiv\ \ell \geq\mp\ell_{\gamma}^{\pm} $,     only equilibrium torus  $\cc_3$ exists--see Fig.\il\ref{Fig:Plotaaleph1II}.
The toroidal surfaces are characterized by $K_{\pm}\in [K^{\pm}_{min}, K^{\pm}_{Max}[ \subset]K_{mso}^{\pm},1[\equiv \mathbf{K0}$. Otherwise,  there can be  funnels of  material, associated to matter jets, along an open configuration   $O^{\pm}_x$ with   $K^{\pm}_{Max}\geq1$ ($\mathbf{K1}^{\pm}$).

Constraints in  this model are provided by the  conditions  of no-penetration of  matter ($\bar{\lambda}\neq0$-absence of collision) and by the geometric constraint for the equilibrium configurations determined by the geometric properties of the Kerr background reflected by  the geodesic structure.
We distinguish   four  types of unstable couples of orbiting configurations  (\emph{states} of the macro-configurations): the
\emph{proto-jet-proto-jet} (\textbf{J-J}) systems,  corresponding to couples of open cusped surfaces,
the \emph{proto-jet-accretion} (\textbf{J-A}) systems, where the proto-jet can follow or precede the accretion point, and finally the
 \emph{accretion-accretion} (\textbf{A-A}) systems,  where   matter    can accrete onto the attractor from  several instability points.
 Finally we consider also  the case of the
 \emph{proto-jet-equilibrium} (\textbf{J-C}) systems,  and the  \emph{accretion-equilibrium}  (\textbf{A-C}) systems.
  We   prove that states depend on the  dimensionless spin of the attractor, the relative rotation of the tori with respect to the attractor, the relative rotation of the fluids in the tori, strongly differentiating between  $\ell$corotating and  $\ell$counterrotating fluids.
After studying  five fundamental states,  we shall consider the possible combination of these states, reorganizing the investigations for ringed disks consisting of more than two rings,  i.e. with configuration of order greater then two.
An interesting task of  this investigation is the search for a  possible proto-jet-accretion correlation  in the states of the ringed disks.
Thus, we  introduce the concept of  \emph{geometrical correlation} between two configurations of a state, when the two surfaces may be in contact, in accordance with the constraints of the system. When a contact between two configurations occurs,   feeding or collision phenomena happen, leading eventually to a topological transition of  the ring state and, in the end, of  the entire macro-configuration.
 Consequentially we  face the problem of the \emph{state  evolution}: an initial couple  of configurations   (starting state) could evolve towards    a   transition of the surface topologies following an evolutive line from the initial state. We show that in  some cases equilibrium configurations  can only lead to proto-jet configurations and not to the accretion.

 In the following, we  will use also  the  symbols $\lessgtr$; we intend  the ordered sequence of  maximum points of the pressure, or $r_{\min}=r_{cent}$, minimum of the effective potential which corresponds to  the configuration center. Therefore, in relation to a couple  of rings,   the terms ``internal'' (inner-$i$) or ``external'' (outer-$o$), will always refer, unless otherwise specified,  to the   sequence ordered according to the center location.
  $r_{cent}$;
then  with
symbols $\mathbf{\succ}$ and $\mathbf{\prec}$,  we refer  to the sequentiality according to the  location of the \emph{minimum} points of the pressure,   or $r_{Max}$, maximum point  of the effective potential   $r_{x}^-<r_{x}^+$.

We organize our analysis   dividing   discussion in the  section (\ref{Sec:criticalII}) for the $\ell$corotating sequences and Section\il(\ref{Sec:ell-cont-double}) for the $\ell$counterrotating sequences. For easy of reference we listed   in Table\il(\ref{Table:nature-Att}) major  Kerr \textbf{BH} spin values defined   during this  analysis.
 In \ref{App:location}
 we provide  proof of the some  assertions used  in  Secs\il(\ref{Sec:criticalII}) and  (\ref{Sec:ell-cont-double}), and a more general  discussion of  some results.
\begin{table*}[ht]
\scriptsize{\tiny}
\centering
\caption{
Major  Kerr \textbf{BH} spin values defined   during the analysis. Details on the relevance of the 21 selected values of the \textbf{BH} spins  can be found in the text. }\label{Table:nature-Att}
\begin{threeparttable}[hp]
\resizebox{1\textwidth}{!}{%
\begin{tabular}{|l|l|l|}
\hline
$a_{\aleph_2}\equiv 0.172564M:-\ell_{mso}^+=\ell_{mbo}^-$&
$a_o^* = 0.201697M:\,\ell_{mso}^-=\ell^-(r_{mbo}^+)$&
  $a_{\iota}\equiv0.3137M:r_{mbo}^-=r_{\gamma}^+$
  \\
  $a_{\iota_a}\equiv0.372583M:r_{mso}^-= r_{mbo}^+ $ &
 $a_{\aleph_1}=0.382542M\in]a_{\iota_a},a_{\aleph_a}[:\ell_{\gamma}^-=-\ell_{+}(r_{mso}^-)$&
    $a_{\aleph_0}\equiv0.390781M:\ell_{\gamma}^-=-\ell_{mbo}^+$
    \\
    $\breve{a}_*\equiv 0.401642 M:\breve{\ell}_*=\ell_{\gamma}^-$&
    $ \tilde{a}_{\aleph}\approx 0.461854M:\ell^-(r_{mso}^+)=\ell_{mbo}^-$&
    $a_{\aleph}\approx0.5089M:-\ell_{mso}^+=\ell_{\gamma}^-$
    \\
    $a_{\iota}^* =0.618034 M:\breve{\ell}_{+}=\breve{\ell}_{2_+}^-$&
     $a_{\gamma_-}^{\beta}\equiv0.628201 M:\ell_{\beta}^-=\ell_{\gamma}^-$&
      $a_{\gamma_+}^-\equiv 0.638285 M:r_{\gamma}^+=r_{mso}^-$
      \\
${a}_1\approx0.707107M:r_{\gamma}^-=r_{\epsilon}^+$&
 $ a_o = 0.728163M\in]a_1, \breve{a}_{\aleph}[:\ell_{mbo}^-=\ell^-(r_{mbo}^+)$&
 $\breve{a}_{\aleph}=0.73688M:\ell^-(r_{mso}^+)=\ell_{\gamma}^-$
 \\
  $a_{\gamma_-}^{\Gamma}\equiv0.777271M: \ell_{\Gamma}^-=\ell_{\gamma}^-$&
   ${a}_b^-\approx0.828427M:r_{mbo}^-=r_{\epsilon}^+$&
$ a_o^{\gamma} = 0.867744M\in]a_b^-,a_{\mathcal{M}}^-[:\ell_{\gamma}^-=\ell^-(r_{mbo}^+)$
 \\
    $a_{\mathcal{M}}^-\equiv 0.934313M:\ell_{\gamma}^-=\ell_{\mathcal{M}}^-$&
    $a_2\approx0.942809M: r_{mso}^-=r_{\epsilon}^+$&
$\breve{a}\equiv 0.969174M:\breve{\ell}^-=r_{\gamma}^-$
\\
\hline
\end{tabular}}
\end{threeparttable}
\end{table*}
We study inclusions \footnote{$r_{\bullet} \in \pp$  means the inclusion of a radius $r_{\bullet}$ in the configuration $\pp$ (location of $\pp$ with respect to $r_{\bullet}$) according to some conditions;
  $\non{\in}$  is  non inclusion; in general
$\bowtie !$ intensifier  a reinforcement of a relation $\bowtie$, indicating that this is  a necessary relation which is  \emph{always} satisfied.} of the notable radii $r_{\mathcal{N}}^{\pm} \in ()_{\pm}$ and $r_{\mathcal{N}}^{\pm} \in ()_{\mp}$. This analysis  sets location of the disk inner edge  with respect to the geodesic structure $r_{\mathcal{N}}$, according to the fluid specific angular momentum. 
\section{ $\ell$corotating sequences}\label{Sec:criticalII}
We consider  the case of $\ell$corotating sequences focusing first on a couple  of  $\ell$corotating  configurations  and then  extending our investigation to the multiple configurations of the decomposition of  order $n>2$. Figure \il\ref{Fig:rssi-na}-\emph{third panel} shows a colliding couple of corotating configurations, and   Fig.\il\ref{Fig:rssi-na}-\emph{second panel} shows an inner couple of counterrotating tori in a three tori configuration, the inner rings are in accretion.

We point out that
\bea&&\label{Eq:spe-ci}
\mbox{ if  }\quad ()_i<()_o, \quad \mbox{ where}\quad \ell_i\ell_o>0,
\\\nonumber
&&\mbox{ then}\quad ()_i\succ()_o, \quad \mbox{where}\quad ()\in\{\cc, \cc_x, \oo_x\},
\eea
see Fig.\il\ref{Figs:IPCPl75}.
 For the \emph{ordered} sequences  of surface, with the notation $<$ or $>$,  we intend  the ordered sequence of  maximum points of the pressure or $r_{min}$, minimum of the effective potential and the disk centers for the closed sub-configurations \citep{ringed}. In relation to a couple  of rings,   the terms ``internal'' (equivalently inner) or ``external'' (equivalently outer), will always refer, unless otherwise specified,  to the   sequence ordered according to the location of the  centers.
Then,  if   ${\cc}_{i}<{\cc}_j$ for  $i<j$,  $\cc_i$ is the inner ring, closest to attractor, with  respect to $\cc_j$, and  there is    $r_{cent}^i\equiv r_{min}^{i}<r_{min}^j\equiv r_{cent}^j$. Within these definitions, the rings  $(\cc_i, \cc_{i+1})$ and $(\cc_{i-1}, \cc_{i})$ are \emph{consecutive} as  $\cc_{i-1}<\cc_i<\cc_{i+1}$  \citep{ringed}.
The
symbols $\succ$ and $\prec$  refer instead to the sequentiality between the   ordered location of the \emph{minimum} points of the pressure,   or $r_{Max}$, maximum point  of the effective potential, if they exist, which are  the instability points of accretion, $r_x=r_{Max}$ for $\cc_x$ topologies,  or of launching of proto-jets, $r_{Max}=r_J$ for the open cusped topologies (proto-jets).
Where, {due to definition}, it is always
$r_{min}^i<r_{min}^o$. For an $\ell$corotating sequence, this definition  \textit{implies} also   $r_{Max}^i>r_{Max}^o$, therefore,  for the $\ell$corotating sequences it is always $()_i<()_o$ and $()_i\succ()_o$.

The third inequality in  Eq.\il(\ref{Eq:spe-ci}) makes sense when  the potential function has a maximum point, that is  for   $\ell_i$ and $\ell_o$ in $\mathbf{L1}$ or $\mathbf{L2}$.
The nearest to the source is  the open surface launching point $r_{Max}=r_J$  or the accretion  point $r_{Max}=r_x$, and  largest is the magnitude of the fluid specific angular momentum. The  largest  is the  radius of the  maximum  pressure point and   more stretched on the  equatorial plane is  the configuration, regardless of its  topology.

\medskip

In what follows  we will specify relation  (\ref{Eq:spe-ci}) in  different cases, fixing the topology of the couple $()_i-()_o$: in Sec.\il(\ref{Sec:J-Jc}) we shall consider the   couple of open configurations  $\mathbf{(J^{\pm}-}\mathbf{J^{\pm})}$.
The couple $\mathbf{(A-J)}$, formed  by a configuration in accretion and an open configuration in $\oo_x$ topology,
 are studied in Sec.\il(\ref{Sec:A-J}) for the case where the  opened    cusped topology
is the outer one of the pair.
In Sec.\il(\ref{Sec:innerJ-A}),
the case in which the inner  configuration of the couple has   open topology is  investigated.
 Section\il(\ref{Sec:C-J-clna-v}) describes the couple formed by a disk in equilibrium and an open outer surface (according to the location of minima of the effective potential).
Section\il(\ref{Sec:J-Csmooth}) concentrates on the couple where the open  surface is the inner one.
Some remarks on the couples $\mathbf{(C-A)}$ are addressed in Sec.\il(\ref{Sec:A-C-Asmooth}), closing this section.

This section covers the  $\ell$corotating  sequences of rings. We shall always intend  the  relations between magnitudes of the specific angular momentum,  if not otherwise specified.
\subsection{The $\ell$corotating proto-jet-proto-jet (\textbf{J-J}) systems}\label{Sec:J-Jc}
In this section we consider a couple of    $\ell$corotating  open cusped configurations with P-W instability  points $(r_{Max}^{o},r_{Max}^{i})$ where
\bea\nonumber
&&
\oo_x^i<\oo_x^o\quad\mbox{implies}\quad \oo_x^i\succ \oo_x^o\quad, |\ell_i|<|\ell_o|,\;\ell\in\mathbf{L2}, \\
&& \label{Eq:dim-ana}r_J\in]r_{\gamma},r_{b}],\quad
\mathbf{K}_i\in]1,\mathbf{K}_o[,
\eea
for the   state of the $\ell$corotating couple $O^{\pm}_x-O^{\pm}_x$. Here  we determine  the evolutive lines  for different black hole attractors.

In the sequences of open configurations,  we will mainly deal   with  the  critical points.   It will be  then convenient  to introduce  the \emph{criticality indices}  $\hat{i}$, univocally associated   to the couple ($r_{Max}^i$, $\ell_i$), where $i$ is as usual the configuration index   univocally associated to ($r_{min}^i$, $\ell_i$)\footnote{It is intended that  no confusion  will    arise by a possible overlap  of two  $\ell$counterrotating  configurations.}.
As relation (\ref{Eq:spe-ci}) stands,   we can consider  $\hat{i}(i)$ as a decreasing function of the configuration index $i$ or  $\partial_i \hat{i}(i)<0$. In other words, Eq.\il(\ref{Eq:dim-ana}) can be written in the criticality indexes as
$\oo_x^i\succ \oo_x^o$ and  $\oo_x^{\hat{o}}\succ \oo_x^{\hat{i}}$ as $|\ell_i|=|\ell_{\hat{o}}|<|\ell_o|=|\ell_{\hat{i}}|$.
The function   $\hat{i}(i)$ has inverse  in a restriction of the variation domain of  the configuration index. In fact, the configuration index set  is far more vast of  the set of the criticality indexes containing, for example, the surfaces with momentum in \textbf{L3}. However  we might say that the function $\hat{i}(i)$ could be inverted in $\mathbf{L1}$ and  $\mathbf{L2}$, taking into account that the disk  does not include necessarily the  critical points  when  $K<K_{Max}$ (in other words, the rank $\mathfrak{r_x}\in[0,n]$).

In principle, there is an infinite number of $\ell$corotating critical open funnels, i.e. $n^{Max}_{J}=\infty$.
As mentioned in \cite{ringed}, we can introduce  definition of configuration density $\delta n$, i.e., the density of maximum points of pressure in a
 fixed orbital range $\Delta r$,  $ \delta n\equiv \left.n\right|_{\Delta r}/\Delta r$.  The configuration density naturally depends on the specific angular momentum parameter and particularly on the matrix of displacements, here considered at constant step $\kappa$\footnote{For example, by adopting the   antisymmetric displacement matrix $\epsilon_{[i, i+\kappa]}=\kappa \epsilon$ (with  $\epsilon=$constant  and step of the decomposition $\kappa\geq 1$),  introduced in  \cite{ringed}.}. The density is generally a decreasing function of the step $|\kappa|>0$.  Similarly, we can introduce a  \emph{criticality density} $\delta n_{Max}$ as the density of  P-W  points. Specifically we now consider the density $\delta n_J$ as  the density of P-W launching points of proto-jets in $\Delta r_J$. In the particular case of  $\ell$corotating matter, the following relations hold:
\bea&&\nonumber
\partial_{\delta \mathbf{Q}} \delta n<0,\quad\partial_{\delta \mathbf{Q}} \delta n_J<0\quad\mbox{ with}\quad\delta\mathbf{Q}\in\{\delta |\ell|, \delta K\},
\\
&&\partial_{\delta \mathbf{Q}}\Delta r_J^{i,o}>0\quad \Delta r_J^{i,o}\equiv r_{J}^i-r_J^o>0,
\eea
 where
$\delta K$ and   $\delta |\ell|$ give  the magnitude of the difference  of the $K$ parameter and the specific angular momentum $|\ell|$ of the $\ell$corotating rings,
and $\Delta r_J^{i,o}$ is the distance between the two launching  points--see Fig.\il\ref{Fig:CblaPlot} ($\delta |\ell|$  is   constant displacement $|\kappa|$ in accordance with \cite{ringed}). Here and in the following, by  $\partial_{\mathbf{B}}\mathbf{Q}>0$ we mean  the quantity $\mathbf{Q}$ increases as the quantity $\mathbf{B}$ increases and viceversa.

The increase of the criticality  (configuration) density corresponds to an increase, at fixed orbital range, of the   $ \oo_x $ configurations  number in  the  $\ell$corotating  sequence. This corresponds to an  increase in the magnitude of the specific angular momentum as the launch point $r_J$ is closer to the   surface  $\oo_x^{\gamma_{\pm}}$, cusped surface with angular momentum $\ell_{\gamma}^{\pm}$--Fig.\il\ref{Fig:CblaPlot}:
additional specific angular  momentum  is required  to the orbiting matter for  the formation of a new   $\ell$corotating launching point of proto-jets, inner with respect to  the first $O^{\hat{1}}_x$ of the  $\{O^{\hat{i}}_x\}_{\hat{i}}$ sequence,  increasing thus inwardly the  criticality density  $\delta n_J$,   while approaching the threshold  $\oo_x^{\ell}=\oo_x^{\gamma}$.
In other words, the fluid specific angular  momentum magnitude  decreases \emph{outwardly} in the decomposition.

However, the specific angular  momentum  that has to be provided  for a shift of  the critical point inwardly, is  not constant with the dimensionless spin  of the black hole. It  is  strongly diversified for the two $\ell$counterrotating  sequences, and  does not in general grow regularly with the configuration index (or decreases with the criticality index) \citep{ringed}. In fact, there exists a maximum value with respect to configuration index (or equivalently with respect to the radius) for configurations with  center of maximum hydrostatic pressure in
 $r_{\mathcal{M}}^{\pm}>r_{mso}^{\pm}$,   and specific angular  momentum $\ell_{\mathcal{M}}^{\pm}>\ell_{mbo}^{\pm}$ respectively.

In terms of the  criticality indices  $\hat{i}$,  one has
$\partial_{\hat{i}}\ell <0$, with
$\partial^2_{\hat{i}}\ell=0$ for a configuration with specific angular  momentum magnitude   $\ell_{\mathcal{M}}>\ell_{mbo}$
 and centered in  $r_{\mathcal{M}}$. Indeed, since
 $r_{\mathcal{M}}^{\pm}>r_{mso}^{\pm}$, the radii $r_{\mathcal{M}}^{\pm}$ can be   minimum points of the effective  potential, but \emph{not} maximum points.  Then $\ell_{\mathcal{M}}$ is a minimum (maximum) value for   the function $\partial_{\hat{i}} \ell_{\hat{i}}$ ($\partial_i\ell_i=\partial_{i} \hat{i}\partial_{\hat{i}} \ell_{\hat{i}}$).

The two $\ell$counterrotating   sequences of $\ell$corotating open configurations with specific angular  momenta  $\ell_{\mathcal{M}}^{\pm}$,  have generally different topologies associated with their critical phase. Indeed,
 for the counterrotating fluids in the geometries $a\in]0,M]$, we have
$-\ell_{\mathcal{M}}\in]-\ell_{mbo}^+, -\ell_{\gamma}^+[\equiv \mathbf{L2}^+$; therefore,  the critical configurations with  $\ell=\ell_{\mathcal{M}}^+$ always correspond to the proto-jets $\oo_x^+$.

Conversely, this is not always the case for the corotating fluids
 where, at higher spin  of the attractor, i.e.,  $a\geq a_{\mathcal{M}}^-\equiv 0.934313M$, there is  $\ell_{\mathcal{M}}^-\in \mathbf{L3}^-$ when there are no critical topologies. Whereas in the geometries with   $a\in[0,a_{\mathcal{M}}^-[$, we have $\ell_{\mathcal{M}}^-\in \mathbf{L2}^-$,  and only critical configurations $\oo_x^-$ are possible. Moreover, this also means that  the supplying  specific angular  momentum to be provided for a further inner launch point of proto-jet (according to the criticality  index $\hat{i}$), decreases constantly  with the criticality index   (or constantly increases with the configuration index) in the geometries of the faster attractors\footnote{For \emph{fast} (\emph{slow}) attractors we intend Kerr attractors with  high ({small}) values of the dimensionless spin with  respect to some  reference values of $a/M$, fixed considering the  geodesic spacetime structure.}, as in those spacetimes, there is no minimum of $\partial_{\hat{i}}\ell^-$ in   $\mathbf{L2}^-$.
 One can conclude that the specific angular  momentum to be  supplied in  the disk for an  inner (corotating) proto-jet launch point grows uniformly with  the radius (and  uniformly with the configuration  index)- $\partial^2_i \ell_J> 0$ for very fast attractors.
But the surplus of specific angular  momentum needed to locate the proto-jets in the inner  regions (moving the center outwardly)   increases more and more slowly in the far away regions\footnote{In  $r\gg r_{\mathcal{M}}(a)  (i\gg i_{\mathcal{M}})$, as discussed in \cite{ringed}.}, where the Newtonian limit  could be considered and, as asymptotically    $\lim_{r\rightarrow\infty}\ell'=0$ with $\lim_{r\rightarrow\infty}\ell^{\pm}=\pm\infty$, the quantity  $\delta \ell_{i+j,i}$ is approximately constant with the radius.

The existence of the couple $(r_{\mathcal{M}},\ell_{\mathcal{M}})$ is a  relativistic effect, also present in the static case $a=0$, but for a rotating attractor this is  strongly differentiated from an   albeit minimal intrinsic rotation of the gravitational source.
In the static limit,
there exists a maximum of  $\partial_{r}\ell$ for  $r_{\mathcal{M}}=(6+4 \sqrt{3})M$,
the spread between the two $\ell$counterrotating  cases appears only when $a\neq0$, and becomes obviously more and more pronounced with increasing spin of the attractor.
In general, however, the increase of specific angular  momentum decreases with  increasing of $r$ and   $R=r/a$ where, at the limit of very large $R$, there is no distinction between the two $\ell$corotating sequences.
This is therefore a feature of the toroidal rotating fluids strongly affected by the
dragging of the spacetime. Indeed, the difference $\partial_r(\ell^++\ell^-)$ goes to zero  as $a=0$  and $r$ goes to infinity \footnote{
However, for $a/M\approx 0$, it is $\approx{2 (a/M) (3 r/M-2)}/{(r/M-2)^3}$.}.
In terms of  the configuration and criticality  density
we have:
\bea&&\label{Eq:densi}
\mathbf{1.}\quad\partial_r\left.\delta n(r)\right|_{\bar{\delta}_{\ell}}\gtrless0\quad\mbox{for}\quad r\lessgtr r_{\mathcal{M}}\quad\mbox{and}
\\\nonumber
&& \mathbf{2.}\quad \partial_r\left.\delta n_J(r)\right|_{\bar{\delta}_{\ell}}\lessgtr0\quad\mbox{for}\quad {r}\lessgtr \bar{r}_{\mathcal{M}}.
\eea
The first relation of Eq.\il(\ref{Eq:densi}) is  always verified for all specific angular  momentum ranges  $\mathbf{Li}$, whereas  the second of  Eq.\il(\ref{Eq:densi}) makes sense only  for $\ell\in\{\mathbf{L1}, \mathbf{L2}\}$.

For these specific angular  momenta   we can consider the  radii to be $\bar{r}_{\mathcal{M}}\equiv r_{Max}(\ell_{\mathcal{M}})$  the critical radii associated with the specific angular  momentum $\ell_{\mathcal{M}}$,   coupled with the center ${r}_{\mathcal{M}}$.  The plot of $\bar{r}_{\mathcal{M}}$ as function of $a/M$, will be analogue to the plot of ${r}_{\mathcal{M}}$ as function of $a/M$,  but   rotated along an axis parallel to $r=$constant and located in the  orbital range according
to  $\ell_{\mathcal{M}}$ in the ranges $\mathbf{Li}$ \footnote{ The couple $({r}_{\mathcal{M}},\bar{r}_{\mathcal{M}})$)  corresponds to the couple  of indices  $(i, \hat{i})$.}.

Similarly to the case of the equilibrium ringed disks,  we can explain  the significance of the presence of the $r_{\mathcal{M}}$ points  and of  Eq.\il(\ref{Eq:densi})
  in terms of density of critical points.
 Assuming that  increase of the specific angular  momentum in magnitude within the  macro-configuration is adjusted for constant displacement ($\kappa$ constant), and  $\partial_i \hat{i}<0$ (we mean to say  that one quantity  decreases where the other  increases, repositioning  the labels  such that  $i\rightarrow \hat{o}$ and viceversa $o\rightarrow \hat{i}$), one has $\epsilon_{[\hat{i}\hat{j}]}=-\epsilon_{ij}=\epsilon_{ji}$ and  $\epsilon_{ \hat{i}+\hat{\kappa},\hat{i}}=\hat{\kappa} \epsilon>0$.
We note that the configuration density, directly related to $\kappa$,  is maximal at $r_{\mathcal{M}}$.
Let  $\bar{r}_{\mathcal{M}}=r_J<{r}_{\mathcal{M}}=r_{cent}$  be  the (proto-jet) instability point associated  with ${r}_{\mathcal{M}}$. This point is always present in the counterrotating case  and for corotating fluids orbiting attractors  with   $a<a^-_{\mathcal{M}}$. The couple of critical points    $(\bar{r}_{\mathcal{M}},{r}_{\mathcal{M}})$  is  an intrinsic property of the specific geometry and it is   function  of $a/M$ only, as such it is unique  for each attractor.
 It follows that the disk centers are  more spaced in the region $r>{r}_{\mathcal{M}}$ (lower density of the index configuration), and viceversa at  $r<{r}_{\mathcal{M}}$; the rings are closer together   as they approach  ${r}_{\mathcal{M}}$.  In a neighborhood of  ${r}_{\mathcal{M}}$, the configuration density $\left.\delta n\right|_{\kappa}\approx\delta n_{\mathcal{M}}$ is  highest-- Fig.\il\ref{Figs:IPCPl75}.
Similarly, rewriting all in terms of the criticality density $\delta n_J$, we would say that, under the conditions given by the displacement matrix with step  $\kappa$ constant\footnote{It is indeed immediate to argue the relation in the assumptions $\kappa=$ constant and  $\epsilon=$constant.},  as the launching points are more spaced (lower criticality density)  in  $]r_{\gamma},\bar{{r}}_{\mathcal{M}}[$,  and decreasing as they move towards the inner $\oo_x^{\gamma}$ surface at $r<\bar{{r}}_{\mathcal{M}}$. At  $r_J>\bar{{r}}_{\mathcal{M}}$, the launching points are getting closer, as  they  approach $\bar{r}_{\mathcal{M}}$  from the outer  regions, or  $r_J\in]\bar{{r}}_{\mathcal{M}},r_{mbo}[$. It follows that, in a neighborhood of ${\bar{r}}_{\mathcal{M}}$, the  corresponding criticality density  $\left.\delta n_J\right|_{\kappa}\approx\delta n_{\mathcal{M}}$ is \emph{minimal}.

This means that, in a ringed  model   where the specific angular  momentum varies (almost) monotonically with a constant step $\kappa$, two remarkable points in the distribution of matter appear:  the $r_{\mathcal{M}}$, where the density of stable configurations (or density of the maximum hydrostatic pressure) reaches maximum, and  the corresponding point $\bar{r}_{\mathcal{M}}$  where the density of proto-jet launch shall be   at minimum.
We recall that  $(r_{\mathcal{M}},\bar{r}_{\mathcal{M}})$, are uniquely  fixed by the  attractor geometry and   the  sense  of the fluid  rotation with respect to the geometry. Therefore, it would be possible to deduce both  the spin of the attractor and the sign of rotation of the fluid relative to this from the knowledge of one of the points of  $(r_{\mathcal{M}},\bar{r}_{\mathcal{M}})$ and the  surfaces with specific angular momentum $\ell\in \{\ell_{\gamma},\ell_{mbo},\ell_{mso}\}$.  

For very fast attractors with  $a>a_{\mathcal{M}}^-=0.934313M$, there is $\ell_{\mathcal{M}}^-=\ell_{\gamma}^-$, and  no maximum of density of the corotating fluid in open topologies exists; then, in the  conditions provided by the assumption of $\kappa$ constant, the set of critical points constantly increases approaching the critical $\gamma$-surface configuration (with specific angular momentum $\ell\in\{\ell_{\gamma},\ell_{mso},\ell_{mbo}\}$).
We  specify that  the assumption of constant step $\kappa$   has been   here adopted as the most  elementary and illustrative example of displacement matrix, the extension of these considerations to a general displacement law, with  a generic $\epsilon_{ij}$ matrix, could  be developed in a rather straightforward manner.

All the  considerations involving  very fast attractors for corotating fluids   can  be interpreted  as an indication of the role played by the intrinsic rotation of the attractor in the formation of  the corotating proto-jets.
In this respect, it is worth noting that the following relations hold
\bea
&&\nonumber
Q_{\mathcal{M}}^+>Q_{\mathcal{M}}^-,\quad \Delta r_J^{\pm}<\Delta r_x^{\pm},\quad \delta r_J^{\pm}<\delta r_x^{\pm},
 \quad
\delta\ell_J^{\pm}>\delta\ell_x^{\pm},
\\
&&\nonumber \delta\ell_s^{-}<\delta\ell_s^{+},\; s\in\{J, x\}, \;  Q_{\mathcal{M}}\in\{r^{\pm}_{\mathcal{M}},\mp\ell_{\mathcal{M}}^{\pm},K_{\mathcal{M}}^{\pm}\},\;a_*\equiv a/M
\\
\label{Eq:ban-appl-cho}
&&
\partial_{a_*}\delta Q_s^{\pm}\gtrless0,\quad \delta Q^{\pm}\in\{\delta r^{\pm},\delta\ell^{\pm}\},
\quad\partial_{a_*}Q_{\mathcal{M}}^{\pm}\gtrless 0.
\eea
Relations (\ref{Eq:ban-appl-cho}) highlight various properties of the  $\ell$counterrotating sequences and the role of the black hole spin in the formation of  the decompositions.

However, before moving to the analysis of the two isolated subsequences\footnote{The definition of  isolated and mixed  subsequences were introduced in  \cite{ringed}, here we remind that $\ell$counterrotating sub-sequences of a  decomposition of the  order $n=n_++n_-$, {are    \emph{isolated} $ \overbrace{\mathbf{C}_s}$   if  $\cc_{n_-}<\cc_{1_+}$ or $\cc_{n_+}<\cc_{1_-}$  and \emph{mixed}  $ \overbrace{\mathbf{C}_m}$  if $\exists\; {i_+}\in[1_+, n_+]:\; \cc_{1_-}<\cc_{i_+}<\cc_{n_-}$ or viceversa  $\exists\; {i_-}\in[1_-, n_-]:\; \cc_{1_+}<\cc_{i_-}<\cc_{n_+}$.}},   we need to clarify  the role of the two orbits $\bar{r}^{\pm}_{\mathcal{M}}$.
Considering Eq.\il(\ref{Eq:ban-appl-cho}), we have  $Q_{\mathcal{M}}^+>Q_{\mathcal{M}}^-$ that means
a greater  specific angular  momentum and  $K$-parameter   is required for the matter  in stable orbit    to reach a maximum pressure point, see also \cite{ringed}.
In general, there is    $r_{min}^+>r_{min}^-$. Then for $\ell_{\pm}\in \mathbf{L2}^{\pm}$, the solution
$\mp\partial_r \ell^{\pm}=0$ has only one maximum point, i.e.,
$\bar{r}_{\mathcal{M}}$ is not a solution of $\mp\partial_r \ell^{\pm}=0$, but  every critical point of maximum  is associated to a minimum,
 and in this sense we consider $\bar{r}_{\mathcal{M}}$ to be a minimum point as associated to the critical point ${r}_{\mathcal{M}}$.

More  generally, without restricting the analysis to the $\ell_{\pm}\in \mathbf{L2}^{\pm}$ case, if
${r_{min}^-<r_{min}^+}$,    then {  \emph{necessarily} }  $\ell^-<-\ell^+$, but  the relation between the maximum points has still to be established. However,  if   ${\ell^-\in]\ell_{mso}^-,-\ell^+[}$, then it can be either   $r_{min}^-<r_{min}^+$ \emph{or}  $r_{min}^-\in]r_{min}^+,\bar{r}_-[$ where $\bar{r}_-:\;\ell^-(\bar{r}_-)=-\ell^+$.

If $\ell^->-\ell^+>\ell_{mso}^+$, then
{\emph{necessarily}  }
$r_{Max}^-<r_{Max}^+<r_{mso}^+<r_{min}^+<r_{min}^-$.
But if  ${\ell^-<-\ell^+}$, then
neither the maximum or the minima are fixed and it can be either $\cc^-<\cc^+$ or $\cc^+<\cc^-$  for each  $r_{Max}^-<r_{Max}^+$ or $r_{Max}^+<r_{Max}^-$.

To summarize, considering also Fig.\il\ref{Figs:IPCPl75}:
if  $r_{min}^+>r_{min}^-$, then   $-\ell^+>\ell^-$.
But it is simple to see that we can obtain $r_{Max}^+>r_{Max}^-$ or $r_{Max}^->r_{Max}^+$. For example, the last case occurs for $r_{min}^-$ being very close to $r_{mso}^-$ and a great separation in  $(\ell^++\ell^-)$. Thus if
$
r_{min}^-\in[r_{mso}^-, r_{min}^+]$ and $\ell^-\in]\ell_{mso}^-, \ell^-(r_{min}^+)[$,
if
$r_{Max}^-: \ell_{mso}^-<\ell^-<\ell^-(r_{min}^+)$
 there is  $\partial_{r_{min}^+}r_{Max}^->0$, see also \cite{ringed,pugtot}.
However, this would make sense  for matter orbiting around sufficiently slow attractors $\mathbf{A}_{\iota}^<\equiv [0,a_{\iota}[$, where $a_{\iota}\equiv0.3137M$.

In other words, knowing the relation between the  points of maximum  pressure, we could ignore  the sequentiality according to the minimum points of pressure.
This information is indeed  important to determine the relative position of the two  $\ell$corotating sequences  also in the case of consecutive sequences.
Part of these considerations will  be resumed in \ref{App:location}.

For attractors  with higher  spins, at $r>r_{\mathcal{M}}$, for the  corotating sub-configurations   the Newtonian limit is  reached in regions closer to the source, and with lower specific angular  momenta magnitude  with respect to the counterrotating ones. This is in agreement with the fact that  the spacetime spin   clearly distinguishes the two types of fluids,
 where the   relativistic effects are essentially determined by the rotation of the Kerr attractor.
In the first place, as discussed  above, the radius corresponding to the  minimum critical density  $\delta n_J$ is, for corotating fluids, always closer to the source than for the counterrotating proto-jets. For corotating matter, this point approaches the source as the attractor spin increases, and the specific angular  momentum  required for the launch of a  corotating proto-jet decreases with this. With increasing black hole spin, the magnitude of the   specific angular  momentum required  for a corotating proto-jet decreases and is decreasing in relation to those corresponding to the  counterrotating case.

The situation is indeed just the opposite for the $\oo_x^+$ surfaces where,  with  increase of the black hole spin, the point of minimum proto-jet density moves outwards, confining  these configurations in regions more and more distant from the source, and  requiring also   increasing  magnitude of the specific angular  momentum. It is then  possible to prove that  the  $K_{\mathcal{M}}$-parameter,  associated   to the momenta $\ell_{\mathcal{M}}$, shows an analogue behavior. This would suggest that the black hole  spin   distinguishes the two types of matter, by favoring the formation of corotating proto-jets with respect to the counterrotating ones.

In order to fully characterize the  role of the   dragging effects with respect to the $(J...J)$ $\ell$corotating sequences, it is  important to analyze the  ranges of the variation for the parameters    of  the proto-jet configurations.
Higher specific angular  momentum in magnitude is required to set the matter in open funnels than for  the accretion.
In any Kerr geometry, the orbital region   where  the open funnels $\oo_x$ are possible is the  closest  to the attractor, being inner with respect to the accretion regions (at  $\mathbf{L1}<\mathbf{L2}$)  and in general it is smaller: the orbital range  allowed for an accretion point is larger then that  where the proto-jet launch ($r_J$) can be formed. The extension of the specific angular  momentum  range possible for the accretion, i.e.  the measure of $\mathbf{L1}$, is less then  the measure of the  $\mathbf{L2}$  range for the jets.
These properties are important in the characterization of the   critical points sequences, and the determination of the criticality density: we could conclude that the accretion phases are favored with respect to the    instable proto-jets.

We could suppose that the  proto-jet of open funnels  arises at the final stage of the formation of an accretion disk which increases  its elongation, approaching  the source, or the $\oo_x$ surface could also  arise  being non--correlated in any way by the accretion  phase.
In the model we are considering, this can be achieved  keeping $\ell\in \mathbf{L2}$ fixed for a disk $\cc_2$, or fixed in the range $\mathbf{L2}$, with increasing $K$ (growing up  of  the density due,   for example, to the  interaction with the surrounding material, with the consequent increase in the disk size and elongation), or decreasing  the specific angular  momentum  from a starting configuration  $\cc_3$ in  \textbf{L3}, losing specific angular  momentum (in magnitude) and  then moving inwards. Finally, starting by a model in  \textbf{L1}, with topology $\cc_1$ or $\cc_{x}^1$, an open critical surface could be  the consequence of an increase of  specific angular  momentum magnitude (due feeding   matter for example, or even for  a direct interaction with the source \citep{pugtot}, as the disk $\cc_1$, or in the critical topology $\cc_x^1$, lies in  the orbital region  closer to the source). A possible mechanism for the feeding of matter and increasing  momentum  for  the  $()_1$ disks, with the  consequent shift of the critical point from  $\Delta r_x$ to $\Delta r_J<\Delta r_x$,  could take place  in these  multiple ringed systems,  for feeding from external surface to $()_1$, for example in the $\mathbf{C}_{\odot}^n$ ringed disks.

The  case  of  starting data in $\mathbf{L1}$ is particularly interesting as the proto-jet launch is related in a direct way  to the accretion phase. On the other side,  from starting data  in $\mathbf{L2}$ or $\mathbf{L3}$, the emergence of the instability points $r_J$ are not directly related  with the accretion, and then the chronological lines   have different evolutionary histories.

However it is worth to note that the configurations $\cc_1$  could evolve directly,   without an accretion stage, towards an  $\oo_x^2$ topology.
A transition from $ \cc_1$ or $\cc_x^1$, requires an increase  of the disk specific angular   momentum, shifted accordingly from $\mathbf{L1}$ to $\mathbf{L2}$. The disk  surface will increase the elongation (but also the density) by the  growing of the $K$-parameter  for the shift  from $\mathbf{K0}$ to $\mathbf{K1}$. It is therefore necessary, for the configurations in  $\mathbf{L1}$,  to provide a mechanism  able to explain   the growth of both the angular momentum  and  the density. On the other hand, for a starting  equilibrium $\cc_3$ disks, with angular  momentum in $\mathbf{L3}$, an evolution towards the launch of an  $\oo_x^2$ proto-jet implies a transition  $\mathbf{L3}$ to $\mathbf{L2}$, therefore a loss of the specific angular  momentum  and   increase in density (because $K$ increases) and size.  This can occur  due  to for example to some feeding  by  embedding material or also a further outer disk of the configuration. For initial data in   $\mathbf{L2}$,  only an increase of   $K$ from  $\mathbf{K0}$ to $\mathbf{K1}$ is required. From this the formation of $\oo_x^2$ proto-jets would seem to be favored   starting from an equilibrium disk   $\cc_2$ or   $\cc_3$ and  not from an accretion phase of a $ \cc_x^1$ surface.

Further considerations concerning the  ``transitions'' to the final state of open funnels, are  given in  \ref{App:location} where different aspects of the geometric correlation  are addressed,  analyzing  the location of the inner and outer edges of the disks in the geodesic structure.
It is important to point out that these considerations are not based on the analysis of the geometrical correlations, but on the  analysis of  variation of the specific angular  momentum:  the orbital region for the  proto-jet is  internal with respect to orbital range of the accretion.

For corotating fluids orbiting the  faster Kerr attractors,   the reduced range  of possible specific angular  momentum (and correspondingly the  orbital range)  is associated  in the \textbf{RAD}  to a reduced  possibility of the  multiple proto-jets formation,  see Eq.\il(\ref{Eq:densi}),  implying  clearly a smaller critical density $\delta n_J$,  implying also a reduced  distance between two consecutive proto-jets $\delta r_J\propto \delta\ell_J$ (close together according to a fraction $\delta r_J/n_J\propto \delta\ell_J/n_J$), with a proportionality factor that can be easily assessed,  being  in general a function of  radius.
 Multiple co-rotating proto-jets shall be then very close to the horizon, approaching this with  increasing  dimensionless  spin of the attractor.

The $\ell$corotating  sequences  of corotating proto-jets  would be favored (and more spaced) in the case of low spin, due to the greater extension of orbital range $\Delta r_J$, and
 for the greater range  of the specific angular  momentum $\mathbf{L2}^-$. As we shall see in Sec.\il(\ref{Sec:ell-cont-double}), in these geometries mixed  $\ell$counterrotating sequences  of proto-jets  are  possible. Further analysis of the situation for the  fast Kerr attractors  suggests that   the  accretion  orbital regions, greater in measure  than the  proto-jet regions in general,  tend to have the same extension as the  proto-jets orbital regions.
 Then, for the  faster  Kerr attractors, we could say that  the probability that a slight change of specific angular  momentum  generates a transition  between the two critical ($\ell$corotating) topologies $(\oo_x, \cc_x)$   increases, inducing therefore  a  possible causal correlation between proto-jet and accretion that would be thus  characteristic of the attractors with large spin. For  fast  attractors,  the orbital distance $\Delta r_J$ is extremely small,  and $\Delta r_J$  would be  negligible (however, in those situations the role of increasing proper distance could be relevant). A special class of  ``fast''  attractors corresponds to the spins   $a>a_{\mathcal{M}}^-$.  In these geometries, where the dragging effects are significant, the critical density  $\delta n_J$, at the constant step $\kappa$,    decreases uniformly in the  outward direction, but the orbital range  is very narrow with  the proto-jet funnels  very close and  eventually indistinguishable.

 We also note that in such spacetimes, this region is entirely contained in the ergoregion  $\Sigma_{\epsilon}^+$;
 further details will be discussed
 in Sec.\il(\ref{Sec:ell-cont-double}) where we will compare the two $\ell$counterrotating sequences of  proto-jets considering the influence of the dragging effects.

This situation is totally reversed for the counterrotating case. Following arguments similar to  the corotating case, we would say in general that the multiple surfaces of counterrotating  proto-jets  appear in regions further away from the attractor with respect to the  sequence of corotating proto-jets.

For $a\in \mathbf{A}_{\iota}^>\equiv]a_{\iota},M]$, only isolated $\ell$counterrotating subsequences of proto-jets  may exist, and the maximal distance between the two subsequences, assessed as the distance between the  outer open surface $\oo_x^{\hat{o}_-}$  of the corotating inner sequence and the first  $\oo_x^{\hat{i}_+}$  of the outer  counterrotating one,  increases with the spin. Whereas in the geometries of    sufficiently slow attractors $\mathbf{A}^<_{\iota}$, \emph{mixed} sequences are possible  and they will be investigated  in Sec.\il(\ref{Sec:J-J-cont}) where  the $\ell$counterrotating couples are addressed.
The geometric separation has some effects on the geometrical and causal  correlations  between the two sequences. It should be noted that, if the  corotating proto-jets density increases for the slow attractors, just the opposite  holds for the counterrotating fluids. The counterrotating sequences are clearly favored at large distances from the black hole, increasing the  orbital range $\Delta r_J$ and the range of the   specific angular  momentum $\Delta \ell_J$. Then we should expect the multiple surfaces of  counterrotating  proto-jets to be  separated and spaced apart for  a broad  differential rotation with orbiting the faster attractors.
The possibility of a transition between $(\oo_x^+, \cc_x^+)$ topologies is reduced then for the counterrotating matter.

In the  sequences  of  counterrotating proto-jets, a minimum  of the criticality density   is  always present, farther   away from the source in comparison with  the corotating  fluids.  Where the differential rotation, as defined in \cite{ringed}, is approximately  equally spaced in the macro-configuration, the presence of a minimum density  would be evident in the multiple sequences of proto-jets, being  more significant with decreasing  step $\kappa$ of the sequence.
\subsubsection{\tb{Final notes  on the  $\ell$corotating proto-jets $(\mathbf{J}^{\mathbf{\pm}}-\mathbf{J^{\pm}})$}}
Considering the  $\ell$counterrotating subsequences made up by   $\ell$corotating proto-jets, it is necessary  to distinguish between two types of attractors:  one set including    slow rotating black holes, at $\mathbf{A}_{\iota}^<$, detailed  in Sec.\il(\ref{Sec:ell-cont-double}), where the two subsequences can be mixed, and the fast attractors, for  $a\in  \mathbf{A}_{\iota}^>$, where the  $\ell$counterrotating subsequences  \emph{must} be separated.
A possible geometric correlation can occur inside each subsequence, or also among the two subsequences. In this last case, we should  consider the confinement of the two separated subsequences  in the geometries  $\mathbf{A}_{\iota}^>$, and  particularly for  $\oo_x^{\hat{o^-}}\prec \oo_x^{\hat{i}}$.
The minimum separation between the two sequences becomes significant with the increase of the spin, until the minimum distance has  maximum value  $r_J^{\hat{o}_-}-r_J^{\hat{i}_+}\approx 3M$
for  very fast attractors with  $a\lesssim M$.
Eventually, the states
$\textbf{J-A}$ and  $\textbf{J-C}$, addressed in the  next sections,  could be seen as  precursors of the  $\ell$corotating $\textbf{J-J}$  decompositions, induced by a P-W instability for the initial closed  topology  evolving towards the open cusped one, or possibly  for collision and then geometrical correlation.
So far we have considered  in the  study of the multiple  open cusped  surfaces  the only criterion  of  the density of critical points $r_J$.
Yet another aspect to be considered, is the collimation of the $\oo_x$ funnels along the rotation axis, which could be related to the formation of a collimated proto-jet. The funnels of matter, in the case of very fast attractors
and corotating fluids, have an opening angle relative to the axis of rotation that is  smaller in comparison with those related to  the slower attractors, thus favoring a stronger collimation along the axis.  Then,  according to  the motion of the test particles,
a product $ \ell a> 0$ is associated to   stabilizing  effects for the rotating matter, being  `` attractive'' with respect to the proto-jets, because for increasing $a/M$, or the specific angular momentum $\ell>0$, the point $r^-_J$ moves inwards.
The inverse occurs  for the counterrotating case, $ \ell a <0$, which  would act  `` repulsively'', in the sense of
  favoring the instability of the orbiting matter, since for  increasing $a/M$, but  decreasing  magnitude of $\ell^+$, the point $r_J^+$ moves outwards.
\subsection{Outer proto-jet: the $\ell$corotating  proto-jet-accretion (\textbf{A-J}) systems}\label{Sec:A-J}
We consider a couple of $\ell$corotating critical  configurations,  formed by an open surface and an accreting configuration with $r_J<r_x$.  We will refer to the scheme $\oo_x-\cc_x$, 
 for a $\ell$corotating couple. We have
\bea&&\nonumber
\cc_x^i<\oo_x^o\quad\mbox{implies }\quad \cc_x^i\succ \oo_x^o\quad |\ell_i|<|\ell_o|\quad \ell_o\in\mathbf{L2},
\\\label{Eq:22}
&&|\ell_i|\in\mathbf{L1},\quad K_i\in \mathbf{K0},\quad K_o\in \mathbf{K1}.
\eea
We note that
the conditions
in   (\ref{Eq:22}) are always verified for the   $\ell$corotating  couples.
In terms of the   criticality indices,   we can express Eq.\il(\ref{Eq:22}) with
$\oo_x^{\hat{i}}\prec \cc_x^{\hat{o}}$
where $ \ell_{\hat{i}}\in \mathbf{L2}$  and  $\ell_{\hat{o}}\in \mathbf{L1}$.

In an  evolutive interpretation, the  $O^o_x$ configuration could model  the final stage of a  $\cc_x$ disk and, as pointed out in Sec.\il(\ref{Sec:J-Jc}), the evolution towards an $\oo_x$ surface from a $\cc^1_x$ one  requires    an  increase of the specific angular  momentum  magnitude  during the time. This could be the consequence, for example, of  feeding of the disk $\cc_x^i\in \mathbf{C}_x^{n}$ in a ringed  structure $ \mathbf{C}_x^{n}$, from a consecutive and outer ring  of the  decomposition  \citep{ringed}.

A different situation for the critical couple in Eq.\il(\ref{Eq:22}) could   occur when  the two P-W instability points  are   very close,  $r_{x}^{\hat{o}}\gtrapprox r_{J}^{\hat{i}}$, or  even   coincident. However, this  last case can be  possible   \emph{only} for an $\ell$counterrotating couple giving rise to a possible geometrical and causal proto-jet-accretion correlation.

As specified before,
in general a geometrical correlation in the $\mathbf{A^{\pm}-J^{\pm}}$ couple  could be favored   when the distance  between the specific angular  momenta ranges   is small, and  then a slight change of  $\ell$ in one of the configuration could lead to correlation or to  a topological  transition,  for loss of specific angular  momentum with the  formation of a  $\cc^1_x$ disk,
 or for  increase of $|\ell|$,  with the formation of an  $O^2_x$ surface.
However, it  should be noted that a  small  step $\kappa$, element of the displacement matrix $\epsilon_{oi}\gtrapprox0$, always corresponds  to a small difference
  $r_{min}^o-r_{min}^i=f_{min}(\epsilon_{oi})\gtrapprox0$ and, for $\ell\in \mathbf{L1}$ or $\mathbf{L2}$,  it corresponds also  to  the difference  $r_{Max}^o-r_{Max}^i=f_{Max}(\epsilon_{oi}) \lessapprox0$, where $(f_{min}, f_{Max})$
are  not independent functions of the step $\kappa$  \citep{ringed}. This is  not always true in the $\ell$counterrotating case.

More specifically,
according to  Eq.\il (\ref{Eq:ban-appl-cho}),  a correlation for {corotating} fluids, $\mathbf{A^{-}-J^{-}}$, is more likely  to occur in the geometries of  very fast attractors,  where   the ranges of possible specific angular  momentum,  $\ell_{Max}^s-\ell_{min}^s$   for $s\in\{J, K\}$ (correspondingly  the  differences $\ell_{Max}^J-\ell_{Max}^x-(\ell_{min}^J-\ell_{min}^x)$), are very small. Conversely, in the spacetimes of   slower attractors, this difference is   highest.
One could conclude  that the  topology of the corotating  orbiting matter is more stable for slow attractors, at least for sufficiently low specific angular  momentum (the surface topology remains unaffected by sufficiently small change in the specific angular  momentum), where $r^-_{cent}\gtrsim r^-_{mso}$.

If the  disk center moves outward, then it is  inevitable that also for  the slower attractors  the   $\cc_x^-$ disk expands in the equatorial plane (increasing elongation at almost constant difference $K_--K_{min}^-$), and it eventually acquires a  critical morphology,
  see also \ref{App:location}.

The situation  is just the opposite for the  counterrotating fluids: the configurations  orbiting   the slower attractors are  more likely to give rise to a proto-jet-accretion correlation, while in the geometries of  faster attractors,  a clear geometrical separation among the configurations can occur.
The range $\ell_{Max}^{x_+}-\ell_{min}^{x_+}$ remains approximately constant with increasing spin  of the attractor, even if $\ell_{min}^{x_+}$ increases.
On the other hand, the range of the specific angular  momentum  for a counterrotating proto-jet increases, and therefore, the  difference $\ell_{Max}^{J^+}-\ell_{Max}^{x_+}-(\ell_{min}^{J^+}-\ell_{min}^{x_+})$ increases, and analogously for the respective orbital ranges.
As a consequence of this, the orbital distance between the two surfaces   remains generally significant, even for very slow  attractors with  $a \approx 0$ (indeed the minimum range  for counterrotating fluids is approximately the maximum range for the corotating ones).

In conclusion, a correlation among the $\ell$corotating $\mathbf{A}^--\mathbf{J}^-$ configurations  is facilitated in the geometries of  faster attractors, while a correlation   in  the counterrotating $\mathbf{A}^+-\mathbf{J}^+$ configurations is in general less likely to occur  with increasing spin.
Therefore. the $\ell$corotating proto-jet-accretion correlation  should be   more evident for   corotating matter orbiting the  fast attractor, and  an increase of the black hole spin should favor a possible correlation.

To properly characterize a possible causal correlation, it is important to characterize  the distance between the two critical points of the couple.
The $\cc_x^1$ configuration  could be   close to the inner proto-jet point $r_J$,  with flooding of material towards   the inner critical point $r_J$, and   supply of  specific angular  momentum. These  phenomena would lead to an increasing  separation among the two surfaces $\oo_x$ and $\cc_x$.  In fact,  the proto-jet point $r_J$ shifts  inwards. Conversely,  the maximum pressure point $r_{min}^{x_1} $ of the accreting disk loosing its specific angular  momentum, will approach the radius $r_{mbo}$,  reducing its  elongation by decreasing  $K=K_{Max}^{1_x}$. However, the minimum point of hydrostatic pressure, $r_{Max}^{1_-}$, will move outwards.
This could eventually  lead to a   negative-feedback effect  which, after  certain time,  might even stop the  feeding (see also \ref{App:location}).
The final result of this very  simplified scenario would turn in a couple made by a  small  outer  disk, eventually in equilibrium, and an inner configuration in proto-jet with  increased specific angular  momentum.

As discussed  in Sec.\il(\ref{Sec:J-Jc}),
in principle it is possible that  an infinite number of $J...J$ couples could be formed.
On the other hand, a possible  $\ell$corotating $\mathbf{C}_x-\mathbf{C}_x$ couples  would violate the principle of non-penetration of matter (at  the first Roche lobes, as the  two $\ell$corotating configurations  would contain the same $r_{mso}$).
Therefore, in the multiple $\mathbf{A-J}$ couples, there would be a  $\ell$corotating subsequence  of a number $n_J$ of open funnels, as considered in Sec.\il(\ref{Sec:J-Jc}), and one (outer) accreting configuration.
The geometric correlation could be then  between the  $\cc_x^1$  configuration and $
\oo_x^{\hat{i}}\prec \cc_x^{\hat{o}}$.

Therefore, an  $\ell$corotating   proto-jet cannot  be formed  from  the  feeding of material from an outer ($\ell$corotating) surface, through  a P-W point    on a stable inner configuration or on an  accreting  inner one. An open cusped surface  may  be formed instead  from  a $\cc_2$ or $()_1$ configuration, after a (hypothetical) collision with an  outer equilibrium disk at higher specific angular  momentum,  which is  increasing its mass for example due to  interaction with the  embedding environment, or by some other collisional phenomena  occurring in the  macro-configuration.
These situations  are discussed in more details in \ref{App:location}, where  different situations for  corotating or counterrotating  couples orbiting attractors of different classes are explored. In  \ref{Sec:coorr}, the  location of $r_{out}$, outer edge of the disk, is also investigated; here we present  some general considerations based on the results proved  there.
 Firstly, as a disk in equilibrium can contain the marginally stable orbit (but not the marginally  bounded orbit as detailed in \ref{Sec:usua-D-mbo-con},  the inner margin of the  disk in   equilibrium can be close to the inner $\ell$corotating proto-jet point, $r_{in}=r_{mbo}+\epsilon$ with $\epsilon>0$. The inclusion\footnote{The \emph{inclusion} notation, $(\in, \non{\in})$ and $\in!$, will be  widely used in \ref{App:location}. The use of $\bar{r}\in()$,  for the radius $\bar{r}$  and any  surface $()$, means that  there  can be  found  proper $K$ or $\ell$ parameters  such  that this  property is satisfied. The symbol $\in\;!$ is  a reinforcement of this inclusion, indicating that this is  a necessary relation which is  \emph{always} satisfied.  The symbol  $\non{\in}$ (meaning non-inclusion)  does not generally have any  intensifier $(!)$, as this analysis is to underline the  possibility of inclusion and the condition for this to be satisfied.} $r_{mso}^{\pm}\in \cc_i^{\pm}$  implies  some restrictions on the specific angular  momentum of the disk, different for    the  corotating and  counterrotating  fluids, and  ultimately    distinction between different attractors--\ref{Sec:usua-D-mbo-con}. In general, for larger specific angular  momentum magnitude (such that $r_{in}=r_{mbo}+\epsilon$),  the disk would be significantly extended outward,  it would have a very large elongation\footnote{The  definition of elongation of a ring and ringed accretion disk was introduced in  \cite{ringed}, here we recall that in general for an accretion disk  the elongation range is defined as   $\Lambda\equiv[r_{in}, r_{out}]$ and the disk elongation on the equatorial plane as $\lambda=r_{out}-r_{in}$, where $r_{in}$ ($r_{out}$) are the inner (outer) edge of the disk. }  $\lambda$,  with
$r_{cent}-r_{in}\gg r_{out}-r_{in}$,  and the  surface, for increase in size  (also at almost  constant specific angular  momentum) might  become opened, being unstable but not reaching an $\oo_x$ topology.

We finally  note that an  immersion of  a $\cc_i$ disk  in a $\cc^i_x$ one, where  $\pm(\ell_i^{\mp}-\ell_1^{\mp})>0$,  could lead to an increase in size and  in the specific angular  momentum magnitude of the inner disk, and  eventually to  transition of   $\cc^1_x$ to a  $O^2_x$ topology. This can happen in fact  only  with   sufficient supply of specific angular  momentum  for the $\cc_x^{1}$ surface. However, the increase of specific angular  momentum magnitude  is a necessary but not sufficient condition for  the occurrence of a  topological transition. In fact, the disk may be stabilized by moving outwards the center of maximum  pressure  with a $K\in \mathbf{K0}$ (however for  $|\ell_{{o}}|$ very large, also $K_{{o}}$ can be very large, as there is  $\partial_{|\ell|}K_{crit}>0$ and,
for  $\ell_{{o}}$ very close to  $\ell_{mso}$,  the parameter  $K_{{o}}$  will be  close to the minimum $K_{mso}$).
 The specific angular  momentum  cannot be too large,  being  proportional to the distance among the centers or,    $
\partial_{\delta \ell}\delta r_{min}>0$. Thus, as  confirmed in \ref{App:location}, there will be an upper bound on the suitable range of specific angular  momentum magnitude.

\begin{figure}[h!]
\begin{center}
\begin{tabular}{c}
\includegraphics[width=1\columnwidth]{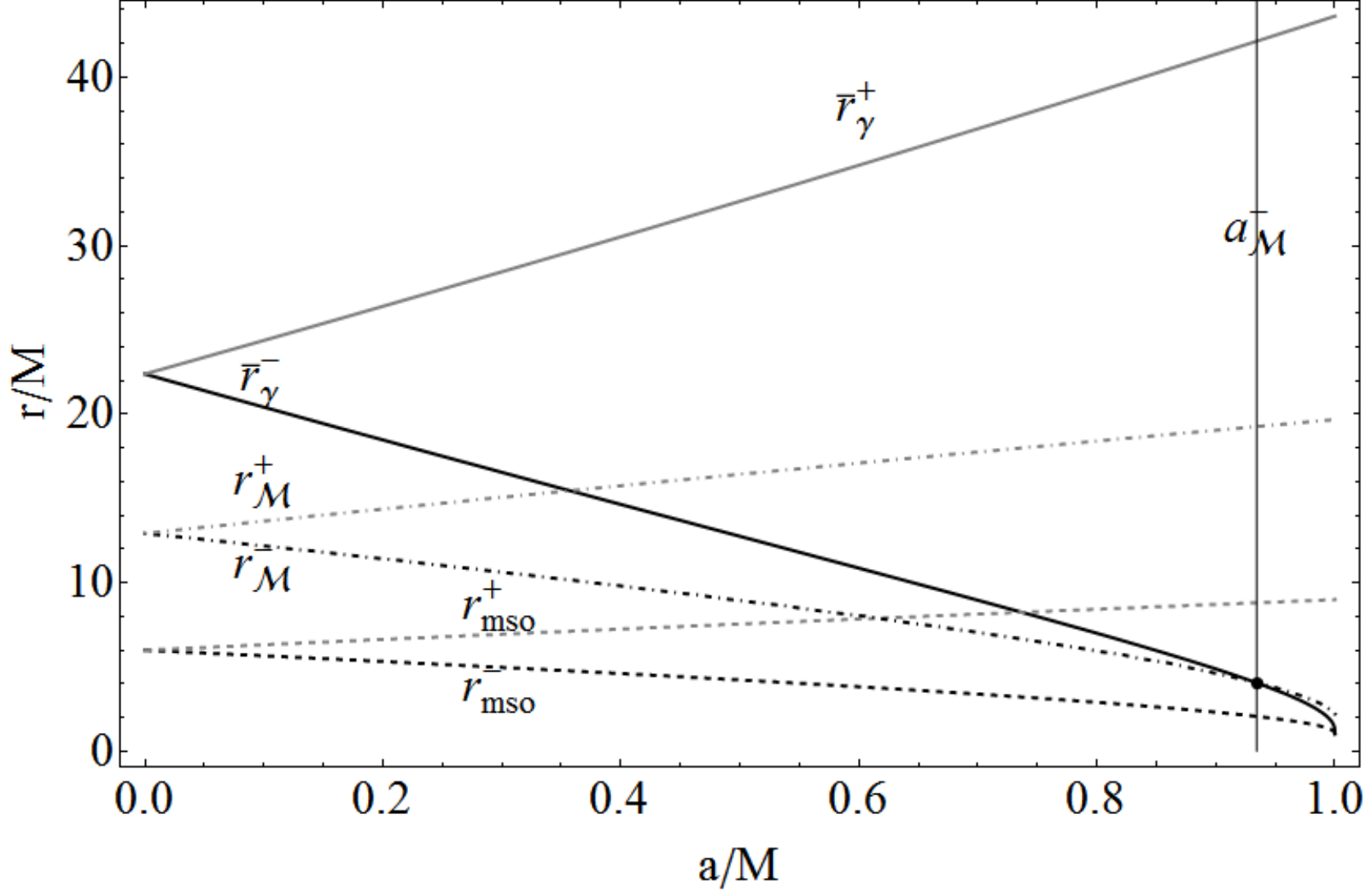}
\end{tabular}
\caption{The panel shows the marginally stable orbits $r_{mso}^{\pm}$, the radii $r_{\mathcal{M}}^{\pm}$ (orbits of maximum growing of the specific angular  momentum magnitude, as function of $a/M$) and the radii $\bar{r}_{\gamma}^{\pm}>r_{\gamma}^{\pm}:\; \ell_{\pm}(\bar{r}_{\gamma}^{\pm})=\ell_{\gamma}^{\pm}$, as functions of the black hole spin-mass ratio $a/M$. For $a=a_{\mathcal{M}}^-\approx 0.9343 M:\;\ell_{\gamma}^-=\ell_{\mathcal{M}}^-$ there is  $r_{\mathcal{M}}^-=\bar{r}_{\gamma}^-$.}\label{Fig:WayveShow}
\end{center}
\end{figure}

We conclude this analysis  describing the configurations in   the Newtonian limit.
The specific angular  momentum  of the equilibrium configuration could  be $\ell\in \mathbf{L3}$ for  a stable $\cc_3$ disk centered  in $r>\bar{r}_{\gamma}\gg r_{mso}$, with $\bar{r}_{\gamma}> r_{\gamma}:\; \ell(\bar{r}_{\gamma})=\ell_{\gamma}$,
 where  there is  also
$\partial_{\delta |\ell|}\delta K_{crit}>0$, and $
\partial_{\delta r_{critc}}\delta K_{crit}>0$.
However, according to the discussion in \cite{ringed}, even if the inequality  $\bar{r}_{\gamma}> r_{mso}$ is always verified in every geometry, for  the corotating  fluids orbiting fast  attractors, this  cannot be considered necessarily as an indication of the location of this radius  in a region where the Newtonian limit could be  applied.
This situation can be inferred from  Fig.\il\ref{Fig:WayveShow},  where the radii $\bar{r}_{\gamma}^{\pm}$,  have been plotted for different attractors, emphasizing their location with respect to the geodesic structure of the spacetime and  the radii $r_{\mathcal{M}}^{\pm}>r_{mso}^{\pm}$.
Fig.\il\ref{Fig:WayveShow} confirms  that  $\bar{r}_{\gamma}^{\pm}\gg r_{mso}^{\pm}$  for the countrorotating fluids in any geometry, and  for the corotating  case only  at $a\ll M$.
Then  it is clear that $\bar{r}_{\gamma}^+\gg r_{\mathcal{M}}^+$ for any attractor spin,  while  the situation is different in the corotating case.
Considering an attractor with   $a=a_{\mathcal{M}}^-:\; \ell_{\gamma}^-=\ell_{\mathcal{M}}^-$, we have $\bar{r}_{\gamma}^-=r_{\mathcal{M}}^-$, and $a_{\mathcal{M}}^-$  could be seen as the infimum  of the  spin range where it is not possible to  consider  the inequality  $\bar{r}_{\gamma}\gg r_{mso}^-$, and then the region where  a Newtonian limit could be considered.

The radii  $\bar{r}_{\gamma}^{\pm}$    (points of maximum hydrostatic pressure) are  the solutions of the algebraic fourth degree equation:
\bea\nonumber
&&\left.\partial_{r}V_{eff}\right|_{\ell_{\gamma}^{\pm}}=0,\quad
\bar{r_{\gamma}}^{\pm}:\;\left.a^2 (a-\ell)^2-4 (a-\ell)^2 r+\right.\\\label{Eq:compl-ell-ky}
&&\left.2 (a-2 \ell) (a-\ell) r^2-\ell^2 r^3+r^4\right|_{\ell_{\gamma}^{\pm}}=0,
\eea
{in dimensionless quantities.}
We note that actually the second equation, solved for each $r$ in terms of the parameters $\ell$ and $a$, is able to provide all the extremes of the effective potential and thus  also the minimum and maximum points of the hydrostatic pressure.
Specifically we can write the solutions  $\bar{r}_{\gamma}^{\pm}$ as  follows\footnote{Notice that this different behavior with respect to the spin suggests, as indeed  can be immediately verified, that the equation associated to  the problem (\ref{Eq:compl-ell-ky}) is not completely re-parameterizable as a function of the dimensionless quantities  $R\equiv r/a$ and  $\bar{\ell}\equiv\ell/a$, but the equations   explicitly depend on   $a/M$. The couples $(R,\bar{\ell})$ are used for example in  \cite{pugtot},  to identify the  Newtonian limit and underline the emergence of  the properly relativistic and dragging effects. }:
\bea\nonumber
&&
\bar{r}_{\gamma}^{\pm}\equiv\frac{1}{12} \left[\sqrt{3} h_3+3 \ell^2+\right.
\sqrt{6} \left(-2^{2/3} h_2-2^4 (a-2 \ell)
 (a-\ell)\right.
\\
&&\nonumber+3 \ell^4-\frac{ 2^{10/3} \left(2 a^2-3 a \ell+\ell^2\right)^2}{h_2}+
\\
&&\left.\left.\frac{3 \sqrt{3} \left[2^5 (a-\ell)^2-8 (a-2 \ell) (a-\ell) \ell^2+\ell^6\right]}{h_3}\right)^{1/2}\right]
\\\nonumber
&&
h_3\equiv\left( 2^{5/3} h_2-2^{4} (a-2 \ell) (a-\ell)+3 \ell^4+\right.
\\
&&\nonumber\left.\frac{ 2^{13/3} \left(2 a^2-3 a \ell+\ell^2\right)^2}{h_2}\right)^{1/2};
\\&&\nonumber
h_2\equiv\left[h_1+\sqrt{h_1^2-2^8 \left(2 a^2-3 a \ell+\ell^2\right)^6}\right]^{1/3};
\\\nonumber
&&h_1\equiv-(a-\ell)^2 \left[2^7 a^4-5.2^6 a^3 \ell+16 \ell^2 \left(\ell^2-3^2\right)+\right.
\\
&&\left.8 a \ell \left(3.2^4+13 \ell^2\right)-9 a^2 \left(48-8 \ell^2+3 \ell^4\right)\right].
\eea
\subsection{Inner proto-jet: the $\ell$corotating  proto-jet-accretion (\textbf{J-A}) systems}\label{Sec:innerJ-A}
The  $\ell$corotating couples  $\oo_x^i<\cc_x^o$,  
are  \emph{not}  possible.  In fact their existence  should imply that   $|\ell_i|<|\ell_o|$, where $\ell_o\in \mathbf{L2}$  and, on the other hand it has to be  $\ell_i \in\mathbf{L3}$  that is contradictory.

\medskip

We conclude   the discussion on the $\ell$corotating couples  with an inner open surface $J_x$, noting  that  in a  $\ell$corotating couple  with an accretion point and a proto-jet, the launching  point  $r_J$ must be the  \emph{inner} with respect to the accretion point $r_x$, implying that  it has to be a  \textbf{A-J} couple,  with a possible correlation between the two surfaces.

We note finally that, given a correlated  couple  $\oo_x$ or  $\cc_x$,  an  overflow of matter occurs, with the consequent decrease of specific angular  momentum magnitude  and the value of $K$ parameter.  This   may lead to a series of accretion stages with a progressive ``drying'' of the disk\footnote{ Suppose that a  $\cc_x$  disk is characterized by  the couple of parameters  $(\ell_a,K_{Max}^a)$   with $\ell_a\in \mathbf{L1}$
 and
 $K_{Max}^a<1$, and critical points  $(r_{min}^a,r_{Max}^a)$ of the hydrostatic pressure. Decreasing the specific angular  momentum in magnitude  and also $K_{Max}$, the disk might return to an equilibrium phase with topology $\cc$, here labeled by the superscript  $b$, with $r_{min}^b<r_{min}^a$,
$r_{Max}^b>r_{Max}^a$,   and also $K_b\leq K_{Max}^b<K_{Max}^a$ (it can be also $K_b=K_{Max}^b-\epsilon\lesssim K_{Max}^b$). This transition, from a cusped $\cc_x$ to a $C$ topology, should be a continuous process.
The initial topology  could also involve an   open surface $\oo_x$, resulting therefore in a  proto jet-accretion transition, as decreasing of specific angular momentum  magnitude may involve a transition from $\mathbf{L2}$ to $\mathbf{L1}$, or proto-jet-equilibrium configuration transition with angular momentum in $\mathbf{L2}$ or $\mathbf{L1}$.  A second new  phase of instability may be  induced by the  interaction  of the new surface with the surrounding matter, present in the macro-configuration as it was in its  initial phase, prior the transition, or by some other mechanism.}.
\subsection{Outer proto-jet: the $\ell$corotating equilibrium disk-proto-jet  (\textbf{C-J}) systems}\label{Sec:C-J-clna-v}
We consider  the $\ell$corotating  \textbf{C-J} couple,  with  only one P-W instability point $r_J$, associated to  the outer $\oo_x^o$ proto-jet. 
We obtain
\bea&&\nonumber
\cc_i<\oo_x^o\quad \ell_i<\ell_o\quad
r_{min}^i<r_{min}^o\quad K_i<1\quad \ell_o\in \mathbf{L2},
\\\label{Eq:fina-j-metri-one}&&
\mbox{thus}\quad \oo_x^o\prec \cc_i,\quad r_{J}^o= r_{in}^+-\epsilon_r
\\
&&\nonumber\mbox{where}\quad \epsilon_r>0,\;\ell_i\in \{\mathbf{L1},\mathbf{L2}\}.
\eea
In fact, there is  $ r_{Max}^o<r_{Max}^i<r_{mso}<r_{min}^i<r_{min}^o$.
This last relation  is a property of  any    $\ell$corotating subsequence, and it also assures  that, since $r_{Max}^o$ must necessarily exist being  associated to   $\oo_x^o$, and as $\ell_i<\ell_o$, for  definition, then  $\ell_i\in\{\mathbf{L1}, \mathbf{L2}\}$.
The inequality      $\oo_x^o\prec \cc_i$    of Eq.\il(\ref{Eq:fina-j-metri-one}), and then the location of the inner edge of the inner disk,  is guaranteed  by the fact that,  considering the  topologies  $(\oo_x^o, \cc_i)$,  we find    $ r_{Max}^o<r_{mbo}<r_{mso}<r_{min}^i<r_{min}^o$. However, as Eq.\il(\ref{Eq:beha-dire}) and Eq.\il(\ref{Eq:the-ag}) hold, then  the inner edge of the $\cc_i$ disk has to be external to the launching point of the proto-jet, $r_J$. It other words,  the disk in regular topology  is  \emph{entirely} contained in the orbital region  $r>r_J$, which finally validates Eq.\il(\ref{Eq:fina-j-metri-one}).

Under these circumstances, a viable correlation among these configurations can take place, for example, for possible impact  of the funnels\footnote{We stress  that this statement requires the study of proto-jet surfaces along the rotation axis. } $\oo_x^o$ on the internal surface $\cc_i$.
Or a geometrical correlation may occur by action of the  surface $\cc_i$  approaching  the critical point $r_J$, for loss of specific angular  momentum $\ell_i$ and  increasing  of the  parameter $K_i$. However,  as discussed in  \ref{App:location}, by the analysis of the possible inclusion relation $r_{mso}\in \cc_i$, the investigation of this case  may require  additional restrictions on the specific angular  momentum $\ell_i$. In fact,  for particular  specific angular  momenta  of the rings, and depending of the kind of attractor they are orbiting, some rings are \emph{entirely}  confined   in the region  $r>r_{mso}$, and then the two surfaces in Eq.\il(\ref{Eq:fina-j-metri-one}) are necessarily geometrically (and causally) separated. Particularly the situation is summarized in   Eq.\il(\ref{Eq:in-raff}) for the  $\cc_1^{\pm}$ disk,  in Eq.\il(\ref{Eq:ali-06},\ref{Eq:curv-n},\ref{Eq:font-cu})  for the corotating $\cc_2^{-}$ disks  and  in  Eq.\il(\ref{Eq:states-energies-Y}) for the  $\cc_2^+$ counterrotating disk. This study has being done in all details in   \ref{Sec:graph-def}, here we report the general results based on the parameter conditions for the confinement of the two surfaces in separated orbital regions and the main idea under  these results.

Then, for  $\ell_i\in \mathbf{L1}$  Eq.\il(\ref{Eq:in-raff})  holds,  thus  for sufficiently large (in magnitude)  $\ell_i\in \mathbf{L1}$  and sufficiently large density  $\mathbf{K0}$,  the  inner margin of the $\cc^{\pm}_i$ disks   can approach the orbits  $r_{mbo}^{\pm}$  (the gap between $r_{in}^i$ and $r_{mbo}$ can be in fact exactly evaluated).
Conversely, for $\ell_i\in \mathbf{L2}$, the situation is much more complicated as it  essentially depends on the range of specific angular  momentum  and, for the corotating disks $\cc_i^{2_-}$, also on the different classes of  attractors: we may be able, from the couple  (\ref{Eq:fina-j-metri-one}), to establish,  if  the ringed disk is  made up by counterrotating or corotating  fluids and, in this last case, also to identify the class of  the attractor.

The case   $\cc_i^{2_-}$ is   described by Eq.\il(\ref{Eq:ali-06},\ref{Eq:curv-n},\ref{Eq:font-cu}).
For attractors with sufficiently hight spin, as in  Eq.\il(\ref{Eq:font-cu}),  the situation is  similar to the  $\cc_i$  disks with $\ell_i\in\mathbf{L1}$,  and a geometric  correlation between the couple may  occur.
In the geometry of   the slower attractors, in order to have a correlation, it is necessary to balance the dragging effects due to the  slower spin of the attractor by  a sufficiently low  disk specific angular  momentum (in magnitude), namely in the range \textbf{II}  as specified by Eq.\il(\ref{Eq:curv-n}). Whereas, for  high enough specific angular  momenta, i.e.,  $\ell_i\in \mathbf{L2}$, the two  disks cannot be  geometrically correlated,  see Eq.\il(\ref{Eq:ali-06}).

For the counterrotating fluids of  the couple $\cc_i^{2_+}<\oo_x^{2_+}$,  orbiting  any Kerr black hole attractor,  Eq.\il(\ref{Eq:states-energies-Y}) holds. The situation is, in general,  analogous to the case of  $\cc_2^-$ disks but, in the counterrotating case, for specific angular  momentum  sufficiently high in magnitude, the   two surfaces are geometrically separated, while for the lower values of $|\ell_i^+|$,   there can be a  geometrical correlation in any Kerr geometry.
It has to be  specified that, for the $\cc_i$  disk, one has to  consider the role of the elongation   parameter $K_i>K_{min}^i\in \mathbf{K0}$, which indeed constitutes an additional free parameter of the system. In the unstable surface (in  open or closed topology), this is  uniquely determined by the specific angular  momentum of the fluid, thereby reducing, in  the cusped topologies, the number of free parameters to the only angular momentum $\ell$.

If the  two surfaces are geometrically separated, and therefore no correlation is possible,  then no feeding of  matter, or any matter penetration after collision, can occur from a $\cc_i$ disk to
an $\oo_x^o$ configuration.

As for the multiple  $\ell$corotating  configurations with topology $C$ and $\oo_x$, regulated by Eq.\il(\ref{Eq:fina-j-metri-one}), there can certainly be two $\ell$corotating subsequences   formed by surfaces with  equal topologies,  respectively, with  configuration of order   $n_J$  and   decreasing magnitude of specific angular  momentum with $r_J$,   and of order $n_c$ for the closed topologies  with the specific angular  momentum increasing with the orbit $r_{min}$. However, for the considerations outlined before,  the subsequences will be isolated and separated by the  orbit  $r_{mbo}$. Further considerations, of  the  corotating or counterotating nature   of the fluids and different classes of attractors are provided in \ref{App:location}.
\subsection{Inner proto-jet: the $\ell$corotating  proto-jet-equilibrium disk (\textbf{J-C}) systems}\label{Sec:J-Csmooth}
We close  this part on  $\ell$corotating  systems with an inner open cusped configuration, by considering the couple  $\oo_x^i<\cc_o$. But it is easy to see that
  existence of such a couple, similarly to the  $\oo_x^i<\cc_x^o$ one,  leads  to a contradiction, and we can finally conclude that  any $\ell$corotating couple \textbf{(J-C)}  must be  described by Eq.\il(\ref{Eq:fina-j-metri-one}).
\subsection{The $\ell$corotating accretion-equilibrium disk  \textbf{(A-C)} systems}\label{Sec:A-C-Asmooth}
We add here some notes on the $\ell$corotating couples made up by a  disk in  accretion and a configuration  in equilibrium. This case was discussed in detail in \cite{ringed}.
Let us suppose that the disks are  $\cc_a^{x}$ and $\cc_b$, then  it has to be
\bea&&\nonumber
K_a\in\mathbf{K0},\quad K_b\in \mathbf{K0}\cc_b\non{<}\cc_x^a \quad \ell_a \ell_b>0\quad
\pm\ell_b^{\mp}<\pm\ell_a^{\mp},
\\
&&\label{Sai-everx}
 \ell_a \in \mathbf{L1},\quad \ell_b\in \mathbf{L1},
\\\label{Sai-ever}
&& \cc_x^a<!\cc_b \; \ell_a \ell_b>0\;
\pm\ell_a^{\mp}<\pm\ell_b^{\mp},\; \ell_a \in \mathbf{L1},\; \ell_b\in \mathbf{Li}.
\eea
If the  equilibrium surface $\cc_b$   would be  the  inner one of the couple, in contradiction with Eq.\il(\ref{Sai-everx}),  it is immediate to prove that it would violate the non-penetration of matter, as shown below.
We note that
 $r_{mbo}<r_{Max}^{a}<r_{mso}^-$.
 But we know, from the assumption on the specific angular  momentum $\ell_{b/a}<1$,  that $r_{Max}^{a}<r_{mso}^-<r_{min}^b<r_{min}^a$,  implying that this configuration is not possible because there would be a penetration  of the first Roche lobe. Therefore   $\cc_x^a\non{>}\cc_b $
and $\cc_x^a<!\cc_b $, that is  reflected in Eq.\il(\ref{Sai-ever}).

We now focus on Eq.\il(\ref{Sai-ever}): the specific angular  momentum $\ell_b$ of the configuration in equilibrium can be, in principle, in any range $\mathbf{Li}$. However, in order to establish if   the condition of non-penetration of matter is really preserved,  it is necessary to establish the location of the outer edge of  $\cc_x^a$ and of the inner edge of $\cc_b$ disk. This analysis has been addressed in all detail in \ref{Sec:usua-D-mbo-con}. This  discussion  points out   significant distinctions  between the $\ell$corotating  couples made up respectively by corotating  and  counterrotating  fluids.

In general,  the two surfaces will be separated  for sufficiently large momentum $\ell_b$ as compared to  $\ell_a$ (in magnitude), and for  $K_b$ small enough at fixed $\ell_b$ (while  $K_a=K_{Max}^a$  is uniquely determined by the specific angular  momentum $\ell_a$).
The specific angular  momentum   $\ell_b$  should  be sufficiently  high  for $r_{mso}\non{\in} \cc_b$ at any $K_b$.
Obviously, this inclusion relation will also be determined by the   $K_b$  value, at least for some specific angular  momenta: the situation does   indeed  depend   of the  specific angular  momentum  range $\ell_b\in \mathbf{Li}$ and the class of the  attractor.  In some cases, it   will depend especially on the corotating or counterrotating  nature  of the disk.
This issue is detailed
in Eq.\il(\ref{Eq:in-raff}), for  specific angular  momentum $\ell_b^{\pm}\in \mathbf{L1}^{\pm}$, in
Eqs\il(\ref{Eq:ali-06}--\ref{Eq:font-cu}) for the corotating disk $\cc_b^-$ with  $\ell_b^-\in \mathbf{L2}^-$. Whereas for  counterrotating disks
Eq.\il(\ref{Eq:states-energies-Y}) holds.

 Finally, for counterrotating disks, with $\ell_b^+\in \mathbf{L3}^+$, Eq.\il(\ref{Eq:En-Dir}) holds, and then one can always find a $ K_b$
  such that the two configurations are separated.
  Viceversa,  for the  $\cc_3^{a_-}$  disks, Eq.\il(\ref{Eq:imply-Col}) and  Eq.\il(\ref{Eq:grounds}) apply,
distinguishing   fast  and slow  attractors.

We conclude this section by  noting that the couple in   (\ref{Sai-ever}) could be seen perhaps as a precursor of the $\ell$corotating  couple $\cc_x-\cc_x$, briefly discussed on the sidelines of Sec.\il(\ref{Sec:A-J}). But in fact, considering the   topology of the couple, and taking account of the requirement of non-penetration   of matter, a possible evolution of a $\cc_b$  disk towards the $\cc_x$ phase, which would  imply a variation of one or both $K_b$ and $\ell_b$ parameters, would be non correlated to the inner configuration.
Indeed, the inner $\cc_x^a$  disk would be correlated to the  onset of an   accretion phase of the $\cc_b$ configuration only by increasing  its own  specific angular  momentum  $\ell_a$   and therefore the
parameter  $K_{Max}^a$, which is not  expected.
A final note regards the decompositions  of order greater than two with a seed couple  $\cc^a_x<\cc_b$.
From the former analysis we conclude that the ``additional'' configurations  would be  in equilibrium  and therefore in the   outer  orbital regions with respect  to $\cc_x^b$.
\section{ $\ell$counterrotating sequences}\label{Sec:ell-cont-double}
The situation for  a   $\ell$counterrotating  couple, with a   critical configuration,  is  determined by the two families of the notable radii $r_{\mathcal{N}}^{\pm}$  of the     geodesic structure of the spacetime    and by the associated specific angular  momenta $\ell_{\mathcal{N}}^{\pm}$. The discussion of this case turns to be more articulated  than the $\ell$corotating case investigated  in Sec.\il(\ref{Sec:criticalII}).
 Some of the results considered here will be  discussed  more  deeply    in \ref{Sec:coorr},    where the  location of the notable radii $r_{\mathcal{N}}^{\pm}\in ()_{\mp}$ is considered with  respect to the configurations $()_{\mp}$.

Here  we will first consider  the  couples of  fixed topology and rotation  with  respect to the central attractor,  providing   rather stringent constraints on the decomposition. Then,  by considering the couple of configurations as a  seed for   a  decomposition of  order $n>2$, we investigate the configurations made up by more disks. 
To simplify the notation we introduce the total angular momentum $\mathcal{L}$ and total $\mathcal{K}$ parameter of the couple.
Examples of $\ell$counterrotating tori are in
Figures \il\ref{Fig:rssi-na}-\emph{Upper} and \emph{Bottom},
 also the outer couple of non accreting tori in \emph{Second} panel.
\subsection{The $\ell$counterrotating proto-jet-proto-jet \textbf{(J-J)} systems}\label{Sec:J-J-cont}
We consider  a couple of $\ell$counterrotating  opened-crossed  configurations.
It has to be
\bea&&\nonumber
\mathcal{L}=\mathbf{L2}^-\cup \mathbf{L2}^+,\quad \mathcal{K}=\mathbf{K1}^-\cup \mathbf{K1}^+ \mbox{for}\; a >a_{\iota}\; \oo_x^+\succ \oo_x^-
		\\\label{Eq:lagespicn}
		&&
		 (\mathbf{J}^+-\mathbf{J}^-),\quad a_{\iota}: \; r_{mbo}^-=r_{\gamma}^+,\;\mbox{for}\; a <a_{\iota}\; \oo_x^+\succ \oo_x^-
\\\label{Eq:rela-SusER-PR}
&&
 (\mathbf{J}^+-\mathbf{J}^-)\quad\mbox{\emph{or}}\quad \oo_x^+\prec \oo_x^-  \quad (\mathbf{J}^--\mathbf{J}^+),
\eea
see Fig.\il\ref{Fig:Plotaaleph1II}. For  the smaller attractor spin values, Eq.\il(\ref{Eq:rela-SusER-PR}) holds and, as we have $r_{mbo}^->r_{\gamma}^+$,   a  $(J-J)$  couple can be in the state
 $\oo_x^+\prec \oo_x^-$  or  also  $\oo_x^+\succ \oo_x^-$. But in the geometries determined by the larger spin values, where $r_{mbo}^-<r_{\gamma}^+$,   Eq.\il(\ref{Eq:lagespicn}) stands, and  the inner proto-jet \emph{must} be corotating i.e.  the only possible state  for this couple is  $\mathbf{J}^+-\mathbf{J}^-$.

In the following  discussion we will refer to the results  in Eq.\il(\ref{Eq:ban-appl-cho}) for the $\ell$corotating couples, which also emphasizes some properties of the  density $\delta n_J$ of   unstable points  $r_J^{\pm}$.

The separation between the $\ell$counterrotating  subsequences of  launching  points increases with $a/M$.
This could indicate that the black hole  spin  favors the launch of corotating material. For   large   vales of the  spin, the inner  proto-jets should be mainly regulated by the dragging effects  of the Kerr spacetime, and more generally by the curvature  effects, while the outer and counterrotating sequence shall be mainly regulated   by the centrifugal effects (where $\ell\in \mathbf{L2}^+$), and especially by the  PW instability (in $r_J^+$) due to the high  values of the  $K$  parameter (indeed $K=K_{Max}^2$ is fixed by the specific angular  momentum $\ell_2$), regulating both the elongation on the equatorial plane and  the disk density\footnote{Then one should certainly  consider, especially for these sequences, the role played in the unstable states  by other factors typically characterizing the evolution of the accretion disks, such as electromagnetic effects, which were not included in the  model adopted here for each ring.}.

In a possible  evolutionary scheme, where the attractor is not meant to be isolated but interacting with the surrounding material, a possible increase of its  dimensionless spin $a/M$ should have a stabilizing effect for  the corotating material, eventually ``separating''  the two $\ell$corotating  sequences, as the counterrotating one could fill, according to the discussion in \cite{pugtot,ringed}, the regions  far away  from the source.

As mentioned also in Sec.\il(\ref{Sec:J-Jc}), matter in the critical $\oo_x^-$ topology  penetrate the ergoregion $\Sigma_{\epsilon}^+$ in the equatorial plane for sufficiently fast Kerr attractors.  The funnels of material will eventually cross  the static limit with an initial velocity  $\dot{\phi}>0$,  following a possible energy extraction process. The static limit $r_{\epsilon}^+$ is, on $\theta=\pi/2$, independent of  $a$-but not of $M$.
More precisely, we have
 \bea\label{Eq:symb-comple-W}
 && r_{J}^-\in \Sigma_{\epsilon}^+ \quad\mbox{for }\quad a\in]a_1, a_b^-[,\quad \mbox{and}\quad  r_{J}^-\in ! \Sigma_{\epsilon}^+
 \\ \nonumber
&&\nonumber\mbox{for }\quad a\in[a_b^-, M],
\\
&&\nonumber
\mbox{
 where}\quad {a}_1/M\equiv1/\sqrt{2}\approx0.707107;,\quad
 {a}_b^-/M\equiv 2 (\sqrt{2}-1)\approx0.828427 ;
 \\
 &&\nonumber
a_1:\;r_{\gamma}^-=r_{\epsilon}^+;
\quad a_b^-:\;r_{mbo}^-=r_{\epsilon}^+
\eea
see Fig.\il\ref{Fig:Plotaaleph1IIa} and \cite{ergon}. At $a>a_{2}$, the maximum of the  hydrostatic pressure will be in $\Sigma_{\epsilon}^+$ where $a_2/M\equiv {2 \sqrt{2}}/{3}\approx0.942809$ and  $ a_2:\; r_{mso}^-(a_2)=r_{\epsilon}^+$.

This fact
confirms  that the dragging effects are dominant in the corotating case,  while  the centrifugal component of the effective potential would become predominant in the counterrotating fluids with respect to the corotating case.

Considering   Eq.\il(\ref{Eq:ban-appl-cho}), we can draw  first conclusion by saying that the unstable points $r_{J}^{\pm}$ in $ \mathbf{L2}^{\pm} $   cannot  be geometrically correlated  in the geometries of fast attractors  ($\mathbf{A}_{\iota}^>:\; a>a_{\iota}$), but a geometric correlation may occur in the geometries of the slower attractors  ($\mathbf{A}_{\iota}^<:\; a<a_{\iota}$), where the launching points may also be  coincident ($r_{J}^+\approx r_J^-$).
In fact
the orbital region $\Delta r_J^{\pm}\equiv\Delta r_J^-\cap\Delta r_J^+$ is rather narrow, $\Delta r_J^{\pm}\lesssim M$,  and  such that  $\partial_a(\Delta r_J^{\pm})<0$.

The considerations outlined here relate primarily to the   orbital ranges eligible for sequences of cusped open  $\ell$counterrotating topologies.
It is worth to note that, as the $\oo_x^{\pm}$  configurations are not closed, the characterization of such multiple decomposition is indeed  a one-dimensional problem, reduced to
 the location of the $r_J^{\pm}$ points  in  a  bounded but continuous range of variation of $\Delta r_J^{\pm}$. In other words, this is not an  extended matter problem  and, as also discussed in Sec.\il(\ref{Sec:J-Jc})  for the  $\ell$corotating proto-jets,  for every  $\ell$corotating subsequences,  the order  $n_{j}^{\pm}$   can be infinite in principle.

Obviously, considering the  geodesic structure of the spacetime the possible multiple  $\ell$counterrotating  decomposition of    proto-jets, orbiting fast attractors with $a>a_{\iota}$, are necessarily isolated,  as it is clear from   Eq.\il(\ref{Eq:symb-comple-W}). Then  there is an inner  corotating  sequence of $(J...J)^-$  configurations, and an outer one  made up by  the counterrotating proto-jets  $(J...J)^+$.
The  $\ell$counterrotating  sequences   may be characterized by a more or less wide spacing,  and the counterrotating  sequence    will be more or less  extended, considering the distance   $[r_{J}^{\hat{n}_+},r_{J}^{\hat{1}_+}]\in\Delta r_{J}^+$, with respect to the corotating sequence which is  confined in an orbital region  increasingly smaller as the attractor dimensionless spin increases--Fig.\il\ref{Fig:Plotaaleph1IIa}.
The minimum separation between the two subsequences is  $\Delta r_{J}{}_+^-\equiv r_{\gamma}^+-r_{mbo}^-$, which states evidently   maximum in the case of extreme geometry.
A possible geometric correlation between two $\ell$counterrotating proto-jets will depend  on the distance $\Delta r_{J}{}_+^-$.

Mixed configurations, on the other hand, are possible for  sufficiently slow  attractors i.e.  $a<a_{\iota}$; this is  quite small  class of attractors, compared to the  class of black hole sources at  higher spin.
Small values of $a/M$ favor  mixed configurations, as the orbital region  $\Delta r_{J}{}_-^+\equiv-\Delta r_{J}{}_+^-= r_{mbo}^--r_{\gamma}^+$ (and the corresponding  specific angular  momentum range) is  minimum in the geometry $a_{\iota}$, where the mixed sequences are possible,   and  maximum  in the static Schwarzschild geometry. We can write, with regards with the   orbital location:
\be
  (J...J)^-<(J...J)_-^+<(J...J)^+.
 \ee
The  outer, $\ell$corotating sequence  $(J...J)^+$  of counterrotating proto-jets  is bounded in the orbital region  $\Delta r_{J}^+- \Delta r_{J}{}_-^+$, with  maximum measure in the geometry $a=a_{\iota}$  and  null in  $a=0$. This  is  followed by   the inner region of mixed sequences  $(J...J)_-^+$ in  $\Delta r_{J}{}_-^+$, and the finally the  $\ell$corotating  inner  isolated sequence  $(J...J)^-$.

We now address the problem of a possible initial state for the  couples in the state  (\ref{Eq:lagespicn}) or
 (\ref{Eq:rela-SusER-PR}).  A more detailed discussion   can be found in \ref{App:location}.
 For each $\ell$corotating subsequence, the considerations outlined   in Sec.\il(\ref{Sec:J-Jc}) apply.
The evolution from a closed  crossed  topology $\cc_x$ of an  accretion configuration   to the topology $\oo_x$ of the open   funnels, requires an increase (in magnitude) of  the specific angular  momentum to ensure the transition  $\mathbf{L1}$ to $\mathbf{L2}$  and an increase of   the  $K$-parameter  with a shift $\mathbf{K0}$ to $\mathbf{K1}$. Whereas, starting from a closed topology, an  analogue   transition  occurs  if  the initial data on the specific angular  momentum are  in $\mathbf{L1}$ or   $\mathbf{L2}$.  Conversely, if the starting data for the closed configurations are in
 $\mathbf{L3}$, then  the   $K$ parameter  increases from $\mathbf{K0}$ to $\mathbf{K1}$, but the specific angular  momentum $\ell$ decreases with a  transition from $\mathbf{L3}$ to $\mathbf{L2}$.
 However, as any equilibrium disk $\cc_{\pm}$ can never contain the marginally bound orbit $r^{\pm}_{mbo}$, as stated in  Eqs\il(\ref{Eq:choral},\ref{Eq:beha-dire},\ref{Eq:the-ag}); this transition implies  a stretching of the disk towards the attractor, moving inward the point  of  maximum of the hydrostatic pressure, if the disk is located in the outer region with larger centrifugal barrier ($\ell\in\mathbf{L3}$), or viceversa,  the shifting  of the disk center outwards, from the  initial data in
  $\mathbf{L1}$.  The  initial configuration   with specific angular  momentum in  $\mathbf{L2}$  will not   necessarily give rise to   a shift of the center of maximum pressure, but it will change morphology and topology  as consequence of the transition $\mathbf{K0}$ to $\mathbf{K1}$.

These considerations   derive mainly from the analysis of the geodesic structure of spacetime. Correspondingly one can observe  arrangement of the eligible  specific angular  momenta  for the $\ell$counterrotating couple of proto-jets in $\mathbf{L2}^{\pm}$, as  summarized in Eq.\il(\ref{Eq:ban-appl-cho}). The two  $\ell$counterrotating  subsequences of  proto-jets  have been discussed extensively in Sec.\il(\ref{Sec:J-Jc}).

We  mention that  associated to the orbital region $\Delta r_{J}{}_-^+$, where the mixed decompositions are possible, is also the  region of common specific angular  momentum
 $\Delta \ell_{j}{}_-^+\equiv -\mathbf{L3}^+\cap \mathbf{L3}^-$. This has the remarkable implication that in the mixed decomposition,   it will tend to have a ratio  $\ell_{+/-}\sim-1$, for sufficiently close points $r_J^{\pm}$,   up to the extreme situation where the attractor has spherical symmetry. This fact is particularly relevant when one considers that  in this  orbital region  a geometric correlation between  the elements of the  mixed sequence   is possible. One can also  see this situation directly  through the analysis of the curves  of specific angular momentum $\mp\ell^{\pm}$ which should approach  in the spacetimes   where $a<a_{\iota}$. This means that, when  correlated, the $\ell$counterrotating  open cusped surfaces  have  specific angular  momentum approximately close in magnitude.

We remind also  that there are no solution $\bar{r}:\, \ell_{+/-}=-1$, but certainly there is an appropriate couple $\bar{r}_{\pm}:\,\ell^+(\bar{r}_+)=-\ell^-(\bar{r}_-)$ where, for  $a<a_{\iota}$,  we have $\bar{r}_-<\bar{r}_+<r_{mso}^-<r_{mso}^+$ or  $r_{mso}^-<r_{mso}^+<\bar{r}_+<\bar{r}_-$.
 Finally, we refer to the discussion of  Sec.\il(\ref{Sec:J-Jc}),
  for the characterization of these structures  in  the Newtonian limit, as  defined through  the couples $r_{\mathcal{M}}^{\pm}$ and  the  corresponding instability points  $\bar{r}_{\mathcal{M}}^{\pm}$.
\subsection{The $\ell$counterrotating  proto-jet-accretion \textbf{(J-A)} systems}\label{Sec:L-A-system}
 %
The $\ell$counterrotating couples of cusped   topologies, made up  by   a proto-jet and  a accreting closed disk, can be analyzed by considering the  following four states:

%
\subsubsection{
\textbf{State I:} $\cc_x^+\succ \oo_x^-$}

First, we focus on  the couple  $\cc_x^+\succ \oo_x^-$: a counterrotating  closed configuration   with an outer accretion point  and an inner instability point $r_J^-$, with  open  configurations corotating with the black hole.
It has to be:
\bea\nonumber&&
 \mathcal{L}= \mathbf{L1}^+\cup \mathbf{L2}^- \quad \Delta r_J^-\cap\Delta r_x^+=\emptyset, \quad\mbox{then } \quad \cc_x^+\succ \oo_x^- 	 \\\label{Eq:la-II-WW}	
 &&\mbox{ and } \quad  \cc_x^+\nprec \oo_x^-  \quad\mbox{as there is} \quad \Delta r_J^-<\Delta r_x^+.
\eea
We shall focus on a possible geometrical  correlation  between the two configurations for action of  the outer counterrotating disk in accretion  to the inner    corotating open proto-jet configuration.

We note that the critical points are located in the regions  $\Delta r_x$ and  $\Delta r_J$: in the first case, the P-W  accretion point $r_x^+$  corresponds to
  the inner margin of the outer Roche lobe. The minimum distance between  the points $r_x^+>r_J^-$ is $\inf(r_{x}^+-r_{J}^-)=r_{mbo}^+-r_{mbo}^->0$ for $a>0$. Even if  the critical points $(r_x^+, r_J^-)$ are geometrically separated, when the outer counterrotating  configuration  reaches the critical topology, the matter falls towards the attractor
and, as the state is
$\oo_x^-\prec \cc_x^+$, this  leads to a possible interesting scenario   with  the    counterrotating  matter accreting with super-Eddington luminosity
on the  $\oo_x^-$ corotating  configuration.

However, associated to the $\oo_x^-$ configuration,   in the region  $r>r_{\epsilon}^+$, there  will be an inner surface (second and inner Roche lobe) embracing  the black hole,  which therefore  could match the second lobe of the outer $\cc_x^+$ one, with the consequent collision     of counterrotating material with  specific angular  momentum, generally, greater in magnitude
 than the specific angular  momentum $\ell^-$.

The  macro-configurations of order $n>2$, with seed couple as in Eq.\il(\ref{Eq:la-II-WW}), will be  made by the isolated sequences of  couples in Eq.\il(\ref{Eq:la-II-WW})  with $n_x^+=1$ and the criticality order $n^-_J$ up to infinity.
For the discussion on the inner  $(J...J)^-$ sequences, we refer to Sec.\il(\ref{Sec:J-Jc}).

Clearly, both the parameters $K_{Max}^{\pm}$ are uniquely fixed by the momenta $\ell_{\pm}$ respectively (we recall then that the lines of constant  $K_{crit}^{\pm}$,  provide exactly the inner and outer edges of the closed critical configurations).

Considering  the radial distances $\Delta r_J^-$,
 and  $\Delta \ell_J^-$,
  we could conclude that the  multiple corotating proto-jets are  favored
  at lower spin, requiring however in general also a  larger specific angular  momentum, see Eq.\il(\ref{Eq:ban-appl-cho}).  Therefore, a larger centrifugal component  gives rise to   launches  of open funnels (in  \cite{pugtot} it has been proposed   in terms of the rationalized specific angular  momentum  $\ell/a$, which emphasizes  the relevance of the ratio between the ``orbital'' specific angular  momentum and the attractor spin in regulating the disk morphology and evolution).

At high values of  spin, the multiple corotating proto-jets are disadvantaged,
   while they are  possible with  lower specific angular  momentum, as discussed in Sec.\il(\ref{Sec:J-Jc}).
For the counterrotating matter in closed (cusped) topology, we find     $\inf{(r_{x}^+- r_J^-)}
  \in[M , 6 M[$ as the black hole spin  $a\in [0,M]$. It is then   $\partial_{-\ell_2^+} K_{Max}^{+}>0$ for $\ell_2^+\in \mathbf{L2}^+$, see Eq.\il(\ref{Eq:ban-appl-cho}).

In this respect, the more relevant effects  for the interaction  between the  $\cc_x^+\succ \oo_x^-$ systems  could occur  for lower vales of spin,  where an  increase of the criticality density $\delta n_J$ at constant step $\kappa$ occurs, and the orbital range $\Delta r_J^-$ is larger.
Concerning the specific angular  momentum
   at $a>a_{\aleph}$, there is   $-\mathbf{L1}^+\cap \mathbf{L2}^-=\emptyset$ and  $-\mathbf{L1}^+> \mathbf{L2}^-$, with  orbital range $\delta r_x^+>\delta r_J^-$, which increases  with increasing attractor spin. Moreover, according to Eq.\il(\ref{Eq:symb-comple-W}), there is  $r_J^-\in\Sigma_{\epsilon}^+$, being inaccessible to any possible  contact with the counterrotating matter.

For attractors with spin $a<a_{\aleph}$, the situation is less articulated
	 and the separation between the orbital regions decreases, being    never zero,  and  the specific angular  momentum    $-\mathbf{L1}^+$ decreases while    $\mathbf{L2}^-$ increases.
As discussed above, in these spacetimes  possible  interaction between the two configurations could be  more relevant.

Then one has   $ -\mathbf{L1}^+\cap \mathbf{L2}^-\neq \emptyset$ in $]a_{\aleph_0}, a_{\aleph}[$, and $a_{\aleph_0}:\; \ell_{\gamma}^-=-\ell_{mbo}^+$,
 where    $-\ell_1^+ <\ell_2^-$  is possible; in this case, the critical points $r_x^+$ and $r_J^-$ are closer, favoring therefore a correlation.
 Finally, at $a\in]a_{\aleph_2}, a_{\aleph_0}[$,  there is  $-\mathbf{L1}^+\subset \mathbf{L2}^-$, while for $a <a_{\aleph_2}$, very close to  $a=0$, there is a range of counterrotating specific angular  momentum  lower in magnitude then the corotating $\ell^-$ one--see Fig.\il\ref{Fig:Plotaaleph1II} and Fig.\il(\ref{Fig:Plotaaleph1IIa}).

\subsubsection{
\textbf{State II:} $\cc_x^+\prec \oo_x^-$} This case is ruled out by Eq.\il(\ref{Eq:la-II-WW}) and this  couple is not allowed. Therefore,  corotating open funnels of matter and  counterrotating accreting configurations must be regulated according to Eq.\il(\ref{Eq:la-II-WW}).

\subsubsection{\textbf{State III:} $\cc_x^-\prec \oo_x^+$}

 We consider the possible presence of  counterrotating funnels of matter, launched from a point $r_J^+$, with  an ``inner'' closed configuration in accretion $\cc_x^-$, or
 \bea&&\nonumber
 \cc_x^-\prec \oo_x^+\quad r_{x}^-<r_J^+,\quad \ell^-\in \mathbf{L1}_-\quad \ell^+\in \mathbf{L2}_+,\\\label{Eq:calc-parti}
 &&
 r_{mbo}^-<r_x^-<r_{mso}^-<r_{out}^-<r_J^+<r_{mbo}^+<r_{mso}^+.
 \eea
The conditions  (\ref{Eq:calc-parti}) imply a condition on the location of the outer edge $r_{out}^-$ of the accreting $\cc_x^-$ disk, which should satisfy the last sequence of inequalities  in (\ref{Eq:calc-parti}).
 On the basis of these considerations only, we could say that this  couple would be possible in the geometries of the fastest attractors, however at higher spins this is the \emph{only} state possible for  the elements of the couples $(\cc_x^-, \oo_x^+)$. Therefore we could say
 \be\label{Eq:Eigen-VV}
 \mathbf{A}_{\iota_a}^>:\quad a>a_{\iota_a}\quad \cc_x^-\prec! \oo_x^+;\quad \mathbf{A}_{\iota_a}^<:\quad a<a_{\iota_a}\quad \cc_x^-\non{\prec} \oo_x^+,
 \ee
 where the spin $a_{\iota_a}$ is introduced in Eq.\il(\ref{Eq:disti-g}).
One could consider this state  as the  opposite with respect to the state $\cc_x^+\succ \oo_x^-$, analyzed in point \textbf{I} above.

As made explicit in Eq.\il(\ref{Eq:Eigen-VV}), the situation depends on the attractor characteristics and  one could say that the couple $\cc_x^-\prec \oo_x^+$  is a characteristic of the geometry of the  fast attractors.
At lower spins the problem of the location of the  outer margin of the corotating disk in accretion
 with respect to the launching point becomes relevant, and   an overlapping of material with the outer configuration is possible.
However, considering the last inequality of Eq.\il(\ref{Eq:calc-parti}), we recall that since the inner configuration
 is closed, it makes sense  to enquire if the \emph{outer} edge  of the disk in accretion, that is  $r_{out}^{-}$, can be near  the outer critical point $r_J^+$, where this kind of situation would lead to collisions.   This issue is partially discussed  in \ref{Sec:coorr}.
In any case, a  possible correlation
  could emerge   from an  interaction between
 the outer edge  $r_{out}^{1_-}$ and the outer critical point $ r_{J}^+$.

In order to provide some constraints on the parameters of $\cc_x^{1_-}$, we should investigate   the inclusion relation $r_{mbo}^+\in \cc_x^{1_-} $. This problem is fully addressed  in the last point of \ref{Sec:scale-ques}, fully responding  to this problem by providing the relations in  Eq.\il(\ref{Eq:A-nimos}) and   following discussion.
Since  the last circular orbit is the  lower boundary of the orbital range $\Delta r_J^+$, we should consider  the inclusion   $r_{\gamma}^+\in ()^-$, as discussed in \ref{Sec:photon-C}.

We  close this section  with    some general considerations.
The configurations $(\cc_x^-,\oo_x^+)$, regulated by Eq.\il(\ref{Eq:calc-parti}), can evolve  independently  without geometrical contact. On the other  hand, from the analysis  of the total  specific angular  momentum  and the orbital ranges,  we can infer a classification in different sets of attractors where it could be more likely to observe such critical configurations.
The first set
 is defined for geometries $a >a_{\gamma_+}^-$, introduced and characterized  in Eq.\il(\ref{Eq:set-identity}),
where at $a =a_{\gamma_+}^- >a_{\iota_a}$ we have $ r_{mso}^-=r_{\gamma}^+$, and the critical points $(r_x^-, r_J^+)$
 are  geometrically separated by the distance $r_{\gamma}^+-r_{mso}^-$ increasing with the spin, see Fig.\il\ref{Fig:Plotaaleph1IIa}. This is a subset of the class of attractors $\mathbf{A}_{\iota_a}^>$.

Increasing
 the  dimensionless spin, the geometric separation  increases, and the minimum distance is at most $ \approx  4 M$ for
$ a\approx M$.

We can then read  the situation  in terms of the specific angular  momentum. The analysis of the specific angular  momentum range $\mathbf{L2}^+\cup \mathbf{L1}^-$ shows  an interesting situation;  there is
$-\mathbf{L2}^+\cap \mathbf{L1}^-=\emptyset$ and   $-\mathbf{L2}^+>\mathbf{L1}^-$, $\mathbf{L1}^-\prec-\mathbf{L2}^+$, and $-\mathbf{L2}^+$ increases with the spin increasing, while $\mathbf{L1}^-$ decreases with increasing $a/M$--
 see  Eq.\il(\ref{Eq:symb-comple-W}).

As a consequence of this fact, if there is a  collision   between these two systems,
then the specific angular  momentum of the open configuration  in  funnels is always larger in
 magnitude then the counterrotating one in accretion.
From these results one can easy draw some general conclusions on the  multiple surfaces generated from the seed $\cc_x^-\prec \oo_x^+$ in Eq.\il(\ref{Eq:calc-parti}), considering that one might have a $\ell$corotating  sequence  of $(\mathbf{J}^+...\mathbf{J}^+)$ proto-jets. 

\subsubsection{\textbf{State IV:} $\cc_x^-\succ \oo_x^+$}

The couple
$\cc_x^-\succ \oo_x^+$, formed by an inner proto-jet of counterotating matter  and  an outer accretion  corotating ring, can exist only in the geometries of  the slow attractors, i.e., for  $a\lesssim0.6 M$   where $r_{mso}^->r_{mbo}^+$. In fact,  as stated in Eq.\il(\ref{Eq:Eigen-VV}), the couples  $\cc_x^-\succ \oo_x^+$ are forbidden  at higher spins:
\be\label{Eq:Q-E-D}
\mathbf{A}_{\iota_a}^>:\quad a>a_{\iota_a}\quad \cc_x^-\non{\succ} \oo_x^+;\quad \mathbf{A}_{\iota_a}^<:\quad a<a_{\iota_a}\quad \cc_x^-{\succ}! \oo_x^+,
\ee
see also Fig.\il\ref{Fig:Plotaaleph1IIa}. The results of this section are expanded and deepened in the last point of  \ref{Sec:scale-ques} and in \ref{Sec:photon-C}, where the inclusion relations $r_{mbo}^+\in \cc_x^-$  and $r_{\gamma}^+\in \cc_x^-$ will be analysed.   At  $a<a_{\iota_a}$, it is always possible to find
$\ell_1^-$  small enough  for  $r_{min}^-\approx
r_{mso}^-$, and then  $K_{Max}^{1_-}<1$ small enough to get a small critical elongation $\lambda_x^{1_-}$, having
 $
 r_{mso}^->r_{mbo}^+$.
 We have then $r_x^{1_-}\in!]r_{mbo}^+,r_{mso}^-
[$, while   $r_J^+ \in ] r_{\gamma}^+,r_{mbo}^+[$ that is  Eq.\il(\ref{Eq:Q-E-D}). For the exact conditions we refer to the study of the inclusion relations of   Eq.\il(\ref{Eq:A-nimos}) and Fig.\il\ref{Fig:Pgreat0}.

The geometric separation between these configurations increases with increasing  spin  of the attractor. Then an  interaction is possible
due to a geometric correlation and  favored  for low spins and large (great elongation and great density) corotating disks $\cc_x^{1_-}$, whereas     smaller disks $\cc_x^{1_-}$  are  favored  at higher spin $a\lessapprox a_{\iota_a}$. Indeed, in this last case the outer corotating matter accreting onto the black hole could impact on the inner counterrotating matter, which  is unstable according to the P-W model.

Considering  that  the specific angular  momentum of the couple is    $\mathcal{L}=\mathbf{L1}^-\cup \mathbf{L2}^+$,  it is  clear  that the specific angular  momentum of the counterrotating matter will be  in magnitude greater then the specific angular  momentum of the  inner corotating one.
Therefore, the location of the couples $(\cc_x^-, \oo_x^+)$ clearly distinguishes the two classes of attractors, according to Eq.\il(\ref{Eq:Eigen-VV}) and Eq.\il(\ref{Eq:Q-E-D}).

Considering the decomposition of  order  $n>2$, generated by the  seed couple
 $\cc_x^-\succ \oo_x^+$ in Eq.\il(\ref{Eq:calc-parti}), one could consider couples at fixed topologies with multiple copies of $\cc_x^-\succ \oo_x^+$ and  an outer sequence of $(\mathbf{J}_+...\mathbf{J}_+)$. The orbital region of the open sequence is  therefore  more reduced while approaching the limit $a=0$.

\begin{figure}[h!]
\begin{tabular}{c}
\includegraphics[width=1\columnwidth]{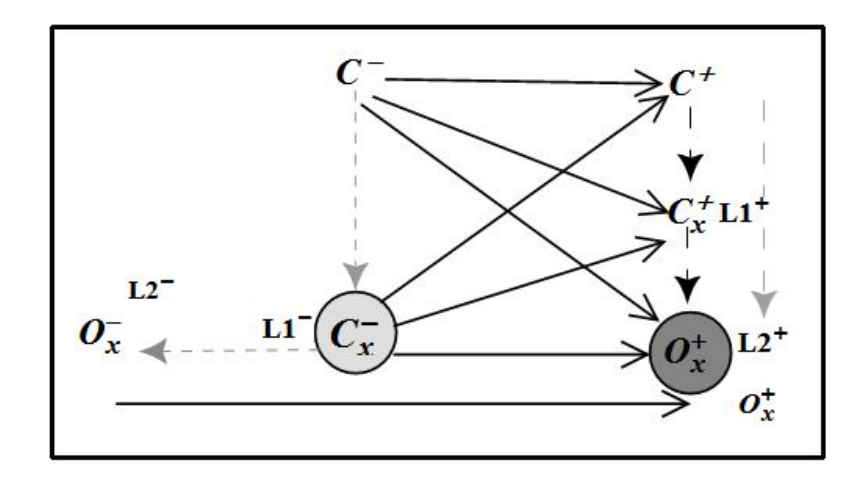}
\end{tabular}
\caption{The $\ell$counterrotating  couple  $\cc_x^-{\succ}! \oo_x^+$ (circled), allowed only in the Kerr geometry with  $a<a_{\iota_a}$. 
Dashed arrow-lines show the evolution of the individual configuration,  starting  from the initial state and pointing  to the final state.  Black continuum  arrow-lines indicate the status of the couple in a specific evolutionary phase, the direction of the arrow is in accordance to the sequentiality relation  $>$ (maximum of the hydrostatic pressure sequentiality), in the order of configuration (and criticality). Where uniquely fixed,  the  region of specific angular  momentum  is indicated  close to each configuration. We assume that  $O^{\pm}_x$ configurations   have an initial state  in closed topology, but it is indeed  possible that the outer  configuration $\cc^-$ could be formed after the formation of  $\oo_x^+$.}
\label{Fig:fig1}
\end{figure}

We now focus on the possible initial states for the couple  $\cc_x^-\succ \oo_x^+$,  making referring  to Fig.\il\ref{Fig:fig1}. As  this is a $\ell$counterrotating couple, the starting  state could be, for example,   $\cc^->()^+$, where the topology of the counterrotating configuration   remains to be fixed, specifically among the topologies $(\cc^-, \cc_x^-)$ for the corotating  fluids  and  $\{\cc^+, \cc_x^+, \oo_x^+\}$ for the counterrotating ones.  
The general scheme  is then  provided by Fig.\il\ref{Fig:fig1}, which does not consider   all the constraints imposed by the geodesic structure of the  $\mathbf{A}_{\iota_a}^<$ spacetimes.
To prevent any penetration of material, each initial couple shall be such that  $()^->()^+$. Being counterrotating,  their formation can occur independently and therefore they can take place at different phases.
All the possible initial  states and   evolutive lines,   leading to the final couple in (\ref{Eq:Q-E-D}), are  shown. However, some of the lines of evolution are to be suppressed due to the dynamical structure  of the $\mathbf{A}_{\iota_a}^<$ geometries and the  conditions of  non-penetration of matter. It is clear  that the couple  with  $\cc^-$ as initial state, with $\ell^-\in \mathbf{L3}^- $, is  always possible, but if  $\ell^-\in \mathbf{L1}^-$, then different    constraints have  to be considered. The  initial state in $\mathbf{L3}^+$  will be disadvantaged by the simultaneous presence of an outer  $()^-$ configuration. The analysis of specific constraints of the evolutive lines of Fig.\il\ref{Fig:fig1} is straightforward and it follows directly the arguments of  \ref{App:location}  and   we will not focus more  on it here.
The analysis of Fig.\il\ref{Fig:fig1} also shows the possible evolutive lines of the corotating configurations  which would bring the system to the state $\oo_x^-\succ \oo_x^+$, reflected in Eq.\il(\ref{Eq:rela-SusER-PR}).
%
\subsection{The $\ell$counterrotating accretion-accretion \textbf{(A-A)} systems}\label{Sec:A-A}
For a couple of $\ell$counterrotating accreting disks it has to be:
\be\label{Eq:result-A-A}
\cc_x^-\prec! \cc_x^+ \quad(\mbox{and}\quad \cc_x^-<! \cc_x^+ )\quad\forall  a\in]0,M], \quad \mathcal{L}=\mathbf{L1}^-\cup \mathbf{L1}^+.
 \ee
 %

In fact, for the attractor spins, $a>a_{\iota_a}$ (at $a_{\iota_a}:\; r_{mso}^-=r_{mbo}^+ $), the inequality in Eq.\il(\ref{Eq:result-A-A})  could not be inverted to create a hypothetical  couple $\cc_x^+\prec \cc_x^-$, as
 $\Delta r_x^-< \Delta r_x^+$ and  $\Delta r_x^-\cap \Delta r_x^+=\emptyset$,
 see also \cite{pugtot,ringed}.

In  the spacetimes of the slower attractors we find  $ r_{mbo}^+<r_{mso}^-$, and therefore   $\Delta r_x^-\cap \Delta r_x^+\neq\emptyset$.

However, in such spacetimes, an hypothetical  couple  $\cc_x^-\succ \cc_x^+ $,  would violate
the condition of non penetration of matter   in  the outer Roche lobes of $ \cc_x^{\pm}$:   in fact $r_{mbo}^-<r_{mbo}^+<r_x^+<r_{x}^-<r_{mso}^-<r_{mso}^+<r_{cent}^+$.

Thus, according to  Eqs\il(\ref{Eq:fORD-LE},\ref{Eq:fORD-LE1}),  in the geometries  $\mathbf{A}_{\iota_a}^<$
a proper specific angular  momentum has to be selected (small enough in magnitude) for the outer  $\cc_x^+$ disk satisfing the condition $r_{mso}^- \non{\in} \cc_1^+$.

In conclusion,   Eq.\il(\ref{Eq:result-A-A}) holds  for the slower attractors $\mathbf{A}_{\iota_a}^<$  under the restrictions  (\ref{Eq:fORD-LE1}) on $\cc_x^+$.

Furthermore, one should consider also the location of the  outer edge $r_{out}^-$ of the corotating disk in accretion with respect to the outer accretion point $r_x^+$.
 For sufficiently small  specific angular  momentum $\ell_1^-$  (corresponding to smaller accretion disks), $K_{Max}^{1_-}$  will be    small enough for the disk $\cc_x^-$  being  not extending to include  $r_{J}^+$. Further details on the conditions on the  $K_1$ parameter  can be found  also in
 \cite{ringed}. We refer also to the analysis and discussion in \ref{Sec:photon-C} and particularly Fig.\il\ref{Fig:Pgreat0}.

In  the  couple $\cc_x^-\prec \cc_x^+ $,  penetration of matter occurs from the material of  the outer accreting ring $\cc_x^+$ to the inner one $\cc_x^-$: the case of two disks in accretion clearly results in a penetration of matter  affecting  the  inner  lobe  of the outer ring  and the outer lobe of the inner ring.
In this case,  the  counterrotating matter is accreting onto the  corotating one: the two configurations  in the $\cc_x^{\pm}$ topology  could have   independent origins and evolutions
up to the collision phase,  giving rise  then to a  geometric correlation, and a consequent  instability    induced on the inner disk from  the outer disk in accretion.

The specific angular  momentum  $-\mathbf{L1}^+$  is certainly  larger that $\mathbf{L1}^-$ for sufficiently high spins      $a \gtrsim a_{\aleph_2}$ therefore in particular also for attractors $a\in]a_{\aleph_2}, a_{\aleph_1}[$--see Fig.\il\ref{Fig:Plotaaleph1II}.
 The outer disk  could therefore be characterized by a  larger elongation,  $\lambda_x^+>\lambda_x^-$, and correspondingly    a greater density.
For the slower attractors, the situation is just the opposite and
	 $ \Delta\ell_x^+ \cup\Delta\ell_x^- \neq \emptyset$.

Regarding the separation between the two configurations in the first Roche lobe, we should consider  the relations
from  Eq.\il(\ref{Eq:A-nimos})--the inclusion relation implies   $r_{mbo}^+\non{\in}\cc_x^+$.  As    $r_{x}^+>r_{mbo}^+$, the two configurations are separated by   the orbit $r_{mbo}^+$.
The counterrotating disk $\cc_1^+$  can approach this value as much as  the specific angular  momentum $-\ell_1^+$ increases in $-\mathbf{L1}^+$.
 For the slower attractors,
this  spacing will be reduced, according to the relation $\Delta r_x^-\cap \Delta r_x^+\neq \emptyset$.
The spacings $\bar{\lambda}_{i+1,i}$  in general increase with the attractor spin and decrease with the increasing  magnitude of the specific angular  momentum  (corresponding to a decrease of their elongation)  see \cite{ringed}).
\subsection{The $\ell$counterrotating  equilibrium disk-accretion \textbf{(C-A)} systems}\label{Sec:C-A-SYStEMS}
In this section we study  the couples $(\mathbf{C}-\mathbf{A}) $ of   $\ell$counterrotating configurations formed by   a closed regular  configuration  and a configuration in accretion.
This case has been   also studied in some detail in \cite{ringed}, where it was considered a single point of instability within the
ringed macro-configuration $\mathbf{C}_x^n$. 

The multiple configurations with a seed couple \textbf{(C-A)}  would be made by  an $\ell$corotating sequence    of  regular closed surfaces  in equilibrium, whose location and rotation with respect to the attractor are fixed in  accordance with the specific state of the couple $(\mathrm{C-A})$.

What  is significant here is  that, in general, in the  decompositions  with an  inner sequence of corotating or   counterrotating fluids in equilibrium, the configuration density  (at $\kappa$ constant)  will be smaller  in the geometries of the   slower attractors for counterrotating disks, and  in the   field of  faster attractors  for the  corotating ones. But the configuration density turns to be  generally lower for  the corotating fluids, predicting therefore  that  the  multiple surfaces of this kind are not favored. However, the conditions from  the geodesic structure   for the existence of these configurations and the condition of no-penetration  of matter  are very stringent and give a clear indication of these limits.

As in this case there are all closed configurations, it will be convenient to make reference to the configuration index.
Consider then the following four states:
\subsubsection{
\textbf{State I}: The $\mathbf{\cc^+-A^-}$ systems} This  {state} of the decomposition is easily ruled out.  Indeed, its existence  would imply:
\bea&&\nonumber
\cc_x^->\cc^+ \quad (r_{min}^{1_-}>r_{min}^+)\quad r_{x}^-\geq r_{out}^{+},
\\\label{Eq:pic-sus}
&& {\mathbf{\mathcal{L}}}=\mathbf{Li}^+\cup \mathbf{L1}^-\quad {\mathbf{\mathcal{K}}}={K}_+^{<}\cup K_{Max}^{1_-}.
\eea
This configuration is \emph{not} possible:
\bea&& \label{Eq:re-pre-plus-minus}
 r_{mso}^->r_{x}^->r_{mbo}^-,\quad  r_{x}^-\geq r_{out}^{+}>r_{mso}^\\
 &&\nonumber\mbox{but it is}\quad r_{mso}^-<r_{mso}^+		 \quad\mbox{therefore}\quad \cc_x^-\non{>}\cc^+.
\eea
Considering Eq.\il(\ref{Eq:re-pre-plus-minus},\ref{Eq:Eigen-VV},\ref{Eq:Q-E-D}) and Eq.\il(\ref{Eq:result-A-A}), we conclude that   the  $\cc_x^-$ configuration of   $\ell$counterrotating closed couple (\ref{Eq:pic-sus}) \emph{must} be the inner one.

\subsubsection{\textbf{State II:} The $\mathbf{A^--\cc^+}$ systems}   This  second  state of  the  $\ell$counterrotating  couple $(\mathbf{A-C})$ is  formed by
a corotating  disk in accretion and a counterrotating  one in equilibrium. According  to Eq.\il(\ref{Eq:re-pre-plus-minus}), we have
\bea&&\label{Eq:pic-sus-orto}
\cc_x^-<\cc^+ \quad (r_{min}^{1_-}<r_{min}^+)\quad r_{x}^{1_-}\leq r_{in}^{+},\\\nonumber
&& {\mathbf{\mathcal{L}}}=\mathbf{Li}^+\cup \mathbf{L1}^-\quad {\mathbf{\mathcal{K}}}={K}_+^{<}\cup K_{Max}^{1_-}.
\eea
Therefore, the outer counterrotating $\cc^+$ disk   is  in equilibrium and  the inner corotating one $\cc_x^-$ is in accretion.

These two configurations can be geometrically separated and   they could   evolve  independently, at lest from some time, unless the outer one is not reaching its unstable mode,  according  to (\ref{Eq:Eigen-VV}),  for open counterrotating proto-jets, leading then  to the topological transition $\cc^+\rightarrow \oo_x^+$.

Therefore, a topological shift towards the  couple $\cc_x^--\oo_x^+$ is  possible, with $\mathbf{Li}^+=\mathbf{L2}^+$, only in the geometries of  higher   $a>a_{\iota_a}$.

Or, as Eq.\il(\ref{Eq:result-A-A}) holds, the couples $\cc_x^-<\cc^+$ could term in a  final $\mathbf{A-A}$ system  with $\mathbf{Li}^+=\mathbf{L1}^+$.  The couple (\ref{Eq:pic-sus-orto})  then,  representing  an initial stage for the  states $\cc_x^--\oo_x^+$, when $a>a_{\iota_a}$, or for the states $\cc_x^--\cc_x^+$ for $a\in]0,M]$.
This closes the analysis of a possible interaction, with consequent  change in the topology of  the outer configuration, in the initial state  $\cc^+$.

However, a geometric correlation between the elements of the couple  would be possible also at fixed topologies, for collisions and   penetration of matter due, for example, to an increasing elongation (magnitude of specific angular  momentum $-\ell^+$ in any of $-\mathbf{Li}^+$, or   growing of  $K_+$ at fixed $\ell^+$),  or  increasing the specific angular  momentum of the accreting  matter  $\ell_1\in \mathbf{L1}^-$,  increasing therefore its critical elongation (also) outwards (we recall  that for  each cusped configuration  the  elongation $ K $ is uniquely determined by the matter specific angular  momentum).
These couples are detailed in \ref{Sec:coorr}.
It could be significant therefore, to discuss the existence of possible constraints for the spacing $\bar{\lambda}_{+,-}$ between this couple.

 According to  Eq.\il(\ref{Eq:choral},\ref{Eq:beha-dire},\ref{Eq:the-ag}) it has to be $r_{in}^{+}\geq r_{out}^{1_-}>r_{mso}^-$ and  there is  $r_{in}^{+}>r_{mbo}^+$.

We should now consider the situation with respect to the outer margin $r_{out}^{1_-}$.
Under the conditions provided  by the latest relation  of Eq.\il(\ref{Eq:fORD-LE1}) we can find that, for   slow attractors ($a<a_{\iota_a}$), counterrotating configurations satisfy the condition
 $r_{mso}^-\in \cc_1^+$. Then a geometric correlation with an  overlapping of material from the outer to the inner configuration of the couple $\cc_x^-<\cc^+$ can certainly occur, if the  magnitude of the specific angular  momentum $\ell^+$ is sufficiently high, accordingly to the constraints provided by  Eq.\il(\ref{Eq:fORD-LE1}).

On the other hand, we note  that  $\mathbf{L1}^-<-\mathbf{Li}^+$ for $a>a_{\aleph_2}$, although the  $K_+$ parameter of the counterrotating disk, can be significantly lower than the $K_-$ parameter of the  inner disk.

At lower spins, $a<a_{\aleph_2}$ the situation is not uniquely determined and one needs to distinguish  between different specific angular  momentum regimes   $\mathbf{Li}$:  in fact at $a<a_{\aleph_2}$ we have   $\Delta \ell_x^1 \cap (-\Delta \ell_1^+)\neq\emptyset$,
 and consequently $-\ell^-_1/\ell^+_1>1$, while  the measures of $- \mathbf{L2}^+$ and $-\mathbf{L3}^+$  are larger then that of $\mathbf{L}_1^-$.

To complete this analysis,  we consider the location of the radius
$ r_{mbo}^+$ with  respect to the configuration $\cc_x^-$, and  of
$r_{mso}^+$ with  respect to  $\cc^+$.

One would  think that it may be always possible to find a  sufficiently high specific angular  momentum to locate, for example, the center of a \emph{corotating} accreting disk,  $\cc^-_x$, on the orbit $r_{mbo}^+$, as proved in
  \ref{Sec:coorr} and stated in Eq.\il(\ref{Eq:05L}) and the following\footnote{We can explicit the general idea of this  argument as   follows: for the couple $\cc_x^-<\cc^+$, there is
$ r_{mbo}^-<r_x^-<  r_{mso}^-<r_{out}^{-}<r_{in}^{+}<r_{min}^{+}<r_{out}^+$,
 having  to make sure there is non-null distance
  $r_{in}^{+}-r_{out}^{-}$ where  $r_{in}^{+}>r_{mbo}^+$.
  Then, if $r_{mbo}^+\in \cc_x^{-}$, which   means $r_{out}^{-}>r_{mbo}^+ $, it is possible to choose a specific angular momentum $\ell^-$  and $\ell^+$  large enough in magnitude, because the distance is zero or negative.
  On the other side, there is
  $r_{mbo}^+>r_{mso}^-$ only for spacetimes of class $\mathbf{A}^>_{\iota_a}$,   which implies that in these geometries  we have $r_{mbo}^-<r_x^-< r_{mso}^-<(r_{mbo}^+<r_{out}^{-})<r_{in}^{+}$,
 where the brackets  signify a commutation between the terms $(r_{mbo}^+,r_{out}^{-})$. Then the inclusion  $ r_{mbo}^+\in \cc_x^+$ is clearly  favored  in the case of attractors
  $\mathbf{A}^<_{\iota_a} $ where    $r_{mbo}^-<(r_x^-<r_{mbo}^+)<  r_{mso}^-<r_{out}^{-}<r_{in}^{+}$.
  Thus,  one could inquire if
 $ r_{mso}^-\in \cc^+$   in the spacetimes $\mathbf{A}^<_{\iota_{a}}$, and this is considered in Eq.\il(\ref{Eq:fORD-LE1},\ref{Eq:ri-sp-zion},
\ref{Eq:not-non-crea-1},\ref{Eq:equa-one}). For faster attractors,  $\mathbf{A}^>_{\iota_{a}}$,
there is  $ r_{mso}^- \non{\in} \cc^+ $ implying that these two surfaces are always separated.
 Whereas for slower attractors  the situation is different and collision could take place for  $|\ell|$ sufficiently high, according to Eq.\il(\ref{Eq:L-Time}) and  Eq.\il(\ref{Eq:fORD-LE1}).}.

However, we have proved that  a corotating disk in accretion must have a specific angular  momentum $\ell_1\in \mathbf{L1}^-$. But for these disks, and a fortiori for $\cc_1$ (which has  lower $K_-$), the density and elongation could be  not sufficient to induce a gravo-hydrostatic instability. Therefore these disks are all contained in  $]r_{mbo}^+, r_{mbo}^-[$. In other words, it could be  $r_{out}^{1_-}<r_{mbo}^+$, implying
  that the  supremum of the spacing
is  at higher spin $\mathbf{A}_{\iota_a}^>$, and it increases with the spin (this fact    might be expected also  from general considerations about the influence of the spin of the black hole on the  $\ell$countarrotating couples, discussed in more points here and  in detail in \cite{ringed}).

It  is also possible to provide a rough estimation of the distance  $r_{mbo}^+-r_{mso}^-$
   as minimum spacing, but the infimum of this spacing, for this class of attractors, is reached in the geometry $a=a_{\iota_a}$. Therefore, this property may be used as tracing to identify the  spin class  the attractor  belongs to.

One can also read this situation in the following way: for higher spins, the topologies
	$\cc_{\odot}^x$  of order $n=2$,  with the couple $\cc_x^-<\cc^+ $,
 could not be possible, and therefore a geometric (and consequently a causal) correlation cannot occur in this couple whose sub-configuration will evolve independently,
	at least until the outer counterrotating configuration will not change topology. However, Eqs\il(\ref{Eq:choral},\ref{Eq:beha-dire},\ref{Eq:the-ag})   prohibit the penetration of  $r_{mbo}^-$ in the corotating disk.

For the counterrotating disk in  open topologies, with  $\mathbf{K1}^+\cup \mathbf{L2}^+$,  there is the   instability point $r_{J}^+ \leq r_{mbo}^+$, and  matter collision  can occur as described in the  point $\mathbf{III}$ $\cc_x^-\prec \oo_x^+$.

We can conclude that   at higher spin a correlation between the
outer counterrotating matter and the inner disk in accretion  (as $\lambda_x=\sup{\lambda}$, a
fortiori this will be true also  for a possible inner closed and regular topology) occurs  if the outer configuration is open,  $()^+=\oo_x^+$
  (we note that $r_x^+>r_{mbo}^+$ for the cusped closed  configuration).
Finally this analysis is completed  in \ref{Sec:photon-C} with the discussion of  the location of
$r_{\gamma}^{\mp}$ in the  $()_{\pm}$	 configurations.

\subsubsection{\textbf{State III:} The $\mathbf{\cc^--A^+}$ systems}
These couples are described by the conditions
\bea&&\label{Eq:pic-get}
\cc^-<\cc_x^+ \quad (r_{min}^-<r_{min}^{1_+})\quad r_{x}^+\geq r_{out}^{-},
\\\nonumber
&& {\mathbf{\mathcal{L}}}=\mathbf{Li}^-\cup \mathbf{L1}^+\quad {\mathbf{\mathcal{K}}}={K}_-^{<}\cup K_{Max}^{1_+}.
\eea
It is clear that  in this case  a correlation  in the couple $\cc^-<\cc_x^+$ must exist, for the \emph{outer} disk $\cc_x^+$ is accreting on the attractor and therefore a collision with the \emph{inner} $\cc^-$ corotating  disk will certainly occur.

This couple could   even be considered as a precursor of an  $\ell$counterrotating $\mathbf{A}-\mathbf{A}$ couple, as analyzed in   Eq.\il(\ref{Eq:result-A-A}), where it was proved   that  $\cc_x^-\prec! \cc_x^+$ $\forall a\in]0,M]$, or also  of a $\mathbf{A-J}$ configuration, considered as point \textbf{III},   where  $\cc_x^-\prec \oo_x^+$,  implying the outer configuration opening in proto-proto-jet.

Decreasing   the specific angular  momentum $-\ell^+\in-\mathbf{L1}^+$, or increasing $K_-\in \mathbf{K0}$ or  $\ell^-$ for  the topology $\cc^-$, the spacing between the two surfaces will in general  decrease as  $\partial_{|\ell|} r_{Max}<0$, $\partial_{|\ell|} r_{min}>0$ and  $\partial_{|\ell|} r_{out}^x>0$.

For the conditions in  Eq.\il(\ref{Eq:pic-get}), it has to be
$r_{mbo}^-<r_{in}^-<r_{min}^-<r_{out}^-<r_{x}^+<r_{min}^+$,
with $r_{mbo}^-<r_{mbo}^+<r_{x}^+$.
Therefore it is  important to consider the relative position of the radii
$ r_{mso}^{\pm}$ and  $r_{mbo}^+\in \cc^-$.

This has been partially faced   in \cite{ringed}  and addressed in detail in this article
in \ref{Sec:coorr}  by Eq.\il(\ref{Eq:A-nimos}).
 From  Eq.\il(\ref{Eq:the-ag}) if follows that
   $r_{mbo}^+\non{\in} \cc_x^+$,
 thus we can certainly find separated couples $\cc^-<\cc_x^+$  (meaning here $r_{out}^-<r_x^+$) in the geometries
$\mathbf{A}_{\iota_a}^>$, where in fact  $r_{mbo}^+\in
  ]r_{mso}^-,r_{mso}^+[$, and one can find
  a sufficiently small momentum $\ell^-\in \mathbf{Li}^-$  for
  $r_{cent}\in ]r_{mso}^-,r_{mbo}^+[$.

However in this case we need to deal  also with the location of the  outer boundary of the inner  corotating disk $\cc^-$
  to establish  if  configurations   $\cc^-$ with $r_{out}^-<r_{mbo}^+$  are possible.
  This condition  is sufficient but \emph{not}  necessary
  for the separated disks.
 We addressed his problem in Eq.\il(\ref{Eq:ge-get}),   which ensures   the   validity of this result for fluids with momentum $ \ell^-\in \mathbf{L1}^-$  and, in the other cases, for different classes of attractors.
 Here we introduce the general idea behind the main arguments.

 As discussed in \ref{Sec:coorr}, we need to distinguish the class of attractors $\mathbf{A}_{\iota_a}^< $ at $ a\in[0,a_{\iota_a}]$ and $\mathbf{A}_{\iota_a}^>$  with  $ a\in]a_{\iota_a},M]$ respectively.

In general,  if $\ell=
 \ell_{mso}^-+\epsilon_{\ell}$ then $r_{min}^-= r_{mso}^-+\epsilon_{r_{min}}$, where $\epsilon_{r_{min}}=\epsilon_{r_{min}}(\epsilon_{\ell})>0$ and $\epsilon_{\ell}>0$ (analogously for the maximum points with $\epsilon_{Max}<0$, and in general  $\epsilon_{r_{min}}\neq\epsilon_{r_{Max}} $
    \citep{ringed}, thus ensuring  the disk is small enough or placed far enough from the  marginally bounded orbit,  as $r_{in}^-<r_{mbo}^+$, in the geometries  $\mathbf{A}^<_{\iota_a}$, and  $r_{out}^->r_{mbo}^+$ for $\mathbf{A}^>_{\iota_a}$.

 Thus  $K_-$ could be sufficiently small to consider    $r_{mbo}^+\non{\in}\cc^-$,
 but this  must imply
  $\ell^-\in \mathbf{L1}^-$ (from the  definition of $\mathbf{L1}^-$).

On the other side, increasing  $\ell^-$ towards $\ell^-({r_{mbo}^+})$, see  Fig.\il\ref{Figs:PXasdPXasdP}, the center with maximum hydrostatic pressure approaches
  $r_{mbo}^+$ at $a>a_{\iota_a}$, or  the minimum of the pressure for attractors  $\mathbf{A}_{\iota_a}^<$. The ring has to have a $K_-$ parameter smaller and smaller to ensure the disk does not include $r_{mbo}^+$.
	Therefore, we can say that these disks not including $r_{mbo}^+$, are nor favored at low $K$ for low specific angular  momentum.

	For attractors with  $ a\in]0, a_{\iota_a}[ $, an increase of the spin  acts against the formation of these configurations, that are favored instead for higher spins,
and thus also the   larger $K$ for the multiple configuration, according to the analysis of Eq.\il(\ref{Eq:ge-get}).

Indeed, $ r_{mso}^-<r_{min}^-<r_{out}^-<r_x^+<r_{mso}^+$
and $r_{mbo}^+<r_x^+$.
				Thus one can always find  an  $-\ell^+\in -\mathbf{L1}^+$ sufficiently small,  and a small $\ell^-$, accordingly to allow a separated couples.

The decompositions of higher order than $n=2$ would be possible, for example, with an inner $\ell$corotating  sequence  of \emph{corotating} disks in equilibrium, and one outer with respect to $\cc_x^+$, formed by corotating or counterrotating fluids in equilibrium.


\subsubsection{\textbf{State IV:} The $\mathbf{A^+-\cc^-}$ systems}
These systems are defined by the relations:
\bea\label{Eq:pic-get-o}
\cc_x^+<\cc^- \quad (r_{min}^{1_+}<r_{min}^-)\quad r_{in}^{-}\geq r_{out}^+,
\\\nonumber {\mathbf{\mathcal{L}}}=\mathbf{Li}^-\cup \mathbf{L1}^+\quad {\mathbf{\mathcal{K}}}={K}_-^{<}\cup K_{Max}^{1_+}.
\eea
In this case the \emph{outer} disk is $\cc^-$ in equilibrium, and  the counterrotating  \emph{inner} disk is in accretion.
It has to be
$r_{mbo}^+<r_x^+<r_{mso}^+<r_{min}^+<r_{out}^+\leq r_{in}^-<r_{min}^-$.
These configurations have also been discussed in details in  \cite{ringed}.

A geometric correlation between the two configurations is possible, and therefore a  causal correlation, as a consequence of
a shift inward of the outer disk, due to  lost of specific angular  momentum  or to an  increase of $K_-\in \mathbf{K0}$, with the fixed center but longer elongation. An increase of specific angular  momentum magnitude $-\ell^+\in -\mathbf{L1}^+$, would also produce a shift outward of the outer margin $r_{out}^+$. Eventual    emergence of an instability of the disk in regular topology $\cc^-$,  leading to the couple  $\cc_x^+<\cc_x^-$, cannot occur but following the collision with the inner counterrotating disk $\cc_x^+$ already in accretion (as $r_{mso}^+<r_{mso}^-$).

In fact,
  Eq.\il(\ref{Eq:result-A-A}) holds, indicating that  $\cc_x^-\prec! \cc_x^+$ and therefore  $\cc_x^-\non{\succ} \cc_x^+$,  from which it results that   the couples $\mathbf{\cc^+-A^-}$,  according to the definition  Eq.\il(\ref{Eq:pic-sus}), \emph{could not} evolve into the topology   $\mathbf{A^+-A^-}$.  For the faster attractors,  $a>a_{\aleph_1}$, such a kind of couple is forbidden for the geodesic structure of the spacetime and, for the slower attractors, $a<a_{\aleph_1}$, from the condition of non-penetration of matter.

  As we have mentioned above,
 a further possibility of correlation occurs for the action of the \emph{inner} unstable configuration, due to increasing specific angular  momentum  $-\ell^+\in- \mathbf{L1}^+$ and consequently increasing  $K_+$, with a consequent shift outward of the center of maximum pressure and of the outer boundary.

As can be seen  from Fig.\il\ref{Fig:Plotaaleph1IIa},
 these couples can be  always possible, within  the necessary condition
 $r_{mso}^+\non{\in}\cc_{-}$, if
 $\ell^-$ is sufficiently large and
  $K$ sufficiently close to the minimum (lower density), with $
  \ell^+\in \mathbf{L1}^+$ low enough (in magnitude)  for the non-penetration of matter condition will be satisfied.
  We can specify these limits considering
  Eqs\il(\ref{Eq:evol-suss-corr}, \ref{Eq:evol-suss-corr2}
\ref{Eq:evol-suss-corr3}).

By referring to  Fig.\il\ref{Figs:PXasdPXasdP},
  we need to distinguish between the attractors $\breve{\mathbf{A}}_*^<$  with spin  $a <\breve{a}_*$,
 and the geometries of the faster attractors $\breve{\mathbf{A}}_*^>$ with  $a >\breve{a}_*$.
 Together with the   further restriction, for  increasing  $\ell^-$ and  decreasing  $-\ell^+$, to avoid the condition $r_{out}^{x_+}\in \cc^-$, uniquely fixed by the specific angular  momentum $\ell^+$.
In any case, we still consider the non-penetration of matter from the outer Roche lobes of the two configurations as described by the second relation of Eq.\il(\ref{Eq:pic-get}).
\subsection{The $\ell$counterrotating equilibrium disk-proto-jet  \textbf{(C-J)} systems}\label{Sec:C-J}
We will consider a couple formed by an equilibrium configuration and a configuration opened in proto-jet.

 As specified in  Sec.\il(\ref{Sec:L-A-system}) and Sec.\il(\ref{Sec:C-A-SYStEMS}), multiple surfaces formed by couple seed, shall contain  two $\ell$corotating sequences   at equal topology. In general, at constant  $\kappa$, the  density of the inner sequence with regular  and closed topologies will be small particularly at high spin  $a/M$ for co-rotating fluids.
Below are discussed several limitations and considerations on the possible orbital extension for such sequences.
The following four cases occur:
\subsubsection{\textbf{Case I:} The $\mathbf{\cc^+-J^-}$ systems}
This case is described by the condition
\bea\nonumber
\cc^+<\oo_x^-\quad (r_{min}^+<r_{min}^-)\quad \mbox{with}\quad r_{out}^+\non{\leq} r_{J}^- \quad \mbox{and}
\\ \label{Eq:pro-Ue-UK}r_{in}^+\geq r_{J}^-, \quad \mathcal{L}=\mathbf{Li}^+\cup \mathbf{L2}^-,\quad \mathcal{K}=\mathbf{K0}^+\cup \mathbf{K1}^-,
\eea
that is, in order to avoid any overlap of material, and considering  the geodesic structure of the spacetime as in Fig.\il\ref{Fig:Plotaaleph1IIa}, the equilibrium $\cc^+$ disk   has to be entirely contained  in the region
 $r>r_J^-$,  but with $r_{min}^+<r_{min}^-$, as it comes from the definition  $\cc^+<\oo_x^-$.
 This is in  contrast  with  the state   $\cc^-<\oo_x^+$, analyzed in  Eq.\il(\ref{Eq:MX-en}), where there is still a closed inner configuration.

Therefore $
  r_{J}^-<r_{mbo}^-<r_{mbo}^+<r_{in}^+<r_{min}^+<r_{out}^+$,
 the third inequality  is a consequence of  Eq.\il(\ref{Eq:choral},\ref{Eq:beha-dire},\ref{Eq:the-ag}) and   $r_{cent}^->r_{mso}^+$.

The couples of  Eq.\il (\ref{Eq:pro-Ue-UK})
could  be  always geometrically separated.  However, the closed $\cc^+$ configuration cannot change topology towards the transition $ \cc^+>J^+$ as this  would imply a transition where  the closed   configurationf changes topology creating a couple  $ \mathbf{J^--J^+ }$ or $\mathbf{J^--A^+}$ considered above. In any case, it has to be $
r_{J}^-<(r_{J}^+<r_{min}^+)< r_{min}^-$ or
$r_{J}^-<(r_{x}^+<r_{min}^+)< r_{min}^-$.
The first inequality is to avoid any penetration of matter, starting the initial condition Eq.\il(\ref{Eq:pro-Ue-UK}), the last one
follows the definition of the specific state of the decomposition, more probably to be formed at spins $a >a_{\iota}$ and  $a>a_{\iota_a}$, as it follows from the geodesic structure in Fig.\il\ref{Fig:Plotaaleph1IIa}.

\subsubsection{\textbf{Case II: }The $\mathbf{J^--\cc^+}$ systems}
We consider the couple
\bea&&\label{Eq:wav-freq}
\oo_x^-<\cc^+\quad (r_{min}^-<r_{min}^+)\quad \mbox{with}\quad r_{out}^+\non{\leq }r_{J}^-
 \\&&\nonumber
 \mbox{and }\quad r_{in}^+\geq r_{J}^-, \quad \mathcal{L}=\mathbf{Li}^+\cup \mathbf{L2}^-,\quad \mathcal{K}=\mathbf{K0}^+\cup \mathbf{K1}^-.
\eea
Since $r_J^-<r_{mbo}^-<r_{mso}^-<r_{min}^-<r_{min}^+$,  the counterrotating closed configurations are  separated and remain separated  during its  evolution (at fixed topology) from  the inner (using the criticality index) corotating proto-jet.

Therefore, in both the cases, \textbf{I} for the  $\mathbf{\cc^+-J^-}$ systems, and  \textbf{II}- for the  $\mathbf{J^--\cc^+}$ systems,  the outer disk  is in equilibrium and   the two  configurations are geometrically separated.

In this case it has to be  $
r_{J}^-<(r_{J}^+<r_{min}^-)< r_{min}^+$ or
$r_{J}^-<(r_{x}^+<r_{min}^-)< r_{min}^+$.

For these cases to be fulfilled, it must be guaranteed that the  specific angular  momenta  $\ell_{\pm}$  are ensuring the relations above; we know then   that it  has to be $\ell^-<-\ell^+$, see also \cite{ringed} and  Fig.\il\ref{Fig:Plotaaleph1IIa}.

\subsubsection{\textbf{Case III:} The $\mathbf{\cc^--J^+}$ systems}
This state is described by the conditions:
\bea\nonumber
&&\cc^-<\oo_x^+\quad (r_{min}^-<r_{min}^+)\quad \mbox{with}\quad r_{out}^-\leq r_{J}^+ \quad \mbox{\emph{or}} \\\label{Eq:MX-en}
&&r_{in}^-\geq r_{j }^+\quad \mathcal{L}=\mathbf{Li}^-\cup \mathbf{L2}^+,\; \mathcal{K}=\mathbf{K0}^-\cup \mathbf{K1}^+.
\eea
This couple, as in Eq.\il(\ref{Eq:pro-Ue-UK}), includes by definition  a closed and regular inner surface.  In contrast with  the cases defined in  Eqs\il(\ref{Eq:pro-Ue-UK},\ref{Eq:wav-freq}), which are  bound in   $ r_{out}^-\non{\leq} r_{J}^+$, by the condition of non-penetration of matter.  The state  considered here is instead possible  as an \emph{alternative} to the relation  $r_{in}^-\geq r_{J}^+$. This means that the corotating disk  in equilibrium can be entirely contained in  $r>r_{J}^+$.

This couple can always exist in any geometries  but, as in $\mathbf{A}_{\iota_a}^<$, we have  $r_J^+<r_{mbo}^+<r_{mso}^-<r_{mso}^+<r_{min}^+$. To avoid any penetration of matter,  the configuration should be contained in the region $r>r_{J}^+$. Thus  $r_{out}^->r_{in}^->r_J^+$, but there  could be a geometric correlation and indeed it can even be $r_{in}^-=r_J^+$.

On the other side, for attractors $\mathbf{A}_{\iota_a}^>$, there is $r_{mso}^-<r_{mbo}^+$ and in this case the corotating  ring can be outer or inner to  the region $r<r_{J}^+$. There  could be geometric correlation, and indeed it could be even  $r_{out}^+=r_{J}^+$. The existence of a  geometric correlation should be considered according to the limits provided by the analysis of  \ref{Sec:coorr} and \ref{Sec:photon-C}.

\subsubsection{\textbf{Case IV:} The $\mathbf{J^+-\cc^-}$ systems}
The following conditions hold:
\bea\label{Eq:classic-condi-imm}
&&\oo_x^+<\cc^-\quad (r_{min}^+<r_{min}^-)\quad \mbox{with}\quad r_{out}^-\non{\leq} r_{J}^+ \\\nonumber
&& \mbox{and }\quad r_{in}^-\geq r_{J}^+, \quad \mathcal{L}=\mathbf{Li}^-\cup \mathbf{L2}^+,\; \mathcal{K}=\mathbf{K0}^-\cup \mathbf{K1}^+.
\eea
In this case the corotating ring is  located in the region $r>r_J^+$, and the surfaces of the couple are geometrically separated, generally by the distance $]r_{mbo}^+, r_{mbo}^-[$. The situation is clearly articulated   as it depends on the geodesic structure of spacetime, and therefore it differentiates various classes of attractors.

Using the results of \ref{App:location}, and considering the limiting spins introduced in the analysis and in  Fig.\il\ref{Fig:Plotaaleph1IIa}, we can summarize the situation as  follows:

\textbf{1.} At  $a<a_{\iota}$, we have
$r_{\gamma}^+<r_{mbo}^-<r_{mbo}^+<r_{mso}^-<r_{mso}^+<r_{min}^+<r_{min}^-$.
But   $r_{J}^+\in ]r_{\gamma}^+,r_{mbo}^+]$
and $r_{mbo}^-\in]r_{\gamma}^+,r_{mbo}^+] \non{\in }\cc^-$.

Using
Eqs\il(\ref{Eq:in-raff},\ref{Eq:ali-06},\ref{Eq:curv-n}) and Eqs\il(\ref{Eq:states-energies},\ref{Eq:states-energies-Y}),
we can say that, for  $\ell$ large enough, and then the minimum point $r_{min}^+$ far enough, and  large  $K_-$, the radius $r_{mso}^-$ can approach the inner edge of the disk (considering also $\mathbf{L3}^-$).

This means that they will be separated by the range $[ r_{mbo}^+,r_{mso}^-]$.  This is confirmed by Eq.\il(\ref{Eq:A-nimos}) and  Eq.\il(\ref{Eq:C3C2-par}) where  a similar argument is   carried out for the other cases.

\textbf{2. }For  $a\in]a_{\iota}, a_{\iota_a}[$,
one has $r_{mbo}^-<r_{\gamma}^+<r_{mbo}^+<r_{mso}^-<r_{mso}^+<r_{min}^+<r_{min}^-$
thus, by increasing the spin of the attractor, the constraints should be less stringent  reducing to the  only  inclusion $r _{mbo}^+\in \cc^-$ and $r_{mso}^- \in \cc^-$.  In this case, one has to consider  properly large  specific angular  momenta $\ell^-$ and   $K_-$  parameter, because an inclusion relation  may be then satisfied,  in accordance with the above analysis.

  At $a\in]a_{\iota_a},a_{\gamma_+}^-[$, see Fig.\il\ref{Fig:Plotaaleph1IIa},  we find
$r_{mbo}^-<r_{\gamma}^+<r_{mso}^-<r_{mbo}^+<r_{mso}^+<r_{min}^+<r_{min}^-$.
The only condition to be insured is $r_{mso}^+\non{\in}\cc^-$.
For even higher spin,  $\mathbf{A}_{\iota_a}^>$,   which includes the extreme case, there  is
 $ r_{mbo}^-<r_{mso}^-<r_{\gamma}^+<r_{mbo}^+<r_{mso}^+<r_{min}^+<r_{min}^-$.
  It remains to  establish the condition $r_{mso}^+\non{\in}\cc^-$, but also  $r_{\gamma}^+\non{\in}\cc^-$.

The inclusion condition with respect to the photon orbit have been  investigated  in \ref{Sec:photon-C}, and therefore  we have to adhere to the conditions in \il(\ref{Eq:lengh-distance},\ref{Eq:lengh-distance0}).

This analysis confirms that the geometrical correlation, for  the contact in  $r_J^+ \approx r_{in}^-$,  can occur only in specific circumstances, by narrowing both the set of geometries and the  range of  the fluid specific angular  momentum.

Finally,  the equilibrium surface is subjected to a change of topology, ending in a critical phase, for example in  $\oo_x^-$  or $\cc_x^-$. However, the formation  of such a couple from the initial phase  in Eq.\il(\ref{Eq:classic-condi-imm})
$\oo_x^+<\cc^-$, with $r_{min}^+<r_{min}^-$, and  $r_{x}^-\geq r_{J}^+$ or $r_{J}^-\geq r_{J}^+$,  or equivalently  $\oo_x^+\succ()_x^-$,
  is regulated by the conditions in Eqs\il(\ref{Eq:lagespicn},\ref{Eq:rela-SusER-PR}) and Eq.\il(\ref{Eq:Eigen-VV}) respectively.
\section{Phenomenology and observational evidence of RADs}\label{Sec:P-O-E-RADs}
\textbf{RADs} are agglomerates of several accretion tori orbiting very compact objects, following the possibility   that   several accretion disks can form  around  very compact objects as \textbf{SMBHs} ($10^6-10^9 M_{\odot}$, $M_{\odot}$ being solar masses) in \textbf{AGNs} embedding.  For these very compact objects  the curvature effects are  relevant and the host can provide for a very rich and active  \textbf{BH} environment. \textbf{RAD} may be originated after     different  accretion phases   in some  binary systems or  \textbf{BH} kick-out, or by     local   clouds  accretion.
  Concerning the accretion emergence,   the maximum number of  accreting disks orbiting  around one  central Kerr \textbf{BH} is $n=2$. A double accretion  can be observable  only as a    couple  $C_x^-<C_x^+$ (see Fig.\il\ref{Fig:rssi-na}),  around all Kerr \textbf{BHs} ($a\neq0$). The couple is subjected to  constraints provided here on the fluid angular momentum range ($\ell$-parameter) and density ($K$-parameter)--see also \cite{dsystem}.
However,  some ``screening''-configurations for such accreting couples   can form, constituting a   more articulated \textbf{RAD}  system $\mathrm{\mathbf{{(xx)}}}:\,C_x^-<C^-<...<C_x^+<C^{\pm}$,   with only \emph{corotating} non-accreting disks between the  two accreting tori. Eventually this  demonstrates that a  screening disk must be always a corotating (non-accreting) torus. These special configurations may be  detectable for example  as  X-ray spectra emission obscuration.
On the other hand,  if a counterrotating torus  is accreting onto the central \textbf{BH}, then a \textbf{RAD} with a corotating outer torus towards the accretion (i.e. $C^-_1$), can be observed only as an  aggregate  of the kind
$\mathbf{\mathrm{(x)}}:\,\pp_x^-<C_x^+<C_1^-<C^{\pm}$, and orbiting around  ``slow'' \textbf{BHs} with  $a<0.46M$.
We note that, if the inner torus is $C_{x}^{-}$,  then a  configuration  $\mathbf{\mathrm{(x)}}$, or with a string of configurations $\pp_x^-<C_x^+<C_1^-<C^{\pm}$, reduces to a special case of  $\mathbf{\mathrm{(xx)}}$ \textbf{RAD}, that is of the kind $C_x^-<C^-<...<C_x^+<C^{\pm}$   \citep{dsystem}.
This  also implies that, during the evolution of the outer corotating torus   towards accretion, no such couple can form,  prior the emergence of  tori collision, eventually reducing the  actual possibility to observe  a counterrotating  accreting disk, in the \textbf{RAD} context, and tightening it to the \textbf{BH}-\textbf{RAD} early phases of evolution-see also \cite{dsystem}.

A further relevant aspect of this investigation is to support  the need,     endorsed  also by several other   studies, of a more general framework of analysis  envisaging the  \textbf{BH}-disk system as an integrated whole. Evidences  of this fact are the ongoing debates on the  jet-accretion correlation, the issues of the \textbf{BH} accretion rate-disk luminosity, \textbf{BH} growth - accretion disk, and  \textbf{BH}-spin shift-accretion disk correlation.

Results of our  analysis, moreover, show  the importance of proto-jet-accretion correlation, envisaged here as $\textbf{(J-A)}$-correlation---Secs\il(\ref{Sec:A-J},\ref{Sec:innerJ-A}) and Sec.\il(\ref{Sec:L-A-system}). Proto-jets are narrow, relatively  fast, long  matter funnels
 \citep{McKinneyScience,Romanova,NatureMa,Maraschi:2002pp,Chen:2015cga,Yu:2015lqj,Zhang:2015eka,Sbarrato:2014uxa,
Coughlin:2013lva,MSBNNW2009,Ghisellini:2014pwa,
FragileW2012,Miller,AbramowiczSharp(1983),Sadowski:2015ena,Okuda:2004zv,Ferreira:2003yy,Lyutikov(2009)}.
Findings of  Secs\il(\ref{Sec:J-Jc},\ref{Sec:J-J-cont}) suggest  the possibility to detect
\emph{{structured proto-jets}} as   sequences of jet-like configurations (or \emph{jet-bundles}), constrained in spacings and relative fluids rotation.

\ref{App:location} specifies the more general statement according to which the inner edge of an accreting torus is located in   $r_{x}\in]r_{mbo},r_{mso}[$ \citep{Krolik:2002ae,BMP98,2010A&A...521A..15A,Agol:1999dn,Paczynski:2000tz}. We bounded the location of the accretion disk cusp to   the variation ranges of the  $\ell$-parameter and   in accordance with the torus evolutionary phases and the \textbf{RAD} structure.
In turn,  in this analysis, we
{narrowed down the location of the single accretion torus  inner edge,  showing the  strong  connection between  \textbf{RAD} structures  and  \textbf{BH} spins}--see also \cite{multy}.

A further phenomenological application of these studies, already
mentioned in \cite{ringed}, regards the possible
connection between \textbf{RADs} seismology and \textbf{QPOs} - the pattern of the
possible oscillation modes of the tori aggregate has been provided
and related to the evolution of instabilities in \textbf{RAD}. \textbf{QPOs} are low
and high frequency peaks in the
power density spectra, studied in missions like \textbf{XMM-Newton} (X-ray Multi-Mirror Mission)\footnote{{http://sci.esa.int/science-e/www/area/index.cfm?fareaid=23}}  or \textbf{RXTE} (Rossi X-ray Timing
Explorer)\footnote{  {http://heasarc.gsfc.nasa.gov/docs/xte/xtegof.html}}.
 Each torus of the aggregate considered here, has its
own axisymmetric and incompressible modes which have been
variously associated with the \textbf{QPO} emergence. For global oscillations
of slender tori in the radial and vertical directions, their frequencies are determined by combinations of the geodesic epicyclic frequencies - for details see \cite{Stuchliketal(2013)}. In the \textbf{RADs} case, these modes have to be
combined with the stratified structure of the \textbf{RADs} and its own
modes, finally leading to an alteration of the macrostructures
elongations and spacings.

In any case, the internal structure of the \textbf{RAD} presents a rich multiplicity of situations and different working frameworks. It is clear that in many aspects the physics of \textbf{BHs} and its  host galaxy would be altered  by the relevance of the \textbf{RAD} argument  in support of the  hypothesis of a more complex \textbf{BH}-accretion disk system,  than is commonly considered.
\textbf{RAD} accretion  in galaxy may
 produce
radiative power  outshining  the
host galaxy itself (note that the \textbf{AGNs} accretion {disk}   is encircled
by a thick (outer) torus of gas and
dust). On the other hand,  \textbf{AGN} are generally characterized by very fast jets
(almost speed of
light), which might  be connected with an inner \textbf{RAD} disk  jet launch.

A \textbf{RAD} structure  analysis  demands for an accurate determination  of the processes timescales   to determine the  \textbf{RAD} evolution timeline. This issue was  detailed discussed in  \cite{dsystem}, where   a \textbf{RAD} of the order $n=2$ was considered as an   $\ell$corotating  or  $\ell$counterrotating couple.  This analysis was  performed as part of the   broadest investigation on  the  tori collision emergence.
In fact, the complexity and variety of the  processes   characterizing  the tori  agglomerate can be actually  contained in a few evolutionary patterns, heavily  depending  on the initial data of  the single component of the aggregate.
 The \textbf{RAD} timescales  were  shown then  to strongly depend on  timescale of tori formation more then
from the instability of  each torus. Different situations are   distinguished according to  the relative fluids rotation   and the rotation of the  \textbf{RAD} \emph{inner} torus  with respect to the central \textbf{SMBH}.
These studies  show how,  for  some couples,  an  expected final  collisional phase occurs,  eventually followed  by a tori  merging with a modification of the \textbf{RAD} internal structure, or a drying-feeding effect with rising of oscillatory modes.
Couples   $\mathbf{\mathrm{(\textbf{x})}}$ and  $\mathbf{\mathrm{(\textbf{xx})}}$, are  examples where  one of the  tori approaches the instability, while  the evolution of the outer corotating torus inevitably leads to the collision.  
This situation  ultimately ends up  constraining the  \textbf{BH} evolution itself in its environment.

  Concerning then the disk  \emph{process timescales}  in the  aggregate, these   depend-on and determine the tori model characteristics (as thickness, opacity, accretion rates or instabilities). The determination of the \textbf{RAD} timescales  should be made by combining  the analysis of each \textbf{RAD} component  timescale  with   the \textbf{RAD} internal oscillation.
 For the  geometrically thick configurations considered here, it is generally assumed that
 the timescale of the dynamical processes, $\tau_{dyn}$,  (regulated by the gravitational and inertial forces, the timescale for  pressure to balance the  gravitational and centrifugal force) is much lower than the timescale of the thermal ones, $\tau_{the}$,  (i.e. heating and cooling processes, timescale of  radiation entropy redistribution) that is lower than the time scale of the viscous processes, $\tau_{vis}$, and the effects of strong gravitational fields are dominant with respect to the  dissipative ones and predominant to determine  the unstable phases of the systems  \citep{F-D-02,Igumenshchev,AbraFra}, i.e. $\tau_{dyn}\ll\tau_{the}\ll\tau_{vis}$. Thus  the effects of strong gravitational fields dominate  the  dissipative ones, grounding the assumption of  perfect fluid energy-momentum tensor--see also \cite{AbraFra,Pac-Wii}.
Moreover,   during the evolution of dynamical processes, the functional form of the angular
momentum and entropy distribution depends on the initial conditions of the system and on
the details of the dissipative processes:
the entropy is constant along the flow and, according to the von Zeipel condition, the surfaces of constant angular velocity $\Omega$ and of constant specific angular momentum $\ell$ coincide \citep{M.A.Abramowicz,Chakrabarti0,Chakrabarti,Zanotti:2014haa}, implying  the rotation law $\ell=\ell(\Omega)$,  independently by the equation of state  \citep{Lei:2008ui,Abramowicz:2008bk}.
Eventually, this model describes an opaque and super-Eddington, radiation pressure supported accretion disks cooled by advection with low viscosity, where proto-jet configurations are  funnels of material with highly super-Eddington luminosity.


Despite the fact that the \textbf{RAD} model we used is based on aggregates of  thick tori, actually  the  major significance of \textbf{RAD} presence in \textbf{BH} host environment  should emerge from the  ``macrostructure-scale'',  to be considered in some extents quite
independently of the single torus model. \textbf{RAD} can be made by  aggregate components   with  very different models, according to the different evolution processes advocated for each torus origin.
In this respect, the macrostructure morphology  is   more decisive,  for the point of view of the \textbf{RAD}-\textbf{BH} system phenomenology,     than  the  model for the each  component.

In the following, we briefly consider different   observational spots expected to be associated with the  \textbf{RAD}, and strongly dependent of the \textbf{RAD} morphology.
Firstly, \textbf{RAD} blends the geometry of a thin disk  (the \textbf{RAD} is generally a geometrically thin disk as demonstrated in \cite{ringed}) with the  specific characteristics of a  geometrically thick disk (for example, high accretion rates), together with a stratified  inner structure, a differential relational law and  a  \emph{knobby},
although axial-symmetric, disk surface  \citep{ringed}. In this sense, the macrostructure disrupts the usual  ``disk-model''- ``disk geometry'' correlation,  especially as regard of  the assessment of the  accretion rate.  First important consequence of this  mix of different elements is in the possibility of episodic accretion phases,  with super-Eddington accretion  rates. This  distinctive feature can enter into the debate on the  \textbf{SMBHs}  origin, combining  however with   drying-feeding processes and screening effects. The already mentioned   possibility  of  structured proto-jet bundles and  possible evidences in the  \textbf{QPOs} analysis are other important fields of application.

On the other hand, from the observational  view-point, the need for such multiple systems is actually already stated in the literature,  for example in the  analysis of  screening  effects of  X-ray emission  supposed so far to be induced by some ``bubbles'' of material orbiting  between an accreting disk and its  central attractor  \citep{Marchesi,Gilli:2006zi,Marchesi:2017did,Masini:2016yhl,DeGraf:2014hna,Storchi-Bergmann}. Results of our investigation, therefore,  strongly advocate for  a framework shift in the  {screening X-ray emission  study},
which is here   traced back to the \emph{only} cases $\mathbf{\mathrm{(x)}}$ and $\mathbf{\mathrm{(xx)}}$.
More generally, the X-ray emission investigation can provide an accurate description of  the spectral  features  of the \textbf{RADs} structure. We  expect that the tori spacings ($\bar{\lambda}$) and the \textbf{RAD} knobby surface  would leave traces in a stratified emission spectra.
The  X-ray emission
from \textbf{AGNs} has been  variously  assumed to be  related to accretion disk instabilities and surrounding corona.
This spectra profile should provide  a fingerprint of the ringed disk structure,
 possibly as  a radially stratified emission profile.
The simplest structures of  this kind are thin radiating rings  \citep{Schee:2008fc,Schee:2013asiposs,S11etal}.
Future X-ray spectroscopy may reveal the  \textbf{BH} accretion ring models as relatively indistinct excesses
on top of the relativistically broadened spectral line  profile \cite{S11etal}, arising  in a well-confined
radial distance in the agglomerate-- \citep{ringed,open,dsystem}.
In  \cite{KS10}
extremal energy shifts of radiation from a ring near a rotating \textbf{BH} were particularly  studied:
radiation from a narrow circular ring  was proved to show a  double-horn profile with photons with
energy around the maximum or minimum of the  range.
This energy span of  spectral lines is a function of the observer's viewing angle, the black hole spin and the  ring radius.

 Eventually,   \textbf{RADs}  might represent   an  environment of the  episodic   accretion phases  advocated as  explanation of the  \textbf{SMBHs}   origin   from    (intermediate   or low mass) \textbf{BH} seeds.
Formation and evolution of \textbf{SMBHs}, especially at cosmological distances (redshift  $z\approx 6$), is still an open
topic in High Energy Astrophysics. One of the key issues is the identification of the different processes associated with the   \textbf{SMBHs}  origin with    very large masses \citep{apite1,apite2,apite3,Li:2012ts,Oka2017,Kawa}.
Note that recently another fundamental alternative of direct creation of \textbf{SMBHs} has been proposed in \cite{Stuchlik:2017qiz}, being based on gravitational instability of central region of the so called trapping relativistic polytropes that could model dark matter galactic halos  \citep{Stuchlik:2016xiq}.

It should be noted then, that the evaluation of the \textbf{SMBHs} spin is strictly  correlated with the ``masses-problem''.
The  assessment of the precise value of the spin parameter of the \textbf{BH}  is connected with  the evaluation of  the main features  of the \textbf{BH} accretion disk system,  as the \textbf{BH} accretion rate or the location of the inner edge of the accretion disk.
  Several processes have been proposed and analyzed:  for example,  \textbf{BHs}  characterized  by    long and continuous accretion episodes arising due to    merging, and   involving  a relevant spin-shift process (especially in elliptic galaxy), or   sequences of  small and random accretion
episodes occurred after different situations  as cloud accretion or also   tidal disruption of a  star  companion (especially in  spiral galaxies). These two different situations would, however, lead to two relatively different populations of   \textbf{BHs}  with different masses.
Collapse from stellar-mass
black holes, \textbf{BHs} mergers, accretion of some  gas-clouds  with low radiative efficiency,  are other proposed  evolutive patterns.
\textbf{SMBHs},  originated from  some  ``seeds'' ($10^4-10^2 M_{\odot}$)  with different evolutive patterns, generally depend on the seed initial mass and on  the \textbf{BH} environment, needing
 enough matter for accretion,  proper processes timescales, and  large initial angular momentum of the   accretion disks. An alternative then   consists of a
 succession of accretion episodes  from misaligned disks with  randoml \textbf{BH} spinning-up and spinning-down, 
or also a sequence of   turning-on and turning-off of  super-Eddington accretion phases interspersed with sub-Eddington phases\footnote{A super-Eddington phase may have  very low efficiency
in converting mass into radiation. We note that the efficiency of a
thick torus with $r_x \approx r_{mbo}$ is nearly zero (indeed in the \textbf{RAD} model such a torus  is subjected to special  boundary conditions). This is  a general relativistic effect, the   binding energy  decreases as the inner
boundary of the disk moves inwards inside the marginally stable
orbit towards $r_{mbo}$, where $E_{mbo}=1$.}.

\begin{figure}[h!]
\begin{tabular}{c}
\includegraphics[width=1\columnwidth]{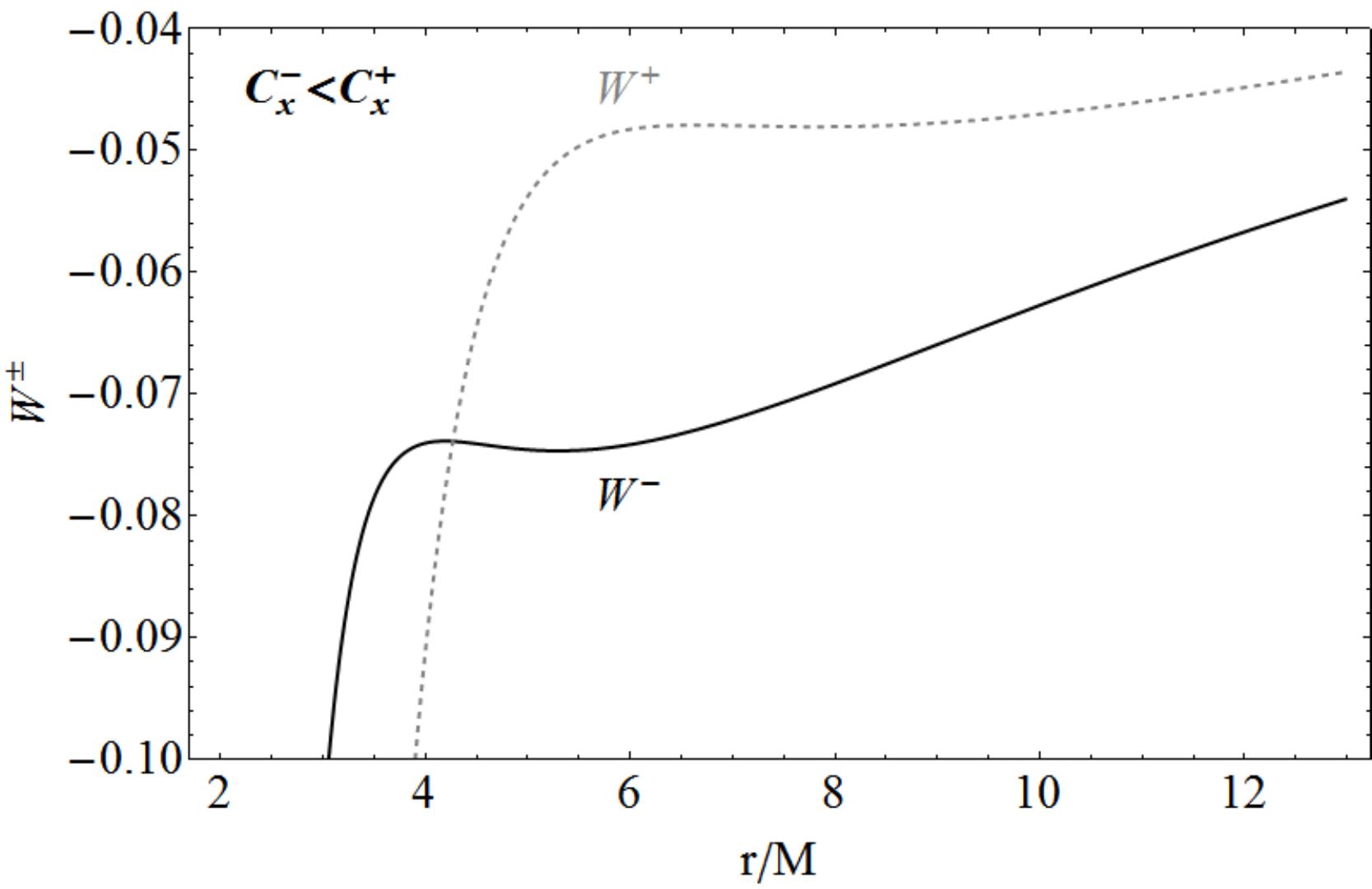}\\
\includegraphics[width=1\columnwidth]{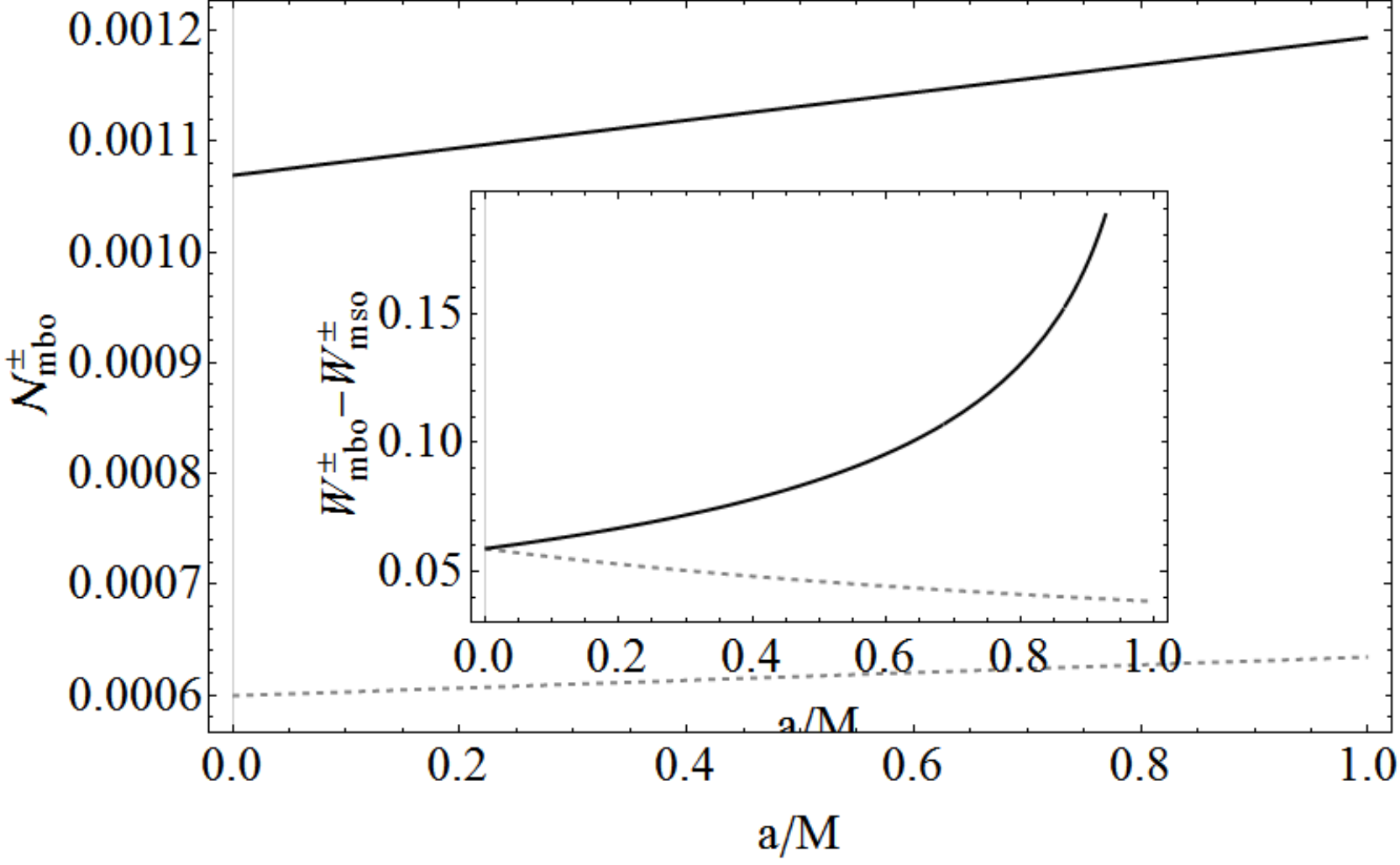}
\end{tabular}
\caption{\emph{Upper panel}: Paczynski-Wiita  (P-W) potential $W^{\pm}$ of Eq.\il(\ref{Eq:scond-d}) as function
  of $r/M$ ($W=\ln V_{eff}$), for the  tori of the couple $C_x^-<C_x^+$   of Fig.\il\ref{Fig:rssi-na} respectively,
  where  $a = 0.38 M$, $\ell^+_o = -3.99$ and $\ell^-_i = 3.32$. \emph{Bottom panel}:  $\mathcal{N}_{mbo}\equiv{r_x} (W_{mbo}-W_{x})^\kappa(\Omega_K(r_x))^{-1} $  as function of  $a/M$. $\Omega_K(r_x)$ is the the Keplerian angular velocity at the cusp $r_x$, $r_{mbo}$ is the marginally bounded orbit.  $\kappa=n+1=4$ has been fixed for an adiabatic fluid with polytropic index $4/3$.  \emph{Inside panel}: difference $(W^{\pm}_{mbo}-W^{\pm}_{mso})$  (maximum location of inner edge is  $r_x\lessapprox r_{mso}$), as function of $a/M$.  Quantities are evaluated at fixed $\ell^+_o $ and $\ell^-_i$. Dashed line is  for the accreting torus $C_x^+$, black line is for the  $C_x^-$ torus.
}
\label{Fig:cannlla}
\end{figure}

\begin{figure*}[ht]
\begin{center}
\begin{tabular}{ccc}
\includegraphics[width=.338\textwidth]{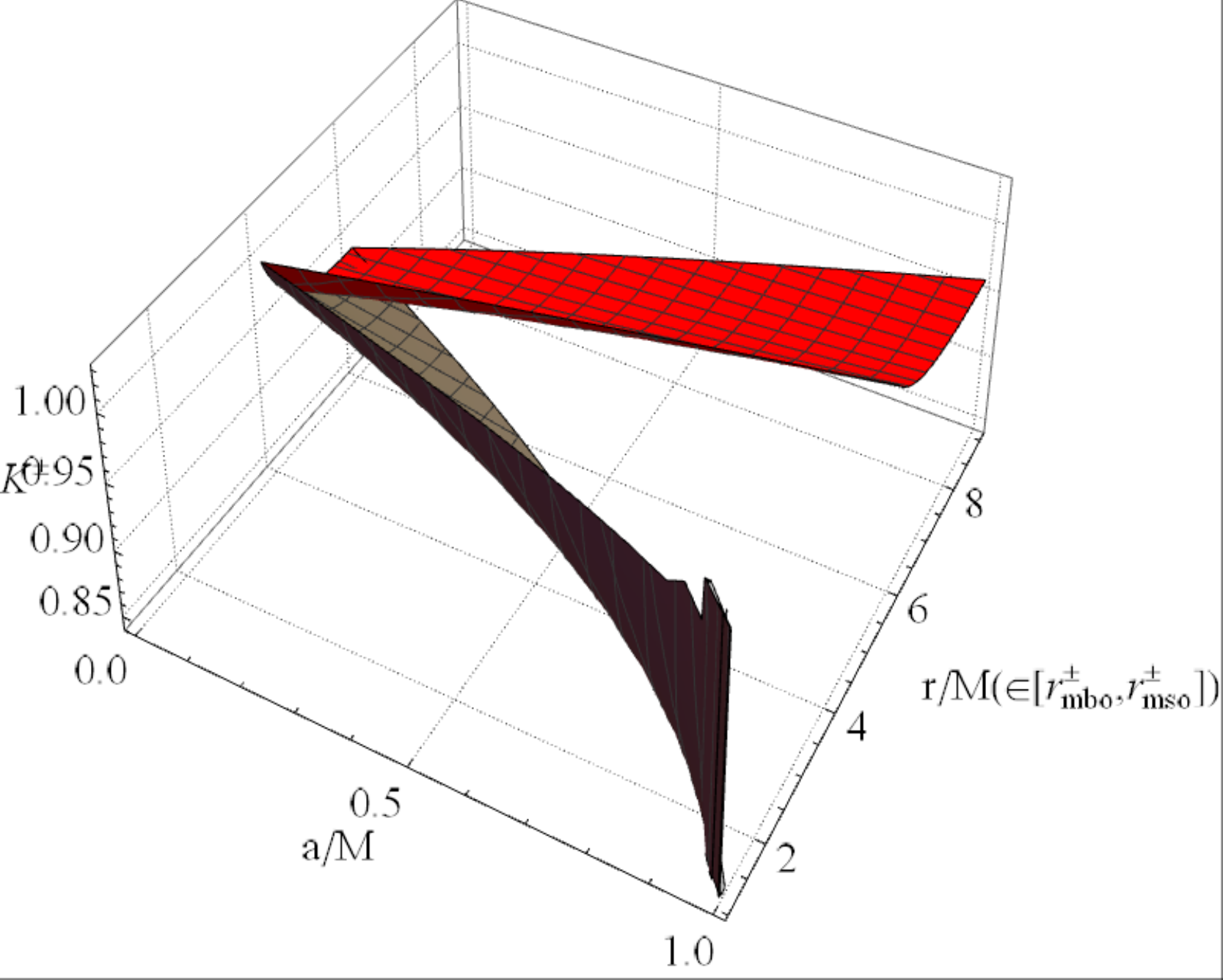}
 \includegraphics[width=.23\textwidth]{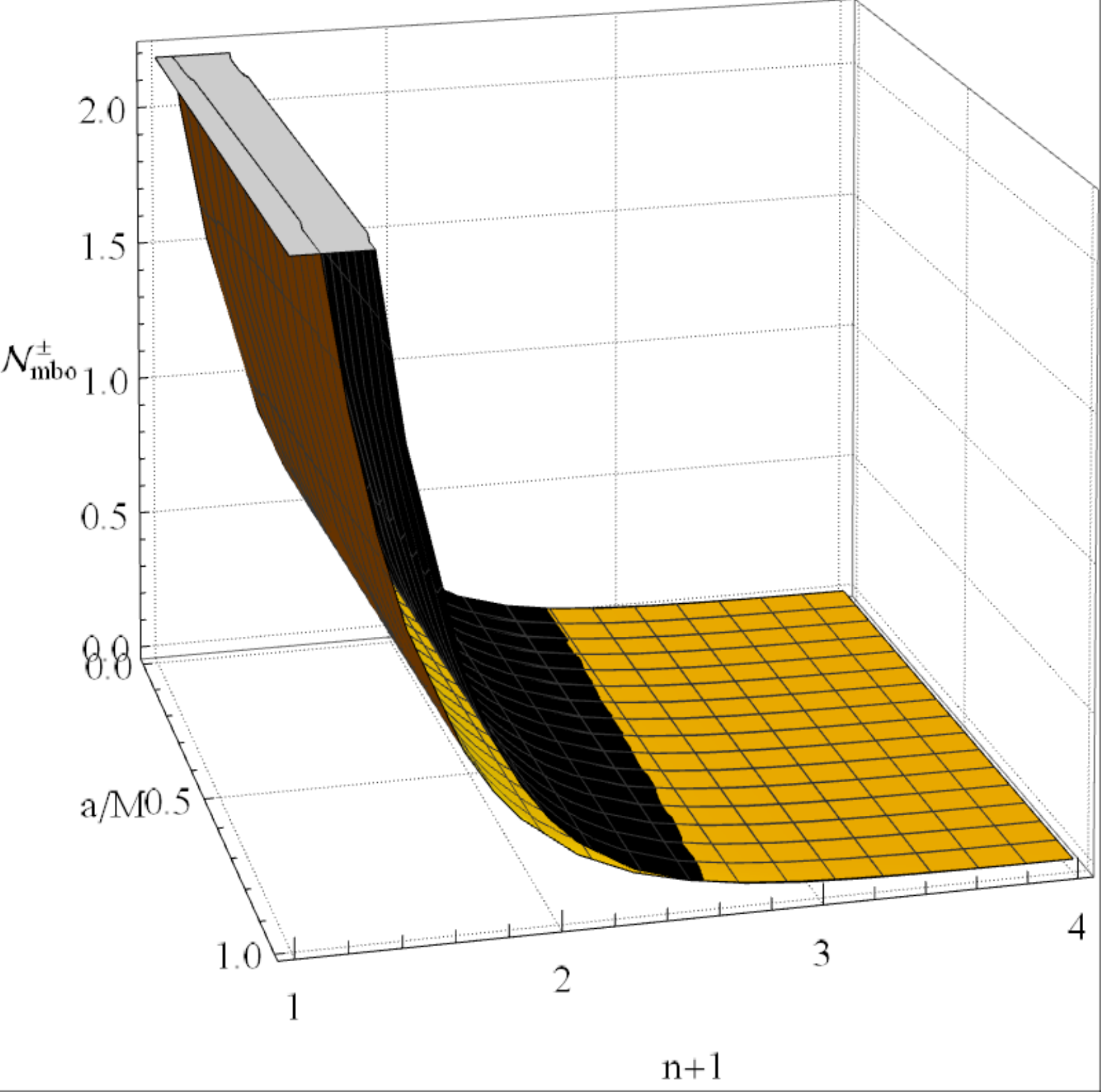}
 \includegraphics[width=.23\textwidth]{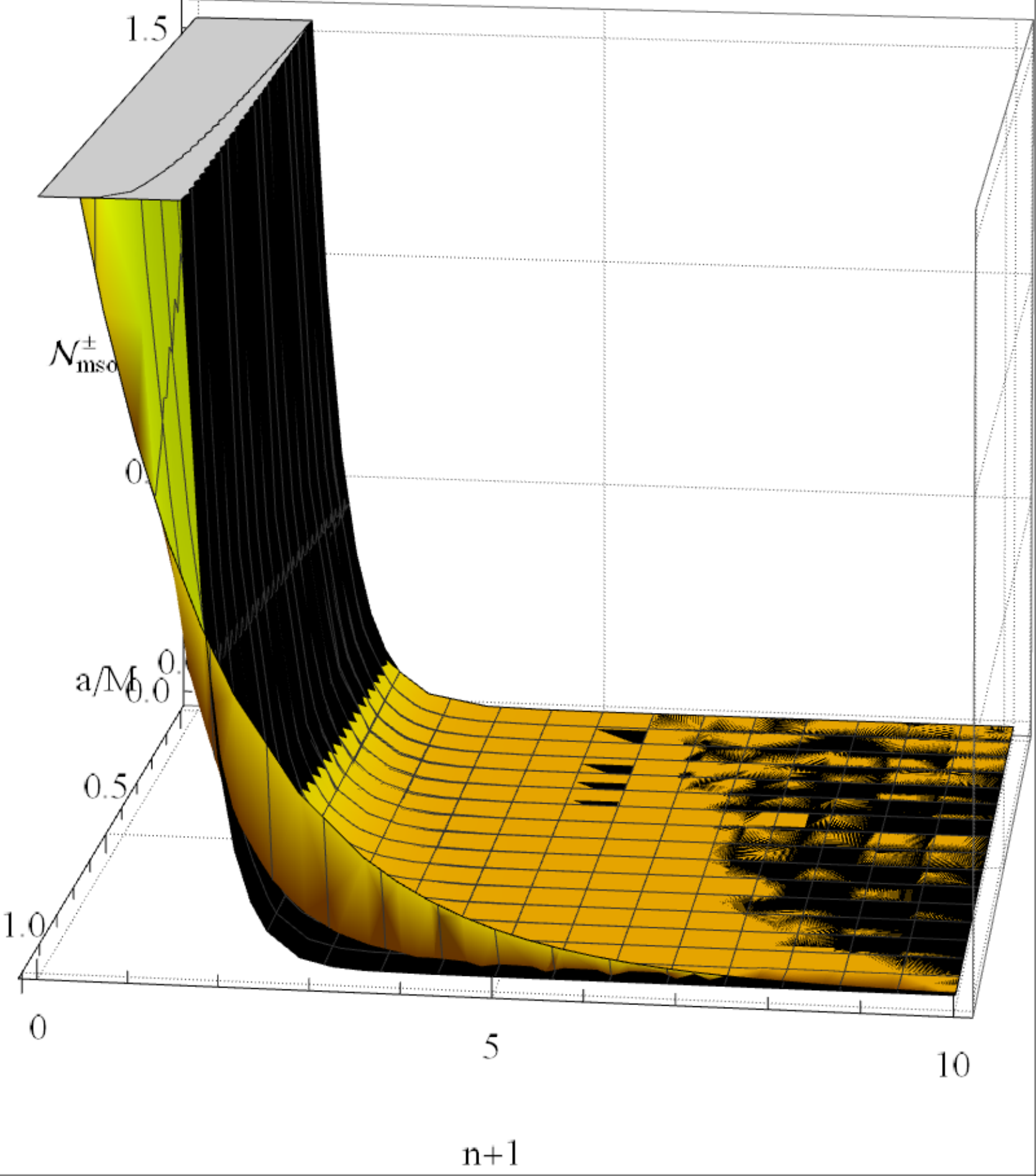}
\end{tabular} \caption{\emph{Left panel}: Parameter $K^{\pm}\equiv\exp{W^{\pm}}$   at the fluid critical  specific angular momentum $\ell^{\pm}$ for the counterrotating  and corotating fluids respectively, as function  of $a/M\in[0,1[$ and the radius $r/M\in[r_{mbo}^\pm,r_{mso}^{\pm}]$.  $W^{\pm}$ is the Paczynski-Wiita  (P-W) potentials.   $\mathcal{N}_{*}^{\pm}\equiv{r_*} (W^{\pm}_{mbo}-W^{\pm}_{*})^\kappa(\Omega_K(r^{\pm}_*))^{-1} $   at  $r_*=r_{x}^{\pm}$ (\emph{Center panel})  and $r_*=r_{mso}^{\pm}$  (\emph{Right panel})  as function of  $a/M$ and $\kappa=n+1$,  where $\gamma=1/n+1$ is the polytropic index, for
   corotating (yellow surfaces)   and for counterrotating (black surfaces)  fluids. $\Omega_K$ is  the Keplerian angular velocity, $r_x$ is the cusp, $r_{mbo}$ is the marginally bounded orbit.  The maximum location of inner edge is  $r_x\lessapprox r_{mso}$.  Quantities are evaluated at fixed  $\ell^+_o = -3.99$ and $\ell^-_i = 3.32$--see also Fig.\il\ref{Fig:cannlla}. }
\label{Fig:canntaz}
\end{center}
\end{figure*}

The \textbf{RAD} agglomerates, due to their stratified inner structure,   are a source  of    episodic accretions,
 combined  with the effects of tori collisions, accretion obscuration and  drying-feeding processes. The macrostructure stands then  as promising  arena of investigation for \textbf{SMBH} formation, following accretion from multi-disks of  $\mathbf{\mathrm{(x)}}$ and $\mathbf{\mathrm{(xx)}}$ configurations.
Efficiency of the \textbf{RAD} and its luminosity are not uniquely determined  by the inner accreting disk, in fact  our  investigation  shows here evidences that the aggregate structure falsifies this hypothesis, by considering  the possibility, albeit restricted to cases $\mathbf{\mathrm{(x)}}$  and $\mathbf{\mathrm{(xx)}}$, of screening effects and alternated phases of accretion and  collision.
Many of the   mentioned aggregate characteristics can be easily evaluated under the hypothesis of the thick torus components--see for example \cite{Abramowicz:2004vi,StraubS,AJS78,ottes,otte0,Abramowicz:2004vi,AbraFra}.    We can provide  a quantification of the key parameters   in a special case of \textbf{RAD}. This will allow us also to evaluate some trends and determine special  aspects of problem space-scales.
Considering   polytropic fluids with pressure $p=\kappa \varrho^{1+1/n}$,
 we can evaluate many  of the  \textbf{RAD}   characteristics, as the  \textbf{RAD} thickness $(h)$ and elongations $(\lambda)$, tori spacings $(\bar{\lambda})$, through an assessment of the  $ K $-parameter only \citep{ringed}.
 For a single torus, calculations of its polytropic structure can be found in \cite{Stuchlik:2009jv}.
 Furthermore,   %
we can also provide an estimate of  the
mass-flux,  enthalpy-flux (evaluating also the temperature parameter),
and  the flux thickness--see for example  \cite{otte0}. All these quantities  have form   $\mathcal{O}(r_x,r_s,n)=q(n,\kappa)(W_s-W_{x})^{d(n)}$, where $q(n,\kappa)$ and $d(n)$ are different functions of the polytropic index\footnote{More precisely we can say that
$\mathrm{{{Enthalpy-flux}}}=\mathcal{D}(n,\kappa) (W_s-W)^{n+3/2}$,
$\mathrm{{{Mass-Flux}}}= \mathcal{C}(n,\kappa) (W_s-W)^{n+1/2}$, while
$\mathcal{L}_{x}/\mathcal{L}= \mathcal{B}/\mathcal{A} (W_s-W_{x})/(\eta c^2)$  stands for  the  fraction of energy produced inside the flow and not radiated through the surface but swallowed by central \textbf{BH}. Efficiency
$\eta\equiv \mathcal{L}/\dot{M}c^2$,    $\mathcal{L}$ representing the total luminosity, $\dot{M}$ the total accretion rate where, for a stationary flow, $\dot{M}=\dot{M}_x$         \citep{otte0}.},
  $W=\ln V_{eff}$ is Paczynski-Wiita  (P-W) potential  of Eq.\il(\ref{Eq:scond-d}),
$W_s\geq W_{x}$ is the value of the equipotential surface, which is taken with respect to the asymptotic value.
Consequently, the   determinant parameter, in this analysis is  $K:\, W_*=\ln K_*$ for any radius $r_*$. Therefore, as the cusp approaches the limiting radius  $r_{mbo}$,  the potential  $W_x\approx0$,  which is the limiting asymptotic value for very large  $r$.
The mass flow rate through the cusp (mass loss, accretion rates)  $\dot{M}_x$  and the cusp luminosity $\mathcal{L}_x$ (and the accretion efficiency $\eta$),
 measuring the
rate the thermal-energy  is  carried at cusp,
 are\footnote{
There is  $\mathcal{L}_{x}={\mathcal{B}(n,\kappa) r_{x} (W_s-W_{x})^{n+2}}/{\Omega_K(r_{x})}$, accretion rate for the disk is   $\dot{m}= \dot{M}/\dot{M}_{Edd}$,
  while  $\dot{M}_{x}={\mathcal{A}(n,\kappa) r_{x} (W_s-W_{x})^{n+1}}/{\Omega_K(r_{x})}$  \citep{otte0}.
} as $\mathcal{P}=\mathcal{O}(r_x,r_s,n) r_x/\Omega_K(r_x)$.
In fact the relativistic frequency $\Omega$  reduces  to the Keplerian  value $\Omega_K$ at the edges of the accretion torus, where  the pressure forces   vanish--see also Figs \il\ref{Fig:cannlla} and \ref{Fig:canntaz}.

 A throughout investigation of these quantities  for the \textbf{RAD} couples,  considered here, is planned for a future analysis,  in connection  with the  analysis of the \textbf{SMBH} accretion rates. We provide here some general considerations for a $\mathbf{\mathrm{(xx)}}$ couple.
For this purpose, we can consider the couple $\mathbf{\mathrm{(xx)}}$ or  $C_x^-<C_x^+<C^-$ of Fig.\il\ref{Fig:rssi-na},
where  there is  ($a = 0.38 M$, $\ell^-_o= 3.99$,  $\ell^+_o = -3.99$, $\ell^-_i = 3.32$). We do not consider the outer corotating non-accreting torus $C^-$ of the triple system.
 Notice that in this triple  $\ell$counterrotating system, the outer  $C^-$ disk  is not constrained to a  higher  height with respect to  the    internal disks, or any equilibrium disk or an $\ell$counterrotating accreting couple has no special {constraints} on the relative height of the tori. This obviously implies a very wide set of  possibilities for  a knobby
 \textbf{RAD} disk. Then, we can evaluate the  tori center ($r_{cent}$), the cusp location $(r_x)$, the outer margin $(r_{out})$, the torus  elongation on the equatorial plane $(\lambda)$, the torus height $(h)$, and the tori spacings  $(\bar{\lambda})$ as follows:
  for the counterrotating  torus, $C_x^+$, there is
($r_{cent}^+= 7.8195M$,
$r_x^+=6.645M$,
$r_{out}^+= 8.60278 M$,
$\lambda^+= 1.95744M$,
$h^+/2= 0.267283M$),
for the corotating torus, $C_x^-$, there is
($r_{cent}^-= 5.29868M$,
$r_{x}^-= 4.1907 M$,
$r_{out}^-= 6.18481M$,
 $\lambda^-= 1.99412 M$,
$ h^-/2=0.364584 M$).
In this specific case,  the  \textbf{RAD}  aggregates a set of small tori. Nevertheless, a $\mathbf{\mathrm{(xx)}}$ couple can be observable  orbiting any  Kerr \textbf{BH}, and the larger is the \textbf{BH} dimensionless spin the bigger the tori can grow,  while a larger spacing is required.  An immediate evaluation  shows  that the maximum spacing  possible for  a double accreting tori is  $\bar{\lambda}_{\max}\lesssim 8M$, in the case of an extreme Kerr \textbf{BH} where  $a\approx M$. Spacing for the   $\mathbf{\mathrm{(xx)}}$ couple  of  Fig.\il\ref{Fig:rssi-na} is $\bar{\lambda}\equiv r_x^{o}-r_{out}^i=0.460526M$  (Note that the space-scales are in units of \textbf{SMBH} masses)-- \citep{ringed}. To realize the significance of this data  we should note that the  spacing parameter $\bar{\lambda}$ is of essential  importance for  determination of   tori  collision, the aggregate oscillation and in the analysis of the  stratified \textbf{RAD}  X-ray emission spectra.  We also recall that only  this case $\mathbf{\mathrm{(xx)}}$  screening effects are only possible for  the corotating tori as  follows $C_x^-<C^-<...<C_x^+<C^{\pm}$.
In the next session, we summarize the final methodological considerations and future perspectives of this work.

\section{Conclusions}\label{Sec:Conclusions}
The systems  investigated here  offer    several   methodological  and observational challenges.  Describing a set of virtually separated  tori orbiting one attractor  as an entire configuration, requires a certain number of assumptions.
We distinguish three periods of \textbf{BH}-ringed accretion disk  life: the first featuring tori formations, a second facing the accretion of one or two tori onto the central \textbf{BH} and the eventually emerging of tori collisions.
Picturing these situations is clearly a risky and complex task.

From  observational view point
 we believe our results may   be  of significance for the High Energy  Astrophysics:  these multi-configurations  may be at the root of phenomena eventually  detectable  by the planed X-ray observatory ATHENA,  such as  the
shape of X-ray emission spectra,  the X-ray obscuration and absorption
by one of the ring, and the extremely energetic  radiative phenomena in quasars  and  \textbf{AGNs}.
The   phenomenology associated with these systems   may   be  very wide. We note that the presence of such structures is capable to substantially modify    the    single disk  scenario, which has been effectively taken  so far  as the common  ground  of the High Energy  Astrophysics connected with the accretion onto \textbf{BHs}.
Explanation of some of the most intriguing and unveiled issues of \textbf{BH} physics interacting with matter may be reset in this new framework. The single torus paradigma would be then  just seen a limit or special case related to an  evolutive  phase of \textbf{BHs} life in their  Host.
The existence of these objects clearly  opens an incredible amount of possibilities to be investigated.
As pointed in \cite{ringed,dsystem},  tori interactions or  oscillations can be associated to a variety of phenomena with relevant energy release.
 The radially oscillating tori of the ringed disk could be related to the high-frequency quasi periodic oscillations observed in non-thermal X-ray emission from compact objects (QPOs), a still obscure feature of the  X-ray astronomy related to the inner parts of the disk.
 The presence of an inner tori may also enter as a new unexpected ingredient in the accretion-jet puzzle, as proposed also in   \cite{S11etal,KS10,Schee:2008fc}.

There are evidences suggesting what  these structures may play a major role in Galaxy dynamics and particularly in \textbf{AGNs}.
In fact, there are studies in  support of  the existence  of \textbf{SMBHs}  characterized by  multi-accretion episodes  during their life-time in Galaxy cores.
Consequently \textbf{SMBHs} life may report  traces  of  their  host Galaxy dynamics as a diversified feeding   of  \textbf{SMBHs}. These processes may involve for example
repeated galaxy mergers
or also interacting
binary \textbf{BHs}, X-ray binaries or \textbf{SMBHs} binary systems.
As consequence of these activities, matter around  attractor could find an equilibrium configuration as
counterrotating  and    misaligned disks-- \citep{Aly:2015vqa,Dogan:2015ida}.

In this analysis we specialize our investigation on  sequences of   toroidal axi-symmetric (ringed accretion disks) configurations  orbiting  in  the equatorial plane of  a  central  Kerr black hole.
From methodological view point, the ringed disks evolutions should arise    from the evolution of    each torus.
Tori in ringed disk   may  collide and merge,  or, eventually
the accreting matter from the outer torus of the couple can  impact on the  inner torus, or    the outer torus may be inactive  with  an active inner torus  accreting  onto the \textbf{BH}, or both tori  may be active.  Our analysis shows the occurrence of these situations  is strictly constrained.
We discussed also  the emergence of  the instability phases for  each torus, identifying   classes  of  central  Kerr attractors   in dependence of   their dimensionless spin.
   Existence and evolution of these structures strongly depends   on the black hole dimensionless spin,  and the relative rotation of the fluids. This aspect has important implications  on the possible observational  effects    providing  a perspective on the  phenomena  emerging from  their dynamics, isolating those situations where actually  these configurations may be chased.
{Finally, the analysis carried out here reduces the range of possibilities in the description of the several possible  \textbf{RAD} configurations to  determination of the $2n$ parameters, where  $n$ is the number of tori in the agglomerate, the  \textbf{RAD} ``order''. The (non-constrained) parameters are the fluid specific angular momentum  $\ell$, and the  $K$-parameter which is related to the torus   density and   morphology.
As explained in Section (\ref{Sec:P-O-E-RADs}), this parameter is directly connected to some of the main phenomenological aspects might be associated with \textbf{RAD}, as the  X-ray emission screening and  the spacings $\bar{\lambda}$, that would eventually emerge in an emission containing a fingerprint of the \textbf{RAD} stratified structure. The $K$-parameter  enters moreover in the evaluation  of the \textbf{SMBH} accretion rate.}
 Accretion or collision constitute possible scenario for the entire ringed  disk instability. We feature the constraints for the emergence of these situations, foreseen in the  occurrence of these disruptive phenomena.
These results can then be used in  any numerical analysis of more complex situations,  sharing  the same symmetry of one at last disk to set up the initial data configurations, as it is generally adopted in many GR-HD or GR-MHD  dynamical integrations for the single accretion disk case.

\appendix
\section{Location of the notable radii $r_{\mathcal{N}}$ in the accretion disks: the location of the inner edge of the disk}\label{App:location}
In this section   we provide  proof of the some  assertions used  in  Sec.\il(\ref{Sec:criticalII}) and  (\ref{Sec:ell-cont-double}), and a more general  discussion of  some results.

We   address the  issue of  the  location of the  notable radii  $r_{\mathcal{N}}\equiv \{r_{\gamma}^{\pm}, r_{mbo}^{\pm},r_{mso}^{\pm}\}$,  defining the geodesic structure of the Kerr spacetime, with respect to  the matter distribution $()_{\pm}$   with momentum in the range  $\mathbf{Li}^{\pm}$.

This is in fact important particularly in the determination of a possible correlation  between  the $\ell$corotating configurations.
We discuss  the   case of $\ell$corotating matter,
 investigating the inclusion of  $r_{\mathcal{N}}^{\pm} \in ()^{\pm}$  respectively in \ref{Sec:usua-D-mbo-con} and \ref{Sec:graph-def}, and of  $r_{\mathcal{N}}^{\pm} \in ()^{\mp}$  respectively in  \ref{Sec:coorr}.

It is worth noting here that this  investigation   actually matches  the  broader  problematic of    the location of
the inner edge of the disk.
Indeed, this investigation will often imply, especially for  the inclusion
$r_{\mathcal{N}}^{\pm} \in ()_{\pm}$, a  discussion of the  location   of  these radii with respect to the inner  margin of the disk, while the location of the outer edge turns to be  important especially for the discussion of the  $r_{\mathcal{N}}^{\pm} \in ()_{\mp}$ case.
This  analysis will eventually turn in a set of   constraints on the    parameters $\ell$ and $K$.

We expect  that  considerations traced here  could be applicable  also for more general   models where
the specific angular  momentum is   not constant along the disk \footnote{For example it could be  $\ell=\ell(r,\alpha_i)$, where $\alpha_i$ is for a  set for  parameters,  \citep{Lei:2008ui}.}.
\subsection{Location of the notable radii $r_{\mathcal{N}}^{\pm} \in ()_{\pm}$}\label{Sec:possi-conc}
In the following, we discuss the location of the marginally bounded orbit $r_{mbo}^{\pm}\in ()_{\pm}$ in \ref{Sec:usua-D-mbo-con}. Location of the marginally stable orbits $r_{mso}^{\pm}\in ()_{\pm}$  is analyzed in \ref{Sec:graph-def}, whereas the location of marginally circular orbit $r_{\gamma}^{\pm}\in ()_{\pm}$ is considered in \ref{Sec:photon-C}.
\subsubsection{Location of the marginally bounded orbits $r_{mbo}^{\pm}$}\label{Sec:usua-D-mbo-con}
We will prove   that the marginally bounded orbits $r_{mbo}^{\pm}$  is \emph{not} included in any disk, for any specific angular  momentum in the range  $\mathbf{Li}^{\pm}$, neither in the  equilibrium configurations $\cc^{\pm}_i$ (with $\ell^{\pm}_i\in \{\mathbf{L1}^{\pm}, \mathbf{L2}^{\pm}, \mathbf{L3}^{\pm}\}$) or in accretion  $\cc_x^{i_{\pm}}$ (with $\ell_i^{\pm} \in \mathbf{L1}^{\pm}$).

There  is indeed  $r_{mbo}^{\pm}>r_J^{\pm}$ for  $\ell^{\pm} \in \mathbf{L2}^{\pm}$,  while  for the  open cusped configuration  $\oo_x$ with  specific angular  momentum $\ell=\ell_{mbo}$ we have   $K_{mbo}=1$ and critical point in located in $r_{J}=r_{mbo}$.

We will assume, for  every  $\ell$, the radial  function $V_{eff}$ to be  well defined in $r_{mbo}$ and in all the  orbital regions considered in this analysis. On the other side, if  the effective potential is not well defined in $r_{mbo}$,  as indeed it is  in  some orbital regions  for  $\ell \in\mathbf{L3}$, this  is sufficient to  prove   that the marginally bounded orbit can not belong to any configuration regulated by that effective potential.
\subsubsubsection{\textbf{Configuration $\cc_3^{\pm}$:} $\ell_3\in \mathbf{L3}$}
We start by observing  that given  a radius $\bar{r}$ located in the orbital range   where there is  $\left.\partial_{|\ell|} V_{eff}\right|_{r}>0$ and  $V_{eff}^2>0$  is well defined then,  being $\ell_2 <\ell_3$ for  any $\ell_2\in \mathbf{L2}$ and $\ell_3\in \mathbf{L3}$, we have  $\left. V_{eff}(\ell_3)\right|_{\bar{r}}>\left. V_{eff}(\ell_2)\right|_{\bar{r}}$.

But since  $\ell_{mbo}=\inf{\mathbf{L2}}=\min{\mathbf{L2}}\leq\ell_2$,  we have $\left. V_{eff}(\ell_3)\right|_{\bar{r}}>\left. V_{eff}(\ell_{mbo})\right|_{\bar{r}}$, then $\left. V_{eff}(\ell_{3})\right|_{{r}_{mbo}}>\left. V_{eff}(\ell_{mbo})\right|_{r_{mbo}}=1$, and if \footnote{Note that we are using, along this work,  the convention introduced in the end of Sec.\il(\ref{Sec:Kerr-2-Disk}): in general,  the  label $(i)$ with $i\in\{1,2,3\}$ indicate  any  quantity $\mathbf{Q}$ relative to the range  of specific angular momentum $\mathbf{Li}$ respectively, thus in this case $r_{in}^3$
is the inner margin of the  regular  configuration with specific angular momentum  $\ell_3\in\mathbf{L3}$.} $r^3_{in}=r_{mbo}$ there is  $K_3=1$ (here in the following we adopt the notation $r^\mathbf{i}$ where $\mathbf{i}\in\{1,2,3\}$ according to the related fluid specific angular momentum $\ell^{\mathbf{i}}$ respectively).

In fact, it cannot be  $r^3_{in}<r_{mbo}$  and therefore $r_{mbo}\in \cc_3$, for this to happen, it should be $K_3>{1}$, as the effective potential increases at $r<r_{in}$ without a maximum point.

Therefore we can conclude  that:
\be\label{Eq:choral}
r^{3_{\pm}}_{in}>r_{mbo}^{\pm}\quad \mbox{implying}\quad r_{mbo}^{\pm}\non{\in} \cc^{\pm}_3.
 \ee
 We note that these arguments are quite  independent from the  corotating  or counterrotating nature of the fluid, but depend mainly  on the variation range   of  the momentum magnitude.
In fact  $\mathbf{L3}$ is the  range  of higher specific angular  momentum  magnitude, and one could assume that the  distinction in the geodesic structure of the  Kerr spacetime between the $\ell$counterrottaing fluids is higher (as essentially determined by the ratio $\ell/a$).
However, it should be noted that for  $\ell\in \mathbf{L3}$,   the centers of the equilibrium disks are placed in   a orbital region  rather distant from the attractor, namely in $r>\bar{r}_{\gamma}\gg r_{mso}$, where $\bar{r}_{\gamma}> r_{\gamma}:\; \ell(\bar{r}_{\gamma})=\ell_{\gamma}$, in the region  $R=r/a\gg0$, with the exception of the  corotating configurations  where, for $a\in[a_{\mathcal{M}}^-,M[$, we find  $\bar{r}_{\gamma}^-\in[ r_{\mathcal{M}}^-,r_{mso}^-[$, see Fig.\il\ref{Fig:WayveShow} and discussion  in Sec.\il(\ref{Sec:A-J}).

\subsubsubsection{\textbf{Configurations $\cc^{\pm}_2$ and $\cc_x^{2_{\pm}}$:} $\ell_2\in \mathbf{L2}$}

  The situation in $\mathbf{L2}$ is as  follows: for $\ell_{mbo}=\inf{\mathbf{L2}}=\min{\mathbf{L2}}$ (where $\sup{\mathbf{L2}}=\ell_{\gamma}$), and $r_{mbo}=\sup{r_{Max}^2}=\max{r_{Max}^2}$ (where $\inf{r_{Max}^2}=r_{\gamma}$), similarly to the  argumentation for the $()_3$ configurations, we consider a specific angular momentum  $\ell_2:\; \ell_{mbo}<\ell_2\in \mathbf{L2}$, thus  $\left. V_{eff}(\ell_2)\right|_{\bar{r}}>\left. V_{eff}(\ell_{mbo})\right|_{\bar{r}}$.
In particular,
   $K_2=\left. V_{eff}(\ell_2)\right|_{r_{mbo}}>\left. V_{eff}(\ell_{mbo})\right|_{r_{mbo}}=1$.

   The unstable phase, expected for the equilibrium disks  $\cc_2$ with specific angular momentum in  $\mathbf{L2}$, is  the open  cusped configuration $\oo_x$.
If the inner edge $r_{in}^2$ of the disk $\cc_2$ in equilibrium would be  $r_{mbo}$ for a  $\ell_2\neq\ell_{mbo}$ then there is  $K_2>1$ (at $\ell_2=\ell_{mbo}$ the disk is open  but this is  a special, unstable   case).
 On the other hand, if  $r_{mbo}\in \cc_2$, that is $r_{mbo}<r_{in}^2$, then the following considerations apply: for  $\ell_2 \in \mathbf{L2}\; \exists\; r_{Max}^2\in]r_{\gamma},r_{mbo}]:\; K_{Max}^2\geq1$  where $r_{mbo}=r_{Max}^2$ only if $\ell_2=\ell_{mbo}$.

  Therefore, at $\ell_2$,   $r_{Max}^2$ being a maximum of the effective potential, the function $V_{eff}$  is increasing in $]r_{Max}^2,r_{mbo}]$ and, being  $\ell_2>\ell_{mbo}$,  we have $V_{eff}(\ell_2,r_{mbo})>V_{eff}(\ell_{mbo},r_{mbo})=1 $.  As a consequence of this, for the equilibrium closed $\cc_2$ disk, there is $K_2\neq V_{eff}(\ell_2,r_{mbo})$  and as the minimum point is $r^2_{min}>r_{mso}>r_{mbo}$, then   we have $K_2=V_{eff}(\ell_2,r_{in})=V_{eff}(\ell_2,r_{out})<1$, where $r_{mbo}<r_{in}^2<r_{min}<r_{out}^2$,  and  therefore $r_{mbo}<r_{in}^2$.

We  finally conclude that
\bea&&\nonumber
r^{2_{\pm}}_{in}>r_{mbo}^{\pm}\quad \mbox{implying}\quad r_{mbo}^{\pm}\non{\in} \cc^{\pm}_2,\quad \mbox{and}
\\\label{Eq:beha-dire}
&&  \quad r_{J}^{\pm}\leq r_{mbo}^{\pm}\quad\mbox{for } \quad \oo_x^{2_{\pm}}\quad\mbox{thus}\quad r_{mbo}^{\pm}{\in} \oo_x^{2_{\pm}}.
 \eea
\subsubsubsection{\textbf{Configurations $\cc_1^{\pm}$ and $\cc_x^{1_{\pm}}$: }$\ell_1\in \mathbf{L1}$}

For the rings with specific angular momentum  $\ell_1 \in \mathbf{L1}$ we will repeat the argument  used for the $()_2$  configurations.

 The unstable configurations for the disks within this data set has topology $\cc_x$. Let $\ell_1\in \mathbf{L1}$, then $\ell_1<\ell_{mbo}=\sup{\mathbf{L1}}$.
Thus in particular there is  $V_{eff}(\ell_1, \bar{r})<V_{eff}(\ell_{mbo},\bar{r})$.   In general one could say that $V_{eff}(\ell_1, r_{mbo})<V_{eff}(\ell_{mbo},r_{mbo})=1$, which does not solve the problem. However as it is  $r_{Max}^1\in]r_{mso},r_{mbo}[$, then  $\partial_r V_{eff}<0$ for $r\in]r_{Max}^1,r_{mbo}[$.  This implies that $r_{mbo}\non{\in}\cc_x^1$, or $r_{mbo}\non{\in}\cc_1$.

One can say that since the  maximum disk orbital extension   occurs  for the  critical configuration $\cc_x^{1}$, then it is sufficient  to say that for no $\ell_1\in \mathbf{L1}$  there is  $r_{Max}^{1}<r_{mbo}$, but  $r_{Max}^{1}\in[r_{mso},r_{mbo}[$. Then it follows that  $r_{mbo}$ cannot belong to any topology associated to the range $\mathbf{L1}$.

We  therefore conclude that
\bea&&\nonumber
r^{1_{\pm}}_{in}>r_{mbo}^{\pm}\quad \mbox{implying}\quad r_{mbo}^{\pm}\non{\in} \cc^{\pm}_1,\quad \mbox{and}
		\\\label{Eq:the-ag}
		&& r^{\pm}_{x}\geq r^{\pm}_{mbo}\quad \mbox{for} \quad \cc_x^{1_{\pm}},\quad\mbox{thus}\quad r_{mbo}^{\pm}\non{\in} \cc_x^{1_{\pm}}.
 \eea
\subsubsection{Location of the marginally stable orbits $r_{mso}$}\label{Sec:graph-def}
An unstable configuration, according to P-W mechanism, \emph{must} contain the marginally stable orbit,  or $r_{mso}\in ()_x$. Therefore
\bea\nonumber
&& r^1_x> r_{mso} \quad r_{mso}\in \cc_x^1\quad\mbox{for}\quad \ell_1\in \mathbf{L1};\quad  r^2_J> r_{mso} \\\label{Eq:comb-with-simm}
 && r_{mso}\in \oo_x^2 \quad\mbox{for}\quad \ell_2\in \mathbf{L2}.
\eea
It remains to  establish the  condition for the marginally stable circular orbit to be contained in a disk of $C$ topology, corresponding to a surface   in equilibrium\footnote{In fact,  as there is  always $r_{min}>r_{mso}$,
   it is possible to select a  value of the $K$ parameter  small enough (i.e. $K=K_{mso}+\epsilon_K$ with $\epsilon_K\gtrapprox0$) for  $r_{mso}<r_{in}$, and consequently   there is    $r_{mso}\non{\in}C$.
  However, if a  disk with $\ell\in\{\mathbf{L1},\mathbf{L2}\}$, admits critical configurations ($r_{Max} <r_{mso}$), then
  $r_{mso}\in \cc_x$ or  $r_{mso}\in \oo_x$. The potential function is monotonically decreasing  in the region $]r_{Max}, r_{mso}[$, or   in a sufficiently narrower left region  of  the orbit $r_{mso}$, for  specific angular momentum  $\ell\in\mathbf{L3}$ where the effective potential  admits no  maximum.
   However, this does not ensure that $V_{eff}(r_{mso})\leq 1$,
  and therefore this condition  does not ensure that  the orbit $r_{mso}$ is  included in the closed disk.
   Essentially, this condition depends on the  specific angular  momentum and also on the location  of  $r_{mbo}$. Moreover, in some cases for
  $ \ell \in\mathbf{L3} $,  it is necessary to assess whether the effective potential is actually well defined.}, considering  the specific angular  momentum   for $\ell\in \mathbf{Li}$.
\subsubsubsection{\textbf{Configurations $\cc_1^{\pm}$:} $\ell_1\in \mathbf{L1}$}

There is :
\bea&&\nonumber
\forall \ell_{1}\in \mathbf{L1}\quad(\mbox{where}\quad r_{min}^1>r_{mso}>r_{Max}^1>r_{mbo}) \quad\mbox{there is}\\
&& \label{Eq:dis-hadr}K_{mso}<V_{eff}(\ell_1, r_{mso})< V_{eff}(\ell_1, r_{Max}^1)<1.
\eea
The first inequality of Eq.\il(\ref{Eq:dis-hadr}) is due to the fact that $\ell_{mso}=\inf{\mathbf{L1}}$ (and the effective potential is in general an increasing function of the specific angular  momentum magnitude   \citep{pugtot}). The second inequality is a consequence of the relation  $\partial_r V_{eff}<0$ in $]r_{Max}^+,r_{mso}]$.  The third and last inequalities \emph{show} that, for any $\ell_1 \in \mathbf{L1}$, there is $K_1=V_{eff}(\ell_1, r_{mso})<1:\; r_{in}^1=r_{mso}$, which constitutes the result of this paragraph.

the inner edge of the equilibrium disk $\cc_1$ is located \emph{on} the stable orbit  $r_{mso}$--Fig.\il\ref{Figs:SupLa}.
Thus,  it is possible to select a set of parameters  $K_1\in]V_{eff}(\ell_1, r_{mso}), K_{Max}[$, for  $r_{mso}\in \cc_1$. This  range of parameters increases as the range $r^1_{Max}-r_{mbo}$ decreases,  along with  the range   $|\ell_{mbo}-\ell_1|$.

\begin{figure}[h!]
\begin{center}
\begin{tabular}{cc}
\includegraphics[width=1\columnwidth]{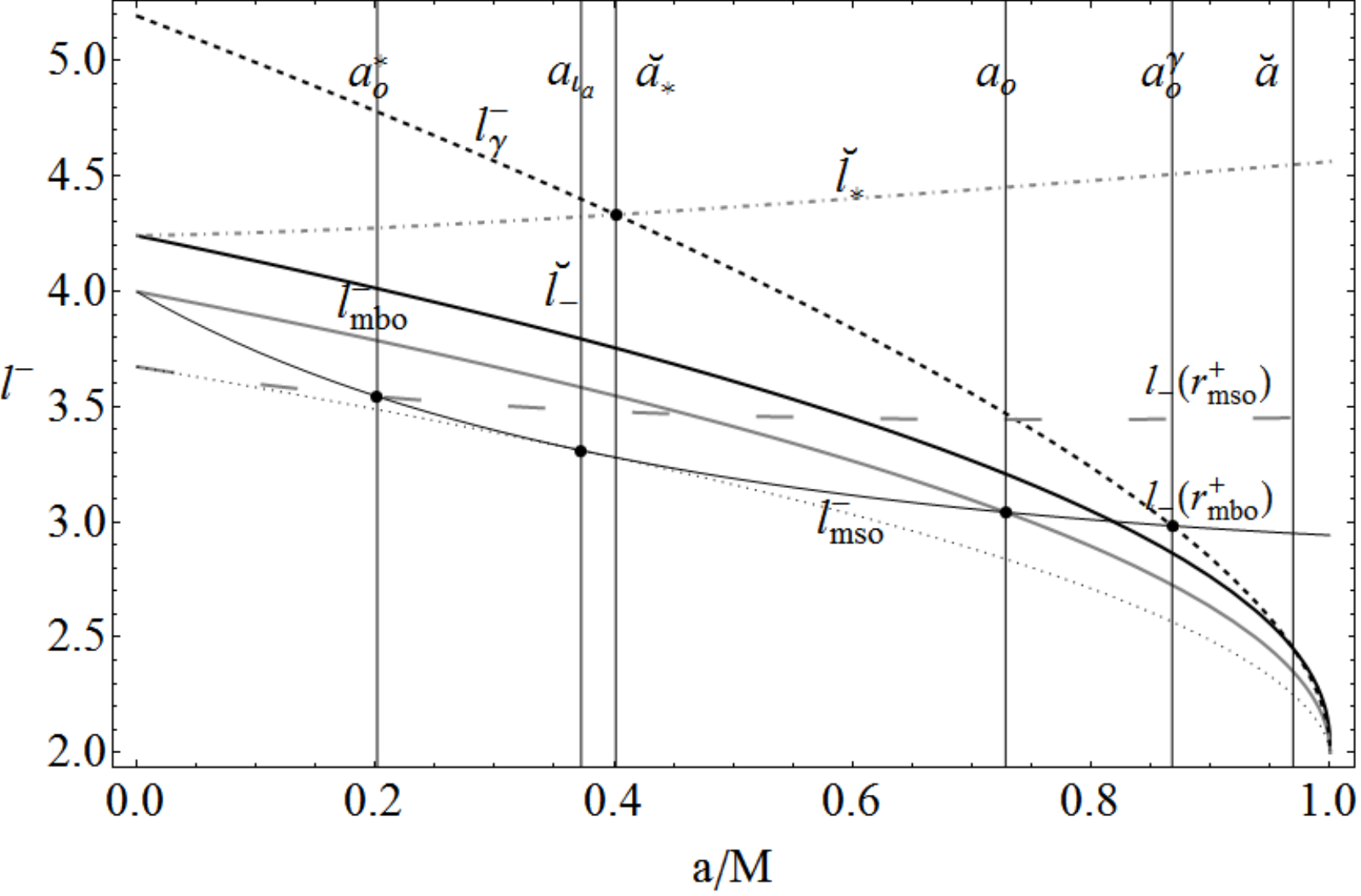}\\
 \includegraphics[width=1\columnwidth]{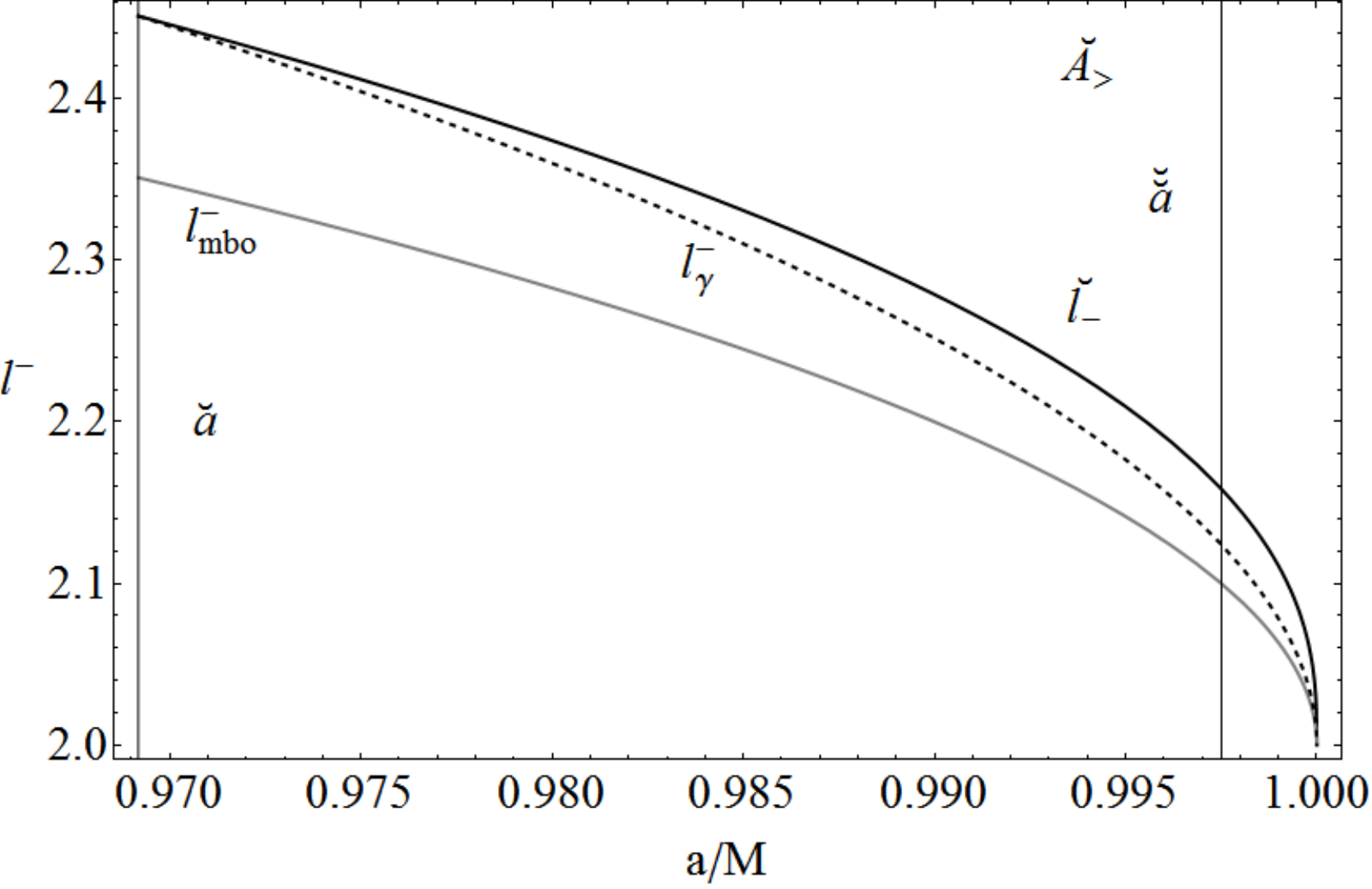}
\end{tabular}
\caption{Corotating fluids: specific angular  momentum $\ell_{mbo}^-\equiv \ell^-(r_{mbo}^-)$, $r_{mbo}^-$, is for the marginally bounded orbit,
$\ell_{mso}^-\equiv \ell^-(r_{mso}^-)$  where $r_{mso}^-$  is the marginally stable
orbit,
$\ell_{\gamma}^-\equiv \ell^-(r_{\gamma}^-)$  and  $r_{\gamma}^-$  is the
 marginally circular
orbit (photon orbit). The angular momentum $\ell^-(r_{mso}^+)$ and $\ell^-(r_{mbo}^+)$,
 $\breve{\ell}_-(a/M):\; V_{eff}(\breve{\ell}_-,r_{mso}^-)=1$ and
 $\breve{\ell}_*:\; V_{eff}(\breve{\ell}_*,r_{mso}^+)=1$ are also plotted--see also Fig.\il\ref{Figs:SupLa}.
  The spins $\{a_o^*,a_{\iota_a},\breve{a}_*,a_o, a_o^{\gamma}, \breve{a},\breve{\breve{a}}\}$ are plotted with black  vertical lines,
   where $\breve{\breve{a}}=0.997508M$--see Eq.\il(\ref{Eq:components}).
   Upper panel shows the the range of spin  $a\in[0,M]$.  Bottom panel shows details of the class of attractors
 $\breve{\mathbf{A}}_>$, as defined  in Eq.\il(\ref{Eq:AmaJ.ge}).
}\label{Figs:PXasdPXasdP}
\end{center}
\end{figure}

\begin{figure}[h!]
\begin{center}
\begin{tabular}{cc}
\includegraphics[width=1\columnwidth]{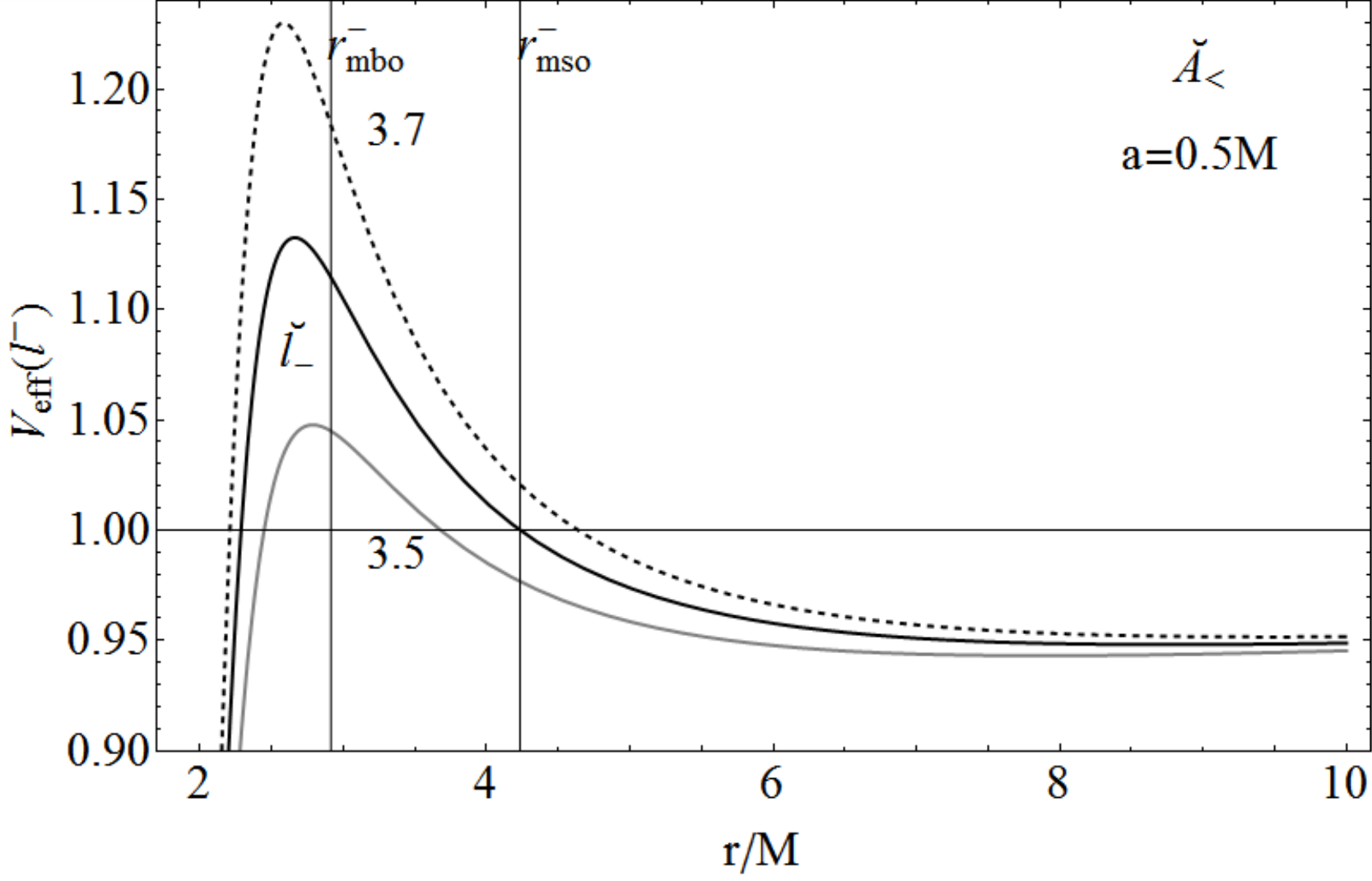}\\
 \includegraphics[width=1\columnwidth]{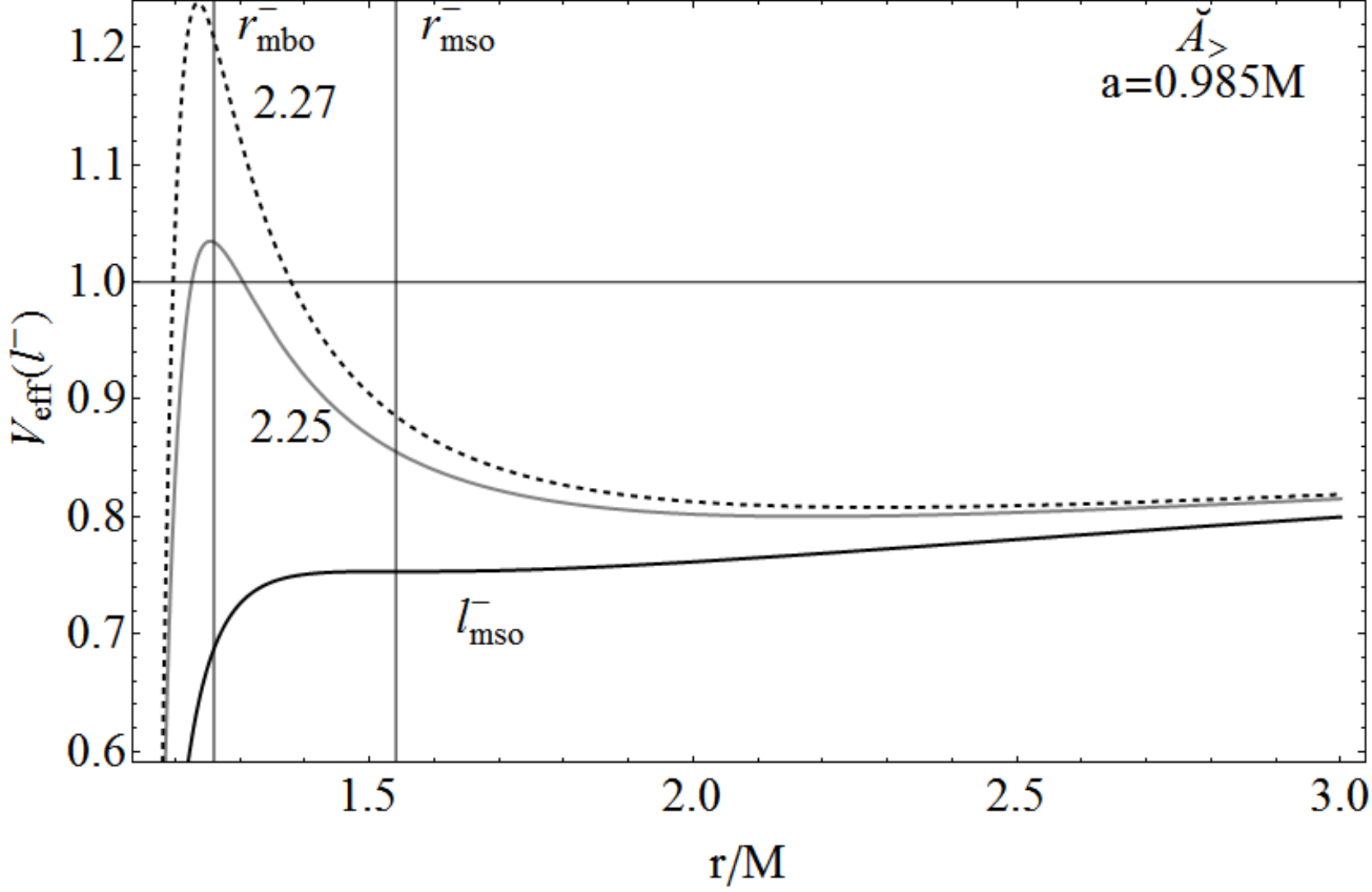}
\end{tabular}
\caption{Corotating fluids $\cc_2^{-}$. Upper panel: For spacetime spin $a=0.5M\in\breve{\mathbf{A}}_<$, the outer horizon is $r_+=1.86603M$ and  $\breve{\ell}_-=3.61088$ and $\ell_{\gamma}^-=3.41421$ with $\ell_{mbo}^-=4.09627$.  Bottom panel: For spacetime spin $a=0.985M\in\breve{\mathbf{A}}_>$, the outer horizon is $r_+=2.16614M$,  $\breve{\ell}_-=2.33082$ and $\ell_{mbo}^-= 2.24495$ with $\ell_{\gamma}^-=2.3102$.}\label{Figs:SupLa}
\end{center}
\end{figure}

We can summarize the situation by saying that:
\be\label{Eq:in-raff}
r_{mso}^{\pm}\in \cc_1^{\pm},\quad r_{mso}^{\pm}\in\;! \cc_x^{1_{\pm}} \quad \forall \ell_1^{\pm}\in \mathbf{L1}^{\pm},
\ee
see also Fig.\il\ref{Figs:PXasdPXasdP}.

Therefore for  Eq.\il(\ref{Eq:in-raff}),  the critical configuration must include the marginally stable orbit.

\subsubsubsection{\textbf{Configurations $\cc_2^{\pm}$:} $\ell^{\pm}_2\in \mathbf{L2}^{\pm}$}

\begin{figure}[h!]
\begin{center}
\begin{tabular}{c}
\includegraphics[width=1\columnwidth]{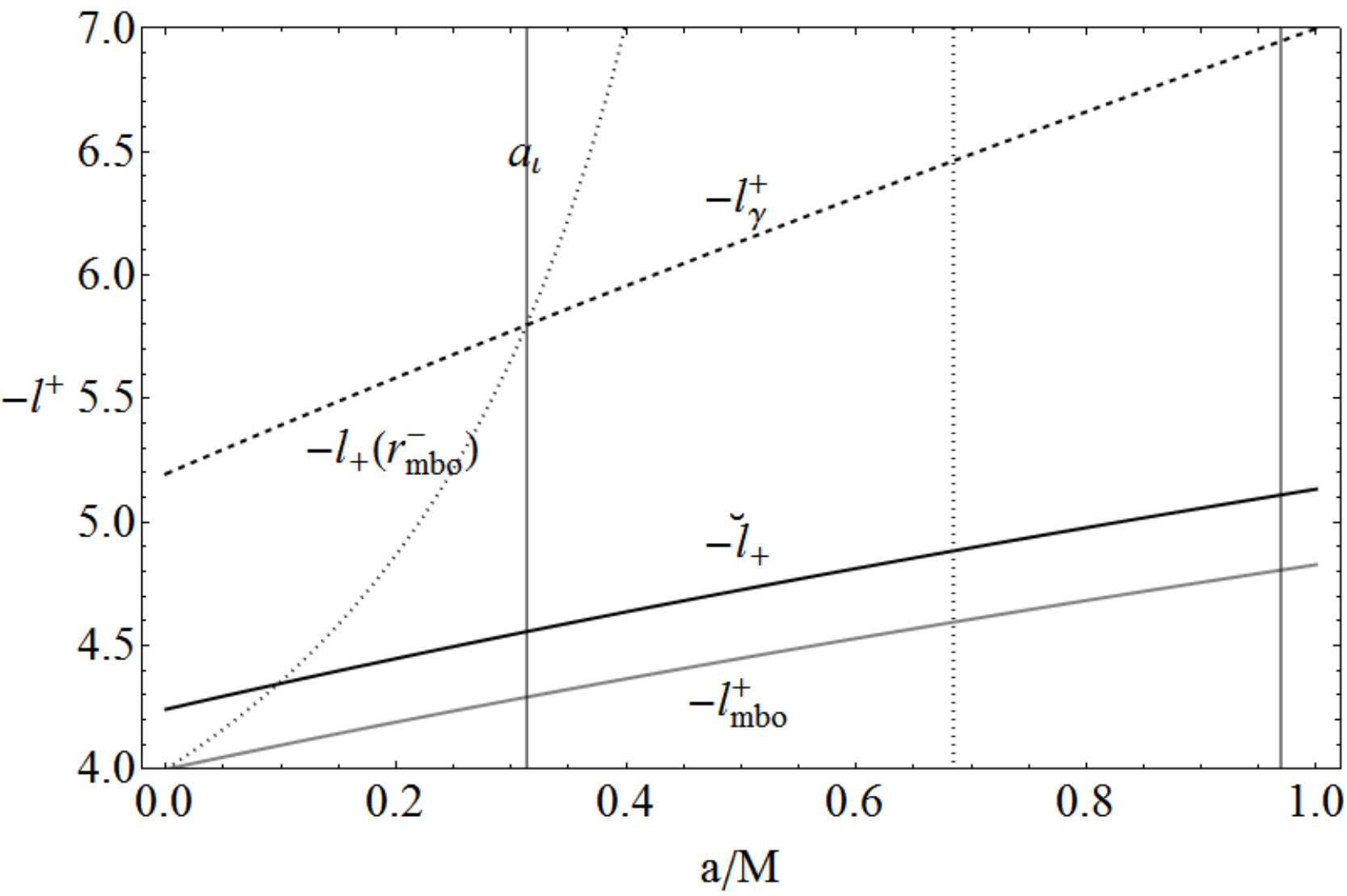}
\end{tabular}
\caption{Counterrotating fluids: the specific angular  momentum $\ell_{\gamma}^+$  and $\ell_{mbo}^+$, boundaries of $\mathbf{L2}^+$, as function of the dimensionless spin $a/M$.
The specific angular  momentum $\breve{\ell}_+:\; V_{eff}(\breve{\ell}_+,r_{mso}^+)=1$. The spin  $a_{\iota}= 0.3137M $ is also signed where $r_{mbo}^-=r_{\gamma}^+$ and then $\ell_{\gamma}^+=\ell^+(r_{mbo}^-)$.}\label{Figs:PXasdcont}
\end{center}
\end{figure}

This case is  rather well articulated and  enables to analyze    deeply  a possible correlation  between critical $\ell$counterrotating  sequences     of corotating or  counterrotating fluids.
Firstly, suppose  there exists a couple of  corotating or counterrotating fluid configurations such that there is a special specific angular  momentum  $\breve{\ell}(a/M):\; V_{eff}(\breve{\ell},r_{mso})=1$. The exact expression of   $\breve{\ell}(a/M)$  can be easily found as a solution of  a quadratic  equation for the variable  $\ell$-- see also Figs\il\ref{Figs:PXasdPXasdP},\ref{Figs:PXasdcont}.

We analyze below the different situations for corotating  and counterrotating  disks decreases.

\textbf{(1) The corotating  disk $\cc_2^-$}

There is a  wide class of rotating attractors defined as
\bea&&\nonumber\label{Eq:Achecl}
\breve{\mathbf{A}}_<\quad\mbox{ at}\quad a/M\in[0, \breve{a}[\quad\mbox{ where}
\\
&&\nonumber\breve{a}\equiv 0.969174M: \;\breve{\ell}_-=r_{\gamma}^-,
 \eea
 which includes  the limiting  static case $a=0$ of the  Schwarzschild solution, where  $\breve{\ell}_-\in \mathbf{L2}^-$. Accordingly,  we consider the following two  ranges of values of the specific angular  momentum
 \bea\label{Eq:Achecl-index}
 &&\mathbf{I}\quad\ell^-_2\in]\breve{\ell}_-, \ell_{\gamma}^-[\quad\mbox{ when}\quad  r_{in}^{2_-}>r_{mso}^- \;\mbox{and}\
 \\
 &&\mathbf{II}\; \ell_2^-\in]\ell_{mbo}^-, \breve{\ell}_-[,
 \eea
--see Fig.\il\ref{Figs:PXasdPXasdP}. We analyze the configurations in the two ranges of angular momentum in the following.

$\mathbf{I}$  For  specific angular  momentum sufficiently high, i.e.  $\ell^-_2\in]\breve{\ell}_-, \ell_{\gamma}^-[$, when  $r_{in}^{2_-}>r_{mso}^-$,  the disk in equilibrium can  never contain the marginally stable orbit or  $r_{mso}^-\non{\in}\cc^-_2$.

$\mathbf{II}$  For  the configurations with lower specific angular  momentum, i.e.  $\ell^-_2\in[\ell_{mbo}^-,\breve{\ell}_-[$ for $V_{eff}(\ell^-_2, r^-_{mso})<1$,  it is possible to select, for the equilibrium $\cc^-_2$ disk, a
 $K^-_2:\; r^{2_-}_{in}=r^-_{mso}$ or $r^{2_-}_{in}<r_{mso}^-$, and therefore  $r_{mso}^-\in \cc^-_2$, being located in the region with extremes $(r_{in}^{2_-},r_{min}^{2_-})$-- see Figs\il\ref{Figs:PXasdPXasdP}.
However, in the     case $\mathbf{I}$, in an evolutive scheme where there is a possible time evolution of the disk morphology and topology, the increase of the $K_-$ parameter, with specific angular momentum in $\mathbf{L2}^-$,
 does not necessarily correspond   to a final stage of P-W  instability with  $O^{2_-}_x$ topology,  but it will pass through the $O^-_{in}$ phase
and then a situation similar to the  case with  specific angular momentum in   $\mathbf{L3}^-$ occurs. Although,  increasing   $K_-$,  the disk will finally reach the    $\oo_x$ topology, passing through $r_{mso}$ in $\oo_{in}$.
In fact, increasing $K_2^-$ for $\ell_2^-\in]\breve{\ell}_-,\ell_{\gamma}^-[$,  the sequence of configurations on $\Sigma_K$ will be $\left.\mathcal{B_{\ell}}^{2_-}\right|_{\mathbf{I}}=\{\cc_2^-, \oo_{in}^-, \oo_{x}^-\}$ see Fig.\il\ref{Figs:SupLa}-Upper. We note that the only way to make $r_{mso}$ a P-W point, inducing therefore a gravo-hydrostatic instability in the disk, is to reach $\ell=\ell_{mso}$ where $r_{mso}$ is a cusp point \citep{pugtot}.
 Consider now  the  lower values of specific angular  momentum:   $\ell_2^-\in]\ell_{mbo}^-, \breve{\ell}_-[$. In order to reach an instability of the disk, if it does not include $r_{mso}^-$ for its density, it is not sufficient to provide a proper elongation, and it has to pass through
  the point  $r_{mso}^-$ maintaining its equilibrium topology.
 Then it will include in any case a stage where it acquires an open, not cusped  $\oo_{in}^{2_-}$ topology, as  discussed for  the case of the marginally bounded orbit  in \ref{Sec:usua-D-mbo-con}.

  There  is indeed $r_{Max}^{2_-}<r_{mbo}^-<r_{mso}^-$, therefore, the sequence of configurations on   different $\Sigma_K$ will be  $\left.\mathcal{B_{\ell}}^{2_-}\right|_{\mathbf{II}}=
  \left.\mathcal{B_{\ell}}^{2_-}\right|_{\mathbf{I}}=\{\cc_2^-, \oo_{in}^{2_-},O^{2_-}_x\}$.
Thus,  the configuration   $O^{2_-}_x$  cannot emerge as ``direct'' consequence of the accretion  in an evolutionary  model (by increasing $K$).  The  correlation in $\mathbf{L2}^-$ can occur only from  $\cc_2^-$ and $O^{2_-}_x$, and the matter cannot pass through a continuum evolution in $\mathbf{L2}^-$.
 We specify better this statement in the  following, making reference to the  sequence of the  effective  potentials  in Fig.\il\ref{Figs:SupLa}.

First, with  $\ell_2^-=$constant in $\left.\mathbf{L2}^-\right|_{\mathbf{I}}$ or  $\left.\mathbf{L2}^-\right|_{\mathbf{II}}$, the disk starting from a regular topology $\cc_2^-$ cannot reach the P-W point configuration $\oo_x^{2_-}$, so far as it passes through $\oo_{in}^{2_-}$.
Therefore, in order to  get a transition (through the surfaces $\Sigma_t$) from $\cc_2^-$ to $ \oo_x^{2_-}$, it is necessary to change $\ell_2^-\in \mathbf{L2}^-$ only, or also  change   the  $K_-$ parameter.
But it is immediate to see that, starting from a closed topology, it is not possible to reach  such a transition, and, on the other side, an initial  phase of $\oo_{in}^{2_-}$  has no meaning here.
Therefore, the specific angular  momentum $\ell_2^-\in \mathbf{L2}^-$ has to be changed together with $K_-$  shift from $\mathbf{K0}$ to $\mathbf{K1}$. However, such a transition, with $\ell^-\in \mathbf{L2}^-$ and $K: \; \mathbf{K0}^-\rightarrow \mathbf{K1}^-$,  has to be continuous  and the only possible solution is the one where the final state is  $\ell_2^-=\inf{\mathbf{L2}^-}=\ell_{mbo}^{2_-}$, $K=1$ and $r_{Max}^{2_-}=r_{mbo}^-$.

The only possible evolution in this scheme, starting from a configuration in equilibrium with $\ell\in\mathbf{L2}^-$, is  the one  leading, for a decrease of the specific angular  momentum and  the concomitant increase of  $K_-$, to the final configuration with cusp in $r_{mbo}^-$.
Finally, concerning  the correlation between the two
configurations,  $\oo_x^{2_-}$ and $\cc_2^-$, the considerations outlined in Sec.\il(\ref{Sec:criticalII}) for the (\textbf{C-J}) $\ell$corotating systems apply and  in  particular  Eq.\il(\ref{Eq:fina-j-metri-one}).

We close this discussion  noting that for $\ell^-=\breve{\ell}_-$ we have $V_{eff}(\breve{\ell}_-, r_{mso}^-)=1$. For $K_-=V_{eff}(\breve{\ell}_-, r_{mso}^-)$, this cannot give rise to a P-W point, but  it should correspond to an $\oo_{in}^-$ surface, see Fig.\il\ref{Figs:SupLa}-bottom.

Then for fast  attractors, i.e.,
\be\label{Eq:AmaJ.ge}
 \breve{\mathbf{A}}_>\quad\mbox{at}\quad a/M\in]\breve{a}, 1],
  \ee
 which  include the extreme Kerr spacetime ($a=M$),  the situation is different with respect to the geometries of $\breve{\mathbf{A}}_<$, defined in Eq.\il(\ref{Eq:Achecl}), as  $\breve{\ell}_->\ell_{\gamma}^-$--Fig.\il\ref{Figs:PXasdPXasdP}. Then $V_{eff}(\breve{\ell}_-, r_{mso}^-)<1$, providing a situation  analogous to the case of slower attractors $\breve{\mathbf{A}}_<$, with slow specific angular  momentum $\ell^-_2\in]\ell^-_{mbo},\breve{\ell}_-[ $.

\medskip

We summarize the situation as  follows:
\bea\label{Eq:ali-06}
&&\mbox{for}\; a\in \breve{\mathbf{A}}_<\equiv[0,\breve{a}[\quad\mbox{at} \quad\mathbf{I}\quad \ell^-_2\in]\breve{\ell}_-, \ell_{\gamma}^-[\\\nonumber
&&\mbox{when}\;  r_{in}^{2_-}>r_{mso}^-\quad r_{mso}^-\non{\in}\cc_2^-\quad r_{mso}^-\in ! \oo_x^{2_-}
\\\label{Eq:curv-n}
&& \mbox{at}\;\mathbf{II}\quad \ell_2^-\in]\ell_{mbo}^-, \breve{\ell}_-[\;  r_{mso}^-{\in}\cc_2^-\quad r_{mso}^-\in ! \oo_x^{2_-}
\\\label{Eq:font-cu}
&&\mbox{for} \; a\in \breve{\mathbf{A}}_>\equiv[\breve{a},M],\;  r_{mso}^-{\in}\cc_2^-\quad r_{mso}^-\in ! \oo_x^{2_-}.
\eea

\textbf{(2)  The counterrotating  disk $\cc_2^{+}$}

Figure\il\ref{Figs:PXasdcont} sketches the situation for the  counterrotating  fluids:
 \be \label{Eq:states-energies}
\cc_2^{+}: \quad \mbox{for \emph{any} attractors}
-\breve{\ell}_+\in]-\ell^+_{mbo}, -\ell^+_{\gamma}[,
 \ee
  therefore:
 \bea&& \nonumber
\cc_2^{+}:\; r_{mso}^+ \in! \oo_x^{2_+};\quad \mbox{and}\quad r_{mso}^+ \in \cc_2^{+}\quad \mbox{for}
\\\label{Eq:states-energies-Y}
&&
-\ell_2^+\in]-\ell^+_{mbo},-\breve{\ell}_+[,\\
&&  r_{mso}^+ \non{\in} \cc_2^{+}
\quad \mbox{for}\quad
-\ell_2^+\in]-\breve{\ell}_+,-\ell_{\gamma}^+[.
 \eea
 The situation is  similar to the corotating fluids orbiting  $\breve{\mathbf{A}}_<$ attractors--Fig.\il\ref{Figs:PXasdPXasdP}-Upper.

For specific angular  momentum sufficiently  low  in magnitude, there always exists a region,  where   the equilibrium disks  $\cc_2^+$ contain  the marginally stable circular orbit $r_{mso}^+$. While for higher specific angular  momentum, i.e. $-\ell_2^+>-\breve{\ell}_+\in \mathbf{L2}^+$,  the disk cannot include the radius $r_{mso}^+$; see also discussion  in Sec.\il(\ref{Sec:C-J-clna-v}) and  in  particular  Eq.\il(\ref{Eq:fina-j-metri-one}).

\textbf{(3) Comments on the $\cc_2^{\pm}$ disks}

We can compare the cases of corotating $\cc_2^-$ disks and counterrotating $\cc_2^+$ disks through  Figs\il\ref{Figs:PXasdPXasdP},\ref{Figs:PXasdcont}.

For the analysis of the corotating disks we refer  to  Figs\il\ref{Figs:PXasdPXasdP} and \il \ref{Figs:SupLa}. First, while  the situation for the counterrotating disks is uniform for attractors $a\in[0,M]$--Fig.\il\ref{Figs:PXasdcont},  for the corotating disks it was necessary to distinguish between the two classes  of attractors $\breve{\mathbf{A}}_{\lessgtr}$.

For the inclusion  $r_{mso}^-\in \cc_2^-$ to be satisfied,
  the ratio $\ell^-/a$ has to be  sufficiently low. In fact, at high spin, $a\in \breve{\mathbf{A}}_>$, the  situation is similar to  $\mathbf{L1}^-$,  and the equilibrium disks $\cc_2^-$ can always contain $r_{mso}^-$.

In order to simplify the presentation of the results,  we denote here  by $\mathbf{Q}_{\in}$ any  quantity $\mathbf{Q }$ which satisfies the   inclusion relation,   and, respectively,   by  $\mathbf{Q}_{\not{\in}}$ any quantity  where the   inclusion relation is {not} satisfied;
 $\Delta \mathbf{Q}$ is  the measure of the maximum range of variation for the quantity $\mathbf{Q}$, thus:
\bea&&\nonumber
\cc_2^-:\quad\ell_{\not{\in}}^{2_-} >\ell_{{\in}}^{2_-},\quad  \Delta\ell_{\not{\in}}^{2_-} >\Delta\ell_{{\in}}^{2_-}\quad\mbox{for}\quad  a\lesssim 0.8 M,\\
&&\nonumber
  \Delta\ell_{\not{\in}}^{2_-} =0 \;{for}\quad  a=  \breve{a},  \quad\nexists \Delta\ell_{\not{\in}}^{2_-}\quad\mbox{for}\quad a\in \breve{\mathbf{A}}_{>}\;\partial_a \mathbf{Q}_{2}^-<0,
   \\\label{Eq:ham-0-c2}
  &&   \mbox{for}\quad \mathbf{Q}_2^-\in\{\ell^{2_-}_{\in},\ell^{2_-}_{\not{\in}}, \Delta \ell^{2_-}_{\in}, \Delta\ell_{\not{\in}}^{2_-}, (\Delta \ell^{2_-}_{\in}- \Delta\ell_{\not{\in}}^{2_-})\},
 \eea
--see Fig.\il\ref{Figs:PXasdPXasdP}.
For the counterrotating  disk $\cc_2^+$, the situation is just reversed:
\bea&&\nonumber
\cc_2^+:\quad\ell_{\not{\in}}^{2_+} >\ell_{{\in}}^{2_+}\quad \Delta\ell_{\not{\in}}^{2_+}>\Delta\ell_{{\in}}^{2_+},\quad
  \partial_a \mathbf{Q}_2^+  >0, \\
  && \nonumber\mbox{for}\quad \mathbf{Q}_2^+\in\{\ell^{2_+}_{\in}, \ell_{\not{\in}}^{2_+}, \Delta \ell^{2_+}_{\in},\Delta\ell_{\not{\in}}^{2_+}, (\Delta \ell^{2_+}_{\in}- \Delta\ell_{\not{\in}}^{2_+})\},
  \\\label{Eq:new-A-g}
 \eea
 see Fig.\il\ref{Figs:PXasdcont}.

 However we can still say,  that  a high value of the ratio $|\ell^2_+|/a$  acts to disadvantage  the cases where  $r_{mso}$ is included in the equilibrium disks   (see also \cite{pugtot} for the analysis of the disk equilibrium in terms of the rationalized momentum).

Reminding the relation between  the specific angular  momentum and the  position of the  pressure maximum points,
we may say in general that disks  $\cc^{2_{\pm}}_{\in}$,  including $r_{mso}^{\pm}$,
are  localized in a narrow region of the specific angular  momentum values and
orbital range. The disks  $\cc^{2_{-}}_{\in}$  approach the attractor by increasing  the spin  of the black hole, while $\cc^{2_{+}}_{\in}$ moves away for increasing $a/M$.

There are several differences in the $\ell$counterrotating couples  of   equilibrium disks $\cc^{2_{\pm}}_{\LARGE{\not{\in}}}$.
The counterrotating disk $\cc^{2_+}_{\LARGE{\not{\in}}}$ demonstrates behavior, related to the spin, that is  very similar to
$\cc^{2_+}_{\LARGE{{\in}}}$.
However, the elongation of the disks $\cc^{2_+}_{\LARGE{{\in}}}$   can be   lower in general  than   those of  $\cc^{2_+}_{\non{\in}}$,  as the configuration density   is characterized by lower specific angular   momentum available, and specific angular   range--see Fig.\il\ref{Figs:PXasdcont}.
The corotating case  $\cc^{2_-}_{\LARGE{\not{\in}}}$ presents an articulated morphological characteristic,  different for   different classes of attractors.  For attractors  $\breve{\mathbf{A}}_<$, the trend with the spin-mass ratio  is similar to $\cc^{2_-}_{{\not{\in}}}$.
However, for $a<0.8M$, the extension and the spacing (or conversely the configuration density) of the $\cc^{2_-}_{{\not{\in}}}$ disks  is greater then of the  $\cc^{2_-}_{{{\in}}}$ disks.

The situation is reversed for higher spin, until at $a\in\breve{\mathbf{A}}_>$    the formation of a $\cc^{2_-}_{{\not{\in}}}$ disk is impossible  (it is not possible to find  $K_2^{\pm}:\; r_{mso}^{\pm}\in \cc^{2_{\pm}}_{{\not{\in}}}$).
In fact, if it always  possible to find a set of parameters  $K_2^{\pm}:\; r_{mso}^{\pm}\in \cc^{2_{\pm}}_{{\in}}$,
and indeed the equilibrium disks $ \cc^{2_{\pm}}_{{\in}}$ must contain  $r_{mso}^{\pm}$, as they extend towards  the maximum elongation at  instability $\lambda_x$.
On the other side, the extension of the regions $\Delta\ell_{\in}^{2_{\pm}}$ increases or decreases  with the spin
 more slowly than  $\Delta\ell_{\not{\in}}^{2_{\pm}}$. Thus one could say that the morphological characteristics of the
 $\cc_{\in}^{2_{\pm}} $ case  are less affected by a change of the spin that those of  the  $\cc_{\not{\in}}^{2_{\pm}} $ case.

\subsubsubsection{\textbf{Configurations $\cc_3^{\pm}$:} $\ell_3\in \mathbf{L3}$.}

For the investigation of this case, we will refer to  equations (\ref{Eq:Achecl},\ref{Eq:AmaJ.ge}) and Figs\il\ref{Figs:PXasdPXasdP},\ref{Figs:PXasdcont}. There are no critical configurations for fluids at specific angular  momentum $\ell_3\in \mathbf{L3} $.
Following  arguments similar to those discussed in  the previous cases, one  can see that for the corotating case,
there are no solutions of  the problem  $\breve{\ell}(a/M)\in \mathbf{L3}^-:\; V_{eff}(\breve{\ell},r_{mso})=1$.

Firstly, Figs\il\ref{Figs:PXasdPXasdP},\ref{Figs:PXasdcont} confirm  the results  for $\ell_1^{\pm}\in \mathbf{L1}^{\pm}$.

As  proved  earlier,  it is always possible to find a  proper $K_1^{\pm}$ for the  closed configurations  in equilibrium $\cc_1^{\pm}$, containing respectively $r_{mso}^{\pm}$. In other words, the disk with a specific angular  momentum in $\ell_1\in \mathbf{L1}$,  containing the marginally stable orbit, must be in equilibrium, and in order  to accrete into the black hole, it must extend far beyond the marginally stable orbit.

Similarly, Figs\il\ref{Figs:PXasdPXasdP},\ref{Figs:PXasdcont} provide an immediate description of the situation for the equilibrium disks  $\cc_3^{\pm}$ with $\ell^{\pm}_3\in \mathbf{L3}^{\pm}$.
We detail the results as follows:

\textbf{(1) The corotating disk $\cc_3^-$}

Analogously  to the fluid configurations  with   specific angular  momentum $\ell_2\in \mathbf{L2}$, it will be convenient to consider first the corotating case, as illustrated  in  Fig.\il\ref{Figs:PXasdPXasdP}.

Equations (\ref{Eq:Achecl}, \ref{Eq:Achecl-index},\ref{Eq:AmaJ.ge}) hold.  Therefore we still need  to distinguish the situation for the two classes of attractors $\breve{\mathbf{A}}_{\lessgtr}$.

For slow  attractors, $\breve{\mathbf{A}}_<:\;a\in[0,\breve{a}[$, we have  $\breve{\ell}_-\in]\ell_{mbo}^-,\ell_{\gamma}^-[$, this implies:
\bea&&\label{Eq:imply-Col}
\mbox{for} \quad a\in\breve{\mathbf{A}}_<\quad\mbox{there  is }\; V_{eff}(\ell^-_3,r_{mso}^-)\nleq 1
\\
&&\nonumber\mbox{and}\; r^-_{mso}\non{\in} \cc_3^-.
\eea
However, concerning  the  first  inequality of Eq.\il(\ref{Eq:imply-Col}), we should consider  that the function $V_{eff}(\ell^-_3,r)$ may  not be well defined  in  $r_{mso}^-$. As mentioned at the beginning of this section, if the potential function is not defined in a point $\bar{r}$, this constitutes here evidence of the fact that the disk cannot exist at $\bar{r}$, and  therefore Eq.\il(\ref{Eq:imply-Col})  is then sufficient to prove that if the effective potential is well defined in $r_{mso}^-$, then  it is not contained  in the equilibrium disk\footnote{ The conditions for which the function  $V_{\ell}(\ell_3, r)$ is  not well-defined could  be  easily provided.}.

The situation is different for the geometries of  faster attractors, $\breve{\mathbf{A}}_>\in]\breve{a},M]$, where
\bea&&\label{Eq:grounds}
\mbox{for} \quad a\in\breve{\mathbf{A}}_>\quad\mbox{there  is }\quad r^-_{mso}>r_{\gamma}^-,\\\nonumber
&& \exists\; K^-_3< V_{eff}(\breve{\ell}^-_3,r_{mso}^-)=1:\; r_{mso}^-\in \cc_3^-
\eea
and $K^-_3$ will be clearly  bounded  from below by $ V_{eff}(\ell_{\gamma}^{_-}, \bar{r})$, where $\bar{r}>r_{mso}^-$, corresponding to the condition $\ell_3^-\in \mathbf{L3}^-$.
The similarities with the $\mathbf{L2}$ case   are evidently in the differentiation between the two classes of attractors $\breve{\mathbf{A}}_{\lessgtr}$,
but the situation is very different with respect to the role of the specific angular  moment.
The case of very fast attractors, seen in Fig.\il\ref{Figs:SupLa}, is  particularly interesting. We will analyze  deeply the morphology of these regions for the  case of   the $\breve{\mathbf{A}}_>$  attractors later, comparing  them with the counterrotating case.

\textbf{(2) The counterrotating disk $\cc_3^+$}

The situation for the counterrotating disks at  $\ell_3^+\in \mathbf{L3}^+$ is illustrated in Fig.\il\ref{Figs:PXasdcont}.

As  Eq.\il(\ref{Eq:states-energies}) holds,  then
\be\label{Eq:En-Dir}
r_{mso}^+\non{\in}\cc_3^+.
\ee
Therefore the equilibrium counterrotating disks $\cc_3^+$ \emph{cannot} contain the marginally stable circular orbit, and the disks are  entirely  contained in an outer orbital region $r>r_{mso}^+$.
This situation is indeed similar to the corotating disks orbiting  the slower attractors $\breve{\mathbf{A}}_<$.

\textbf{(3) Comments on the $\cc_3^{\pm}$ disks}

\begin{figure}[h!]
\begin{tabular}{cc}
\includegraphics[width=1\columnwidth]{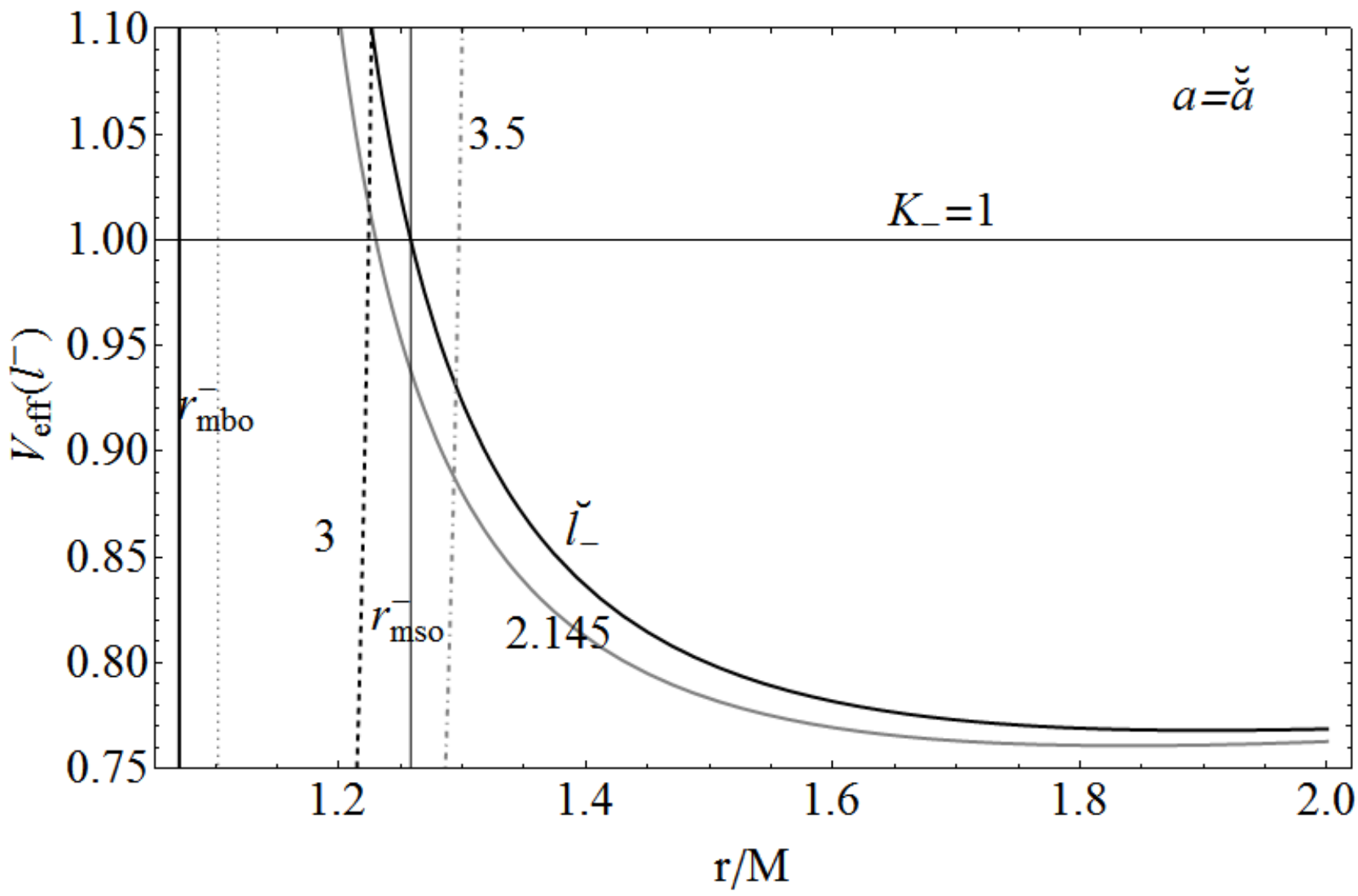}\\
 \includegraphics[width=1\columnwidth]{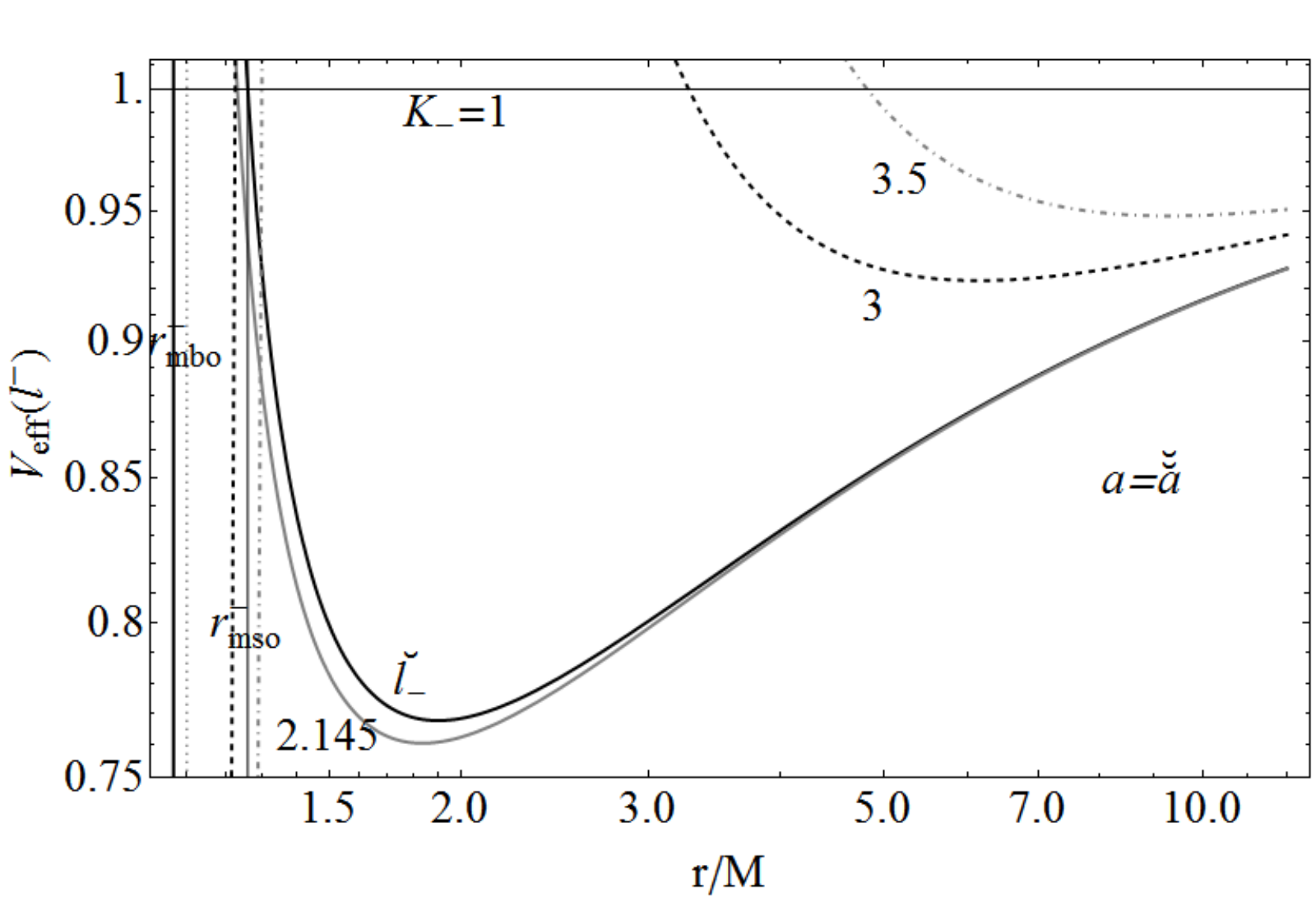}
\end{tabular}
\caption{Corotating fluids $\cc_3^{-}$ in a  geometry with  spin-mass ratio $\breve{\breve{a}}=0.997508M\in\breve{\mathbf{A}}_<$, the outer horizon is $r_+=1.07055M$, $\breve{\ell}_-=2.15829$ and $\ell_{\gamma}^-=2.12395$ with $\ell_{mbo}^-=2.09984$, see also Fig.\il\ref{Figs:PXasdcont}. The plots show the effective potential $V_{eff}(\ell^-)$ versus $r/M$, at different specific angular  momenta, signed on the curves: the black thick curve is for $\ell^-=\breve{\ell}_-$. Upper panel: orbital range $r\in[r_+,r_{\epsilon}^+]$  where $r_{\epsilon}^+=2M$ is the static limit on the equatorial plane.  Bottom panel:  orbital range $r>r_{\epsilon}^+$. Curves $\ln{V_{eff}}$ against $\ln{r/M}$, but the labels on the axes  indicate the original values of $V_{eff}$ and $r/M$.}\label{Figs:C3SupC3La}
\end{figure}

In conclusion, for specific angular  momentum  $\ell_3\in \mathbf{L3}$ only a special   class  of rotating equilibrium disks,   namely $\cc_3^{-}$,   orbiting very fast attractors,  may contain the marginally stable orbit.
The morphology of the range  of specific angular  momenta   for these fluids, is very special.
The more relevant aspect probably is that the  $\Delta \ell_{\in}^{3_-}$ region is   vanishing  at its  extremes, i.e., the region $\Delta \ell_{\in}^{3_-}$ of specific angular  momentum $\ell_3^-$,  whose measure is the distance $(\ell_{\gamma}^--\breve{\ell}_-)$,  has a minimum, with  vanishing  extension, for the two special geometries  $a=\breve{a}$ and  $a=M$--
Figs\il\ref{Figs:C3SupC3La}.
The absolute maximum  of this region occurs  for the  geometry associated with spin $\breve{\breve{a}}=0.997508M$.

Analogously to  Eqs\il(\ref{Eq:ham-0-c2},\ref{Eq:new-A-g}),  we can summarize the situation  for the  $\cc_3^{\pm}$ cases  as  follows:
\bea\nonumber
\cc_3^-: && \quad \mbox{ for}\quad \breve{\mathbf{A}}_<\quad r_{mso}^-\not{\in}\cc_3^-,
\\\nonumber
&& \quad \mbox{ for}\quad \breve{\mathbf{A}}_>\quad r_{mso}^-{\in}\cc_3^-; \quad \inf{\Delta\ell_{{\in}}^{3_-}}=0\\
&&\mbox{for}\quad  a=\{\breve{a},M\},
\\&&\nonumber  \max{\Delta\ell_{{\in}}^{3_-}}=0\quad\mbox{for}\quad a=\breve{\breve{a}}\in]\breve{a}, M[.
\\\label{Eq:components}
&&
 \partial_a \ell^{3_-}_{\in}<0, \; \partial_a \Delta \ell^{3_-}_{\in}\gtrless0 \;\;{for}\; a\lessgtr \breve{\breve{a}},
 \\\nonumber
 \\
 \cc_3^+: && r_{mso}^+\not{\in}\cc_3^+.
\eea
Possibly the region  of specific angular  momentum $\Delta\ell_{\non{\in}}^{3_-}$, and  the corresponding orbital region, is infinite extending up to the regions where the Newtonian limit can be considered.
\subsection{Location of the notable radii $r_{\mathcal{N}}^{\pm}\in ()_{\mp}$}\label{Sec:coorr}
In this section we examine the location of the radii   $r_{\mathcal{N}}^{\pm}\in ()_{\mp}$  with respect to the \emph{closed} (cusped or regular) sub-configurations  $ ()_{\mp}$respectively . This issue is   relevant for example  for the analysis of the  mixed $\ell$counterotating sub-sequences of a decomposition.
As a consequence of this analysis we  will distinguish different classes of attractors,   deriving  conditions on the parameters of the macro-configuration.
First,  we provide several  orbital constraints   which locate   more precisely  the inner  and   the outer edge of the  ring of a decomposition.
Then  we  clarify the possible interaction between two $\ell$counterrotating sub-configurations through the geometrical correlation.

\medskip

In the following discussion, we  use  the arguments and results of   \ref{Sec:possi-conc}  and we  proceed in analogy to the investigation of the  $r_{\mathcal{N}}^{\pm}\in ()_{\pm}$  problem.

In \ref{Sec:geo-Hull} we  consider the counterrotating configurations, studying the problem $r_{\mathcal{N}}^{-}\in ()_{+}$, while  \ref{Sec:scale-ques} focuses on  the corotating  disks, investigating the problem  $r_{\mathcal{N}}^{+}\in ()_{-}$. The location of photon circular orbits $r_{\gamma}^{\pm}$ is investigated in \ref{Sec:photon-C}. 
\subsubsection{Counterrotating disks: $r_{\mathcal{N}}^{-}\in ()_{+}$}\label{Sec:geo-Hull}
\subsubsubsection{
\textbf{Marginally bounded orbit $r_{mbo}^{-}\in ()_{+}$}}

In general, the following propriety holds: as  $r_{mbo}^-<r_{mbo}^+$ in any geometry, then, if
$r_{mbo}^+\non{\in} ()^+$,  it  follows that $r_{mbo}^-\non{\in} ()^+$.

Therefore, as Eq.\il(\ref{Eq:choral}) stands, we have
\be
r_{mbo}^-\non{\in} \cc_3^+.
\ee
We remind then that the $\cc_3$ configurations do not allow critical topologies with $\ell\in \mathbf{L3}$.

Similarly, for an accretion configuration in $\mathbf{L2}^+$, Eq.\il(\ref{Eq:beha-dire}) holds. It follows
$
r_{mbo}^-\non{\in} \cc_2^+
$.
Regarding the location of  $ r_{mbo}^-$ with respect to the open counterrotating configuration $\oo_x^+$,
  we should consider two different classes of attractors.
  In fact, the general considerations  introduced at the beginning of this section  apply strictly only  for a closed (cusped or regular) topology while  for the  open topologies $\oo_x$,
  we need to consider the   geometries
  where
	$r_{mbo}^-\in\Delta r_J^+$,  i.e.  $a\in\mathbf{A}_{\iota}^< \equiv[0,a_{\iota}[$, and  the geometries  with  $ a\in\mathbf{A}_{\iota}^> \equiv]a_{\iota},M]$,    where $r_{mbo}^-<r_{\gamma}^+ $.

Then
we could summarize the results  as follows:
\bea\nonumber&&
r_{mbo}^-\non{\in} \cc_2^+,\quad \mbox{and  we have }\quad r_{mbo}^-\in \oo_x^{2_+} (r_{mbo}^-\succ r_J^{2_+}) \\
&&\nonumber\mbox{ only if}\quad
a<a_{\iota} \quad\mbox{for}\quad -\ell^+(r_{mbo}^-)\in[-\ell_{mbo}^+,-\ell^+(r_{mbo}^-)].
\\
&&\label{Eq:spec-spo-Mix}
  \eea
  We specify that for   the slower attractors,
 $ a\in[0,a_{\iota}[$, \emph{also} the   configurations  with $r_{mbo}^-\in \oo_x^{2_+}$  are possible, while
such a situation is forbidden in the spacetimes of  the faster attractors.
Accordingly,  the  notation $\in$ (not $\in !$)   has been used  in  Eq.\il(\ref{Eq:spec-spo-Mix}). This is because for    $a<a_{\iota}$, there  is $r_{mbo}^->r_{\gamma}^+$, and the open surfaces can include the marginally bounded orbit $r_{mbo}^-$, i.e $r_{J}^+<r_{mbo}^-$, or not include $r_{mbo}^-$, i.e.  $r_{J}^+>r_{mbo}^-$.
This distinction
 clearly depends on the specific angular  momentum, and one should analyze the condition for which $-\ell_2^+\in[-\ell^+(r_{mbo}^-),-\ell_{\gamma}^+[$ where, for  $\ell=-\ell^+(r_{mbo}^-)$, the launch point\footnote{In fact this follows from the general behavior of the curves $\mp\ell_{\pm}$ as function of $r/M$ demonstrated in Fig.\il\ref{Figs:IPCPl75} and  trends of the criticality indices   as described in  Eq.\il(\ref{Eq:dim-ana}).}  is exactly $r_{J}^{2_+}=r_{mbo}^-$, as pictured in  Fig.\il\ref{Figs:PXasdcont}, which also confirms that $-\ell^+(r_{mbo}^-)\in[-\ell_{mbo}^+,-\ell^+(r_{mbo}^-)]$ for $a<a_{\iota}$.

In the  geometry  with $a=a_{\iota}$,  introduced  in Sec.\il(\ref{Sec:criticalII}), there  is  $r_{\gamma}^+=r_{mbo}^-$, and then  it follows that  this case represents a limit,  defining the  $\gamma$-surface $\oo_x^{\gamma_+}$, that can be never reachable also for the $\cc_2^+$ configurations (alternatively one can see this fact also as a consequence of the first   inequality in Eq.\il(\ref{Eq:beha-dire})).

The situation for  $\cc_1^+$ can be  described using  Eq.\il(\ref{Eq:the-ag}), and  we obtain
\be\label{Eq:non-c1mo}
r_{mbo}^-\non{\in} \cc_1^+\quad\mbox{and}\quad r_{mbo}^-\non{\in} \cc_x^{1_+}.
\ee

\subsubsubsection{\textbf{Marginally stable orbit $r_{mso}^{-}\in ()_{+}$}}

We focus now on the location of the marginally stable orbit: $r_{mso}^{-}\in ()_{+}$.

Using  arguments similar to those used to locate the   marginally bounded orbit, we note that since $r_{mso}^+>r_{mso}^-$ in any Kerr geometry,  it  follows  that $r_{mso}^-\non{\in} ()^+$, if
$r_{mso}^+\non{\in} ()^+$  (in this case $r_{in}^+>r_{mso}^+$). Therefore, we  first concentrate our analysis on the cases where  $r_{mso}^+{\in} ()^+$.

In fact, the  critical topologies  $()_x^+$ are  good candidates  for  counterrotating configurations including the orbit $r_{mso}^-$,  as these rings always contain the marginally stable orbit $r_{mso}^+$--see  Eq.\il(\ref{Eq:comb-with-simm}).

However, we point out  that the inclusion   $r_{mso}^+{\in} ()^+$ is  a necessary but not sufficient condition for the inclusion of $r_{mso}^-$ in the disks $()^+$, as  the  investigation of  $r_{mso}^+\non{\in} ()^+$   provides a condition to rule out a series of situation where    $r_{mso}^-$ is certainly not included in $()^+$.

\subsubsubsection{\textbf{On the configurations $\cc_1^+$ and $\cc_x^{1_+}$}}

Consider first  the configurations $()_1^+$. Then, as  Eq.\il(\ref{Eq:in-raff}) holds (which is  necessary condition for $r_{mso}^-\in ()^1_+$),   one has to distinguish  the following two classes of geometries:
\bea\label{Eq:disti-g}
&&\mathbf{A}_{\iota_a}^>: \; a>a_{\iota_a}\quad (r_{mso}^-< r_{mbo}^+),
 \quad \mbox{and} \\\nonumber
 && \mathbf{A}_{\iota_a}^<:\; a<a_{\iota_a}\quad (r_{mso}^-> r_{mbo}^+), \quad\mbox{ where }\\\nonumber
 && a_{\iota_a}\equiv0.372583M.
 \eea
In fact, we  consider only configurations where
$r_{mso}^+\in ()_1^+$, ensured by Eq.\il(\ref{Eq:in-raff}).

Because of Eq.\il(\ref{Eq:the-ag}),  the inner edge of the  $\cc_{x}^{1_+}$ disk  is  $r_{in}^{1_+}\in \Delta r_x^{+}\equiv]r_{mbo}^+,r_{mso}^+[$
(we exclude the case where $r_{in}^{1_+}=r_{mso}^+$).  Then,  from  Eq.\il(\ref{Eq:non-c1mo}), for the  disk   $\cc_{1}^{+}$, there   is $r_{in}^{1_+}\in \Delta r_x^{+}\cap[r_{mso}^+,r_{min}^{1_+}[$.

If  $r_{mso}^-<r_{mbo}^+$,  this implies $r_{mso}^-\non{\in}\cc_1^+$ and  $r_{mso}^-\non{\in}\cc_x^{1_+}$, and this occurs for  sufficiently fast attractors $\mathbf{A}_{\iota_a}^>$--Eq.\il(\ref{Eq:disti-g}).
Therefore, we will consider  only these lower attractors  $\mathbf{A}_{\iota_a}^<$.

Note that these attractor spins also include the  spin $a_{\iota}$: in these geometries, where $r_{mso}^-\in \Delta r_x^+$, one can always find a $K_1^+:\; r_{mso}^-\in \cc_{x}^{1_+}$, or also $K_1^+:\; r_{mso}^-\in \cc_{1}^+$, for sufficiently high magnitude of the specific angular  momentum (we recall that $\lambda^{\pm}_x=\sup{\lambda^{\pm}}$ and $\Lambda^{\pm}_x\supset\Lambda^{\pm}$).

It can be important in some circumstances  to fix the topology of the  counterrotating (outer) configuration $()_1^+$ of a couple,   especially if one assumes that a possible  gravo-hydrostatic instability  may lead to  destabilization on the  inner (corotating)  configuration, as considered for example in Secs\il(\ref{Sec:criticalII},\ref{Sec:ell-cont-double}).

The  situation where   $r_{mso}^-\in \cc_x^{+}$  is ensured by  the condition on the  specific angular  momentum $-\ell_1^+\in[-\ell_1^+(r_{mso}^-),\ell_{mbo}^+[\subset \mathbf{L1}^+$.
In fact, if $\ell_1^+=\ell_1^+(r_{mso}^-)$, then   $r_{Max}^{1_+}=r_{mso}^-$. The limiting values  $\ell_1^+(r_{mso}^-)\in \mathbf{L1}^+$ vary with the spin, see Figs\il\ref{Fig:Plotaaleph1II},\ref{Fig:Plotaaleph1IIa}.

 For    $K_1^+=K_{Max}^{1_+}$, there is  $r_{in}^{1_+}=r_{Max}^{1_+}=r_{mso}^-=r_x^+$,  thus with  increasing $|\ell_1^+|$ the maximum point $r_{Max}^{1_+}$ decreases\footnote{ Since for $a\in]0,a_{\iota_a}[$ we have $r_{mso}^-\in \Delta r_x^{+}$, where the function $-\ell_{1}^+$ is monotonically decreasing, then there  is always $\ell_1^+(r_{mso}^-) \in \mathbf{L1}^+$. } and there is $r_{Max}^{1_+}<r_{mso}^-$.

In conclusion:
 \bea\label{Eq:fORD-LE}
&&\mathbf{A}_{\iota_a}^>:\; a>a_{\iota_a}:\;  r_{mso}^-\non\in \cc_1^+\quad\mbox{and}\quad r_{mso}^-\non\in \cc_x^{1_+},
 \\\nonumber
 &&\mathbf{A}_{\iota_a}^<:\;  a<a_{\iota_a}:
 -\ell_1^+\in]-\ell_{mso}^+,-\ell_1^+(r_{mso}^-)[,\\
 &&  r_{mso}^-\non{\in}  \cc_x^{1_+},\quad r_{mso}^-\non{\in} \cc_1^{+},
 \\\label{Eq:fORD-LE1}
 &&
 \ell_1^+=\ell_1^+(r_{mso}^-),\;   r_{in}^{1_+}=r_{Max}^{1_+}=r_{mso}^-=r_x^+,\\
 && r_{mso}^-\in \cc_x^{1_+},\quad  r_{mso}^-\non{\in} \cc_1^{+},
 \\\nonumber
 &&
 -\ell_1^+\in[-\ell_1^+(r_{mso}^-),-\ell_{mbo}^+[,\quad  r_{mso}^-\in! \cc_x^{1_+},\; r_{mso}^-\in  \cc_1^{+}.
 \eea
\textbf{(1) On the configurations $r_{mso}^-\in()_2^+$}

We consider  now the configuration $()_2^+$. The necessary condition for $r_{mso}^-\in ()_2^+$ in
 Eq.\il(\ref{Eq:states-energies-Y}) (see also Eq.\il(\ref{Eq:new-A-g}))   showed the presence of a specific angular  momentum threshold $\breve{\ell}_2^+$ (we recall that $\breve{\ell}(a/M):\; V_{eff}(\breve{\ell},r_{mso})=1$, solution of  a quadratic  equation for the variable $\ell$-- see also Figs\il\ref{Figs:PXasdPXasdP},\ref{Figs:PXasdcont}, introduced  in
\ref{Sec:graph-def}; here we pointed out the belonging of the specific angular  momentum in $\mathbf{L2}$ with the subscript). Following this analysis, we can state that for any Kerr attractor there is
\bea\label{Eq:not-non-crea}
r_{mso}^-\non{\in}\cc_2^+\quad\mbox{if}\quad -\ell_2^{+}\in]-\breve{\ell}_2^+,-\ell_{\gamma}^+[.
\eea
Thus, for the closed  (and regular)  configurations  $\cc_2^+$, we  shall focus  on  the range of specific angular  momenta  $-\ell_2^{+}\in]-\ell_{mbo}^+,-\breve{\ell}_2^+[$ where $r_{mso}^-{\in}\cc_2^+$.

 The photon circular orbit $r_{\gamma}$ is the  upper boundary  of the orbital range associated with the momentum range  $\mathbf{L2}$, and  the lower boundary is   $r_{mbo}$.

Equation \il(\ref{Eq:spec-spo-Mix}) holds,  which means, for a closed regular topology of a $\cc_2^+$ disk, that  $r_{mbo}^-<r_{in}^{2_+}$ for $a<a_{\iota}$ but, as  $r_{mbo}^-<r_{mso}^-$. This is not sufficient to rule out the condition $r_{mso}^-\in \cc_2^+$ in these geometries.

However, for the critical configuration $\oo_x^{2_+}$,  one could say that as  $r_{mbo}^-\in \oo_x^{2_+}$  only for  $a<a_{\iota}$ then, as  $r_{mbo}^-<r_{mso}^-$, for proper values of the specific angular momentum,  $r_{mso}^-$  must be included in $\oo_x^{2_+}$.

We could see, from Fig.\il\ref{Fig:Plotaaleph1IIa}-Upper that at  $a<a_{\iota_a}$, where $a_{\iota_a}:\;r_{mso}^-=r_{mbo}^+$, we have
$r_{mso}^->r_{mbo}^+$, and  in conclusion
\bea\label{Eq:hanf-s}
r_{mso}^-\in !\oo_x^{2_+} \quad \mbox{for}\quad \mathbf{A}_{\iota_a}^<,\quad -\ell_2^+\in \mathbf{L2}^+,
\eea
see Fig.\il\ref{Figs:PXasdcont} and also  Figs\il\ref{Fig:Plotaaleph1II},\ref{Fig:Plotaaleph1IIa}.
 We note that in the relation (\ref{Eq:hanf-s}), there is  the intensifier $\in !$, meaning  that \emph{all} the  configurations $\oo_x^{2_+} $, in the spacetimes where   $ a<a_{\iota}$,  \emph{must} contain $r_{mso}^-$ i.e. $r_{J}^{2_+}<r_{mso}^-$, because in those spacetimes there is  $r_{mso}^->r_{b}^+$.

Conversely, this does not imply $r_{mso}^-\non{\in}!\oo_x^{2_+}$ for $a\in\mathbf{A}_{\iota_a}^>$ (here we used the intensifier $(!)$ to emphasize that the non-inclusion relation does  need to be always  verified).

As  $r_{J}^{2_+} \in ]r_{mbo}^+,r_{\gamma}^+[$,
it follows that  the radius  $r_{mso}^-<r_{\gamma}^+$ is not included even in the open topology, or
\bea\nonumber
\mbox{for}&&\quad a>a_{\gamma_+}^-\equiv 0.638285 M> a_{\iota_a}:\;r_{\gamma}^+=r_{mso}^-, \\
&&\label{Eq:set-identity}\mbox{there is}\quad
r_{mso}^-\non{\in}\oo_x^{2_+},\quad r_{mso}^-\non{\in}\cc_2^{+},
\\\label{Eq:we-eha}
\mbox{for}&&\quad a\in ]a_{\iota_a}, a_{\gamma_+}^-[\quad\mbox{there is}\quad
r_{mso}^-{\in}\oo_x^{2_+}.
\eea
We stress  that in  Eq.\il(\ref{Eq:we-eha}) there  is $\in$ and not $\in !$,
on the other hand, for this relation to be satisfied, it has to be
$-\ell_2^+\in[-\ell_2^+(r_{mso}^-),-\ell_{\gamma}^+[$ (one can see the general behavior of the curves $\mp\ell_{\pm}$ versus $r/M$  in Fig.\il\ref{Figs:IPCPl75}), see also Fig.\il\ref{Fig:Plotaaleph1II}.

Thus one can specify Eq.\il(\ref{Eq:we-eha}) as  follows:
\bea
\label{Eq:we-eha-I}
&&r_{mso}^-{\in}!\oo_x^{2_+}\quad \mbox{for}\quad a\in ]a_{\iota_a}, a_{\gamma_+}^-[\quad
\mbox{and}\\\nonumber
&&  -\ell_2^+\in[-\ell_2^+(r_{mso}^-),-\ell_{\gamma}^+[.
\eea

The situation is  more structured for the closed  $\cc_2^+$ configurations--see Fig.\il\ref{Fig:Plotaaleph1IIa}-bottom.

In order to establish the location of the $r_{mso}^-$ in the counterrotating disks $\cc_2^+$, we can consider the  following three cases:\footnote{We  note that in dealing with multiple surfaces formed by $\ell$counterrotating fluids,  the static limit  represents an important limitation.  On the equatorial plane, this is  placed on the orbit $r_{\epsilon}^+=2M$, which is invariant with respect to the  change of the attractor spin. The static limit and the inner region $\Sigma_{\epsilon}^+\equiv]r_+,r_{\epsilon}^+[$ (\emph{ergoregion}),  have very peculiar characteristics;  the static limit, for any spacetime $a\neq0$, acts in some way  as a \emph{semipermeable membrane}, separating materials in  counterrotaing orbits  confined in the outer region,  from matter  corotating with the source. Corotating fluids can penetrate  and possibly also to exit from the static limit. In the processes of energy extraction   from the black hole, for example the Penrose process, matter can go outside $\Sigma_{\epsilon}^+$, crossing $r_{\epsilon}^+$  with greater initial momentum  and energy \citep{ergon}. Note that the efficiency of the energy extraction by the Penrose process in the field of  Kerr black holes \citep{AbraFra} is significantly lower than in the in the field of Kerr naked singularities \citep{SJS2013,Stuchlik1980}}
\bea\label{Eq:tenm-direct}
&&r_{mso}^-\in]r_{mbo}^+,r_{mso}^+[\quad \mbox{ for}\quad a\in]0,a_{\iota_a}[  \\
&&\nonumber
 a_{\iota_a}\equiv 0.372583 M:\; r_{mso}^-=r_{mbo}^+,
 \\\label{Eq:rmso-1-charl}
&& r_{mso}^-\in]r_{\gamma}^+,r_{mbo}^+[
 \quad \mbox{ for} \quad a\in]a_{\iota_a}, a_{\gamma_+}^-[  \\\nonumber
 && a_{\gamma_+}^-=0.638285M:\; r_{mso}^-= r_{\gamma}^+,
 \\\label{Eq:third-case}
&& r_{mso}^-<r_{\gamma}^+ \quad \mbox{ for} \quad a>a_{\gamma_+}^-.
\eea
The case (\ref{Eq:third-case}) was ruled out by Eq.\il(\ref{Eq:set-identity}). Therefore we will investigate the situation for slower attractors, i.e. $a<a_{\gamma_+}^-$, considering  Eq.\il(\ref{Eq:not-non-crea}).

\begin{figure}[h!]
\begin{center}
\begin{tabular}{cc}
\includegraphics[width=1\columnwidth]{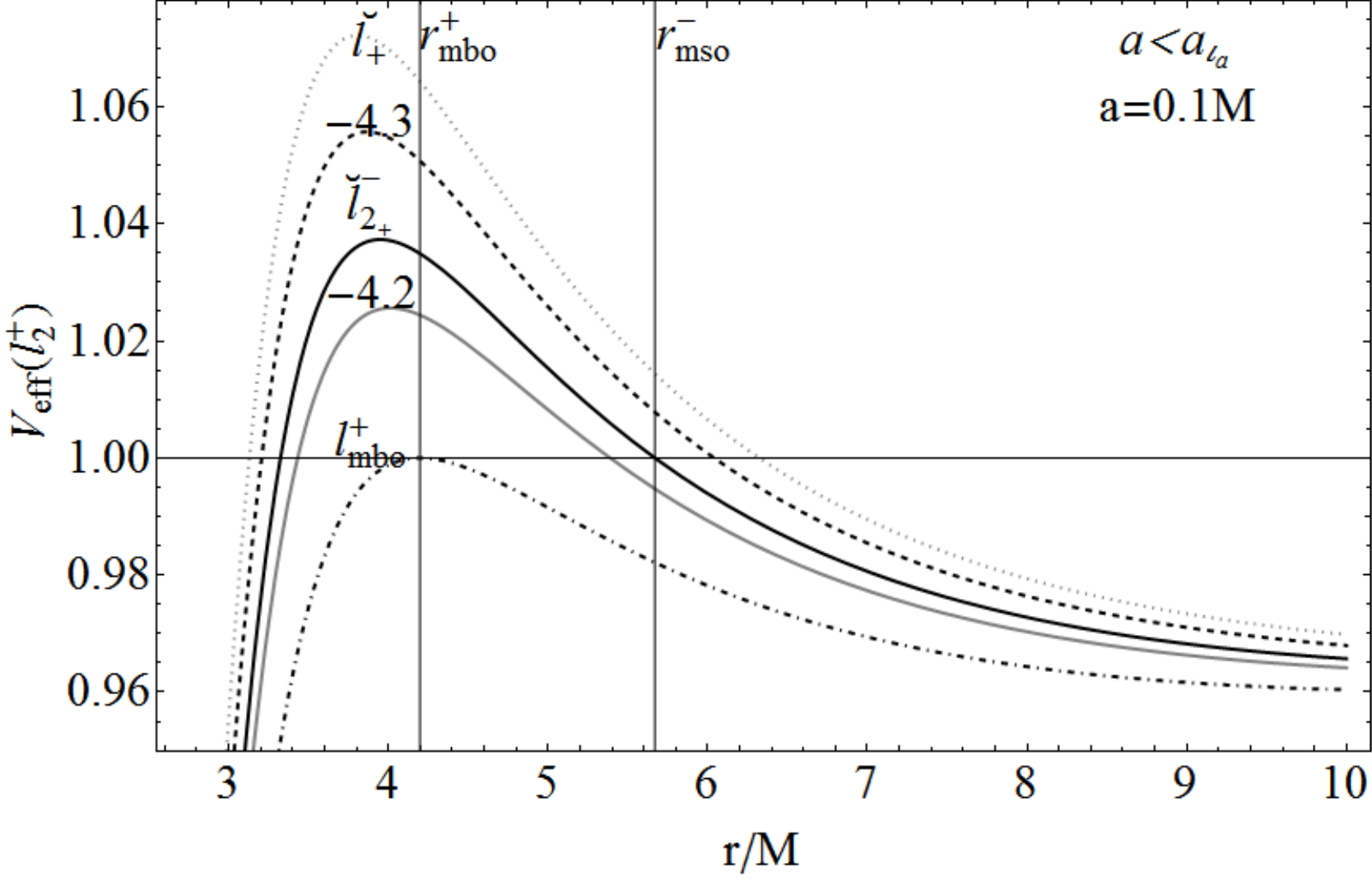}\\
 \includegraphics[width=1\columnwidth]{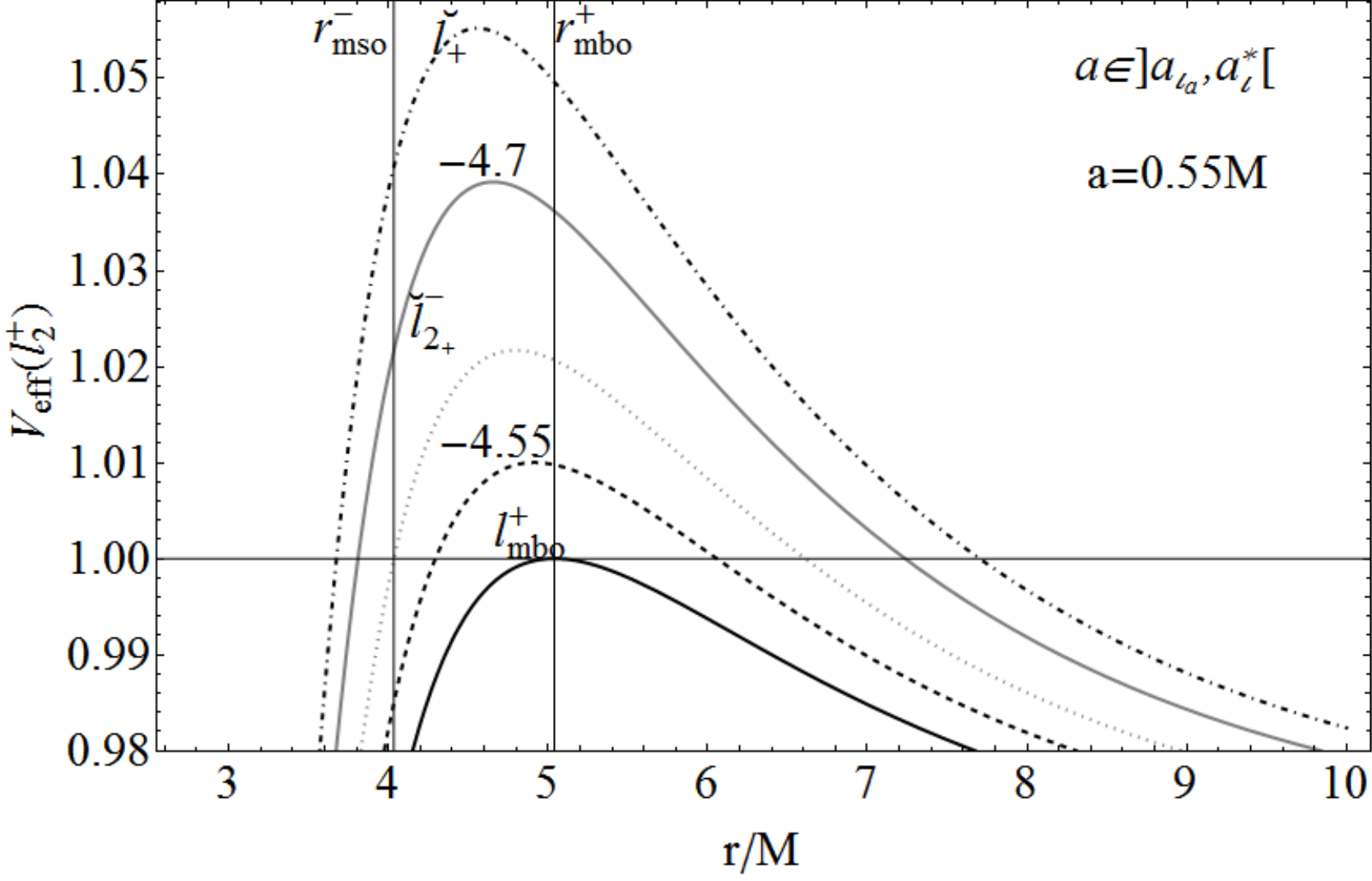}
\end{tabular}
\caption{Counterrotating fluids: disks $\cc_2^+$. Upper panel: spacetime with spin  $a=0.1 M<a_{\iota_a}$, potentials given  for different specific angular  momenta. At $-\ell_2^+\in]-\ell_{mbo}^+,-\breve{\ell}^-_{2_+}]$, with $r_{mso}^-\in \cc_2^+$. The outer horizon $r_+=1.99499M$  and $\breve{\ell}_{2_+}^-=-4.24107$,
$\ell_{mbo}^+=-4.09762$,
$\breve{\ell}_+=-4.3478$. Bottom panel: spacetime with spin $a=0.55 M\in]a_{\iota_a},a_{\iota}^*[$, there are no disk with $r_{mso}^-\in \cc_2^{+}$. In this spacetime  $r_+=1.83516M$, $\breve{\ell}_{2_+}^-=-4.61392$,
$\ell_{mbo}^+=-4.48998$,
$\breve{\ell}_+=-4.77024$.}\label{Fig:liber-Sucx}
\end{center}
\end{figure}

From  Eq.\il(\ref{Eq:rmso-1-charl}) we have $r_{mso}^-<r_{mbo}^+$,  but  we find  also  $r_{mbo}^+\non{\in}\cc_2^{+}$ from Eq.\il(\ref{Eq:beha-dire}). Therefore,  it follows that\footnote{\label{Ref-gfootnote}We address    here  some  general considerations on the arguments that we  use in this section, clarifying particularly certain aspects behind the results given  in  Eq.\il(\ref{Eq:ri-sp-zion}).
The main issue  is to locate the inner edge of the disk, in this specific case a $\cc_2^+$  disk,  in particular with respect to $r_{mso}^-$.  Equation (\ref{Eq:ri-sp-zion}) indicates that the $\cc_2^+$  disk  cannot include  $r_{mso}^-$ for  a very large class of fast attractors, say $a\gtrsim 0.37 M$. In fact, the  inner edge is  $r_{in}^{2_+}>r_{mso}^-<r_{mbo}^+$.  This relation is  trivial when relation  $r_{mso}^-<r_{\gamma}^+$ is satisfied, occurring in the geometries  $a>a_{\gamma_+}^-\gtrsim 0.638 M$. For the geometries   with $a\in]a_{\iota_a},a_{\gamma_+}^-[$,  we have  $r_{mbo}^+>r_{mso}^-$-- see also Fig.\il\ref{Fig:liber-Sucx}-bottom; but the necessary condition for the inclusion of $r_{mso}^-$ into a $\cc_2^+$ configuration  is that there would be $r_{mbo}^+\in \cc_2^+$.  However  previous results in  Eq.\il(\ref{Eq:beha-dire})  had proved that this condition is \emph{never} satisfied, and the reason for this is that the effective potential for the disk with  specific angular  momentum $\mathbf{L2}^+$ is too ``large'' (i.e. $K_{2}^+(r_{mbo}^+)>1)$  as shown in Fig.\il\ref{Fig:liber-Sucx}-bottom. This obviously creates a barrier, essentially due to the centrifugal component of the effective potential, due to which the disk cannot  include,  while remaining in equilibrium,  the orbit $r_{mbo}^+$ and, therefore, a fortiori  the orbit $r_{mso}^+$. In fact this result  is not in contradiction with the results  reflected  in Fig.\il\ref{Fig:Plotaaleph1IIa}-bottom.}
\be\label{Eq:ri-sp-zion}
r_{mso}^-\non{\in}\cc_2^{+} \quad \mbox{for}\quad a>a_{\iota_a} \quad(\mathbf{A}_{\iota_a}^>),
\ee
in accord  with Eq.\il(\ref{Eq:set-identity}).

However we need to discuss the location of the counterrotating  proto-jet point $r_J^+$. Indeed, due to  Eq.\il(\ref{Eq:beha-dire}), we find  $r_{mbo}^+\in \oo_x^{2_+}$. This  task has been completed in  Eq.\il(\ref{Eq:set-identity}) and Eq.\il(\ref{Eq:we-eha-I}).

We focus then on  the toroidal fluids  orbiting with proper specific angular  momentum   $-\ell_2^{+}\in]-\ell_{mbo}^+,-\breve{\ell}_2^+[$, around attractors with $a<a_{\iota_a}$,  considered in  Eq.\il(\ref{Eq:tenm-direct}), where there is  $r_{mso}^-\in]r_{mbo}^+,r_{mso}^+[$.

The effective potential function  decreases monotonically in this orbital range,
but it does not reach the maximum (as the range of specific angular  momentum is  \textbf{L2}). Thus,
we consider a ``starting'' configuration ``embedded'' in an effective potential $V_{eff}(\ell_2^+,r_{mso}^+)<1$, where  $-\ell_2^+\in]-\ell_{mbo}^+, -\breve{\ell}_2^+[$, for Eq.\il(\ref{Eq:states-energies-Y}) (that is not in contradiction with Eq.\il(\ref{Eq:not-non-crea})).

However, we have  $r_{mbo}^+\non{\in}\cc_2^+$ for  Eq.\il(\ref{Eq:beha-dire}), which means  $V_{eff}({\ell_2^+,r_{mbo}^+})>1$.  It follows then that   a radius $\bar{r}\in]r_{mbo}^+, r_{mso}^+[:\; V_{eff}(\ell_2^+,\bar{r})=1$ exists.
 Following arguments  similar to the ones  developed in \ref{Sec:graph-def}, in order
to evaluate if  there are actually solutions of this problem under the condition $-\ell_2^{+}\in]-\ell_{mbo}^+,-\breve{\ell}_2^+[$ on the specific angular  momentum, we have  to know the situation for $V_{eff}(\ell_2^+, r_{mso}^-)$. Therefore we look  for  the solutions of the problem   $\exists\;\breve{\ell}^-_{2_+}:\;V_{eff}(\ell_2^+, r_{mso}^-)=1$ in   $-\ell_2^{+}\in]-\ell_{mbo}^+,-\breve{\ell}_2^+[$.

Fig.\il\ref {Fig:Plotaaleph1IIa} is a restriction of Fig.\il\ref{Figs:PXasdcont} and shows the situation for $-\ell^+\in]-\breve{\ell}_+,-\ell_{mbo}^+[$ where, according to
Eq.\il(\ref{Eq:not-non-crea}), configurations  $\cc_2^+$ including  $r_{mso}^-$ are possible.
We  expect therefore that there will be an orbital region included in $\Delta r_x^{2_+}$, and a range of specific angular  momentum for the counterrotating matter in  $\mathbf{L2}^+$ satisfying this condition.
The figure shows  the function $\breve{\ell}_{2_+}^-$,  as evaluated in $r_{mso}^-$.

 In general, one finds  that for $-\ell_2^+\in]-\ell_{mbo}^+,-\breve{\ell}_{2_+}^-[$,  there  is $V_{eff}(\ell_2^+,r_{mso}^-)<1$. Therefore, there  can be
$r_{mso}^-\in \cc_2^+$, while  for $-\ell_2^+\in]-\breve{\ell}_{2_+}^-,-\breve{\ell}_+[$ we find  $r_{mso}^-\non{\in} \cc_2^+$.

 Concluding this paragraph we note that to establish an analogue inclusion relation with respect to the  open surfaces $\oo_x^{2_+}$, we can compare the two analysis.
\medskip

\textbf{(2) Concluding remarks on the problem $r_{mso}^-\in ()_2^+$}

We have shown that the solution of this problem is different for different classes of attractors. This property, as all other cases in which  different classes of attractors were pointed out, turns to be a possible  useful tool for the identification of features possibly  distinguishing   between different gravitational sources.

More specifically, here the class of attractors  with spin in  $]0, a_{\gamma_+}^-[$  is then decomposed in the following sub-classes  $]0, a_{\gamma_+}^-[= ]0, a_{\iota_a}[\cup]a_{\iota_a},a_{\iota}^*[\cup]a_{\iota}^*,a_{\gamma_+}^-[$,   where:
\be\label{Eq:as-def-iota-star}
a_{\iota}^* =0.61834M\in]a_{\aleph_1}, a_{\gamma_+}^-[:\quad\breve{\ell}_{+}=\breve{\ell}_{2_+}^-,
\ee
 and $ a_{\gamma_+}^-$ was introduced in  Eq.\il(\ref{Eq:set-identity}).

However, these considerations are only necessary to ensure the condition  $r_{mso}^-\in \cc_2^+$,  but not sufficient. Indeed, considerations (\ref{Eq:ri-sp-zion}) rule out the geometries
$\mathbf{A}_{\iota_a}^>$, although in those spacetimes   the condition  $V_{eff}(\ell_2^+,r_{mso}^-)<1$ holds.

  Essentially, the (centrifugal)  barrier, provided at  $r_{mbo}^+$, does not  allow   the inclusion of  $r_{mbo}^+$ in the disk and we  recall that the upper boundary of the range of specific angular  momentum is intrinsically related  to $r_{mso}^+$--see Fig.\il\ref{Fig:liber-Sucx}-bottom.
   In fact, for $a<a_{\iota_a}$ see  Fig.\il\ref{Fig:liber-Sucx}-upper, there is  $r_{mso}^-\in]r_{mbo}^+, r_{mso}^+[\equiv\Delta r_{x}^+$ and, if the specific angular  momentum is low enough, i.e., $-\ell_2^+\in]-\ell_{mbo}^+,-\breve{\ell}_{2_+}^-[$, then
   the inner edge $r_{in}^{2_+}$ can be chosen, if the disk is  sufficiently dense or equivalently, if the hydrostatic pressure is sufficiently large (i.e., the difference between the pressure at its maximum and  the pressure at its minimum located at the disk boundary), so  that  the counterrotating disk  can incorporate  $r_{mso}^-$.

 As we have already noted, in many of these issues, the component of the potential that further changes the behavior of the disk  is its centrifugal part: for the specific angular  momentum too low (in magnitude), no disk  will  form. By increasing the specific angular  momentum a disk with low density and very small size arises. At larger angular momenta, the minimum density of the disk, given  as a function of  $K$, increases and the disk, in order to counterbalance this effect, will  move  the point of maximum pressure away from the central attractor, while  its inner edge will more towards the gravitational attractor,  increasing thus  its extension until arriving to an unstable phase.

 Further increasing  of the  specific  angular  momentum  leads to  destruction of the closed  topology, the outer edge being a  P-W instability point.

 Thus, in conclusion for  the  $\mathbf{A}_{\iota_a}^<$ spacetimes, we have  $r_{mso}^-\in \cc_2^+$ only for   $-\ell_2^+\in]-\ell_{mbo}^+,-\breve{\ell}_{2_+}^-[$,
  this situation is reflected  very clearly by    Fig\il\ref{Fig:liber-Sucx}.

With reference to Fig.\il\ref{Fig:Plotaaleph1IIa}-bottom, this  region of the  specific angular  momentum decreases
with  the spin  in $]0, a_{\iota_a}[$ up to  $a_{\iota_a}:\;r_{mso}^-=r_{mbo}^+$, where it vanishes and no  such configurations are possible, see also  Eq.\il(\ref{Eq:tenm-direct})\footnote{ Similarly, regular  $\cc_2^+$ disks,  satisfying this property,   will be constrained   in terms of  the possible orbital range. The situation  is different  for  attractors  with spin in   $]a_{\iota_a}, a_{\iota}^*[$, where the orbital range increases  again with the spin    up to the  upper boundary of the second class, $a_{\iota}^*$-- see also Eq.\il(\ref{Eq:as-def-iota-star}),  where  maximum extension of the orbital region occurs, and  in the third, very restricted class $]a_{\iota}^*,a_{\gamma_+}^-[$.}.
We can summarize this analysis by saying that:
\bea\label{Eq:not-non-crea-1}
&& r_{mso}^-\non{\in}\cc_2^+\quad \mbox{if}\quad -\ell_2^{+}\in]-\breve{\ell}_2^+,-\ell_{\gamma}^+[\in \mathbf{L2}^+,
\\
&& r_{mso}^-\non{\in}\cc_2^{+},\;
r_{mso}^-\non{\in}\oo_x^{2_+},\;\mbox{for}\; a>a_{\gamma_+}^-> a_{\iota_a},
\\
&&
r_{mso}^-\non{\in}\cc_2^{+} \quad \mbox{for}\quad a>a_{\iota_a},
\\
&&r_{mso}^-{\in}!\oo_x^{2_+}\quad \mbox{for}\quad a\in ]a_{\iota_a}, a_{\gamma_+}^-[\quad
\mbox{and}
\\&&\nonumber  -\ell_2^+\in[-\ell_2^+(r_{mso}^-),-\ell_{\gamma}^+[,
\\
&&r_{mso}^-\in !\oo_x^{2_+} \quad a<a_{\iota_a},\quad -\ell_2^+\in \mathbf{L2}^+,
\\\nonumber
&&
r_{mso}^-{\in}\cc_2^{+} \; \mbox{for}\; a<a_{\iota_a} \; \mbox{and}\;-\ell_2^+\in]-\ell_{mbo}^+,-\breve{\ell}^-_{2_+}].
\\
\label{Eq:L-Time}
\eea

\textbf{(3) On the configurations $r_{mso}^-\in()_3^+$}

We can now focus on a $\cc_3^+$ disk, considering the marginally stable orbit $r_{mso}^-$. Equation (\ref{Eq:En-Dir}) holds. As  $r_{mso}^-<r_{mso}^+$, this proves that:
\be\label{Eq:equa-one}
r_{mso}^-\non{\in}\cc_3^+.
\ee
\subsubsection{Corotating  disks: $r_{\mathcal{N}}^{+}\in ()_{-}$}\label{Sec:scale-ques}

\subsubsubsection{\textbf{Marginally stable orbit $r_{mso}^{+}\in ()_{-}$}}

It is convenient to consider first the location of the  marginally stable orbits. We can compare $r_{mso}^+$ with the radius $r_{mso}^-<r_{mso}^+$. Clearly it  is always possible to select the specific angular  momentum $\ell^-(r_{mso}^+)$ such that the disk can be centered on
 $r_{mso}^+$, see Fig.\il\ref{Fig:Plotaaleph1II}.
  This disk can be stable, in regular topology $\cc^-$, or in accretion  for the cusped topology  $\cc_x^{-}$,
 depending on the specific angular  momentum  $\ell^-(r_{mso}^+)\in \mathbf{Li}$, see Fig.\il\ref{Fig:Plotaaleph1II}.

We obtain
 \bea\nonumber
&& \ell^-(r_{mso}^+)\in \mathbf{L1}^- \; \mbox{for}\; a<\tilde{a}_{\aleph}\approx 0.4618544 M\in]a_{\aleph_0},a_{\aleph}[
\\
&& \tilde{a}_{\aleph}:\;
 \ell^-(r_{mso}^+)=\ell_{mbo}^-,\; r_{mso}^+=r_{cent}^-\in\{\cc_1^-, \cc_x^{1_-}\},
 \\ \label{Eq:dist-mon-1} &&
 \\\label{Eq:dist-mon-2}
&&  \ell^-(r_{mso}^+)\in \mathbf{L2}^- \; \mbox{for}\; a\in]\tilde{a}_{\aleph},\breve{a}_{\aleph}[\\
&& \breve{a}_{\aleph}=0.73688M:\;
  \ell^-(r_{mso}^+)=\ell_{\gamma}^- \\
  &&\nonumber  r_{mso}^+=r_{cent}^-\in\{\cc_2^-, \oo_x^{2_-}\},
  \\\label{Eq:dist-mon-3}
&&  \ell^-(r_{mso}^+)\in \mathbf{L3}^- \; \mbox{for}\; a>\breve{a}_{\aleph} \;  r_{mso}^+=r_{cent}^-\in \cc_3^-,
 \eea
 We now enquire on the situation where the disk is not exactly centered in $r_{mso}^+$.
This analysis {is} specially relevant in the investigation of a  possible interaction between rings orbiting with   specific angular  momenta {in}  different ranges $\mathbf{Li}$.

Equations\il(\ref{Eq:dist-mon-1},\ref{Eq:dist-mon-2},\ref{Eq:dist-mon-3}) show that  the relation $r_{mso}^+\in ()^-$ is invariant for  a slight change of specific angular  momentum around the value $\ell^-(r_{mso}^+)$. However, the orbit $r_{mso}^+$  may be non included in a corotating ring
with lower specific angular  momentum   (with $r_{cent}^-<r_{mso}^+$), or larger specific angular  momentum (with $r_{cent}^->r_{mso}^+$): the first case for   $r_{out}^-<r_{mso}^+$,  the second for $r_{in}^->r_{mso}^+$.

We should compare the value of the maximum $K_{Max}^-$, with the value of the potential at  $r_{mso}^+$. In fact, it is immediate to see that,  for $\ell^-<\tilde{\ell}^-\equiv\ell^-(r_{mso}^+)$, the effective potential   $V_{eff}(\ell^-,r_{mso}^+)<1$, but $r_{Max}^-> \tilde{r}_{Max}^-\equiv r_{Max}^-(\tilde{\ell})$.

If $V_{eff}(\ell^-,r_{mso}^+)>K_{Max}^-$, then  $r_{out}^-<r_{mso}^+$,
 where there is an angular  momentum  $\ell^-<\tilde{\ell}^-$ such that a maximum   $r_{Max}$ exists. In fact, the condition   $V_{eff}(\ell^-,r_{mso}^+)>K_{Max}^-$ can be verified, for definition of maximum point,  only in the asymptotic region $r>r_{min}^-$, as one  assumes $K_{Max}^-<1$--see also the analysis in Fig.\il\ref{Fig:Pgreat0}.

On the other hand,
 we note that the condition
 $V_{eff}(\ell^-,r_{mso}^+)<K_{min}^-=V_{eff}(\ell^-, r_{min}^-)$ cannot be fulfilled for the definition of minimum. Then:
\be\label{Eq:schem-Math}
K_{min}^-<V_{eff}(\ell^-,r_{mso}^+)<K_{Max}^-=V_{eff}(\ell^-, r_{Max}^-),
 \ee
 if   $ r_{Max}^-$ exists, and then $\ell^-$ is  in $\mathbf{L1}^-$ or $\mathbf{L2}^-$ (we will analyze later the case  $ \mathbf{L3}^-$).
  Thus  $r_{mso}^+\in()^+$ can be,  but we can  always  verify (sufficient condition)  that if  $V_{eff}(\ell^-, r_{Max}^-)>1$, then  $\ell^-$ is  in $\mathbf{L2}^-$ and  $r_{mso}^+\in \oo_x^{2_-}$.

The general argument, summarized at the end of this section, is the following: as  we know that  $r_{Max}^-<r_{mso}^-<r_{min}^-$ then\footnote{In fact: if $\ell^-<\tilde{\ell}^-$, then  $r_{Max}^->\tilde{r}_{Max}^-$. But if $r_{Max}^->\tilde{r}_{Max}^-$ then  there is $K_{Max}^-<\tilde{K}_{Max}^-$-- see Fig.\il\ref{Figs:IPCPl75}-bottom. Therefore,  we have $\tilde{r}_{out}^{x_-}>r_{out}^{x_-}$ (we recall that in the case of critical configurations the outer cross point of the curves  $K_{crit}=$constant  in  Fig.\il\ref{Figs:IPCPl75}-bottom is exactly the outer edge of the critical accretion disk). On the other side, we know that  $\tilde{r}_{out}^{x_-}>r_{mso}^+$, because $r_{mso}^+=\tilde{r}_{min}^-$,
 from the  definition of $\tilde{\ell}^-$, as  this fact implies that $\tilde{K}_{Max}^->K_{mso}^+$--see  Fig.\il\ref{Figs:IPCPl75}-bottom. However,  it has to be $K^-_{Max}\in
 ] K_{mso}^-,K^-_{crit}(r_{mso}^+) [ \cup ]K^-_{crit}(r_{mso}^+),\tilde{K}_-^{Max}[$. We recall that $\ell^-<\tilde{\ell}^-=\ell^-(r_{mso}^+)$, because there is $r_{mso}^+=\tilde{r}_{min}^-$. Now, if  $K_{min}^-<K^-_{Max}\in
 ] K_{mso}^-,K^-_{crit}(r_{mso}^+) [$, then $r_{out}^{x_-}<r_{mso}^+$, as  immediate to see by  considering the curves $K_-$ in Fig.\il\ref{Figs:IPCPl75}-bottom, and this finally proves the result in Eq.\il(\ref{Eq:inter-BH}). The condition above implies that there is the  specific angular  momentum $\ell^-\in]\ell_{mso}^-,\bar{\ell}_{mso}^{+}[,$ where $\bar{\ell}_{mso}^{+}>0:\; V_{eff}(\bar{r}_{Max},\bar{\ell}_{mso}^{+})=K^-_{crit}(r_{mso}^+)$.
 In the second case, Eq.\il(\ref{Eq:sho-H-Delta}), where  $r_{mso}^+\in \cc_x^{1_-}$,
 we know that $ r_{min}^-\in]r_{mso}^-, r_{mso}^+[$ for assumption, and then  the specific angular  momentum
 $\ell^-\in]\ell_{mso}^-, \tilde{\ell}=\ell^-(r_{mso}^+)[$,  and therefore $r_{Max}\in]\tilde{r}_{Max}^-,r_{mso}^-[$. However,  $\tilde{K}_{Max}^->K_{mso}^+$ if $ \tilde{K}_{Max}^->K_{Max}^->K_{mso}^+$, and then we have
$\tilde{r}_{Max}^-<r_{Max}^-<\bar{r}_{Max}$.}
\bea&&\nonumber
 \mbox{  if}\quad \ell^- <\tilde \ell^-\quad \mbox{ there is }\\
 &&\nonumber\tilde{r}_{Max}^-<r_{Max}^-<r_{mso}^-<r_{min}^-<r_{mso}^+,\\
 &&\label{Eq:inter-BH}\mbox{and thus}\quad
r_{mso}^+\non{\in}\cc_x^{1_-}.
\eea
So far we have considered
  $\ell^-\in \mathbf{L1}^-$, therefore it holds for any Kerr attractor according to the Equations \il(\ref{Eq:dist-mon-1},\ref{Eq:dist-mon-2},\ref{Eq:dist-mon-3}), also for
fast attractors where there are no  $\tilde{r}_{Max}$,  see Eq.\il(\ref{Eq:dist-mon-3}).
Then
\bea&&\nonumber
\mbox{  if}\quad \ell^->\tilde{\ell}^-\quad \mbox{ there is }\\
&&\nonumber r_{Max}^-<\tilde{r}_{Max}^-<r_{mso}^-<r_{mso}^+<r_{min}^-\\
&&\label{Eq:sho-H-Delta} \mbox{and thus }\quad  r_{mso}^+\in!()^-_x.
\eea
In terms of  the maximum points, conditions in Eq.\il(\ref{Eq:sho-H-Delta}) hold where maximum points of the effective potential exist. Precisely, we can always select a  $K$, where  $K_{Max}^-<1$ (which occurs for $\ell^-\in\mathbf{L1}^-$). The situation is different for proper specific angular  momentum in the ranges  $\mathbf{L2}^-$ and $\mathbf{L3}^-$  where,  if $V_{eff}(\ell^-, r_{mso}^+)$ is well defined, it  should be $V_{eff}(\ell^-, r_{mso}^+)<1$, and there would be an upper  limit on the specific angular  momentum  $\ell^->\tilde{\ell}^-$  depending  on the black hole spin.

Let us consider then Eq.\il(\ref{Eq:sho-H-Delta}) for the minimum points:
$V_{eff}(\bar{\ell}_-,r_{mso}^+)<1$
for $\bar{\ell}_-<\breve{\ell}_*$, where $\breve{\ell}_*:\; V_{eff}(\breve{\ell}_*,r_{mso}^+)=1$, see Fig.\il\ref{Figs:PXasdPXasdP}.
We can now trace easily come conclusions:
\bea
&&\nonumber  r_{mso}^+\in \cc^-\;:
\mbox{for}\quad a\in[0,\breve{a}_*[\quad V_{eff}({\ell}_-,r_{mso}^+)<1 \\
&&\label{Eq:evol-suss-corr}
 \mbox{in}\quad \mathbf{L1}^-\cup[\ell_{mbo}^-,\breve{\ell}_*[\subset \textbf{L2}^-,
\\\label{Eq:evol-suss-corr2}
&&\mbox{for}\quad a\in]\breve{a}_*,M]\quad V_{eff}({\ell}_-,r_{mso}^+)<1 \\
&&\nonumber \mbox{in}\quad \mathbf{L1}^-\cup \mathbf{L2}^-\cup ]\ell_{\gamma}^-,\breve{\ell}_*[\subset \textbf{L3}^-,
\\\label{Eq:evol-suss-corr3}
&&\mbox{where}\quad\breve{a}_*\equiv 0.401642 M:\; \breve{\ell}_*=\ell_{\gamma}^-.
\eea
We recall that   Eqs\il(\ref{Eq:evol-suss-corr}--\ref{Eq:evol-suss-corr3}) are indeed  necessary but not sufficient,  for   it is  always possible to find an appropriate  $K<1$  such that  $r_{mso}^+$ is included in the disk, and then in particular   $r_{mso}^+<r_{out}^-$.

It should be ensured that the maximum of the potential,  being   located  at $r_{Max}^-<r_{mso}^+$ while it  exists,  satisfies  the relation $K_{Max}^-\geq V_{eff}(\ell^-,r_{mso}^+)$.

In the case $\mathbf{L2}^-\cup \mathbf{L3}^-$, the conditions in Eqs\il(\ref{Eq:evol-suss-corr}--\ref{Eq:evol-suss-corr3}) are also sufficient.
While this is not immediate  for the disks with momentum in  $\mathbf{L1}^-$, for which  $K_{Max}^{1_-}<1$. Therefore, we should consider  the condition $K_{Max}^{1_-}>V_{eff}(\ell_1^-,r_{mso}^+)$, implying restriction on  $\mathbf{L1}^-$, see Fig.\il\ref{Figs:IPCPl75}-bottom.

It is worth to say  that the location of the outer edge of the disk, so far ignored in this analysis, becomes relevant in the discussion of the problem of inclusion for the
  corotating  disk.
In fact,
the position of $r_{\mathcal{N}}$, with respect to the outer margin, is basically determined by the possibility to find out a  proper $K$.
\subsubsubsection{\textbf{Marginally bounded orbit $r_{mbo}^{+}\in ()_{-}$}}

The issue of the location  of the $ r_{mbo}^{+}$   with the respect to a corotating configuration is extremely significant. The situation is  rather complex and here  we will  provide some general considerations in the analysis of   different specific situations.

The corotating configuration  could be located either  in the region $r>r_{mbo}^+$ or $r<r_{mbo}^+$, as detailed in Sec.\il(\ref{Sec:ell-cont-double}).
The investigation of this case  involves the distinction of  two classes of attractors  and the analysis of the location of  $r_{mbo}^+$ with respect to both the inner and outer edge of  the closed configurations.
More specifically, we will need to  compare the situation  for  $r_{mbo}^{+} $ with  that for  $r_{mso}^-$ and $r_{mbo}^-$.  As such we distinguish the two classes of attractors: $\mathbf{A}_{\iota_a}^<:\quad a\in[0,a_{\iota_a}[$, where $r_{mbo}^+<r_{mso}^-$,  and $\mathbf{A}_{\iota_a}^>:\quad a\in]a_{\iota_a},M]$, where $r_{mbo}^+>r_{mso}^-$; at $a=a_{\iota_a}$  we have $r_{mbo}^+=r_{mso}^-$, see Fig.\il\ref{Fig:Plotaaleph1IIa}.

In the second class of geometries, $\mathbf{A}_{\iota_a}^>$, the radius $r_{mbo}^+$ can correspond to a center of the disk,  or also any point of the configuration, but not a critical point.  For $\mathbf{A}_{\iota_a}^<$  attractors, the $r_{mbo}^+$ can be any point of the disk in general, but not the center of the closed configuration, it can be however its critical cusped point.

In other words:
\bea\nonumber
&&\mbox{for}\quad\mathbf{A}_{\iota_a}^<:\quad a\in[0,a_{\iota_a}[;\quad r_{mbo}^+\non{\in}\{r^-_J, r_{cent}^-\}; \\\nonumber
&& \mbox{it could be }
\quad r_{mbo}^+=r_{x}^-\quad\mbox{or}\quad r_{mbo}^+\in \left.()^-\right|_{\Theta(r_{cent}-r_{in})}
\\\label{Eq:A-nimos}
&&
\\\label{Eq:A-minos}
&&\mbox{for}\quad\mathbf{A}_{\iota_a}^>:\quad a\in]a_{\iota_a},M];\quad r_{mbo}^+\non{\in}\{r^-_J, r_x^-\};\\
&& \mbox{it could be } \quad  r_{mbo}^+=r_{cent}^- \quad \mbox{or}\quad r_{mbo}^+\in ()^-,
\eea
where $()^-$, as usually, does not particularize the  topology, $\Theta(r_{cent}-r_{in})$ in Eq.\il(\ref{Eq:A-nimos}) is the Heaviside (step) function  such that $\Theta(r_{cent}-r_{in})=1$  for $r_{cent}>r_{in}$ and $\Theta(r_{cent}-r_{in})=0$ for $r_{cent}<r_{in}$.
We need now to specify the specific angular  momentum $\ell^-$ and the topology of the corotating configuration.
We look at the closed regular corotating  topologies because any critical topology must contain the marginally stable orbit, that is  $r_{mso}^-\in ! ()_x^-$. This is because the  margins of   the critical configurations   (both the inner as well as the outer edges of the  closed cusped topology) are univocally fixed by the specific angular  momentum, see curves $K_{crit}=$constant  in Fig.\il\ref{Figs:IPCPl75}-bottom.  Then  we need essentially to find the appropriate specific angular  momentum for  ${\ell}_{\beta}^-:\; V_{eff}(\ell_{\beta}^-,r_{mbo}^+)<1$.

We can provide some immediate constraints  for the  class of geometries $\mathbf{A}_{\iota_a}^<$  in Eq.\il(\ref{Eq:A-minos}), by considering the location of   $r_{mbo}^+<r_{mso}^-$. For the configurations where  $r_{mso}^-\non{\in}\cc^-$,
  it must be  $r_{mbo}^+\non{\in} \cc^-$.
Therefore, from the former analysis,
in the conditions provided by   Eq.\il(\ref{Eq:ali-06}) for the attractors  $\breve{\mathbf{A}}_<\supset \mathbf{A}^<_{\iota_a}$, we find  $r_{mso}^-\non{\in}\cc_2^-$ (in this case the $r_{in}^{2_-}>r_{mso}^->r_{mbo}^+$)
and from  Eq.\il(\ref{Eq:imply-Col}) for all the  $\breve{\mathbf{A}}_<$  black holes, there is   $r_{mso}^-\non{\in}\cc_3^-$ (in this case, the inner margin of $\cc_3^-$ is far beyond $r_{mso}^-$ and $r_{mbo}^+$). Therefore, one finds:
\bea\nonumber&&
\mbox{for } \mathbf{A}^<_{\iota_a}\quad \mbox{it is}\quad r_{mbo}^+\non{\in}\cc_3^-;\quad \mbox{ and for} \quad \ell^-_2\in]\breve{\ell}_-, \ell_{\gamma}^-[\\
&& \label{Eq:C3C2-par}\mbox{ it is}  \quad r_{mbo}^+\non{\in}\cc_2^- \quad \mbox{as}\quad r_{in}^{2_-}>r_{mso}^->r_{mbo}^+.
\eea
The results in Eq.\il(\ref{Eq:C3C2-par}) explain the situation, for the slow attractors completely for the  $\cc_3^-$  configurations and  for $\cc_2^-$, but not for  the $\cc_1^-$ disks.

\begin{figure*}[ht]
\begin{center}
\begin{tabular}{ccc}
\includegraphics[width=.4\textwidth]{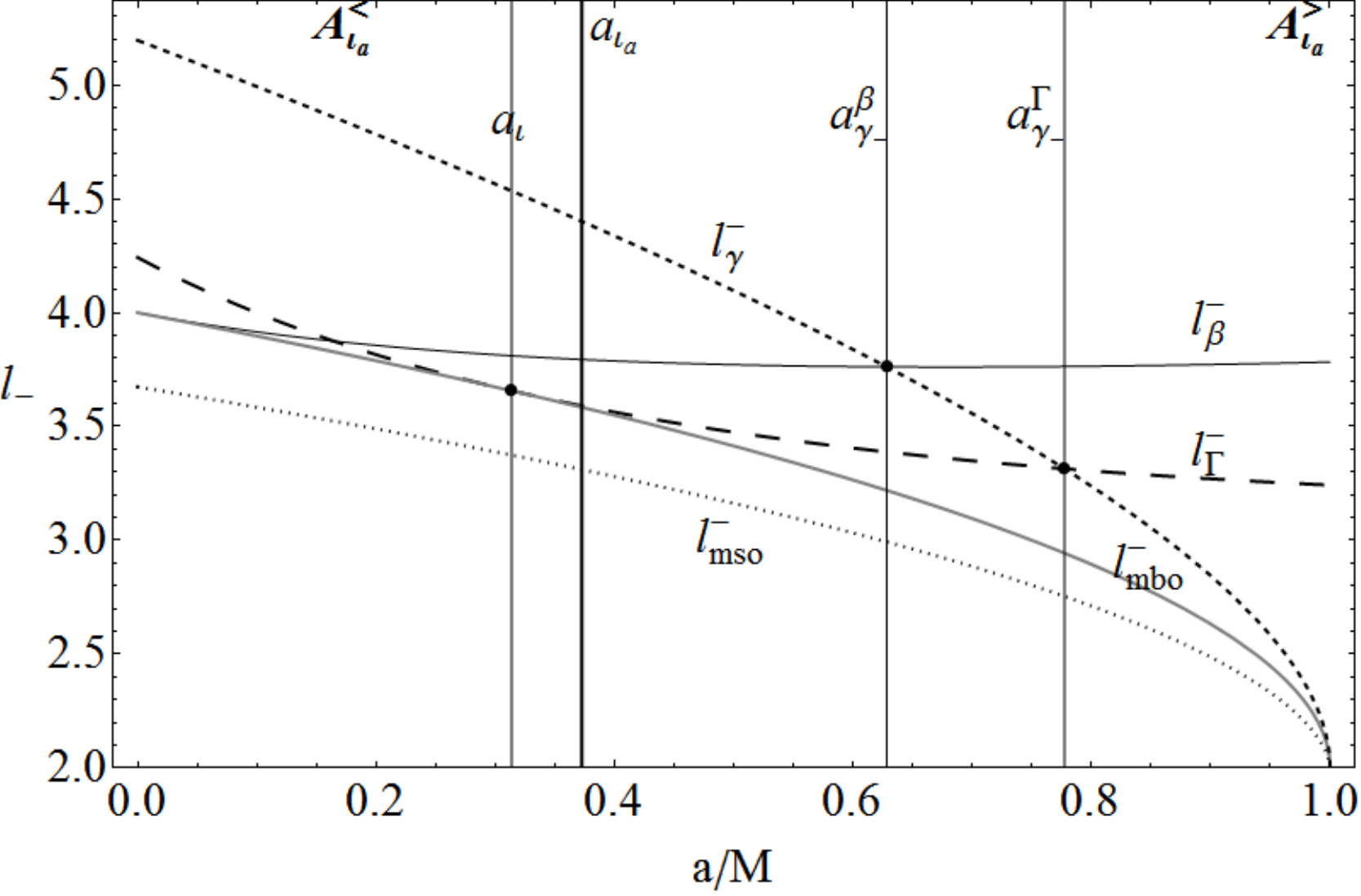}
\includegraphics[width=.4\textwidth]{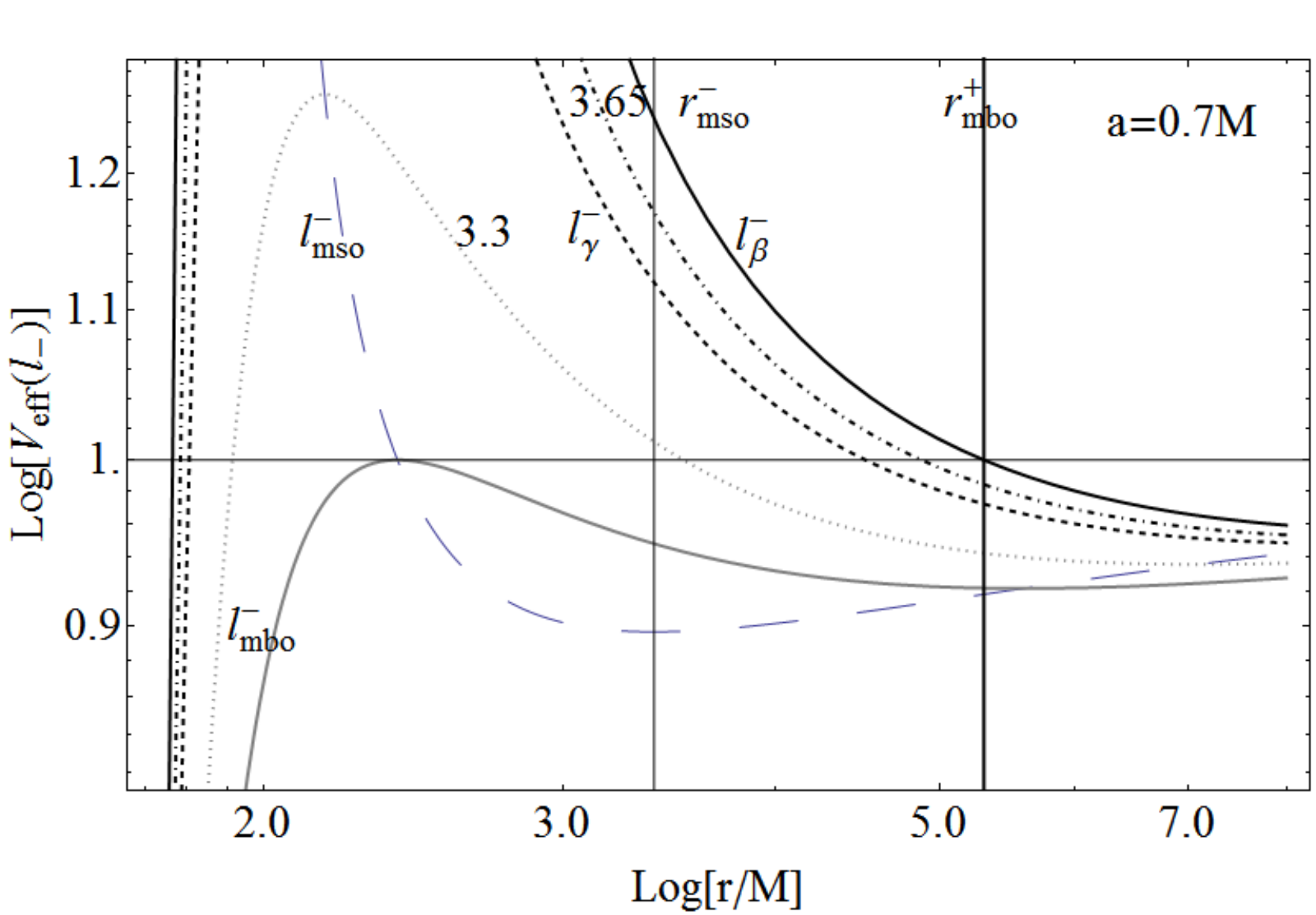}\\
\includegraphics[width=.4\textwidth]{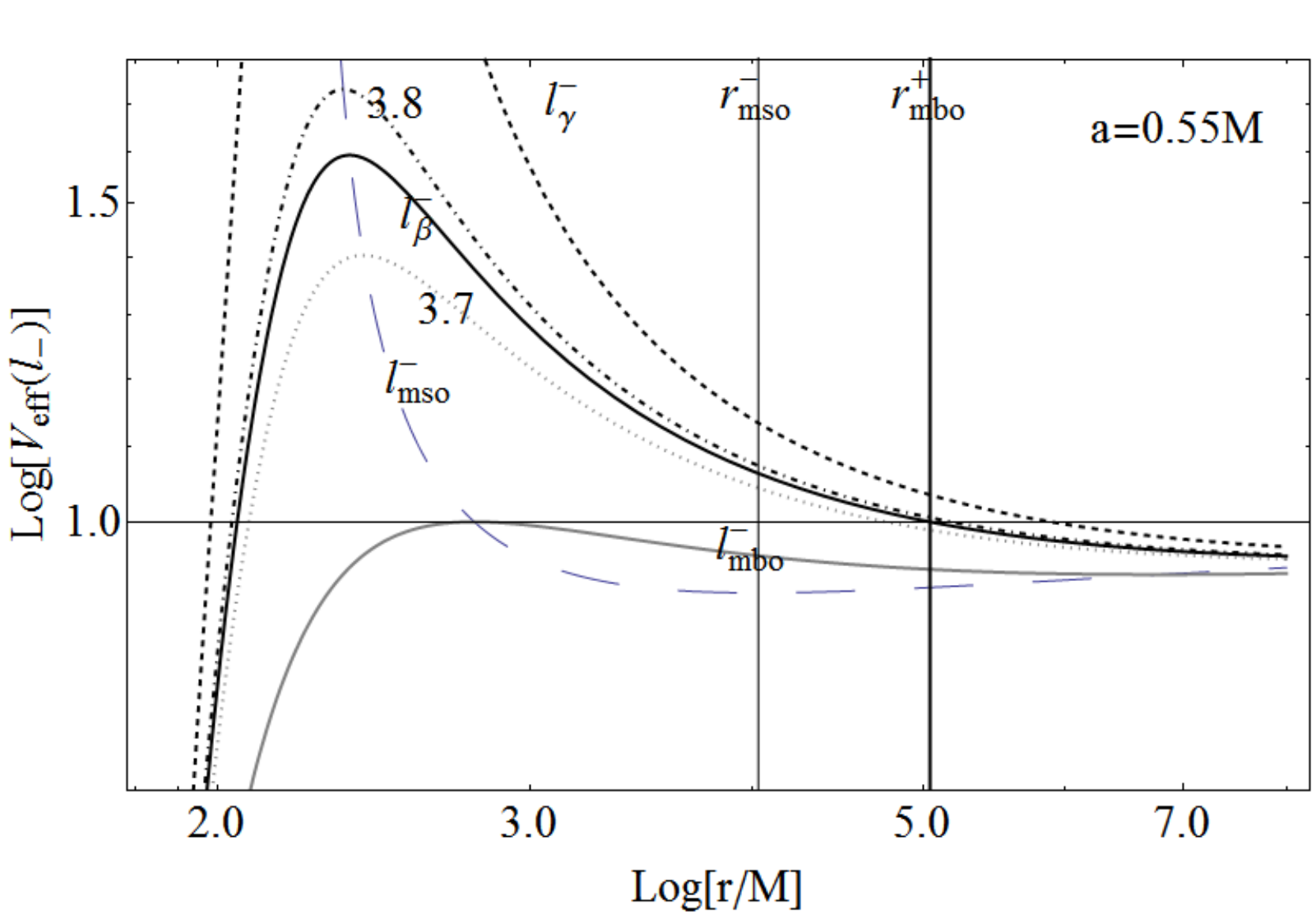}
\includegraphics[width=.4\textwidth]{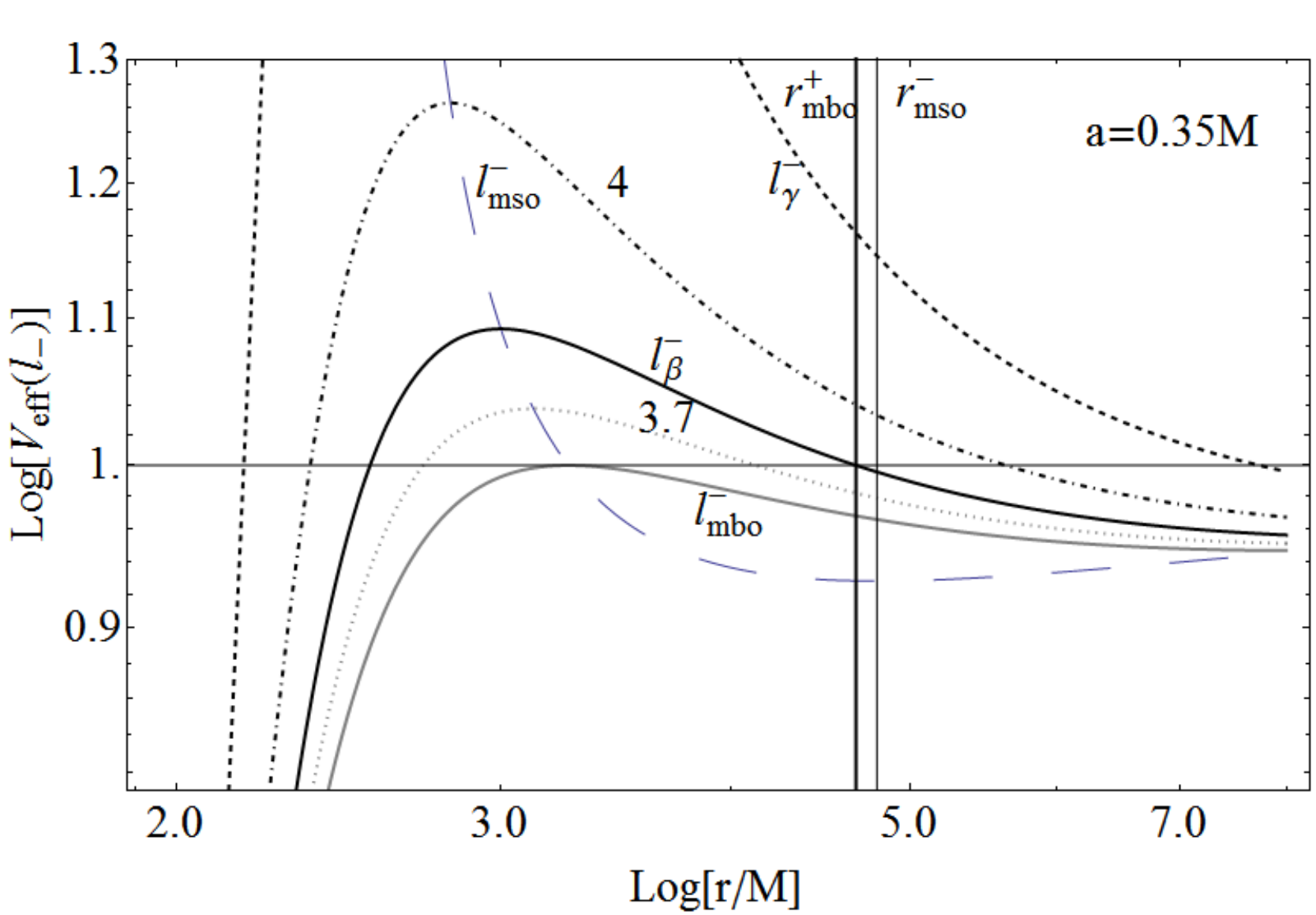}
\end{tabular}
\caption{Corotating  disks: $r_{\mathcal{N}}^{+}\in ()_{-}$: Left-upper-panel: specific angular  momenta $\ell_{mbo}^-\equiv \ell^-(r_{mbo}^-)$, $r_{mbo}^-$ is  the marginally bounded orbit,
$\ell_{mso}^-\equiv \ell^-(r_{mso}^-)$,  $r_{mso}^{\pm}$  is the marginally stable
orbit,
$\ell_{\gamma}^-\equiv \ell^-(r_{\gamma}^-)$,  $r_{\gamma}^{\pm}$  is the
 marginally circular
orbit (photon orbit). Notation $(\pm)$ is for counterrotating and corotating fluids respectively. The specific angular  momentum $\ell_{\beta}^{-}:\; V_{eff}(\ell_{\beta}^{-},r_{mbo}^+)=1$ and $\ell_{\Gamma}^{-}:\; V_{eff}(\ell_{\Gamma}^{-},r_{\gamma}^+)=1$ are also plotted. The spin $a_{\iota_a}$ defines the two classes of attractors $\mathbf{A}_{\iota_a}^<$ and $\mathbf{A}_{\iota_a}^>$ respectively. $a_{\gamma_-}^{\beta}\equiv0.628201 M:\;\ell_{\beta}^-=\ell_{\gamma}^-$ and  $a_{\gamma_-}^{\Gamma}\equiv0.777271M:\; \ell_{\Gamma}^-=\ell_{\gamma}^-$. Right-upper-panel and bottom panels: effective potentials as functions of $r/M$, at different $a/M$ and $\ell$.
}\label{Figs:pePlotre}
\end{center}
\end{figure*}

Similarly to what has been  done in previous cases, we can face this problem introducing  the specific angular  momentum  $\ell_{\beta}^{-}:\; V_{eff}(\ell_{\beta}^{-},r_{mbo}^+)=1$, as shown in Fig.\il\ref{Figs:pePlotre}-left.
From the figure it follows that
\bea\nonumber
&&\ell_{\beta}^-\in \mathbf{L2}^-\quad\mbox{ for}\quad a<a_{\gamma_-}^{\beta}\supset \mathbf{\mathbf{A}}_{\iota_a}^<,\quad\mbox{and}\quad
\ell_{\beta}^-\in \mathbf{L3}^-\quad\mbox{ for}\\
&& a>a_{\gamma_-}^{\beta} \subset \mathbf{A}_{\iota_a}^>
,\quad
a_{\gamma_-}^{\beta}>a_{\iota_a}
\quad \mbox{
for} \quad \ell^-<\ell_{\beta}^- .
\eea
In $ \mathbf{L1}_-$  the effective potential is always less then its  asymptotic limit,  and it follows that it is  possible to locate, with a proper elongation, the disk such that
\bea\nonumber
&&r_{mbo}^+\in ()^-_1;\quad r_{mbo}^+\in ()^-_2  \quad\mbox{for}\quad a>a_{\gamma_-}^{\beta}\quad\mbox{and for }\\
&& a<a_{\gamma_-}^{\beta}\quad\mbox{at}\quad\ell^-<\ell_{\beta}^-,\quad r_{mbo}^+\non{\in} ()^-_2
\\\label{Eq:05L}
&&\mbox{for }\quad\ell^->\ell_{\beta}^-,
\\\nonumber
&&r_{mbo}^+\non{\in} \cc^-_3\quad\mbox{for}\quad a<a_{\gamma_-}^{\beta},\quad \mbox{and for }\\
&& a>a_{\gamma_-}^{\beta}\quad\mbox{and}\quad\ell>\ell^{\beta}_{\gamma_-};\quad r_{mbo}^+{\in} \cc^-_3
 \\\label{Eq:05L1}
 && \mbox{ for } a>a_{\gamma_-}^{\beta}\quad\mbox{and}\quad\ell<\ell^{\beta}_{\gamma_-};
 \eea
these results are in fact verified in Fig.\il\ref{Figs:pePlotre} and they confirm   the conclusion of  Eq.\il(\ref{Eq:C3C2-par}).
Equations (\ref{Eq:05L},\ref{Eq:05L1})  close  discussion of the problem of inclusion.

However, in order to fully characterize this situation we can consider the specific angular  momentum $\ell^-(r_{mbo}^+)$, plotted as function of $a/M$ in Fig.\il\ref{Figs:PXasdPXasdP}-upper.
In fact, as a general premise  we note that for  $\ell^-=\ell^-(r_{mbo}^+)$,  a critical point must be located in  $r_{mbo}^+$, according to Eqs\il(\ref{Eq:A-nimos},\ref{Eq:A-minos}).

In the   $\mathbf{A}_{\iota_a}^<$ geometries, for fluids with  momentum $\ell_{mbo}^+$, the orbit $r_{mbo}^+$ has to correspond to an unstable point, but  it is not  $r_J$ because,  for  possible   $\ell^-(r_{mbo}^+)\in \mathbf{L3}^-$, it could happen according to the discussion in  \ref{Sec:graph-def}  that the potential is not well defined, but in this case, for these attractors, Eq.\il(\ref{Eq:C3C2-par}) holds.

Once the center $r_{min}^-$ is shifted with  respect to $r_{mbo}^+$, such that  $r_{min}\lessgtr r_{mbo}$  (for $\ell^-\neq \ell^-(r_{mbo}^+)$),  one can always select a $K_-$ so that the elongation of the $\cc^-$  configuration  is small enough  to ensure $r_{mbo}^+\non{\in}\cc^-$.
Besides, a more specific  constraint  on $K_-$  could be provided in dependence on $\ell^-$, according to the analogue analysis suggested in     \citep{ringed}, whereas in this section we will  provide constraints on the rotation parameters    $(a/M, \ell)$ for  the location of the center.

We know that the gap  $\bar{\Delta}_{\ell^-}^{mbo}\equiv\bar{\ell}_-\pm \ell^-(r_{mbo}^+)>0$ between the specific angular  momenta (the $\pm$ sign is due to the requirement of positive $\bar{\Delta}_{\ell^-}^{mbo}$ according to the different location of $r_{mbo}^+$ with respect to $r_{cent}$) is proportional to the distance between the radii, or generally $\partial_{\bar{\Delta}_{\ell^-}^{mbo}}(\bar{\Delta}_{r}^{mbo})>0$, where $\bar{\Delta}_{r}^{mbo}$ is the (positive) distance between the radii $(r_{mbo}^+, r_{cent})$. This must imply the possibility of large specific angular  momentum    and then  a  larger $K_-$,   or a larger range of variation for  $K_-$.
 Because   $r_{cent}$ is a minimum of the effective potential  with   $r_{mbo}^+>r_{Max}^-$, then it is always  $V_{eff}(r_{mbo}^+)-V_{eff}(r_{cen})>0$, and this grows with increasing specific angular  momentum (we note that if it  would be  $r_{mbo}^+<r_{Max}^-$,  then the problem would be solved immediately as  $r_{mbo}^+\non{\in}\cc^-$).

Considering  the various locations of $\ell^-(r_{mbo}^+)\in \mathbf{Li}^-$, we  finally  investigate four  classes of attractors,  defined by  the spins
 $\{a_o^*, a_{\iota_a}, a_o, a_o^{\gamma}\}$ as in  Fig.\il\ref{Figs:PXasdPXasdP}-upper:

\textbf{I.}  At
  $a=a_{\iota_a}$ { it is }$ r_{mso}^-=r_{mbo}^+$ and $ \ell^-(r_{mbo}^+)=\ell_{mso}^-$, thus
  \bea
   &&\nonumber \mbox{at}\;  a\in]a_{\iota_a},a_{o}[\subset \mathbf{A}^>_{\iota_a}\;\mbox{it is}\;\ell^-(r_{mbo}^+)\in
	]\ell_{mso}^-,\ell_{mbo}^-[=\mathbf{L1}^-,\\\label{Eq:ge-get}
	&& r_{cent}^{1_-}\lessgtr r_{mbo}^+\quad \mbox{for  }\quad \quad\ell\lessgtr\ell^-(r_{mbo}^+),
\eea
  and then, for $ \ell<\ell^-(r_{mbo}^+)$, the ring density can span in higher values.

\textbf{II.}  At   $a=a_{o}$ we have $\ell_{mbo}^-=\ell_{-}(r_{mbo}^+)$, therefore
	the configuration $\oo_x^-$ has the unstable point $r_J^-=
	r_{mbo}^-$ and  $r_{cent}^-= r_{mbo}^+$.
	\bea&&\nonumber
\mbox{For }\; a \in]a_o,a_{o}^{\gamma}[\subset \mathbf{A}_{\iota_a}^>\;\mbox{it is}\;\ell_{-}(r_{mbo}^+)\in]\ell_{mbo}^-,\ell_{\gamma}^-[= \mathbf{L2}^-\\
&& r_{cent}^{2_-}\lessgtr r_{mbo}^+\quad\mbox{for}\quad\ell\lessgtr\ell^-(r_{mbo}^+),
\eea
and,  for  $\ell=\ell^-(r_{mbo}^+)$,   the center of the disk will be  at
	$r=r_{mbo}^+$.
The region of lower specific angular  momentum,  and the corresponding orbital range, increases with the spin  and thus one could conclude that the separated configurations are favored also with relatively high elongation of the corotating equilibrium disk.
		
\textbf{III.}  Finally:
\bea&&\nonumber
\mbox{for}\quad a\in]a_o^{\gamma},M] \subset \mathbf{A}_{\iota_a}^>\quad \ell^-(r_{mbo}^+) \in \mathbf{L3}^-\\
&& r_{cent}^{3_-}\lessgtr r_{mbo}^+\quad\mbox{for}\quad\ell\lessgtr\ell^-(r_{mbo}^+).
\eea
In this case
	the range of lower specific angular  momentum  is relatively small and it increases with the spin of the attractor.
	At $a=a_o^{\gamma}$,  the center is located at $r_{mbo}^+$ for $\ell^-=\ell^-(r_{mbo}^+)$.

\textbf{IV.} For lower spin, in  the $\mathbf{A}_{\iota_a}^<$ geometries, there is  $r_{mbo}^+\in]r_{mbo}^-, r_{mso}^-[$. Therefore, for  momenta  $\ell_{-}(r_{mbo}^+)$ (for those
in $\mathbf{L1}^-$), the orbit
$ r_{mbo}^+$
is a critical $r_{x}$ point. The  disk center thus will be located   far from $r_{mso}^+$, with maximum distance in the  static case.  For an attractor with $a= a_{\iota_a}$, the center will be located at
$r_{mso}^-$. Therefore, for smaller $a/M$,  at $ \ell<\ell^-(r_{mbo}^+)\in \mathbf{L1}^- $, the disk center will be located close enough
to
the attractor and   the
disk  in its critical (closed) topology will be close to $r_{mso}^-$.

The specific angular   momentum range for  a ring centered in $]r_{mbo}^+,r_{mso}^-[$ is higher at smaller $a/M$ and, at
 $a=a_o^* $, the disk is centered  in $r_{mso}^+$  with  accretion point in  $r_{mbo}^+$.

We note that
 $r_{mbo}^-<r_{mbo}^+$,
 and  for $a<a_{\iota_a}$ we have $r_{mso}^->r_{mbo}^+$.  Therefore no  $\cc^-$ disk can be centered in
 $r_{mbo}^+$. It is necessary then to discuss the location of  $r_{mbo}^+$ with respect to the inner edge of the disk.
For the  larger spin, we find $ r_b^+\in]r_{\gamma}^-, r_{mso}^+[ $  and a $\cc^-$ disk can be centered in  $r_{mbo}^+$.
 Then  we could say  that  for sufficiently high spin, and with  high specific angular  momentum for low spin, it can be $r_{mbo}^+\in \cc^-$, and
 for $a<a_{\iota_a}$, it has to be $r_{mbo}^+\in!()_x^-$. For fast attractors, one has to consider also the outer margin of the disk  and the location of  the marginally stable orbit  as addressed in the previous point --see also the analysis in Fig.\il\ref{Fig:Pgreat0}.
\subsection{On the location of the photon orbits and general consideration on the methods}\label{Sec:photon-C}
We complete our study by considering  the location  of the ring      with respect to the  photon orbits $r_{\gamma}^{\pm}$.
The results of this  investigations  contribute  to possibility to  localize more accurately  a corotating configuration, with respect to the   counterrotating geodesic structure of the Kerr geometry-- Fig.\il\ref{Fig:Plotaaleph1IIa}.

\begin{figure*}[ht]
\begin{center}
\begin{tabular}{ccc}
\includegraphics[width=.338\textwidth]{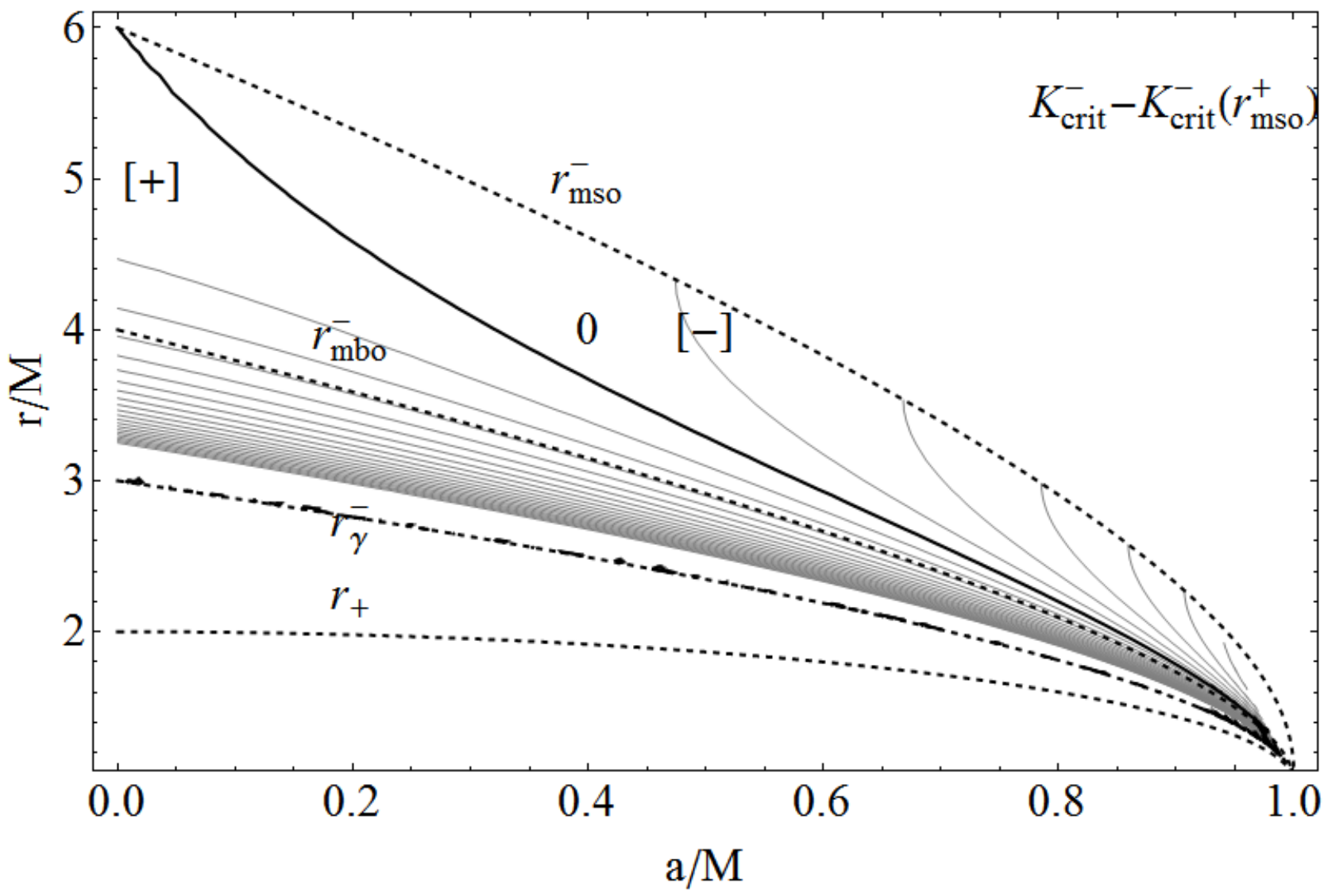}
 \includegraphics[width=.338\textwidth]{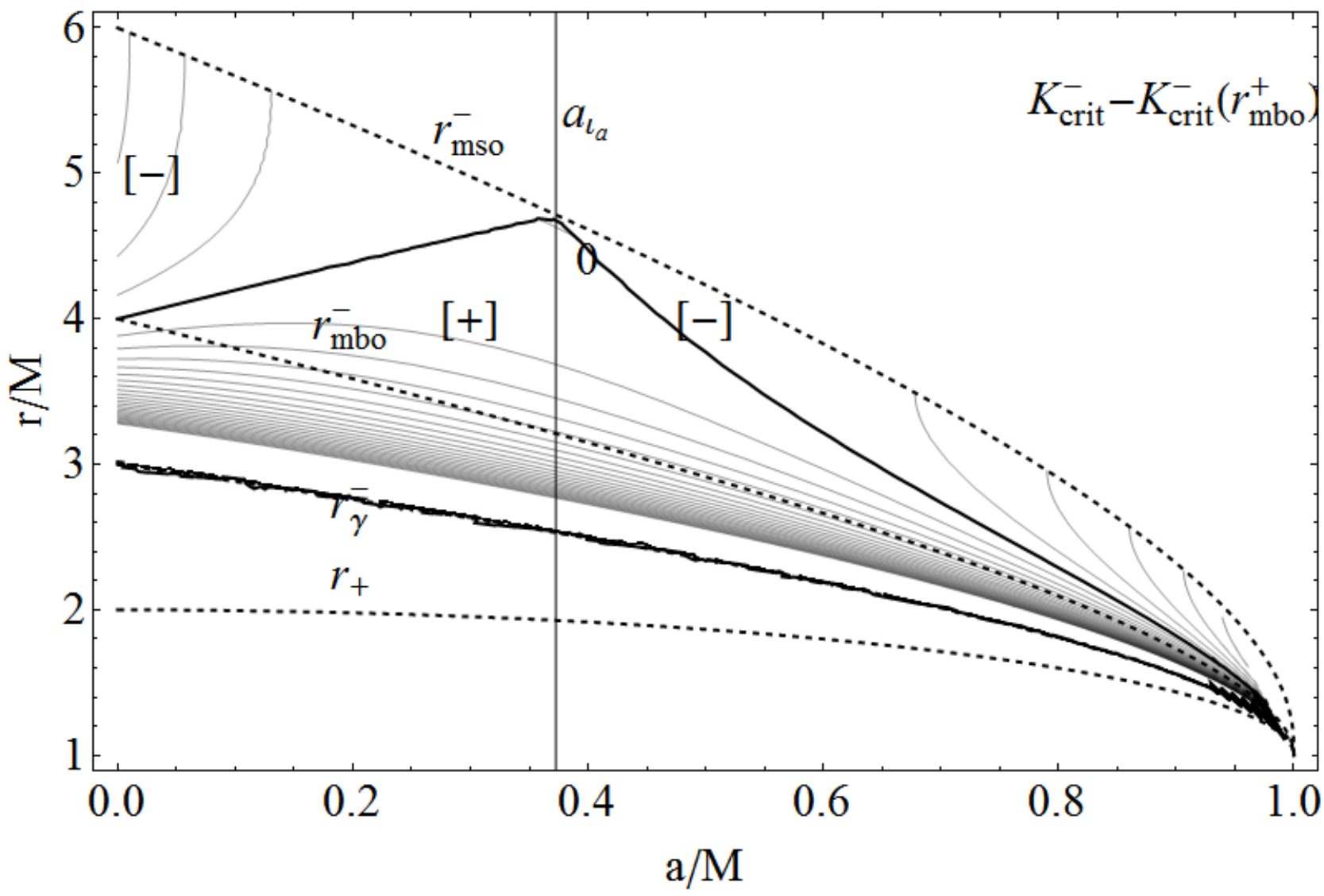}
 \includegraphics[width=.338\textwidth]{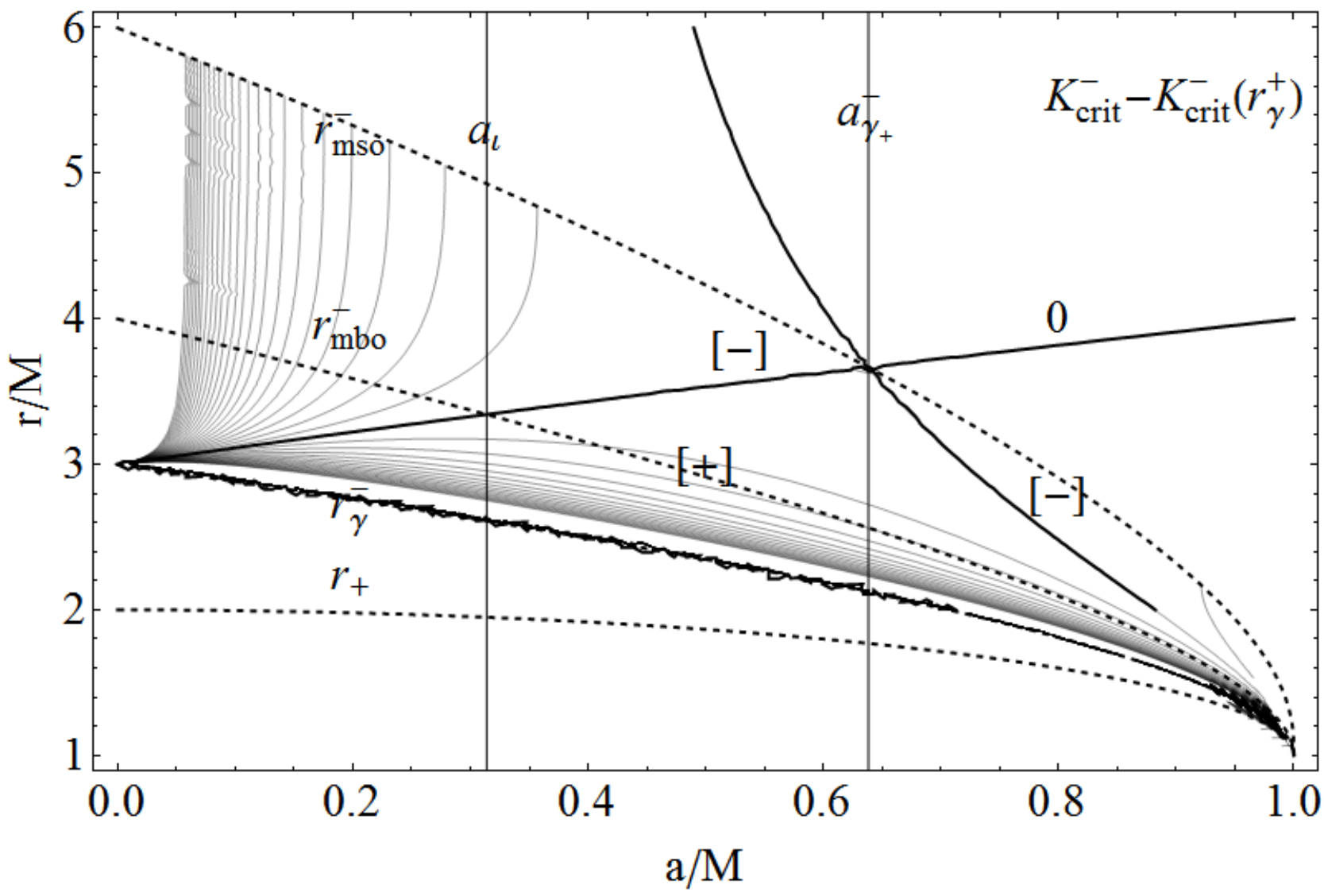}
\end{tabular}
\caption{Location of the notable radii $r_{\mathcal{N}}^{\pm}\in ()_{\mp}$.
Curves $K_{crit}^-(r)-K_{crit}^-(r_{\mathcal{N}}^{+})=$constant
  in the plane $r/M-a/M$. The zero,  $K_{crit}^-(r)=K_{crit}^-(r_{\mathcal{N}}^{+})
  $, is the black thick curve. The region of positive  (negative) value of the constant
  $K_{crit}^-(r)-K_{crit}^-(r_{\mathcal{N}}^{+})$ is indicated with $[+]$ ($[-]$).
  Radii $r_{mso}^-$ (corotating marginally stable orbit),
  $r_{mbo}^-$ (corotating marginally bounded orbit) and
  $r_{\gamma}^-$ (corotating photon orbit) are also plotted,
 $r_{+}$ is the Kerr outer horizon. at $a=a_{\iota_a}$ there is  $r_{mso}^-=r_{mbo}^+$
 see also discussion of  Eq.\il(\ref{Eq:ge-get}). At $a=a_{\gamma_+}^-$ there is  $r_{mso}^-=r_{\gamma}^+$, see also Figs\il\ref{Fig:Plotaaleph1II},\ref{Fig:Plotaaleph1IIa}. }\label{Fig:Pgreat0}
\end{center}
\end{figure*}
It is  clear that the inclusion relations   $r_{\gamma}^{\pm}\non{\in} ()_{\pm}$  as well as $r_{\gamma}^-\non{\in} ()^+$,  due to    $r_{\gamma}^-<r_{\gamma}^+$, can be derived by immediate considerations of the geodesics structure of the Kerr spacetime.
It is now necessary   to discuss  the relation  $r_{\gamma}^+\in ()^-$.
This analysis, together with the considerations drawn on the inclusion relation  $r_{mbo}^+\in ()^-$ at the end of \ref{Sec:scale-ques}, complete the discussion on the $\ell$counterrotating couple $()_{\pm}$ and their possible correlation.

We find:
\bea
&&\nonumber
\mbox{for}\quad a\in{\mathbf{A}}_{\iota}^<\equiv[0,a_{\iota}]\quad\mbox{it is}\quad  r_{\gamma}^+\in]r_{\gamma}^-,r_{mbo}^-]\equiv \Delta r_J^-;\\\nonumber
&&
  \mbox{for}\quad a\in]a_{\iota},a_{\gamma_+}^-]\quad\mbox{it is}\quad r_{\gamma}^+\in]r_{mbo}^-,r_{mso}^-]\equiv\Delta r_x^-,
\\
 &&
\mbox{for}\quad a>a_{\gamma_+}\quad\mbox{it is}\quad r_{\gamma}^+>r_{mso}^-.
\eea
From these relations, and  considering also the results of  \ref{Sec:usua-D-mbo-con} and particularly Eqs\il(\ref{Eq:choral},\ref{Eq:beha-dire}.\ref{Eq:the-ag}), we infer that
  \bea\label{Eq:circum-ved}
 && \mbox{for}\quad a\in[0,a_{\iota}]\quad\mbox{it is}\quad  r_{\gamma}^+\non{\in} (\cc^-,\cc_x^{1_-}),\\\nonumber
&& a\in]a_{\iota},a_{\gamma_+}^-] \quad\mbox{it is }\quad r_{\gamma}^+\non{\in}\cc_3^-;\;r_{\gamma}^+\non{\in}\cc_2^-\\
\nonumber&&\mbox{for}\quad \ell^-_2\in]\breve{\ell}_-, \ell_{\gamma}^-[.
  \eea
Concerning the disks in the geometries
 with $a\in]a_{\iota},a_{\gamma_+}^-]$, certainly   $r_{\gamma}^+\non{\in}()^-$ when $r_{mso}^-\non{\in}()^-$, for example in the different conditions laid down in \ref{Sec:graph-def}, and particularly in
 Eqs\il(\ref{Eq:ali-06},\ref{Eq:states-energies-Y},\ref{Eq:imply-Col}).

 However, to be more precise, we need to refer to  the specific  angular momenta as in Figs\il\ref{Figs:pePlotre}-left. Then we can  introduce the angular  momentum $ \ell_{\Gamma}^-:\;V_{eff}( \ell_{\Gamma}^-,r_{\gamma}^+)=1$ such
  that
 $ \ell_{\Gamma}^-\in \mathbf{L2}^-$ for $a<a_{\gamma_-}^{\Gamma}$,
  while there is  $ \ell_{\Gamma}^-\in \mathbf{L3}^-$ when  $ a>a_{\gamma_-}^{\Gamma}$.
  We have:
  \bea\nonumber&&
  a_{\iota}<a_{\gamma_+}^-<a_{\gamma_-}^{\Gamma}<\breve{a},\quad
a_{\iota}\equiv0.3137M,\;
 a_{\gamma_+}^-\equiv 0.638285 M,\\
 &&
  a_{\gamma_-}^{\Gamma}\equiv0.777271M,\;\breve{a}\equiv 0.969174M.
  \eea
  It follows that:
   \bea\nonumber
&&   \mbox{for}\quad a\in[0,a_{\gamma_-}^{\Gamma}[
\\\nonumber
&&\mbox{it is}\quad
   r_{\gamma}^+\in ()^-_1,\quad r_{\gamma}^+\in ()^-_2 \quad\mbox{for}\quad \ell^-<\ell_{\Gamma}^-,
   \\\nonumber
    &&r_{\gamma}^+\non{\in }()^-_2 \quad\mbox{for}\quad \ell^->\ell_{\Gamma}^-\in \mathbf{L2}^-;\quad r_{\gamma}^+\non{\in}\cc_3^-,
   \\
   &&\label{Eq:lengh-distance}
\mbox{for}\quad a\in]a_{\gamma_-}^{\Gamma},M]^*\quad\mbox{it is}\quad
r_{\gamma}^+\in ()^-_1\quad r_{\gamma}^+\in ()^-_2,
\\
&&\nonumber r_{\gamma}^+\in ()_3^-\quad\mbox{for}\quad \ell^-<\ell_{\Gamma}^-\in \mathbf{L3}^-
\\
&&\nonumber r_{\gamma}^+\non{\in }()_3^-\;\mbox{for}\; \ell^->\ell_{\Gamma}^-\in \mathbf{L3}^-.
\\
\label{Eq:lengh-distance0}
   \eea
We  conclude that  this kind of configuration is favored for corotating matter,  when both rotation parameters,  $(a/M, \ell^-)$, have  sufficiently low values (in accordance with the balance of centrifugal and gravitational component of the effective potential to which the configuration  is subject).
In  case of the closed configurations, this  property  helps  for setting the inner and outer edge of the disk.

We note that the relations in  Eqs\il(\ref{Eq:lengh-distance},\ref{Eq:lengh-distance0}) are \emph{not} in contradiction with  the result of Eq.\il(\ref{Eq:circum-ved}). In fact these two results should be considered  together, and ensure that in this case the potential is lower than its asymptotical limit, but the point, in this case the orbit $r_{\gamma}^+$, is located in a  right  range of the maximum point, where the function is increasing with the radius (in the case of momenta in  $\mathbf{L3}^-$ there is no minimum of the hydrostatic pressure).

\begin{figure*}[h!t]
\begin{center}
\begin{tabular}{ccc}
\includegraphics[width=.341\textwidth]{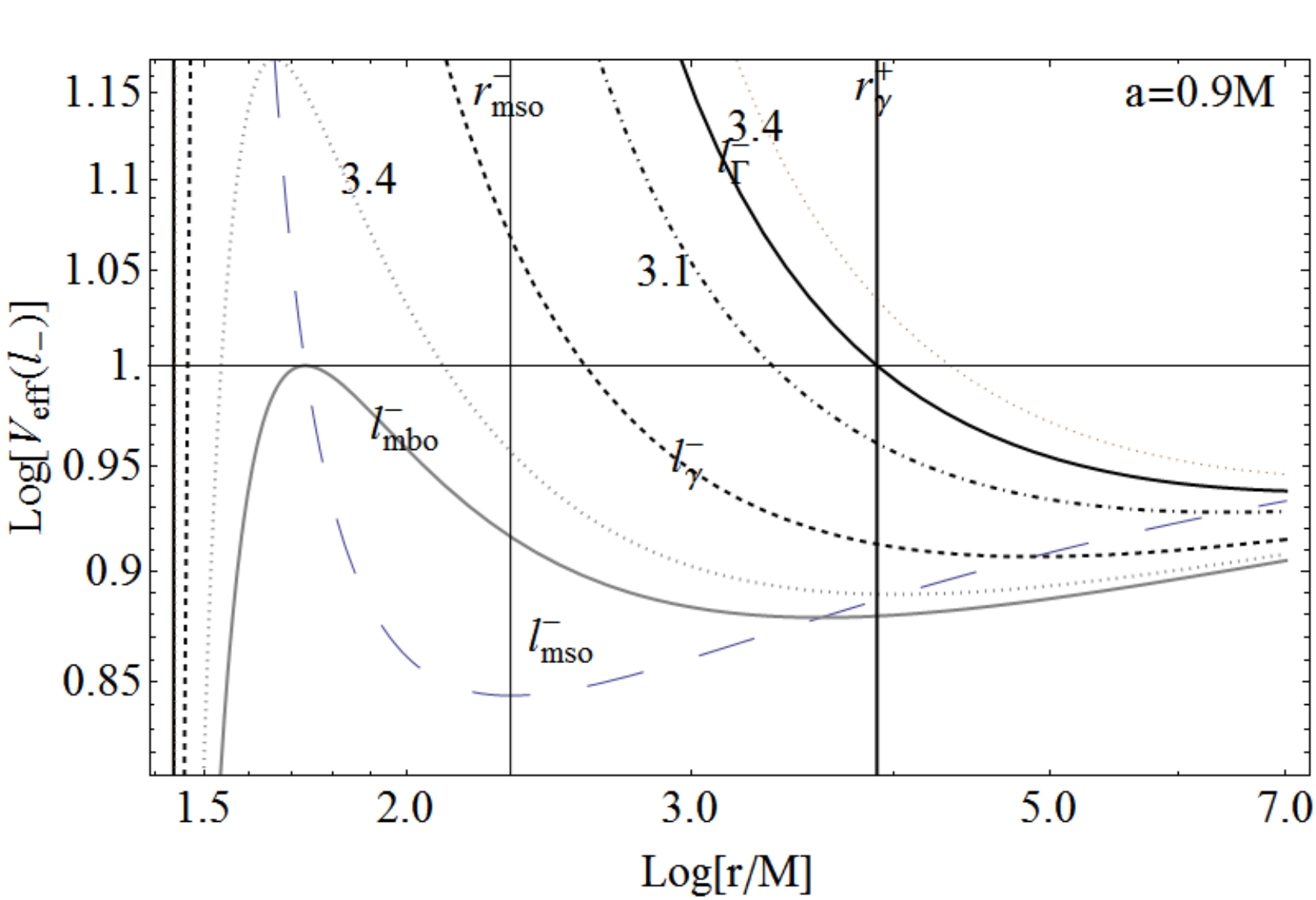}
 \includegraphics[width=.341\textwidth]{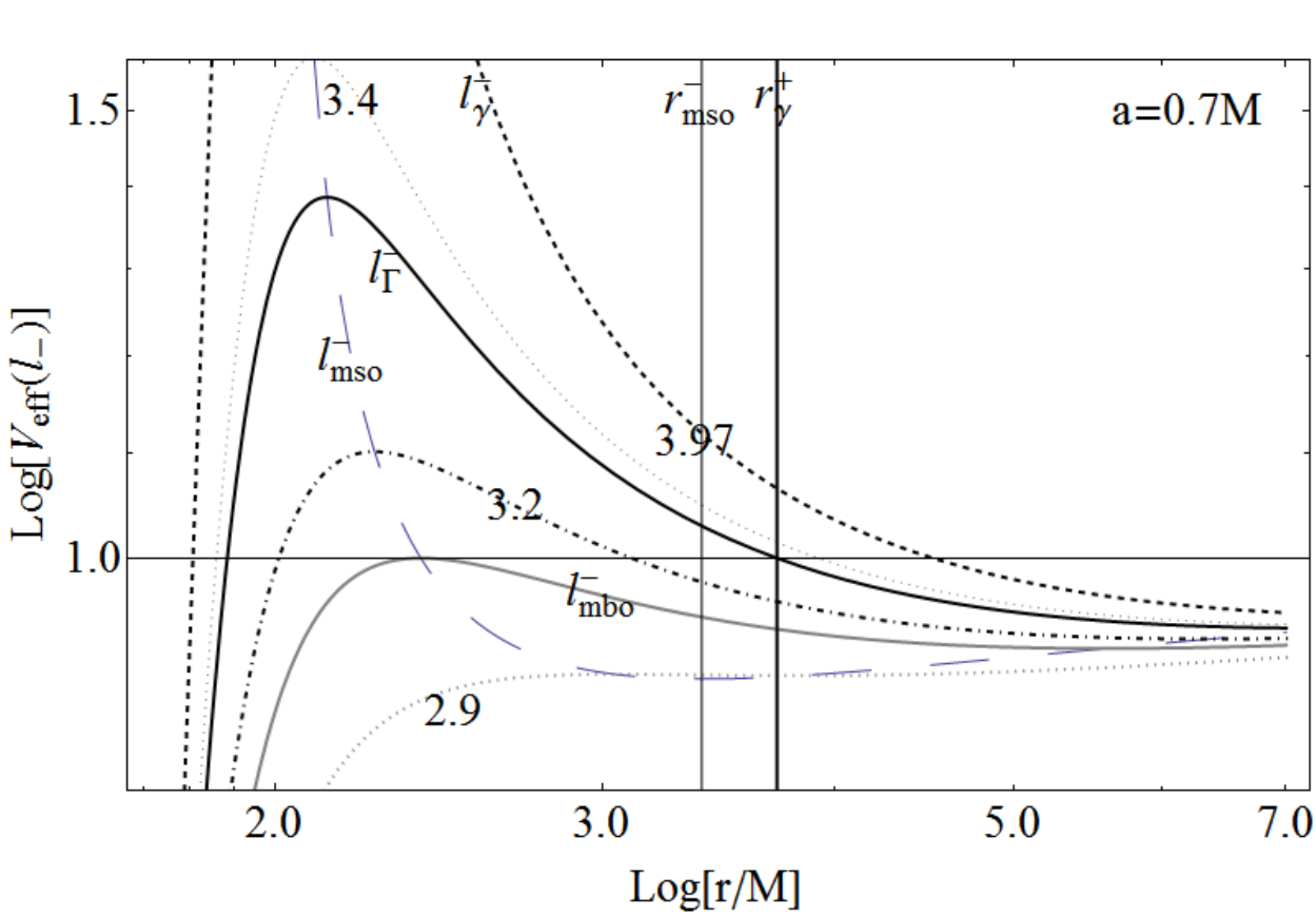}
 \includegraphics[width=.341\textwidth]{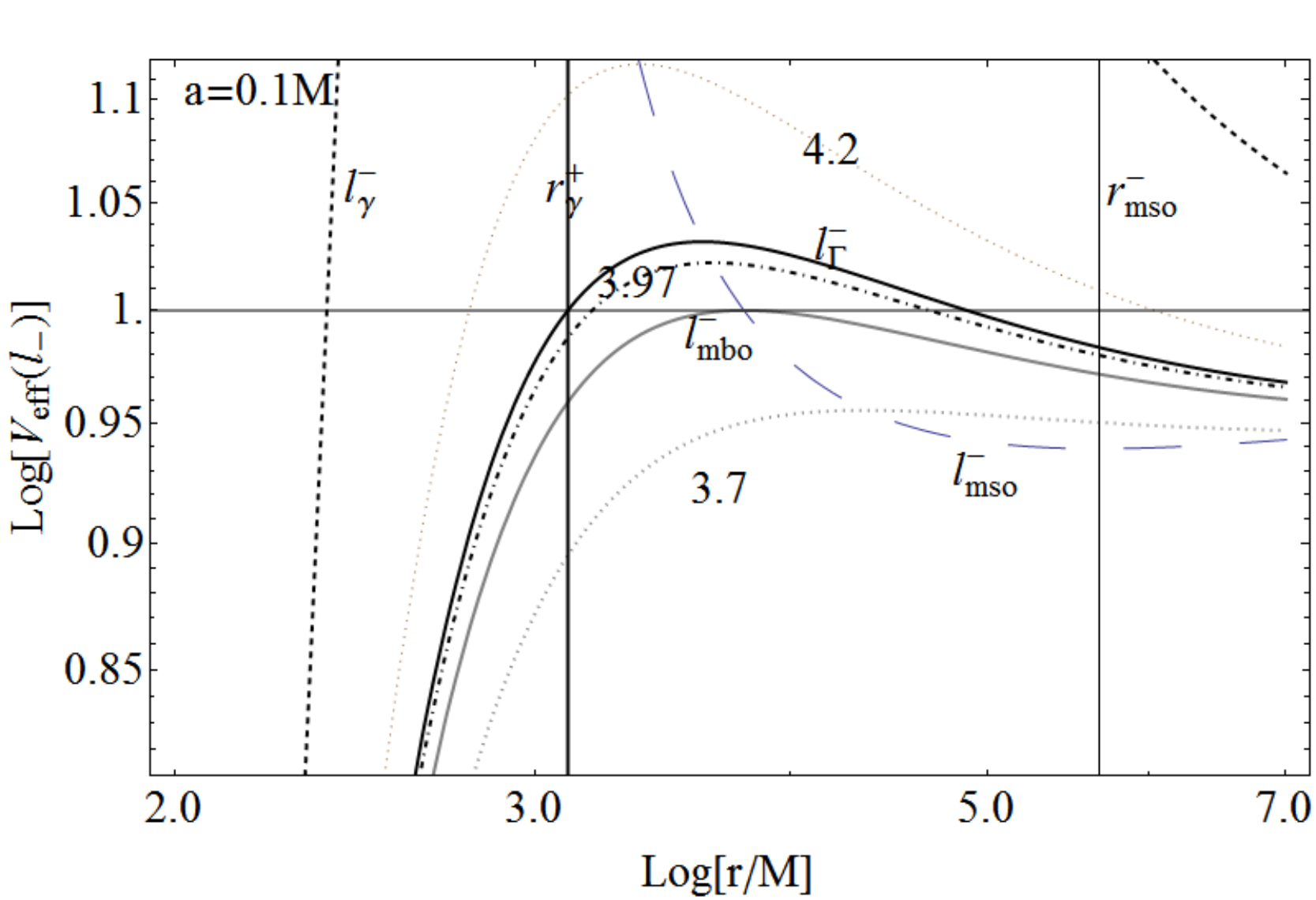}
\end{tabular}
\caption{
Location of the photon orbit $r_{\gamma}^+\in()^-$: Effective potential for corotating fluids versus $r/M$,
for  different spin-mass ratios of the black hole.
For fixed attractor, the effective potential is evaluated at  the specific angular  momentum $\ell_{mbo}^-$ of the marginally bounded orbit $r_{mbo}^-
$, the specific angular  momentum $\ell_{mso}^-$
of the marginally stable orbit $r_{mso}^-,
$,  at $\ell_{\gamma}^-$ for  the corotating photon orbit $
r_{\gamma}^-$ and for  $\ell_{\Gamma}^-\equiv\ell^-(r_{\gamma}^+)$   where $r_{\gamma}^+$ is counterrotating photon
 orbit.
 Numbers close to the curves are further values  of the specific angular  momentum.
Radii $r_{mso}^-$ and $r_{\gamma}^+$ are also plotted.
}\label{Fig:DSic9}
\end{center}
\end{figure*}

These results are  confirmed  by  Fig.\il\ref{Fig:DSic9}, moreover  we note that also  Eqs\il(\ref{Eq:05L},\ref{Eq:05L1}), confirmed by the analysis in  Figs\il\ref{Figs:pePlotre}, are not in contradiction with the constraints provided by the geodesic structure. This is  because for the   orbit  $r_*$ it follows: $V_{eff}^2(\ell_3,r_*)>0$ and
  $K_{Max}^i\equiv V_{eff}(\ell_i,r_{Max})\geq V_{eff}(\ell_i,r_*)$ for $\ell_i\in\{\mathbf{L1}, \mathbf{L2}\}$, where $r_{Max}<r_*$.
This is always true  for $\ell\in\mathbf{L2}$; then for the closed topologies it is necessary to choose a proper elongation and density such that  $K_2<K_{Max}^2$. The situation for the  momenta in $\ell_i\in \mathbf{L1}$, where the maximum of the centrifugal barrier  has  $K_{Max}^1<1$, is more articulated.
The constraint
 $K_{crit}\leq K_{crit}(r_*)$ provides a relation    $r=r(a)$. This analysis is shown in  Fig.\il\ref{Fig:Pgreat0},  which also explains the asterisk  $(*)$ in Eq.\il(\ref{Eq:lengh-distance0}).  On the other hand, this is evident from the shape of the curve $\ell_{\Gamma}^-$ in  Fig.\il\ref{Figs:pePlotre}.
\section*{Acknowledgments}
\noindent
D. P. acknowledges support from the Junior GACR grant of the Czech Science Foundation No:16-03564Y.
Z. S. acknowledges  the Albert Einstein Centre for Gravitation and Astrophysics supported by grant No.
 14-37086G.
 {{The authors  thank }
the anonymous reviewer for the constructive suggestions  which
helped us to improve the manuscript.}
%

\end{document}